\documentclass[11pt]{article}
\pdfoutput=1
\usepackage{geometry}                
\usepackage{graphicx}
\usepackage{amssymb}
\usepackage{amsmath}
\usepackage{epstopdf}
\usepackage{amscd}
\usepackage{mathtools}
\usepackage{slashed}
\usepackage{bbm}
\usepackage{upgreek}
\usepackage[utf8]{inputenc}
\usepackage[matrix,arrow]{xy}
\usepackage{arydshln}
\usepackage[utf8]{inputenc}

\usepackage{stolenstyle}

\usepackage[linktocpage=true]{hyperref}
\hypersetup{
colorlinks=true,
citecolor=blue,
linkcolor=blue,
urlcolor=blue}

\DeclareFontFamily{OT1}{pzc}{}
\DeclareFontShape{OT1}{pzc}{m}{it}{<-> s * [1.10] pzcmi7t}{}
\DeclareMathAlphabet{\mathpzc}{OT1}{pzc}{m}{it}

\newcommand\T{\rule{0pt}{2.6ex}}       
\newcommand\B{\rule[-1.2ex]{0pt}{0pt}} 

\def\sy{\mathpzc{y}}
\def\sf{\mathpzc{f}}

\newcommand{\overbar}[1]{\mkern 1.5mu\overline{\mkern-1.5mu#1\mkern-1.5mu}\mkern 1.5mu}
\def\Gbar{\overbar{G}}
\def\Cbar{\overbar{C}}

\def\Fbar{\overbar{F}}

\def\Gammabar{\overbar{\Gamma}}
\def\Sigmabar{\overbar{\Sigma}}
\def\Psibar{\overbar{\Psi}}
\def\omegabar{\overbar{\omega}}
\def\sbar{\overbar{s}}

\def\rhobar{\overbar{\rho}}
\def\taubar{\overbar{\tau}}
\def\psibar{\overbar{\psi}}
\def\gammabar{\overbar{\gamma}}

\def\kappabar{\overbar{\kappa}}
\def\varepsbar{\overbar{\varepsilon}}
\def\etabar{\overbar{\eta}}

\def\ed{\, \textrm{d}}

\def\ie{{\it i.e.}}
\def\eg{{\it e.g.}}

\def\etc{{\it etc}}
\def\tr{ \, \textrm{tr} \, }
\def\vol{ \, \textrm{vol}}
\def\Tr{ \, \textrm{Tr} \, }
\def\STr{ \, \textrm{STr} \,}
\def\sgn{ \, \textrm{sgn} }

\def\ad{ \, \textrm{ad} }

\def\diag{ \, \textrm{diag}}
\def\uM{\underline{M}}

\def\gm{\gamma_{\rm m}}

\def\ua{\underline{a}}
\def\ub{\underline{b}}
\def\uc{\underline{c}}
\def\ud{\underline{d}}

\def\tp{\tilde{p}}
\def\tq{\tilde{q}}
\def\um{\underline{m}}
\def\un{\underline{n}}
\def\up{\underline{p}}

\def\upsi{\underline{\psi}}
\def\uy{\underline{y}}
\def\uz{\underline{z}}
\def\ur{\underline{r}}

\def\utheta{\underline{\theta}}
\def\uphi{\underline{\phi}}

\def\uA{\underline{A}}
\def\uB{\underline{B}}

\def\uM{\underline{M}}

\def\umu{\underline{\mu}}
\def\unu{\underline{\nu}}
\def\tmu{{\tilde{\mu}}}
\def\tnu{{\tilde{\nu}}}

\def\utmu{{\underline{\tilde{\mu}}}}
\def\utnu{{\underline{\tilde{\nu}}}}

\def\uI{\underline{I}}
\def\uJ{\underline{J}}
\def\uK{\underline{K}}
\def\uzero{\underline{0}}

\def\NN{\mathcal{N}}

\def\half{\frac{1}{2}}

\def\BB{\mathcal{B}}
\def\CC{\mathcal{C}}

\def\FF{\mathcal{F}}
\def\KK{\mathcal{K}}

\def\LL{\mathcal{L}}
\def\VV{\mathcal{V}}

\def\MM{\mathcal{M}}
\def\PP{\mathcal{P}}
\def\DD{\mathcal{D}}

\def\GG{\mathcal{G}}

\def\XX{\mathcal{X}}

\def\pd{\partial}

\def\vareps{\varepsilon}

\def\preprint{ {{MI-TH-1756}}}

\title{Holography for field theory solitons}
\author[a] {Sophia K.~Domokos}
\author[b] {and Andrew B.~Royston}

\abstract{We extend a well-known D-brane construction of the AdS/dCFT correspondence to non-abelian defects.  We focus on the bulk side of the correspondence and show that there exists a regime of parameters in which the low-energy description consists of two approximately decoupled sectors.  The two sectors are gravity in the ambient spacetime, and a six-dimensional supersymmetric Yang--Mills theory.  The Yang--Mills theory is defined on a rigid $AdS_4 \times S^2$ background and admits sixteen supersymmetries.  We also consider a one-parameter deformation that gives rise to a family of Yang--Mills theories on asymptotically $AdS_4 \times S^2$ spacetimes, which are invariant under eight supersymmetries.  With future holographic applications in mind, we analyze the vacuum structure and perturbative spectrum of the Yang--Mills theory on $AdS_4 \times S^2$, as well as systems of BPS equations for finite-energy solitons.  Finally, we demonstrate that the classical Yang--Mills theory has a consistent truncation on the two-sphere, resulting in maximally supersymmetric Yang--Mills on $AdS_4$.}

\affiliation[a]{Department of Physics, New York City College of Technology, 300 Jay Street, Brooklyn, NY 11201, USA}
\affiliation[b]{ George P. and Cynthia W. Mitchell Institute for Fundamental Physics and Astronomy,
Texas A\&M University,
College Station, TX 77843, USA}
\emailAdd{sophia.domokos@gmail.com}
\emailAdd{aroyston@physics.tamu.edu}

\begin{document}

\maketitle
\parskip 7pt

\section{Introduction and summary}

The original AdS/CFT duality \cite{Maldacena:1997re} is nearing its twentieth anniversary. Even AdS/dCFT, which introduces defects in conformal field theories with gravity duals \cite{Karch:2001cw,Karch:2000gx,DeWolfe:2001pq}, is fifteen years old. A small corner of this D-brane universe, however, remains relatively unexplored.

In this  paper, we describe a simple generalization of the D3/D5-brane intersection that forms the basis of the original anti-de Sitter/defect conformal field theory correspondence (AdS/dCFT). Instead of studying a single probe D5-brane in the presence of a large number of D3-branes, we consider the seemingly simple non-abelian generalization, with several parallel D5-branes.

The resulting model, when subjected to Maldacena's low-energy limit and restricted to an appropriate regime of parameters, offers rich physics and rich mathematics, which we begin to uncover here.  Denoting the numbers of D3- and D5-branes by $N_c$ and $N_f$ respectively, and the string coupling by $g_s$, the regime of parameters is $N_c \gg g_s N_c \gg 1$ and $N_f \ll N_c/\sqrt{g_s N_c}$.  The first two conditions are the ones that arise in the usual AdS/CFT correspondence.  They ensure that gravity is weakly coupled and curvatures are small relative to the string scale.  The final condition is a slight refinement of the oft-quoted `probe limit' $N_f \ll N_c$.  We will see that it arises naturally when we demand that corrections from gravity to the $\mathfrak{su}(N_f)$ sector of the D5-brane theory be suppressed.  In this regime, therefore, the effects of closed strings can be neglected relative to the tree-level Yang--Mills interactions.

As in the original AdS/dCFT correspondence, the duality `acts twice' \cite{Karch:2001cw,Karch:2000gx,DeWolfe:2001pq} in the sense that it relates bulk closed strings to operators in the ambient part of the boundary theory, and bulk open strings on the D5-branes to operators localized on a defect in the boundary theory.  Hence the curved-space super-Yang--Mills theory (SYM) describes the physics of operators confined to a defect in the boundary CFT.  As the bulk SYM is dual to a (2+1)-dimensional system, it is potentially relevant to holographic condensed matter applications. Indeed, the bulk SYM admits a zoo of solitonic objects, whose masses and properties are constrained by supersymmetry.  We expect that these  correspond to vortex-like states on the dual defect.  Conversely, holography should provide a new tool for studying SYM solitons in the bulk.

In this paper, however, we focus on the bulk side of the correspondence.  A detailed construction of the dual boundary theory will appear elsewhere, \cite{DR2}.

\subsection{Summary of results}

We begin by constructing a six-dimensional (6D) SYM theory with $\mathfrak{osp}(4|4)$ symmetry from a D3/D5 intersection. We assume that the number of D3-branes ($N_c$) is large, so we can represent them with a Type IIB supergravity solution. We then consider the D5-branes as probes in this background. We arrive at the SYM action by combining and extending D-brane actions that already appear in the literature. For the bosonic theory on the D5-branes, we use the non-abelian Myers action \cite{Myers:1999ps}. We determine the kinetic and mass-like terms for the fermions using the abelian action of \cite{Aganagic:1996pe,Cederwall:1996ri,Bergshoeff:1996tu,Aganagic:1996nn, Marolf:2003ye,Marolf:2003vf,Martucci:2005rb}, and infer the non-abelian gauge and Yukawa couplings via a simple ansatz consistent with gauge invariance and supersymmetry. We then apply the Maldacena low-energy near-horizon limit.

The resulting action is summarized in equations \eqref{SYM}-\eqref{Sbndry}.  While we obtained this action from a D-brane model, it makes sense as a classical field theory for arbitrary simple Lie groups.

We go on to analyze  the vacuum structure, perturbative spectrum, and the BPS equations satisfied by  solitons in the 6D SYM theory.  We  also show that the 6D theory has a nonlinear consistent truncation to maximally supersymmetric YM theory on $AdS_4$. 

Here are a few highlights  from the road ahead:
\begin{itemize}
\item The space of vacua of the 6D theory has multiple components.  There are, in fact, infinitely many when $N_c \to \infty$.  One component is a standard Coulomb branch labeled by vevs of Higgs fields.  The other components are labeled by magnetic charges and are quite complicated: they have roughly the form of moduli spaces of singular monopoles fibered over spaces of Higgs vevs.  A D-brane picture (see Figure \ref{fig4} below) provides some intuition for these vacua.

\item We perform a perturbative mode analysis around a class of vacua that carry magnetic flux. The background fields of this class are Cartan-valued and simple enough to make the linearized equations tractable.  Furthermore, the background fields of any vacuum will asymptote to the same near the boundary, so the results for the asymptotic behavior of fluctuations are robust.  This is important for the holographic dictionary, where one maps modes to local operators in the dual, based in part on their decay properties near the boundary.  

Our analysis of the perturbative spectrum generalizes previous results for the abelian D5-brane defect \cite{DeWolfe:2001pq,Arean:2006pk,Myers:2006qr}, and offers a number of new results. We  display, for instance,  the complete KK spectrum of fermionic modes.  We also observe that a Legendre transform of the on-shell action with respect to one of the low-lying modes, along the lines of \cite{Klebanov:1999tb}, is required for holographic duality.\footnote{The paper \cite{Mezei:2017kmw} appeared when this work was nearing completion.  Its authors make a closely related observation in maximally supersymmetric YM on $AdS_4$.  The consistent truncation of the 6D theory on the two-sphere, explained in section \ref{sec:ct}, shows that these results are in fact describing the same phenomenon.}  

We  identify a set of low-lying non-normalizable modes that can be turned on without violating the variational principle or supersymmetry.  These modes form a natural class of boundary values for soliton solutions in the non-abelian D5-brane theory. In the holographic dual, meanwhile, they source a set of relevant operators---and in one case a distinguished irrelevant operator.

\item Having explored the vacua and perturbative structure of the bulk SYM theory, we then survey various systems of BPS equations. These first order equations arise when we demand that field configurations preserve various amounts of supersymmetry.

Solutions to the BPS equations saturate bounds on the energy functional. These bounds depend on a combination of the fields' boundary values as well as the magnetic and electric fluxes through the asymptotic boundary. 

The BPS systems we obtain house a number of generalized self-duality equations that are well known in mathematical physics, like (translationally invariant) octonionic instantons \cite{Corrigan:1982th}, and the extended Bogomolny equations \cite{Kapustin:2006pk}. All of these equations are defined on a manifold with boundary, where the boundary is the holographic boundary.
\end{itemize}

The paper is structured as follows: In section \ref{sec:Dbranes} we describe the D-brane intersection and take the low-energy limit of the  action to arrive at a curved space SYM theory. In section \ref{sec:susy} we verify the invariance of the action under supersymmetry. In section \ref{boundarystuff} we describe the vacuum structure of the model, and formulate asymptotic boundary conditions on the fields. In section \ref{sec:ct} we derive the consistent truncation of our six-dimensional theory to four dimensions, while in section \ref{sec:monopoles} we derive the BPS equations satisfied by solitons in the system. We conclude and discuss future directions in section \ref{sec:future}. Necessary but onerous details are relegated to a series of Appendices.

\section{Branes and holography}\label{sec:Dbranes}

In this section we describe the brane set-up, the AdS/dCFT picture, and the low-energy limit and parameter regime that isolates six-dimensional SYM as the low-energy effective theory on the D5-branes.

\subsection{Brane configuration}\label{sec:setup}

\begin{figure}
\begin{minipage}[c]{0.4\linewidth}
\begin{center}
\includegraphics[scale=0.5]{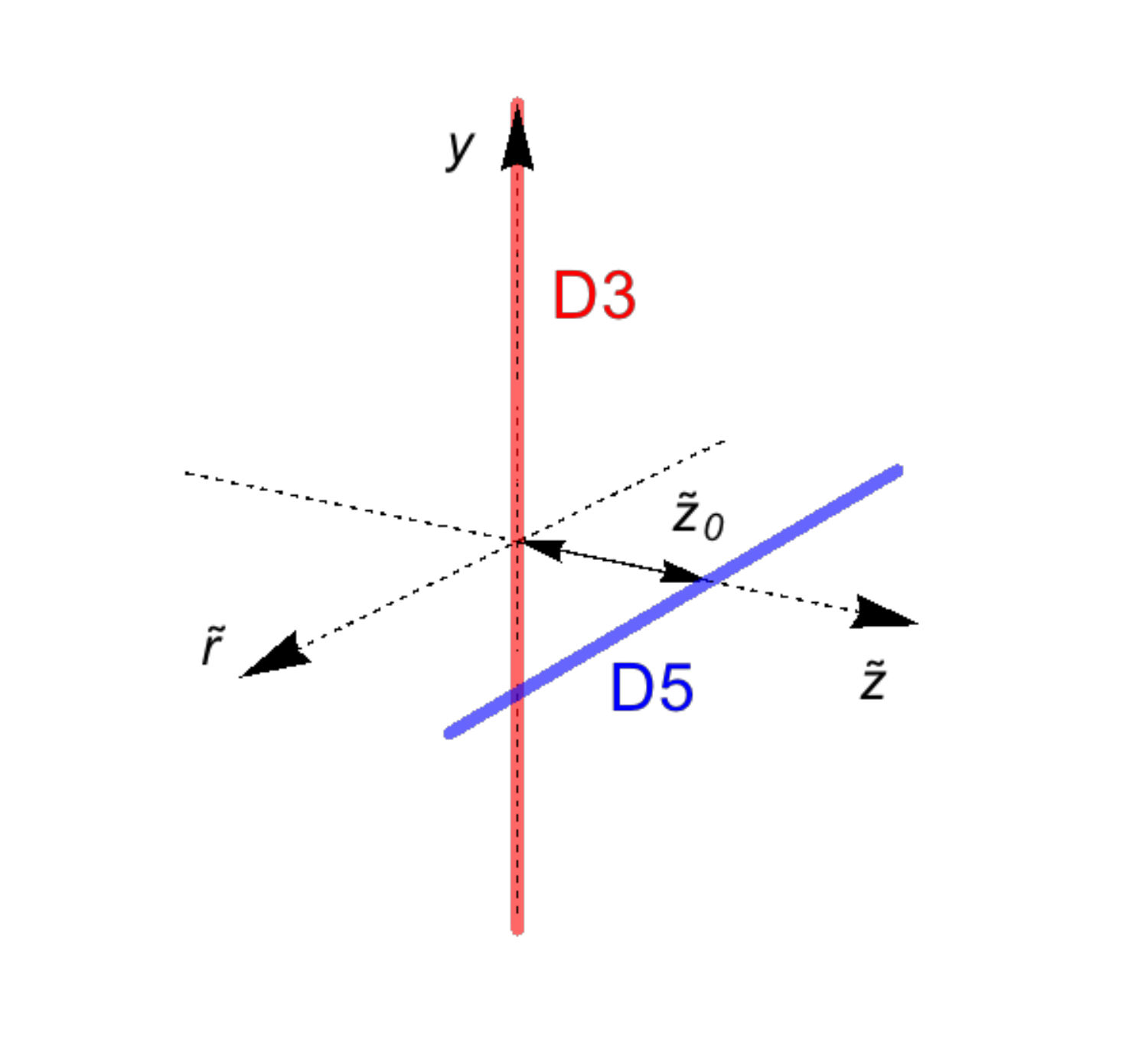}
\end{center}
\end{minipage}
\hspace{0.8cm}
\begin{minipage}[c]{0.4\linewidth}
\begin{center}
\begin{tabular}{c | c c c ;{2pt/2pt} c c c ;{2pt/2pt} c c c ;{2pt/2pt} c}
  & 0 & 1 & 2 & 3 & 4 & 5 & 6 & 7 & 8 & 9  \T\B\B \\
  \hline
 D3 & X & X & X &  & & & & & & X  \T\T\B  \\
 D5 & X & X & X & X & X & X & & & &  \T\B\B \\ \hline
      & \multicolumn{3}{c ;{2pt/2pt}}{$x^\mu$} &\multicolumn{3}{c}{$\tilde{r}_i$} & \multicolumn{3}{;{2pt/2pt} c ;{2pt/2pt}}{$\tilde{z}_i$} & $y$ \T\T
 \end{tabular}
\end{center}
\end{minipage}
\vspace{-10pt}
\caption{The intersecting brane system.  The $x^\mu$ directions common to both types of brane worldvolume are suppressed in the figure on the left.  The D5-branes can be separated from the D3-branes by a distance $\tilde{z}_0$ in the directions transverse to both stacks.}
\label{D3D5system}
\end{figure}

We begin with a non-abelian version of the brane configuration in \cite{DeWolfe:2001pq}. $N_f$ D5-branes and $N_c$ D3-branes in the ten-dimensional IIB theory span the directions indicated in Figure \ref{D3D5system}. Standard arguments \cite{Polchinski:1998rr} show that the intersecting D3/D5 system preserves one quarter of the supersymmetry of 10D type IIB string theory, or eight supercharges.

The ten coordinates, $\tilde{x}^M = (x^\mu; \tilde{r}_i ; \tilde{z}_i;y)$ are divided as follows:  $x^\mu$, $\mu,\nu = 0,1,2$, parameterizes the $\mathbb{R}^{1,2}$ spanned by both stacks; the triplet $\tilde{r}_i = (\tilde{r}_1,\tilde{r}_2,\tilde{r}_3) = (x^3, x^4, x^5)$ parameterizes the remaining directions along the D5-branes; the triplet $\tilde{z}_i = (\tilde{z}_1,\tilde{z}_2,\tilde{z}_3) = (x^6,x^7,x^8)$ parameterizes directions orthogonal to both stacks; and finally $y = x^9$ parameterizes the remaining direction along the D3-branes and orthogonal to the D5-branes.  We reserve the notation $x^M = (x^\mu,r_i,z_i,y)$ for a rescaled version of these coordinates to be introduced below.  We will sometimes use spherical coordinates $(\tilde{r},\theta,\phi)$ to parameterize the $\tilde{r}_i$ directions, and we denote the radial coordinate in the $\tilde{z}_i$ directions by $\tilde{z}$.  We also write $\tilde{x}^a = (x^\mu, \tilde{r}_i)$, $a = 0,1,\ldots ,5$, and $\tilde{x}^m = (\tilde{z}_i,y)$, $m=1,\ldots,4$, for the full set of directions parallel and transverse to the D5-branes respectively. 

The D3-branes are taken to be coincident and sitting at $\tilde{r}_i = \tilde{z}_j = 0$.  The center of mass position of the D5-branes in the transverse $\tilde{x}^m$ space is denoted $\tilde{x}_{0}^m = ( \tilde{z}_{0,i}, y_0)$.  We will allow for relative displacements of the D5-branes from each other, but assume that these distances are small compared to the string scale. In other words, the separation is well-described by vev's of non-abelian scalars in the D5-brane worldvolume theory.  This will be explained in more detail below.  When all D5-branes are positioned at $\tilde{z}_i = 0$ an $SO(1,2) \times SO(3)_{r} \times SO(3)_{z}$ subgroup of the ten-dimensional Lorentz group is preserved.  Nonzero D5-brane displacements in $\tilde{z}$ break $SO(3)_z$.  This can be explicit or spontaneous from the point of view of the D5-brane worldvolume theory, depending on whether the center of mass position $\tilde{z}_{0,i}$ is, respectively, nonzero or zero.

As noted above, eight of the original thirty-two Type IIB supercharges are preserved by the brane setup. From the point of view of the three-dimensional intersection, this is equivalent to $\NN = 4$ supersymmetry.  The $R$-symmetry group is $SO(4)_R = SU(2)_r \times SU(2)_z$ with the two factors being realized geometrically as the double covers of the rotation groups in the $\tilde{r}_i$ and $\tilde{z}_i$ directions.  The light degrees of freedom on the D3-branes and the D5-branes are a four-dimensional $\NN = 4$ $\mathfrak{u}(N_c)$-valued vector-multiplet, and a six-dimensional $\NN = (1,1)$ $\mathfrak{u}(N_f)$-valued vector-multiplet.  Each of these decompose into a 3D $\NN = 4$ vector-multiplet and hypermultiplet.  For those D5-branes intersecting the D3-branes, the $3$-$5$ strings localized at the intersection are massless.  They furnish a 3D $\NN = 4$ hypermultiplet transforming in the bi-fundamental representation of the appropriate gauge groups.  Meanwhile the massless closed strings comprise the usual type IIB supergravity multiplet.

\subsection{Low energy limit and AdS/dCFT}

Let us now consider the low-energy limit of the brane setup, that ultimately yields the defect AdS/CFT correspondence.  This is the famous Maldacena limit \cite{Maldacena:1997re} that, in the absence of D5-branes, establishes a correspondence between 4D $\NN = 4$ SYM and type IIB string theory on $AdS_5 \times S^5$.  To arrive at the AdS/dCFT correspondence one considers the low-energy effective description of the D3/D5 system at energy scale $\mu$ and takes the limit $\mu \ell_s \to 0$, where $\ell_s$ is the string length.  The dynamics of the massless degrees of freedom have two equivalent descriptions in terms of two different sets of field variables. This fact is the essence of the original AdS/dCFT correspondence. 

 To simplify the present discussion we temporarily assume no separation between the brane stacks -- in other words,  $\tilde{z}_0 = 0$.  The first set of variables that describes the D3/D5 intersection is based on an expansion around the flat background: Minkowski space for the closed strings and constant values of the brane embedding coordinates for the open strings.  In this case standard field theory scaling arguments apply.  After canonically normalizing the kinetic terms for open and closed string fluctuations, interactions of the closed strings and 5-5 open strings amongst themselves, as well as the interactions of the closed and 5-5 open strings with the other open strings, vanish in the low-energy limit.  These degrees of freedom decouple from the system. Meanwhile the 3-3 and 3-5 strings form an interacting system described by four-dimensional $\NN = 4$ SYM coupled to a co-dimension one planar interface, breaking half the supersymmetry and hosting a 3D $\NN = 4$ hypermultiplet.  The interface action, which can in principle be derived from the low energy limit of string scattering amplitudes, was obtained in \cite{DeWolfe:2001pq} by exploiting symmetry principles.  The entire theory contains a single dimensionless parameter in addition to $N_f$ and $N_c$---the four-dimensional Yang--Mills coupling---given in terms of the string coupling via $g_{{\rm ym}}^2 := 2\pi g_s$.

The interface plus boundary ambient Yang--Mills theory is classically scale invariant, and it was argued in \cite{DeWolfe:2001pq,Erdmenger:2002ex} to be a superconformal quantum theory.  The symmetry algebra is $\mathfrak{osp}(4|4)$, with bosonic subalgebra $SO(2,3) \times SO(4)_R$ and sixteen odd generators.  $SO(2,3)$ is the three-dimensional conformal group of the interface while the odd generators correspond to the eight supercharges along with eight superconformal generators.  This is the ``defect CFT'' side of the correspondence.  Considering a nonzero separation $\tilde{z}_0$ corresponds to turning on a relevant mass deformation in the dCFT \cite{Karch:2002sh,Yamaguchi:2002pa}.

Our focus here will be mostly on the other side of the correspondence, which is based on an expansion in fluctuations around the supergravity background produced by the $N_c$ D3-branes.  This background involves a nontrivial metric and Ramond-Ramond (RR) five-form flux given in our coordinates by
\begin{align}
\ed s_{10}^2 =&~ f^{-1/2} (\eta_{\mu\nu} \ed x^\mu \ed x^\nu + \ed y^2) + f^{1/2} (\ed \tilde{r}_i \ed \tilde{r}_i + \ed\tilde{z}_i \ed\tilde{z}_i)~, \cr
F^{(5)} =&~ (1 + \star) \ed x^0 \ed x^1 \ed x^2 \ed y \ed f^{-1}~, \qquad\textrm{with} \cr
f =&~ 1 + \frac{L^4}{(\tilde{r}^2 + \tilde{z}^2)^2} ~, \qquad \textrm{where} \quad L^4 = 4\pi g_s N_c \ell_{s}^4~.
\end{align}
The metric is asymptotically flat and approaches $AdS_5 \times S^5$ with equal radii of $L$ when $\tilde{v}^2 \equiv \tilde{r}^2 +\tilde{z}^2 \ll L^2$.  The energy of localized modes in the throat region, as measured by an observer at position $\tilde{v}$, is redshifted in comparison to the asymptotic fixed energy $\mu$ according to $E_v = f^{1/4} \mu \sim (L/\tilde{v}) \mu$, for $\tilde{v} \ll L$.  Hence, while closed string and D5-brane modes with Compton wavelengths large compared to $L$ decouple as before, excitations of arbitrarily high energy can be achieved in the throat region.  The near-horizon limit isolates the entire set of stringy degrees of freedom in the throat region by sending $\tilde{v}/\ell_s \to 0$ in such a way that $E_v \ell_s$ remains fixed.  From the redshift relation it follows that we are sending $\tilde{v}/\ell_s \to 0$ while holding $\tilde{v}/(\ell_{s}^2 \mu)$ fixed.  For fixed 't Hooft coupling $g_s N_c$, this is equivalent to sending $\tilde{v}/L \to 0$ while holding $\tilde{v}/(L^2 \mu)$ fixed.

To facilitate taking this limit we introduce new coordinates
\begin{equation}
r_i = \frac{\tilde{r}_i}{L^2 \mu^2} ~, \qquad z_i = \frac{\tilde{z}_i}{L^2 \mu^2}~,
\end{equation}
and write $(r,\theta,\phi)$ for the corresponding spherical coordinates and $z \equiv \sqrt{z_i z_i}$.  We will also sometimes employ a vector notation $\vec{r} = (r_1,r_2,r_3)$, $\vec{z} = (z_1,z_2,z_3)$.  One finds that with these new coordinates, the metric becomes\footnote{The metric can be brought to the form found in \cite{DeWolfe:2001pq} by first introducing standard spherical coordinates $(z,\zeta,\chi)$ in the $\vec{z}$ directions and then setting $r = v \cos{\psi}$ and $z = v \sin{\psi}$ with $\psi \in [0,\pi/2]$, and $\mu = 1$.  Then $v$ is the $AdS_5$ radial coordinate in the Poincare patch, with $v \to \infty$ the asymptotic boundary, while $(\psi,\theta,\phi,\zeta,\chi)$ parameterize the $S^5$, viewed as an $S^2 \times S^2$ fibration over the interval parameterized by $\psi$.}
\begin{align}\label{nhgeom}
\ed s_{10}^2 \to &~ (L\mu)^2 \bigg\{ \mu^2 (r^2 + z^2) \left[ \eta_{\mu\nu} \ed x^\mu \ed x^\nu + \ed y^2 \right]  + \cr
&~  \qquad \qquad+ \frac{1}{\mu^2(r^2 + z^2)} \left[ \ed r^2 + r^2 \ed \Omega^2(\theta,\phi) + \ed \vec{z} \cdot \ed \vec{z} \, \right] \bigg\} \cr
=: &~ (L\mu)^2 \Gbar_{MN} \ed x^M \ed x^N~,   \cr
F^{(5)} \to &~ 4 (L\mu)^4 \mu^4 (r^2 + z^2) (1 + \star) \ed x^0 \ed x^1 \ed x^2 \ed y \, (r \ed r + z \ed z) \cr
=: &~ (L\mu)^4 \ed \Cbar^{(4)}~,
\end{align}
where we've introduced a rescaled metric and four-form potential, $\Gbar_{MN}, \Cbar^{(4)}$.  $\Gbar_{MN}$ is the metric on $AdS_5 \times S^5$ with radii $\mu^{-1}$.

\begin{figure}
\begin{center}
\includegraphics{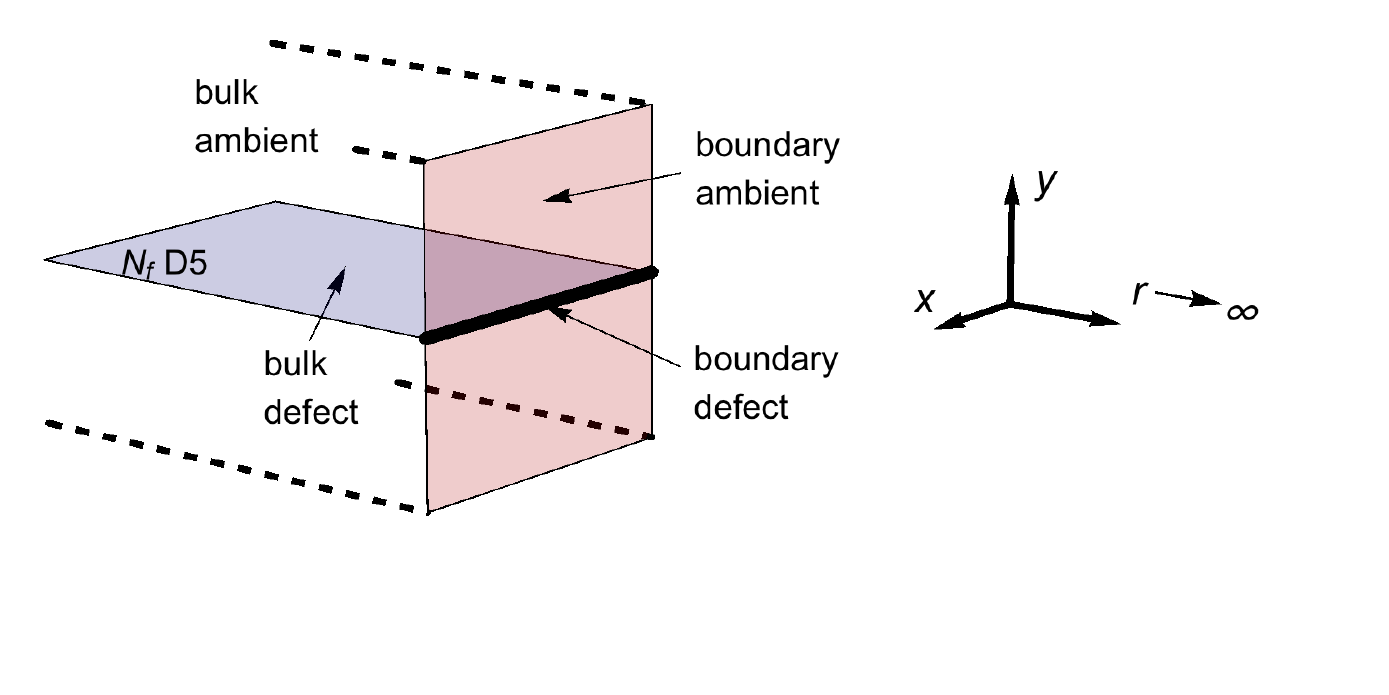}
\vspace{-40pt}
\caption{The defect AdS/CFT correspondence.  The bulk theory consists of an ambient IIB string theory on $AdS_5 \times S^5$, coupled to a defect composed of probe D5-branes.  The boundary theory consists of an ambient $\NN = 4$ SYM on $\mathbb{R}^{1,3}$ coupled to a co-dimension one defect hosting localized modes.  }
\label{fig2}
\end{center}
\end{figure}

The degrees of freedom in the near-horizon geometry include both the closed strings and the open strings on the D5-branes.  String theory in this background is conjecturally dual to the dCFT system, with the duality `acting twice' \cite{Karch:2001cw,Karch:2000gx,DeWolfe:2001pq}.  This means the following: closed string modes in the (ambient) spacetime of the bulk side are dual to operators constructed from the 4D $\NN = 4$ SYM fields in the (ambient) spacetime on the boundary.  Open string modes on the D5-branes, which form a defect in the bulk, are dual to operators localized on the defect in the boundary theory.  These operators are constructed from modes of the 3-5 strings and modes of the 3-3 strings restricted to the boundary defect.  See Figure \ref{fig2}.  

The validity of the supergravity approximation in the closed string sector requires that $N_c \gg g_s N_c \gg 1$.  The first condition suppresses $g_s$ corrections to the low energy effective action, while the second condition is equivalent to $L \gg \ell_s$, ensuring that higher derivative corrections are suppressed as well.  

In subsection \ref{sec:YMleeft} we'll see how this limit suppresses the interactions between closed string and open string D5-brane modes, leading to an effective Yang--Mills theory on the D5-branes.  This extends previous analyses of the D3/D5 system to the case of multiple D5-branes, showing how the non-abelian interaction terms among open strings are dominant to the open-closed couplings, at least in the $\mathfrak{su}(N_f)$ sector of the theory.  In subsection \ref{sec:Myers} we will describe explicitly what these interactions look like (using the  Myers non-abelian D-brane action).

In preparation for that, consider the following redefinition of the relevant supergravity fields. Let $S_{\rm IIB}[G,B,\Delta\upphi,C^{(n)};\kappa]$ denote the type IIB supergravity action in Einstein frame.  Here $B$ is the Kalb--Ramond two-form potential and $\Delta\upphi := \upphi - \upphi_0$ is the fluctuation of the dilaton field $\upphi$ around its vev, $\upphi_0$, with $e^{\upphi_0} \equiv g_s$.  The $C^{(n)}$, $n$ even, are the Ramond-Ramond potentials, and $\kappa$ is the ten-dimensional Newton constant, $\kappa^2 = \half (2\pi)^7 g_{s}^2 \ell_{s}^8$.  Upon rescaling the metric and potentials according to
\begin{equation}\label{closedrescale}
G_{MN} = (L\mu)^{2} \tilde{G}_{MN}~, \quad B_{MN} = (L\mu)^{2} \tilde{B}_{MN}~, \quad C^{(n)} = (L \mu)^{n} \tilde{C}^{(n)}~,
\end{equation}
one finds that
\begin{equation}
S_{\rm IIB}[G,B,\Delta\upphi,C^{(n)};\kappa] = S_{\rm IIB}[\tilde{G},\tilde{B},\Delta\upphi, \tilde{C}^{(n)};\kappabar]~,
\end{equation}
where the new Newton constant is
\begin{equation}
\kappabar = \frac{\kappa}{(L \mu)^4} = \frac{(2\pi)^3 \sqrt{\pi}}{N_c} \mu^{-4}~.
\end{equation}
Thus an expansion in canonically normalized closed string fluctuations, $(h_{MN},b_{MN},\varphi,c^{(n)})$, around the near-horizon background, \eqref{nhgeom}, takes the form
\begin{align}\label{csfluc}
& G_{MN} = (L \mu)^2 \left(\Gbar_{MN} + \kappabar h_{MN} \right)~, \qquad B_{MN} = (L\mu)^2 \kappabar b_{MN}~, \qquad \Delta\upphi = \kappabar \varphi~, \cr
& C^{(4)} = (L\mu)^4 \left(\Cbar^{(4)} + \kappabar c^{(4)}\right)~, \qquad C^{(n)} = (L\mu)^n \kappabar c^{(n)}~,~n\neq 4~,
\end{align}
where $\Gbar_{MN}$ and $\Cbar^{(4)}$ were given in \eqref{nhgeom}, and $n$-point couplings among closed string fluctuations go as $\kappabar^{n-2}$.  

\subsection{The non-abelian D5-brane action}\label{sec:Myers}

The massless bosonic degrees of freedom on the D5-branes are a $U(N_f)$ gauge field $A_a$, $a = 0,1,\ldots,5$, with fieldstrength $F_{ab}$, and four adjoint-valued scalars $X^m = (Z^{1,2,3},Y)$.  The gauge field carries units of mass while the $X^m$ carry units of length.  The eigenvalues of ($-i$ times) the latter are to be identified with the displacements of the $N_f$ D5-branes away from $(\vec{z}_0,y_0)$.  Our conventions are that elements of the $\mathfrak{u}(N_f)$ Lie algebra are represented by anti-Hermitian matrices, so there are no factors of $i$ coming with the Lie bracket in covariant derivatives.  The `$\Tr$' operation denotes \emph{minus} the trace in the fundamental representation, $\Tr := - \tr_{{\bf N}_f}$, with the minus inserted so that it is a positive-definite bilinear form on the Lie algebra.  Later on we will generalize the discussion to a generic simple Lie algebra $\mathfrak{g}$, and then we define the trace through the adjoint representation via $\Tr := - \frac{1}{2h^\vee} \tr_{{\bf adj}}$, where $h^\vee$ is the dual Coxeter number.  This reduces to the previous definition for $\mathfrak{g} = \mathfrak{u}(N_f)$.

The non-abelian D-brane action of Myers, \cite{Myers:1999ps}, captures a subset of couplings between the 5-5 open string and ambient closed string modes.  It takes the form
\begin{align}
S_{\rm D5}^{\rm bos} =&~ S_{\rm DBI} + S_{\rm CS}~, \qquad \textrm{with} \label{Myers} \\
S_{\rm DBI} =&~ \tau_{\rm D5}  \int \ed^6x e^{-\Delta\upphi} \times \cr
&~ \times \STr \bigg\{ \sqrt{ - \det \left( P[E_{ab}] + P[E_{am} (Q^{-1} - \delta)^{mn} E_{nb} ] -i \lambda F_{ab} \right) \det Q^{m}_{\phantom{m}n} } \bigg\}~, \qquad \label{MyersDBI} \\
S_{\rm CS} =&~ -\tau_{\rm D5} \int \STr \bigg\{ P \left[ e^{\lambda^{-1} {\rm i}_X {\rm i}_X} \CC \right] \wedge e^{-i \lambda F} \bigg\} ~, \raisetag{24pt} \label{MyersCS}
\end{align}
where
\begin{equation}\label{MyersQ}
Q^{m}_{\phantom{m}n} := \delta^{m}_{\phantom{m}n} + \lambda^{-1} [X^m, X^k] E_{kn}~,
\end{equation}
$\lambda := 2\pi \ell_{s}^2$, and $\tau_{\rm D5} := 2\pi /(g_s(2\pi \ell_s)^6)$ is the D5-brane tension.  Besides the factor of $e^{-\Delta\upphi}$ in $S_{\rm DBI}$, the closed string fields are encoded in the two quantities
\begin{equation}
E_{MN} := e^{\Delta\upphi/2} (G_{MN} + B_{MN}) ~, \qquad \mathcal{C} = \sum_n C^{(n)} \wedge \exp\left( e^{\Delta\upphi/2} B\right)~.
\end{equation}
The factors of the dilaton are present here because we work in Einstein frame for the closed string fields.  This action generalizes the non-abelian D-brane action of \cite{Tseytlin:1997csa} to the case of a generic closed string background.  

The quantity $P[T_{MN...Q}]$ denotes the gauge-covariant pullback $P$ of a bulk tensor $T_{MN...Q}$ to the worldvolume of the D5-branes. For instance, the pullback of the generalized metric to the brane is
\begin{equation}\label{PofE}
P[E_{ab}] = E_{ab} -i (D_a X^m) E_{mb} -i E_{am} (D_b X^m)  - (D_a X^m) E_{mn} (D_b X^n) ~,
\end{equation}
with $D_a = \pd_a + [A_a,\,\cdot\,]$.  The closed string fields are to be taken as functionals of the matrix-valued coordinates, $E_{MN}(x^P) \to E_{MN}(x^a;-iX^m)$, defined by power series expansion:
\begin{equation}\label{Etaylor}
E_{MN}(x^a,-iX^m) := E_{MN}(x^a,x_{0}^m) + \sum_{n=1}^{\infty} \frac{(-i)^n}{n!} X^{m_1 \cdots m_n} \left(\pd_{m_1} \cdots \pd_{m_n} E_{MN}\right)(x^a,x_{0}^m)~,
\end{equation}
The determinants in the DBI action \eqref{MyersDBI} refer to spacetime indices $a,b$ and $m,n$.  

In the Chern-Simons (CS) action, \eqref{MyersCS}, the symbol ${\rm i}_X$ denotes the interior product with respect to $X^m$.  This is an anti-derivation on forms, reducing the degree by one.  Since the $X^m$ are non-commuting one has, for example, 
\begin{equation}
( {\rm i}_{X}^2 C^{(k+2)})_{M_1 \cdots M_{k}} = \half [X^m,X^n] C_{nm M_1 \cdots M_{k}}^{(k+2)}~.
\end{equation}
See \cite{Myers:1999ps} for further details.

The `STr' stands for a fully symmetrized trace, defined as follows \cite{Myers:1999ps}.  After expanding the closed string fields in power series and computing the determinants, the arguments of the $\STr$ in \eqref{MyersDBI} and \eqref{MyersCS} will take the form of an infinite sum of terms, each of which will involve powers of four types of open string variable: $F_{ab}, D_a X^m, [X^m,X^n]$, and individual $X^m$'s from the expansion of the closed string fields.  The $\STr$ notation indicates that one is to apply $\Tr$ to the complete symmetrization on these variables. 

The precise regime of validity of the Myers action is not a completely settled issue.  First of all, like its abelian counterpart, it captures only tree-level interactions with respect to $g_s$.  Second, if $\mathfrak{F}$ denotes any components of the `ten-dimensional' fieldstrength, $F_{ab}$, $D_a X^m$, or $[X^m,X^n]$,  \eqref{Myers} is known to yield results {\it incompatible} with open string amplitudes at $O(\mathfrak{F}^6)$ \cite{Hashimoto:1997gm,Bain:1999hu}, even in the limit of trivial closed string background.  Finally, the action \eqref{Myers} is given directly in ``static gauge,'' and there have been questions about whether it can be obtained from gauge fixing a generally covariant action.  This could lead to ambiguities in open-closed string couplings at $O(\mathfrak{F}^4)$ according to \cite{deBoer:2003cp}.  However, the results of \cite{Howe:2006rv} suggest that the Myers action can in fact be obtained by gauge-fixing symmetries in a generally covariant formalism where the Chan--Paton degrees of freedom are represented by boundary fermions on the string worldsheet.  As we will see below, none of these ambiguities pose a problem in the scaling limit we are interested in.

\subsection{Yang--Mills as the low energy effective theory}\label{sec:YMleeft}

We now expand the action \eqref{Myers} in both closed and open string fluctuations, where the closed string expansion is an expansion around the near-horizon geometry of the D3-branes, in accord with \eqref{csfluc}.  This was already done in some detail in the abelian case \cite{DeWolfe:2001pq}, but there are some important new wrinkles that arise in  the non-abelian case.  We summarize the main points here and provide further details in appendix \ref{app:fluc}.  

First, the kinetic terms for the open string modes take the form
\begin{equation}
S_{\rm DBI} \supset - \tau_{\rm D5} (L\mu)^6 \int \sqrt{- g_6} \Tr \left\{ \frac{1}{4} \lambda^2 (L\mu)^{-4} F_{ab} F^{ab} + \half (\Gbar_{mn} |_{x_{0}^m}) D_a X^m D^a X^n \right\}~,
\end{equation}
where we recall that $\lambda = 2\pi \ell_{s}^2$.  The factors of $(L \mu)$ arise from writing the background metric in terms of the barred metric.  We have introduced the notation $g_{6} := \det(g_{ab})$, with $g_{ab} := \Gbar_{ab}(x^a, x_{0}^m)$ the induced background metric on the worldvolume.  It takes the form
\begin{equation}\label{6Dmet}
g_{ab} \ed x^a \ed x^b = \mu^2 (r^2 + z_{0}^2) \eta_{\mu\nu} \ed x^\mu \ed x^\nu + \frac{1}{\mu^2 (r^2 + z_{0}^2)} \left( \ed r^2 + r^2 \ed\Omega(\theta,\phi)^2 \right)~.
\end{equation}
When $z_0= 0$ this is the metric on $AdS_4 \times S^2$ with equal radii of $\mu^{-1}$, while $z_0 \neq 0$ gives a deformation of it.  Worldvolume indices will always be raised with the inverse, $g^{ab}$.  We use the notation $|_{x_{0}^m}$ to indicate when other closed string fields are being evaluated at $x^m = x_{0}^m$.

The coefficient of the $F^2$ term determines the effective six-dimensional Yang--Mills coupling:
\begin{equation}
g_{{\rm ym}_6}^2 := \frac{1}{\tau_{\rm D5} \lambda^2 (L\mu)^2} =  4\pi^{7/2} \frac{\sqrt{ g_s N_c}}{N_c} \mu^{-2} ~.
\end{equation}
Note that the dimensionless coupling $(g_{{\rm ym}_6} \mu)$ is small in the regime $N_c \gg g_s N_c \gg 1$.  In order to bring the scalar kinetic terms to standard form we define mass dimension-one scalar fields through
\begin{equation}\label{XtoPhi}
\Phi^m := \lambda^{-1} (L\mu)^2 X^m = \frac{\sqrt{g_s N_c}}{\sqrt{\pi}} \mu^2  X^m~,
\end{equation}
so that $(A_a,\Phi^m)$ carry the same dimension.

Once the closed string fields in the D-brane action are expressed in terms of the rescaled quantities, one finds that $F_{ab}$ is always accompanied by a factor of $\lambda (L\mu)^{-2}$, while $[X^m,X^n]$ is always accompanied by the inverse factor.  After changing variables to $\Phi^m$ for the scalars, all four types of open string quantities appearing in D5-brane action carry the same prefactor:
\begin{align}\label{osev}
& \left( \frac{\lambda}{(L\mu)^2} F_{ab}, D_a X^m, \frac{(L\mu)^2}{\lambda} [X^m,X^n], \mu X^m \right)   = \frac{\sqrt{\pi}}{\sqrt{g_s N_c}} \mu^{-2} \left( F_{ab}, D_a \Phi^m, [\Phi^m,\Phi^n], \mu \Phi^m\right)~,
\end{align}
and this provides a convenient organizing principle for the expansion.  Of course it is $(A_a, \Phi^m)_{\rm c}$, defined by 
\begin{equation}\label{cnorm}
(A_a, \Phi^m) = g_{{\rm ym}_6} (A_a, \Phi^m)_{\rm c}~,
\end{equation}
that are the canonically normalized open string modes.  The open string expansion variables on the right-hand side of \eqref{osev} do not scale homogeneously when expressed in terms of these, and this point must be kept in mind when comparing the strength of interaction vertices below.

Now, let $\mathfrak{C} \in (h_{MN},b_{MN},\varphi, c^{(n)})$ denote a generic closed string fluctuation, let $\mathfrak{O} \in ( F_{ab}, D_a \Phi^m, [\Phi^m,\Phi^n], \mu \Phi^m)$ denote any of the open string expansion variables, and set
\begin{equation}\label{epsop}
\epsilon_{\rm op} := \frac{\lambda}{(L\mu)^2} =  \frac{\sqrt{\pi}}{\sqrt{g_s N_c}} \mu^{-2}~.
\end{equation}
Then the expansion of \eqref{Myers} can be written in the form
\begin{equation}\label{D5expanded}
S_{\rm D5}^{\rm bos} = - \frac{1}{\epsilon_{\rm op}^2 g_{{\rm ym}_6}^2} \int \ed^6 x \sqrt{-g_6} \sum_{n_o,n_c = 0}^{\infty} \epsilon_{{\rm op}}^{n_o} \, \kappabar^{n_c} \, V_{n_o,n_c}~,
\end{equation}
where $V_{n_o,n_c}$ is a sum of monomials of the form $\mathfrak{C}^{n_c} \STr( \mathfrak{O}^{n_o})$, with rational coefficients.  The first few $V_{n_{o},n_{c}}$'s are 
\begin{align}\label{Vocs}
V_{0,0} =&~  N_f ~, \cr
V_{0,1} =&~ N_f \left\{ \half \left( h^{a}_{\phantom{a}a} + \varphi \right) - \frac{1}{6!} \epsilon^{abcdef} c^{(6)}_{abcdef} \right\}~, \cr
V_{1,0} =&~ 0 ~, \cr
V_{0,2} =&~ N_f \left\{ \frac{1}{8} \left( h^{a}_{\phantom{a}a} + \varphi \right)^2 + \frac{1}{4} (b^{ab} b_{ab} - h^{ab} h_{ab}) - \frac{1}{4! 2} \epsilon^{abcdef} c_{abcd}^{(4)} b_{ef} \right\}~, \cr
V_{1,1} =&~ i \Tr \bigg\{ \frac{1}{2} b^{ab} F_{ab} + h_{am} D^a \Phi^m + \half \Phi^m \left. \left( \pd_m (\Gbar^{ab} h_{ab}) + \pd_m \varphi \right) \right|_{x_{0}^m} + \cr
&~ \qquad \quad  -\epsilon^{abcdef} \bigg[ \left( \frac{1}{6!} \Phi^m (\pd_m c_{abcdef}^{(6)}) |_{x_{0}^m} + \frac{1}{5!} (D_a \Phi^m) c_{mbcdef}^{(6)} \right) +  \cr
&~ \qquad \qquad \qquad \qquad  + \frac{1}{3! 2} (D_a \Phi^m) \Cbar^{(4)}_{mbcd} b_{ef} +  \frac{1}{4! 2} c_{abcd}^{(4)} F_{ef} \bigg] \bigg\} ~, \cr
V_{2,0} =&~ \Tr \bigg\{ \frac{1}{4} F_{ab} F^{ab} + \half \Gbar_{mn} D_a \Phi^m D^a \Phi^n + \frac{1}{4} \Gbar_{mk} \Gbar_{nl} [\Phi^m, \Phi^n] [\Phi^k, \Phi^l]  +  \cr
&~ \qquad \quad - \frac{1}{3! 2} \epsilon^{abcdef} (D_a \Phi^m) \Cbar_{mbcd}^{(4)} F_{ef} \bigg\}~,
\end{align}
where $\epsilon^{abcdef}$ is the Levi--Civita tensor with respect to the background metric, $\epsilon^{012345} = (-g_6)^{-1/2}$, and we have used that $\STr$ reduces to the ordinary trace when there are no more than two powers of the open string variables $\mathfrak{O}$.  All closed string fields are to be understood as being evaluated at $x^m = x_{0}^m$ except for those in $V_{1,1}$ that involve taking a transverse derivative before setting $x^m = x_{0}^m$.

\begin{figure}
\begin{center}
\includegraphics[scale=0.3]{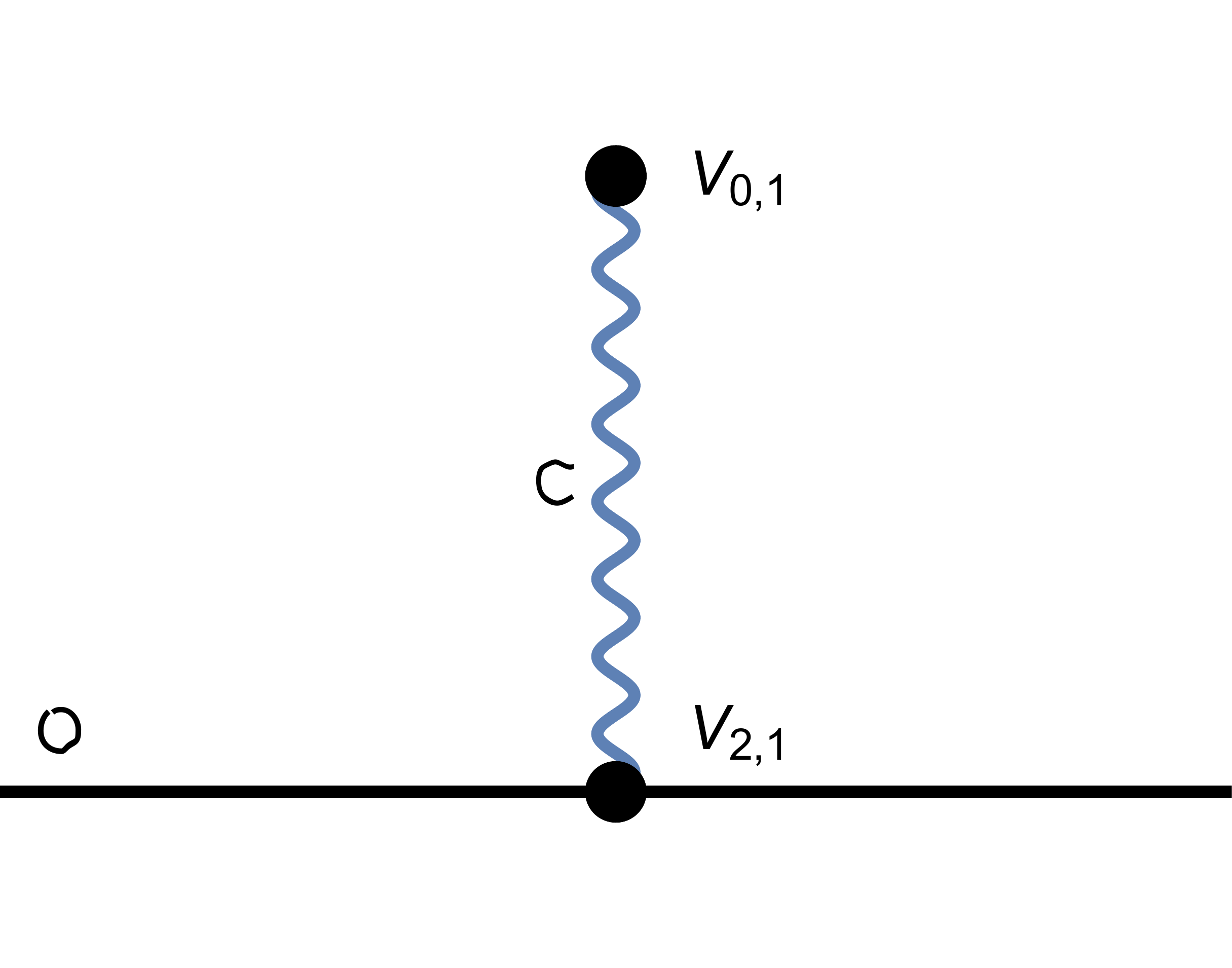}
\caption{A virtual closed string correction to an open string propagator.  The closed string is created from the vacuum by a vertex in $V_{0,1}$.  It propagates to a three-point vertex in $V_{2,1}$.  This gives a correction to the open string propagator that is of the same order as a standard one-loop correction from virtual open string modes.}
\label{fig3}
\end{center}
\end{figure}

There is a great deal of physics in the $V_{n_o,n_c}$'s:
\begin{itemize}
\item $V_{0,0}$ corresponds to the energy density of the background D5-brane configuration.
\item $V_{0,1}$ gives closed string tadpoles for the metric, dilaton, and RR six-form potential.  These are present because we have not included the gravitational backreaction of the D5-branes---\ie\ we have not expanded around a solution to the equations of motion for these closed string fields.  The strength of these tadpoles is $N_f g_{{\rm ym}_6}^{-2} \epsilon_{\rm op}^{-2} \kappabar  \propto N_f \sqrt{g_s N_c} \mu^2$, which is large when $g_s N_c \gg 1$.  However this does not necessarily mean that the probe approximation is bad!  The effects of these tadpoles on open and closed string processes will still be suppressed if the interaction vertices are sufficiently weak.  

Consider, for example, the leading correction to the open string propagators due to these tadpoles.  This corresponds to the diagram in Figure \ref{fig3}.  The correction is proportional to the product of the tadpole vertex with the cubic vertex for two open and one closed string fluctuation.  After canonically normalizing the open string modes via \eqref{cnorm}, the three-point vertex goes as $\kappabar$.  Therefore the product is proportional to $N_f \sqrt{g_s N_c} \mu^2 \kappabar \propto (N_f \sqrt{g_s N_c}/N_c) \mu^{-2} \propto N_f g_{{\rm ym}_{6}}^2 $.  Hence this process acts just like a standard one-loop correction to the Yang--Mills coupling that we would get from open string modes.  As long as $N_f \ll N_c/\sqrt{g_s N_c}$, both the standard one-loop correction and this closed string correction will be suppressed.  Note this is a slightly stronger restriction than the usual  $N_f \ll N_c$ limit when the 't Hooft coupling $g_s N_c$ is large, but nevertheless can be comfortably satisfied for a range of $N_f$ in the regime $N_c \gg g_s N_c \gg 1$.
\item The vanishing of $V_{1,0}$ indicates that open string tadpoles are absent.  This simply validates the fact (already implicitly assumed in the above discussion) that the D5-brane embedding, described by $x^m = x_{0}^m$, extremizes the equations of motion for the open string modes in the fixed closed string background.
\item Only the center-of-mass degrees of freedom corresponding to the central $\mathfrak{u}(1) \subset \mathfrak{u}(N_f)$ participate in $V_{1,1}$ due to the trace.  The strength of these interactions is $g_{{\rm ym}_6}^{-1} \epsilon_{\rm op}^{-1} \kappabar \propto g_{{\rm ym}_6}$, where we have made use of the convenient relation
\begin{equation}
\epsilon_{\rm op} g_{{\rm ym}_6}^2 = 2\sqrt{\pi} \kappabar~.
\end{equation}
Hence they can be treated perturbatively.  Furthermore the $\mathfrak{u}(1)$ and $\mathfrak{su}(N_f)$ degrees of freedom decouple in $V_{2,0}$, so the couplings in $V_{1,1}$ can only transmit the effects of the closed string tadpoles to the $\mathfrak{su}(N_f)$ fields through higher order open string interactions.
\item  The first three terms of $V_{2,0}$ come from the DBI action, and comprise the usual Yang--Mills action on a curved background.  The final term in $V_{2,0}$, meanwhile, comes from the CS action and is non-vanishing because there is a the nontrivial RR flux in the supergravity background.
\end{itemize}

It is also interesting to consider the form of terms in $V_{3,0}$, or higher order open string interactions.  $V_{3,0}$ is nontrivial when $z_0 \neq 0$; $V_{4,0}$ is always nontrivial.  For example, there is an $\STr(\vec{\Phi}^z F^2)$ coupling of the form
\begin{equation}
V_{3,0} \supset \STr \left\{ \frac{\vec{z}_0 \cdot \vec{\Phi}^z}{3(r^2 + z_{0}^2)} \left( F^{r_i r_j} F_{r_i r_j} - F^{\mu\nu} F_{\mu\nu} \right) \right\}~.
\end{equation}
Three- and four-point couplings in $V_{3,0}$ and $V_{4,0}$ come with extra factors of $\epsilon_{\rm op}$ relative to the three- and four-point couplings in the Yang--Mills terms, $V_{2,0}$.  Hence they will be suppressed relative to the Yang--Mills terms for field variations at or below the scale $\mu$.  More precisely, if the fields vary on a scale $\mu'$ we merely require $(\mu'/\mu)^2 \ll \sqrt{g_s N_c}$, in order that these terms be suppressed relative to their counterparts in $V_{2,0}$.

In summary, there is a regime of parameters---namely $N_c \gg g_s N_c \gg 1$ and $N_f \ll N_c/\sqrt{g_s N_c}$---where the leading interactions of the (bosonic) $\mathfrak{su}(N_f)$ open string modes are governed by $V_{2,0}$.  This forms the bosonic part of a six-dimensional super-Yang--Mills theory on the curved background \eqref{6Dmet}.  

We can present this action in two different forms, both of which will prove useful below.  First there is the form we have used to give $V_{2,0}$, in which the scalars carry curved space indices.  In order to be more explicit with regards to the $\Cbar^{(4)}$ term, we have from \eqref{nhgeom} that the relevant components are
\begin{equation}
\Cbar_{012y}^{(4)}(x^a,x_{0}^m) = \mu^4 (r^2 + z_{0}^2)^2~,
\end{equation}
and so the last term of $V_{2,0}$ contributes as follows:
\begin{align}
& \int \ed^6 x \sqrt{-g_6}  \Tr \left\{ \frac{1}{3! 2} \epsilon^{abcdef} (D_a \Phi^m) \Cbar_{mbcd}^{(4)} F_{ef} \right\} = \cr
& \qquad \qquad \qquad = \frac{1}{2} \int \ed^6 x \sqrt{-g_6} \tilde{\epsilon}^{r_i r_j r_k} \mu^4 (r^2 + z_{0}^2)^2 \Tr \left\{  (D_{r_i} \Phi^y) F_{r_j r_k} \right\}~.
\end{align}
Here we have introduced $\tilde{\epsilon}$, which should be thought of as the Levi--Civita tensor on the Euclidean $\mathbb{R}^3$ spanned by $\vec{r}$:  $\tilde{\epsilon}^{r_1 r_2 r_3} = 1$, or if we work in spherical  coordinates $\tilde{\epsilon}^{r\theta\phi} = (r^2 \sin{\theta})^{-1}$.  Then the bosonic part of the Yang--Mills action is
\begin{align}\label{SYMb1}
S_{{\rm ym},b} :=&~  - \frac{1}{g_{{\rm ym}_6}^2} \int \ed^6x \sqrt{-g_6} V_{2,0} \cr
=&~ - \frac{1}{g_{{\rm ym}_6}^2} \int \ed^6x \sqrt{-g_6} \Tr \bigg\{ \frac{1}{4} F_{ab} F^{ab} + \half \Gbar_{mn} D_a \Phi^m D^a \Phi^m + \cr
&~ \qquad + \frac{1}{4} \Gbar_{mk} \Gbar_{nl} [\Phi^m, \Phi^n] [\Phi^k, \Phi^l]  - \half \tilde{\epsilon}^{r_i r_j r_k} \mu^4 (r^2 + z_{0}^2)^2 (D_{r_i} \Phi^y) F_{r_j r_k} \bigg\} ~. \qquad
\end{align}

We can also derive a more standard field theoretic form for the action by rescaling the scalar fields in such a way that their kinetic terms are canonically normalized.  To do this, we make use of a vielbein associated with the background metric $\Gbar_{mn}$:
\begin{equation}\label{canonscalar}
\Phi^{\uy} :=  \mu (r^2 + z_{0}^2)^{1/2} \Phi^y~, \qquad \Phi^{\underline{z_i}} := \frac{1}{\mu (r^2 + z_{0}^2)^{1/2}} \Phi^{z_i} ~.
\end{equation}
Both mass terms and boundary terms arise when we integrate by parts in the kinetic terms.  One can also integrate by parts on the last term of \eqref{SYMb1} and make use of the Bianchi identity, $\tilde{\epsilon}^{r_i r_j r_k} D_{r_i} F_{r_j r_k} = 0$.  We also switch to spherical coordinates, as the only surviving bulk term comes from the derivative of the $(r^2 + z_{0}^2)^2$ prefactor.  This integration by parts also generates a boundary term.  After carrying out these manipulations, the bosonic action becomes 
\begin{align}\label{SYMb}
S_{{\rm ym},b} :=&~ - \frac{1}{g_{{\rm ym}_6}^2} \int \ed^6x \sqrt{-g_6} \Tr \bigg\{ \frac{1}{4} F_{ab} F^{ab} + \half (D_a \Phi_{\um}) (D^a \Phi^{\um}) + \frac{1}{4} [\Phi_{\um}, \Phi_{\un}] [\Phi^{\um}, \Phi^{\un}] + \cr
&~ + \half M_{z}^2 \delta_{ij} \Phi^{\underline{z_i}}\Phi^{\underline{z_j}} + \half M_{y}^2 (\Phi^{\uy})^2 + 2M_{\Psi} \epsilon^{\alpha\beta} F_{\alpha\beta} \Phi^{\uy} \bigg\} + S_{b}^{\rm bndry}~. \raisetag{20pt}
\end{align}
In the last term the indices $\alpha,\beta$ correspond to coordinates $\theta,\phi$ along the two-sphere and $\epsilon^{\theta\phi} = (g_{S^2})^{-1/2} = \mu^2 (r^2 + z_{0}^2)/(r^2 \sin{\theta})$.  The mass parameters are defined as follows:
\begin{equation}\label{SYMmasses}
M_{z}^2 := \mu^2 \left( \frac{r^2}{r^2 + z_{0}^2} - 3\right)~, \quad M_{y}^2 := \mu^2 \left( \frac{r^2}{r^2 + z_{0}^2} +3 \right)~, \quad M_{\Psi} := \frac{\mu r}{\sqrt{r^2 + z_{0}^2}}~.
\end{equation}
As $r \to \infty$ they approach the values $-2,4,1$ in units of the inverse AdS radius.  When $z_0 =0$ they take these values everywhere.  Although the squared mass of the $Z$ scalars is negative, it satisfies the Breitenlohner--Freedman bound \cite{Breitenlohner:1982jf} for $AdS_4$.  The reason for the notation $M_{\Psi}$ will become clear below when we consider the fermionic part of the action.

The boundary terms arise due to the integration by parts and the boundary component $\pd M_6 \cong \{r_b\} \times \mathbb{R}^{1,2} \times S^2$ at $r = r_b \to \infty$.\footnote{We assume the fields are sufficiently regular such that there is no boundary contribution from $r=0$.  This is discussed in some further detail for static configurations later.  See section \ref{sec:Bbound}.}  They are given by
\begin{align}\label{Sbbndry}
S_{b}^{\rm bndry}  =&~ \frac{1}{g_{{\rm ym}_6}^2} \int_{~~\,\mathclap{\pd M_6}} \ed^5 x \sqrt{-g_{(\pd)} }\Tr \left\{ \frac{M_{\Psi}}{2} \left( (\Phi^{\uy})^2 - \delta_{ij} \Phi^{\underline{z_i}}\Phi^{\underline{z_j}} \right) + \half \Phi^{\uy} \epsilon^{\alpha\beta} F_{\alpha\beta}  \right\}~,
\end{align}
where $\ed^5 x \sqrt{-g_{(\pd)}} $ is the induced volume form on the boundary,
\begin{equation}\label{bmeasure}
\sqrt{-g_{(\pd)}} \, \ed^5 x = \mu r_{b}^2 (r_{b}^2 + z_{0}^2)^{1/2} \ed^3 x \ed \Omega~,
\end{equation}
with $\ed^3 x := \ed x^0 \ed x^1 \ed x^2$ and $\ed\Omega := \sin{\theta} \ed \theta \ed \phi$.  If one works with the action in the form \eqref{SYMb} then it is important to keep these terms.  They play a crucial role both in establishing the consistency of the variational principle and in the supersymmetry invariance of the Yang--Mills action.  The limit $r_b \to \infty$ of quantities computed using \eqref{Sbbndry} is understood to be taken at the end of any calculation (when it exists).

\subsection{Fermionic D-brane action}

Ideally, one would like to obtain non-abelian super-Yang--Mills theory on the D5-branes via the limiting behavior of a $\kappa$-symmetric non-abelian super D-brane action for general closed string backgrounds.  While important progress toward constructing such actions has been made ( see \eg\ \cite{Taylor:1999pr,Drummond:2002kg,Howe:2005jz,Bandos:2009xy} and references therein), the subject has not matured sufficiently to be of practical use for our purposes.  

Instead, we will fall back on abelian fermionic D-brane actions that have been discussed extensively, starting with the initial work of \cite{Aganagic:1996pe,Cederwall:1996ri,Bergshoeff:1996tu,Aganagic:1996nn}, and continuing with \cite{Marolf:2003ye,Marolf:2003vf,Martucci:2005rb}.  Here we follow the conventions of \cite{Marolf:2003vf,Martucci:2005rb}.  This will provide the fermionic couplings that are quadratic order in open string fluctuations---kinetic and mass-like terms.  With these and the full set of bosonic couplings in hand, we will be able to deduce the remaining Yukawa-type couplings and the non-abelian supersymmetry transformations via a simple ansatz.

The massless fermionic degrees of freedom on a D5-brane are the same as those in ten-dimensional super-Yang--Mills, and can be packaged into a single ten-dimensional Majorana--Weyl fermion, $\Psi$.  The couplings of $\Psi$ to the IIB closed string supergravity fields are described most conveniently by introducing a doublet of ten-dimensional Majorana--Weyl spinors $\hat{\Psi} = (\Psi_1,\Psi_2)^T$ of the same 10D chirality.  One linear combination will be projected out by the $\kappa$-symmetry projector while the other will be the physical $\Psi$.  The ten-dimensional gamma matrices, satisfying $\{ \Gamma^M, \Gamma^N \} = 2 G^{MN}$, are likewise extended by the doublet structure.  One introduces
\begin{equation}
\hat{\Gamma}^M := \Gamma^M \otimes \mathbbm{1}_2~, \qquad \hat{\Gammabar} := \Gammabar \otimes \sigma^3~,
\end{equation}
where $\Gammabar = \Gamma^{\underline{0123456789}}$ is the ten-dimensional chirality operator and $\sigma^{1,2,3}$ are the Pauli matrices.

The \emph{abelian} fermionic D5-brane action, to quadratic order in $\hat{\Psi}$, takes the form
\begin{align}\label{fermD5}
S_{f} =&~ \frac{\tau_{\rm D5}}{2} \int \ed^{6}x e^{-\Delta\upphi} \sqrt{-\det{ (P[E] -i \lambda F)}} \, \hat{\Psibar} (\mathbbm{1}-\Gamma_{\rm D5}) \left[ (M^{-1})^{ab} \hat{\Gamma}_{b}^{(P)} \hat{D}_a - \Delta \right] \hat{\Psi}~,
\end{align}
where $E_{MN} = e^{\Delta\upphi/2} (G_{MN} + B_{MN})$ as before and the matrix $M$ is
\begin{equation}
M_{ab} = e^{\Delta\upphi/2} P[G_{ab}] +  \FF_{ab} \hat{\Gammabar}~.
\end{equation}
Here we have also introduced the shorthand $\FF_{ab} := e^{\Delta\upphi/2} P[B_{ab}] -i \lambda F_{ab}$.  The idempotent matrix $\Gamma_{\rm D5}$ appearing in the $\kappa$-symmetry projector, $\half (\mathbbm{1} - \Gamma_{\rm D5})$, has a somewhat nontrivial expression\footnote{Our $M_{ab},\Gamma_{\rm D5}$ are denoted $\widetilde{M}_{ab},\widetilde{\Gamma}_{\rm D5}$ in \cite{Marolf:2003vf,Martucci:2005rb}.} in terms of $\FF$:
\begin{align}\label{GammaD5}
\Gamma_{\rm D5} :=&~  \frac{1}{\sqrt{-\det(P[E] -i \lambda F)}} \times \cr
&~ \times \sum_{q+r =3} \frac{\varepsilon^{a_1 \cdots a_{2q} b_1 \cdots b_{2r}}}{q! 2^q (2r)!} (-i)^q \FF_{a_1 a_2} \cdots \FF_{a_{2q-1} a_{2q}} ( \Gamma_{b_1 \cdots b_{2r}}^{(P)} \otimes (-i \sigma^2)) (\hat{\Gammabar})^r ~, \qquad
\end{align}
where $\varepsilon^{012345} = 1$, and the $\Gamma_{a}^{(P)}$ are the pullbacks of $\Gamma_M$ to the worldvolume.  In static gauge, $\Gamma_{a}^{(P)} = \Gamma_a -i (\pd_a X^m) \Gamma_m$.  

The remaining couplings to closed string fields are encoded in the generalized derivative $\hat{D}$ and the mass-like operator, $\Delta$.  We write only the terms that contribute when evaluated on the near-horizon background geometry \eqref{nhgeom}; the full set of couplings can be found in \cite{Marolf:2003vf,Martucci:2005rb}.  In this case
\begin{align}
\hat{D}_a =&~ P[D_a] \otimes \mathbbm{1}_2 + \frac{1}{16\cdot 5!} e^{\Delta\upphi} F^{(5)}_{M_1 \cdots M_5} \left(\Gamma^{M_1 \cdots M_5} \Gamma_{a}^{(P)} \otimes (i\sigma^2) \right) + \cdots~,
\end{align}
where the terms represented by $\cdots$ vanish when closed string fluctuations are switched off, while $\Delta \to 0$ when closed string fluctuations are switched off.  The notation $P[D_a]$ is meant to indicate that one takes the pullback of $D_M \Psi_{1,2}$ to the brane worldvolume, and $D_M$ is the standard covariant derivative on ten-dimensional Dirac spinors.

Now we would like to argue that in the near-horizon geometry \eqref{nhgeom}, the action \eqref{fermD5} has an expansion in closed and open string fluctuations controlled by the same parameters, $\epsilon_{\rm op}, \kappabar$, that appeared in the expansion of the bosonic action \eqref{D5expanded}.  Considering first the rescaling of the closed string fields, \eqref{closedrescale}, there are a few key points:
\begin{itemize}
\item After applying this rescaling under the determinant of \eqref{fermD5} we can pull out a factor of $(L\mu)^6$, and we will have the usual factor of $(L\mu)^{-2} \lambda = \epsilon_{\rm op}$ accompanying $F_{ab}$.
\item The $\Gamma_{b_1 \cdots b_{2r}}^{(P)}$ factor in $\Gamma_{\rm D5}$ rescales according to $\Gamma_{b_1 \cdots b_{2r}}^{(P)} = (L\mu)^{2r} \tilde{\Gamma}_{b_1 \cdots b_{2r}}^{(P)}$, due to the implicit vielbein factors present in it.  Taking into account the $(L\mu)^{-6}$ from the determinant factor out front, $\Gamma_{\rm D5}$ retains its form under the rescaling except that each factor of $F_{ab}$ picks up a corresponding $(L\mu)^{-2}$ prefactor.  This combines with the $\lambda$'s already present so that all $F_{ab}$ in $\Gamma_{\rm D5}$ are accompanied by $\epsilon_{\rm op}$.
\item One can check that $(M^{-1})^{ab} \hat{\Gamma}_{b}^{(P)} \hat{D}_a - \Delta$ gets a net factor of $(L\mu)^{-1}$ when expressed in terms of the rescaled closed string fields, while $F_{ab}$ in $M_{ab}$ acquires an $\epsilon_{\rm op}$ prefactor.
\item Finally, each appearance of $\pd_a X^m$ from pullbacks and $X^m$ from expanding the closed string fields around $x_{0}^m$ is accompanied by a factor of $\epsilon_{\rm op}$ when we express $X^m$ in terms of $\Phi^m$ via \eqref{XtoPhi}.
\end{itemize}
Together, these observations show that all open string interaction vertices between $\Psi^2$ and powers of $F_{ab},\pd_a X^m$, and $X^m$ are controlled by the expected power of $\epsilon_{\rm op}$.  The overall prefactor of the leading $\Psi^2$ term is $\tau_{\rm D5} (L\mu)^5 = (\epsilon_{\rm op}\, g_{{\rm ym}_6})^{-2} (L\mu)^{-1}$.  We can make a rescaling\footnote{The $\hat{\Psi}$ in \eqref{fermD5} must have units of (length)${}^{1/2}$.  It would be natural to include a factor of $\lambda^{1/2}$ out in front of \eqref{fermD5} so that they are dimensionless.  Then the rescaling would be $\tilde{\hat{\Psi}} = \epsilon_{\rm op}^{3/4} \hat{\Psi}$.} of $\hat{\Psi}$ analogous to \eqref{XtoPhi} such that the coefficient of this leading order term is simply $-i /g_{{\rm ym}_6}^{2}$.  We will assume this has been done and continue using the same notation for the fermion.

Hence we write
\begin{align}\label{leadingSf}
S_f =&~ -\frac{i}{2 g_{{\rm ym}_6}^2} \int \ed^6 x \sqrt{ -g_6} \, \hat{\Psibar} (\mathbbm{1} - \Gamma_{\rm D5}^{(0)}) \Gamma^a \hat{D}_{a}^{(0)} \hat{\Psi} \times \left( 1 + O(\epsilon_{\rm op},\kappabar) \right)~,
\end{align}
where
\begin{equation}\label{GammaD5leading}
\Gamma_{\rm D5}^{(0)} := \Gamma_{\underline{012r\theta\phi}} \Gammabar \otimes \sigma^1 ~,
\end{equation}
and
\begin{align}\label{D0hat}
\hat{D}_{a}^{(0)} :=&~ \left( \pd_a + \frac{1}{4} \omegabar_{MN,a} \Gamma^{MN} \right) \otimes \mathbbm{1}_2 + \frac{1}{16 \cdot 5!} \Fbar^{(5)}_{M_1 \cdots M_5} \Gamma^{M_1 \cdots M_5} \Gamma_a \otimes (i \sigma^2) ~.
\end{align}
For $\Gamma_{\rm D5}^{(0)}$ we took the $q = 0$, $r=3$ term in \eqref{GammaD5} and used that $(-g_6)^{-1/2} \varepsilon^{b_1 \cdots b_6} \Gamma_{b_1 \cdots b_6}^{(P)} = 6! \, \Gamma_{\underline{012r\theta\phi}}$ to leading order in open and closed string fluctuations.  In \eqref{D0hat}, $\omegabar_{MN,P}$ are the components of the spin connection with respect to the background metric $\Gbar_{MN}$, evaluated at $x^m = x_{0}^m$, and all gamma matrices with covariant indices are defined using the vielbeine of the background metric.

Let us evaluate \eqref{D0hat} in more detail.  It follows from the background \eqref{nhgeom} that
\begin{align}
& \frac{1}{16 \cdot 5!} \Fbar^{(5)}_{M_1 \cdots M_5} \Gamma^{M_1 \cdots M_5}  = - \frac{\mu}{4} \left\{ \frac{r}{\sqrt{r^2 + z_{0}^2}} \Gamma^{\underline{012ry}}  + \frac{z_{0,i}}{\sqrt{ r^2 + z_{0}^2}} \Gamma^{\underline{012z_i y}} \right\} (\mathbbm{1} - \Gammabar)~, 
\end{align}
where we recall that $(r,\theta,\phi)$ are spherical coordinates for the directions spanned by $\vec{r}$.  But the second term drops out of \eqref{leadingSf} because 
\begin{equation}
\Gamma^a \Gamma^{\underline{012ry}} \Gamma_a = 2 \Gamma^{\underline{012ry}}  ~, \qquad \Gamma^a \Gamma^{\underline{012z_iy}} \Gamma_a = 0~.
\end{equation}
Regarding the ten-dimensional spin connection, there are nonzero components of the type $\omegabar_{bm,a}$ when $z_0 \neq 0$.  (See appendix \ref{app:spinors} for details.)  However, the contribution of these components to $\Gamma^{a} \omegabar_{MN,a} \Gamma^{MN}$ cancels out.  Hence
\begin{equation}\label{D0hatsimp1}
\Gamma^a \hat{D}_{a}^{(0)} = \Gamma^a \left( \pd_a + \frac{1}{4} \omegabar_{bc,a} \Gamma^{bc} \right) \otimes \mathbbm{1}_2 - \frac{\mu r}{2\sqrt{r^2 + z_{0}^2}} \Gamma^{\underline{012ry}} (\mathbbm{1} + \Gammabar)  \otimes  (i\sigma^2)~.
\end{equation}

The projector in the last term of \eqref{D0hatsimp1} will either give the identity or zero when acting on $\Psi_{1,2}$, depending on the 10D chirality of the latter.  The two possibilities distinguish between a D5-brane and an anti-D5, and only one choice will lead to a supersymmetric worldvolume theory on the brane.  We will see that the supersymmetric theory corresponds to
\begin{equation}\label{10Dchirality}
\Gammabar \Psi_{1,2} = \Psi_{1,2}~.
\end{equation}
Thus the coupling to the background $\Fbar^{(5)}$ provides a necessary mass-like term for the fermion.

It is now straightforward to diagonalize the operator $\half (\mathbbm{1} - \Gamma_{\rm D5}^{(0)}) \Gamma^a \hat{D}_{a}^{(0)}$ with respect to the auxiliary doublet structure.  Introducing the unitary transformation
\begin{equation}
U := \frac{1}{\sqrt{2}} \left( \mathbbm{1} - \Gamma_{\underline{012r\theta\phi}} \Gammabar \otimes (i\sigma^2) \right)~,
\end{equation}
one finds
\begin{equation}
U \left[ \half (\mathbbm{1} - \Gamma_{\rm D5}^{(0)}) \Gamma^a \hat{D}_{a}^{(0)} \right] U^\dag = \left\{ \Gamma^a D_a + \frac{\mu r}{2\sqrt{r^2 + z_{0}^2}} \Gamma^{\underline{\theta\phi y}} (\mathbbm{1} + \Gammabar) \right\} \otimes \half (\mathbbm{1} + \sigma^3)~,
\end{equation}
where
$D_a := \pd_a + \frac{1}{4} \omegabar_{bc,a} \Gamma^{bc}$.  Thus, setting $(\Psi, \Psi')^T := U \hat{\Psi}$, one sees that $\Psi'$ is projected out while $\Psi$ encodes the physical degrees of freedom.  Using \eqref{10Dchirality}, and recalling the definition of $M_{\Psi}$ in \eqref{SYMmasses}, the final result for \eqref{leadingSf} takes the form
\begin{equation}\label{leadingSf2}
S_f = -\frac{i}{2g_{{\rm ym}_6}^2} \int \ed^6 x \sqrt{-g_6} \, \Psibar \left\{ \Gamma^a D_a + M_{\Psi} \Gamma^{\underline{\theta\phi y}} \right\} \Psi \times \left( 1 + O(\epsilon_{\rm op},\kappabar) \right)~.
\end{equation}
Note that for a ten-dimensional Majorana--Weyl spinor, the bilinear $\Psibar \Gamma^{M_1 \cdots M_p} \Psi$ vanishes unless $p=3$ (or $7$), so the gamma matrix structure of the mass term is as it had to be.  $\Psi$ contains the degrees of freedom of a single six-dimensional Dirac fermion and we could write \eqref{leadingSf2} in six-dimensional language, but for now it is more convenient to work directly with the `10D' form.

Finally, we will infer from \eqref{leadingSf2} and the bosonic Yang--Mills terms \eqref{SYMb}, the non-abelian analogs of the leading terms in \eqref{leadingSf2} that complete \eqref{SYMb} into a supersymmetric invariant.  Clearly the covariant derivative $D_a$ should be generalized to a gauge covariant derivative, $D_a := \pd_a + \frac{1}{4} \overbar{\omega}_{bc,a} \Gamma^{bc} + [A_a, \, \cdot \,]$.  We will, for convenience, continue to use the same notation for this covariant derivative as we did above.  A natural ansatz that will yield the Yukawa couplings is simply to extend this to a ten-dimensional covariant derivative: $\Gamma^a D_a \Psi + \Gamma^{\um} [\Phi_{\um}, \Psi]$.  Our ultimate justification for this ansatz (detailed below) will be that supersymmetry requires it. 

 Hence we take the fermionic terms of the Yang--Mills action to be
\begin{align}\label{SYMf}
S_{{\rm ym},f} :=&~ -\frac{i}{2 g_{{\rm ym}_6}^2} \int \ed^6 x \sqrt{-g_6} \Tr \bigg\{ \Psibar \left( \Gamma^a D_a + M_{\Psi} \Gamma^{\underline{\theta\phi y}} \right) \Psi + i \Psibar \Gamma^{\um} [\Phi_{\um}, \Psi ] \bigg\} +  S_{f}^{\rm bndry},
\end{align}
where $\Psi$ is now valued in the adjoint representation of $\mathfrak{su}(N_f)$.  

We've included a boundary action for the fermion ,
\begin{align}\label{Sfbndry}
S_{f}^{\rm bndry} :=&~ -\frac{i}{4 g_{{\rm ym}_6}^2}  \int_{~~\,\mathclap{\pd M_6}} \ed^5 x \sqrt{-g_{(\pd)}} \Tr \left\{ \Psibar \Gamma^{\underline{\theta\phi y}} \Psi \right\}~.
\end{align}
The analysis of \cite{Henneaux:1998ch} for fermions on anti-de Sitter space demonstrates that such boundary terms are necessary in order to have a well-defined variational principle.  We will see that the boundary action \eqref{Sfbndry} is also required for supersymmetry.  Without it, the supersymmetry variation of the action would produce boundary terms that do not vanish on their own.  These points are analyzed in sections \ref{sec:cvp} and \ref{sec:bndrySUSY} below.  In principle such boundary terms should have already been present in \eqref{fermD5}, but we are not aware of any previous work on this issue.

\section{Supersymmetry}\label{sec:susy}

As noted previously, the intersecting D-brane system of Figure \ref{D3D5system} preserves eight supersymmetries.  In the near-horizon limit of the D3-brane geometry, the symmetry algebra is enhanced to $\mathfrak{osp}(4|4)$ with sixteen odd generators, provided the D3 and D5-branes have zero transverse separation.  The leading low-energy effective description in the regime $N_c \gg g_s N_c \gg 1$ and $N_f \ll N_c/\sqrt{g_s N_c}$ consists of a six-dimensional Yang--Mills theory on the rigid background \eqref{6Dmet} in which the transverse separation appears as a parameter, (along with decoupled supergravity and $\mathfrak{u}(1)$ sectors).  Thus one expects the Yang--Mills theory to possess eight supersymmetries when $z_0 \neq 0$ and sixteen when $z_0 = 0$.  

In this section we first review the Killing spinors of the background geometry \cite{Lu:1998nu,Claus:1998yw} and the induced Killing spinors on the D5-brane worldvolume \cite{Skenderis:2002vf}.  Then, using the latter as generators, we exhibit the full set of supersymmetry transformations on the Yang--Mills fields and establish the invariance of the action, \eqref{SYMb} plus \eqref{SYMf}, modulo boundary terms.  

\subsection{Killing spinors in the bulk}

$AdS_5 \times S^5$ is a maximally supersymmetric background admitting thirty-two linearly independent Killing spinors---that is, solutions $\epsilon$ to the vanishing of the gravitino variation,
\begin{equation}\label{bulkKSeqn}
\left[ D_{M}^{(0)}  - \frac{i}{16 \cdot 5!} (\Gamma^{M_1\cdots M_5} \Fbar_{M_1 \cdots M_5}^{(5)} )\Gamma_M \right] \epsilon = 0 ~.
\end{equation}
Here $\epsilon$ is complex Weyl and our conventions are that it has positive chirality, $\Gammabar \epsilon = \epsilon$.  Explicit solutions can be found in various references, going back to \cite{Lu:1998nu,Claus:1998yw}.  The form of these solutions depends of course on the choice of coordinate system and frame.  Most references employ a frame adapted to some type of spherical coordinate parameterization of the $S^5$.  This is inconvenient for the applications we have in mind here.\footnote{For example, a naive application of these formulae will lead to expressions for worldvolume Killing spinors on the D5-brane that appear to depend on angular coordinates parameterizing the transverse space that are not well-defined on the brane locus.}  We provide two alternative descriptions that are better-suited to the analysis in subsequent sections; both of them will be useful below.

The first is based on coordinates $(x^\mu,\vec{r},\vec{z},y)$ in which the metric takes the form
\begin{equation}
\Gbar_{MN} \ed x^M \ed x^N = (\mu v)^2 (\eta_{\mu\nu} \ed x^\mu \ed x^\nu + \ed y^2) + \frac{ \ed \vec{r}\,{}^2 + \ed \vec{z}\,{}^2 }{(\mu v)^2}~,
\end{equation}
and a maximally Cartesian-like choice of orthonormal frame:
\begin{align}\label{Cartframe}
& e^{\umu} = (\mu v) \ed x^\mu~, \quad e^{\uy} = (\mu v) \ed y~, \quad e^{\underline{r_i}} = \frac{1}{\mu v} \ed r_i ~, \quad e^{\underline{z_j}} = \frac{1}{\mu v} \ed z_j~,
\end{align}
Here $v$ should be understood as shorthand for $v := \sqrt{ r^2 + z^2}$ and we recall that $\mu$ is the inverse AdS radius.  The equations \eqref{bulkKSeqn} are straightforwardly integrated to yield the solutions $\epsilon \to (\epsilon)_{\rm cart}$, with
\begin{equation}\label{epscart}
(\epsilon)_{\rm cart} = \frac{1}{\sqrt{\mu v}} \left( \frac{r_i \Gamma^{\underline{r_i}} + z_i \Gamma^{\underline{z_i}} }{v} \right) \epsilon_{-}^0 + \sqrt{\mu v} \left[ \epsilon_{+}^0 - \mu (x^\mu \Gamma_{\umu} + y \Gamma_{\uy}) \epsilon_{-}^0 \right]~,
\end{equation}
where $\epsilon_{\pm}^0$ are constant complex Weyl spinors satisfying an additional projection condition:
\begin{equation}\label{eps0conditions}
\Gammabar \epsilon_{\pm}^0 = \pm \epsilon_{\pm}^0  \qquad \& \qquad  i \Gamma^{\underline{012 y}} \epsilon_{\pm}^0 = \pm \epsilon_{\pm}^0 ~.
\end{equation}
We provide some details of the analysis in appendix \ref{app:spinors}.  Each of the $\epsilon_{\pm}^0$ contain eight complex (sixteen real) free parameters, for a total of thirty-two.

The notation $(\epsilon)_{\rm cart}$ is meant to emphasize that these are the components of the Killing spinor with respect to a specific basis of sections (or class of bases) on the Dirac spinor bundle.  The basis is such that the gamma matrices $\{ \Gamma^{\umu}, \Gamma^{\uy}, \Gamma^{\underline{r_i}},\Gamma^{\underline{z_j}} \}$ associated with the frame \eqref{Cartframe} have constant matrix elements.

The second presentation makes use of spherical coordinates $(r,\theta,\phi)$ and the frame
\begin{equation}
e^{\underline{r}} = \frac{1}{\mu v} \ed r~, \qquad e^{\underline{\theta}} = \frac{1}{\mu v} r \ed \theta~, \qquad e^{\underline{\phi}} = \frac{1}{\mu v} r \sin{\theta} \ed \phi~,
\end{equation}
in place of the $e^{\underline{r_i}}$.  The two frames are related by a local rotation, and the components of the Killing spinor with respect to the new frame, which we denote by $(\epsilon)_{S^2}$, are related to the components of the Killing spinor with respect to the old frame, \eqref{epscart}, by the lift of this rotation to the Dirac spinor bundle.  The result is
\begin{equation}\label{bulkKSsol}
(\epsilon)_{S^2} =  \frac{1}{\sqrt{\mu v}} \left( \frac{r \Gamma^{\ur} + z_i \Gamma^{\underline{z_i}} }{v} \right) h_{S^2}(\theta,\phi) \epsilon_{-}^0 + \sqrt{\mu v}\, h_{S^2}(\theta,\phi) \left[ \epsilon_{+}^0 - \mu (x^\mu \Gamma_{\umu} + y \Gamma_{\uy}) \epsilon_{-}^0 \right]~,
\end{equation}
where
\begin{equation}\label{hS2def}
h_{S^2}(\theta,\phi) := \exp\left(\frac{\theta}{2} \Gamma^{\underline{r\theta}} \right) \exp\left(\frac{\phi}{2} \Gamma^{\underline{\theta\phi}} \right)~.
\end{equation}
In this presentation we are implicitly working with respect to a basis where the matrix elements of $\Gamma^{\ur},\Gamma^{\utheta},\Gamma^{\uphi}$ are constant.  $\epsilon_{\pm}^0$ denote the same column vectors of constant entries in both expressions.  Additional details on the relationship between $(\epsilon)_{\rm cart}, (\epsilon)_{S^2}$ and between the gamma matrices $\{ \Gamma^{\underline{r_i}} \}$ and $\{ \Gamma^{\ur},\Gamma^{\utheta},\Gamma^{\uphi} \}$ can be found in appendix \ref{app:spinors}, where we also describe some further transformations that bring the Killing spinors to the form typically found in the literature.

\subsection{Killing spinors on the brane}

The subset of supersymmetries preserved by the D5-brane embedding is generated by those Killing spinors $\epsilon$ that additionally satisfy a $\kappa$-symmetry projection condition \cite{Aganagic:1996pe,Cederwall:1996ri,Bergshoeff:1996tu,Aganagic:1996nn}.  Let 
\begin{equation}
\epsilon = \varepsilon + i \varepsilon'~,
\end{equation}
where $\varepsilon, \vareps'$ are Majorana--Weyl and introduce the doublet $\hat{\vareps} = (\vareps, \vareps')^T$.  Then the condition can be expressed as \cite{Marolf:2003vf,Martucci:2005rb}
\begin{equation}
\Gamma_{\rm D5} \hat{\varepsilon} = \hat{\varepsilon}~,
\end{equation}
where $\Gamma_{\rm D5}$ is given by \eqref{GammaD5}.  If we restrict to the leading-order effective description of the D5-brane, \ie\ the Yang--Mills theory, then it will be sufficient to work with the leading order expression $\Gamma_{\rm D5}^{(0)}$, \eqref{GammaD5leading}.  Since $\Gammabar \varepsilon^{(\prime)} = \vareps^{(\prime)}$, this condition is equivalent to
\begin{equation}\label{kappaconYM}
\vareps'(x^a;x_{0}^m) = \Gamma_{\underline{012r\theta\phi}} \vareps(x^a;x_{0}^m)~.
\end{equation}
Here we also emphasize that the Killing spinors are to be evaluated on the background embedding defined by $x^m = x_{0}^m$.

Let us analyze this condition on the explicit solutions \eqref{bulkKSsol}.  First we extract the Majorana--Weyl components of \eqref{bulkKSsol}.  Let $\vareps_0,\vareps_{0}'$ be constant Majorana--Weyl spinors of positive chirality and $\eta_0,\eta_{0}'$ be constant Majorana--Weyl spinors of negative chirality such that
\begin{equation}
\epsilon_{-}^0 = \eta_0 + i \eta_{0}' ~, \qquad  \epsilon_{+}^0 = \vareps_0 + i \vareps_{0}' + (\mu y_0) \Gamma_{\uy} (\eta_0 + i \eta_{0}')  ~.
\end{equation}
Here $y_0$ is the asymptotic $y$-value of the D5-brane stack; this shift has been included for convenience below.  With these definitions, the second of the conditions \eqref{eps0conditions} is equivalent to
\begin{equation}\label{conMWrel1}
\vareps_{0}' = \Gamma^{\underline{012y}} \vareps_0~, \qquad \eta_{0}' = - \Gamma^{\underline{012y}} \eta_0 ~,
\end{equation}
and the Majorana--Weyl components of \eqref{bulkKSsol} are
\begin{align}\label{bulkKSsolMW}
(\varepsilon)_{S^2} =&~  \frac{1}{\sqrt{\mu v}} \left( \frac{r \Gamma^{\ur} + z_i \Gamma^{\underline{z_i}} }{v} \right) h_{S^2}(\theta,\phi) \eta_0 + \cr
&~ +  \sqrt{\mu v}\, h_{S^2}(\theta,\phi) \left[ \vareps_0 - \mu (x^{\mu} \Gamma_{\umu} + (y-y_0) \Gamma_{\uy}) \eta_0 \right]~,
\end{align}
and $(\varepsilon')_{S^2}$ of the same form with $(\vareps_0,\eta_0)_{S^2} \to (\vareps_{0}', \eta_{0}')_{S^2}$.

Now we impose \eqref{kappaconYM}.  This must hold for all values of $x^a$; in particular, the terms that go as $v^{-3/2}$ and the ones that go as $v^{1/2}$ must match independently.  Matching the $v^{-3/2}$ terms leads to the condition
\begin{equation}
\left( r \Gamma^{\underline{r}} + z_{0,i} \Gamma^{\underline{z_i}} \right) h_{S^2}(\theta,\phi) \eta_{0}' = \Gamma_{\underline{012r\theta\phi}} \left( r \Gamma^{\underline{r}} + z_{0,i} \Gamma^{\underline{z_i}} \right) h_{S^2}(\theta,\phi) \eta_{0}~.
\end{equation}
Since $\Gamma_{\underline{012r\theta\phi}}$ commutes with $\Gamma^{\underline{z_i}}$ but anticommutes with $\Gamma^{\underline{r}}$, this leads to two (additional) conditions on $\eta_0,\eta_{0}'$ that are incompatible with each other when $\vec{z}_0 \neq 0$, unless we set $\eta_0 = 0 = \eta_{0}'$.  However if $\vec{z}_0 = 0$ then we only obtain one additional relation between $\eta_0,\eta_{0}'$, (since  $\Gamma_{\underline{012r\theta\phi}}$ commutes with $h_{S^2}$):
\begin{equation}\label{eta0con2}
\eta_{0},\eta_{0}' = 0 ~~ \textrm{if} ~~ \vec{z}_0 \neq 0~, \qquad  \eta_{0}' = - \Gamma_{\underline{012r\theta\phi}} \eta_0 ~~ \textrm{if} ~~ \vec{z}_0 = 0~.
\end{equation}
Matching the $v^{1/2}$ terms, meanwhile, leads to the requirement
\begin{equation}
\vareps_{0}'  - (\mu x^{\mu}) \Gamma_{\umu} \eta_{0}' = \Gamma_{\underline{012r\theta\phi}} \left( \vareps_{0}  - (\mu x^\mu) \Gamma_{\umu} \eta_{0} \right)~.
\end{equation}
The $x^\mu \Gamma_{\umu}$ terms cancel because of \eqref{eta0con2}.  Thus equality can be achieved by taking
\begin{equation}\label{vareps0con2}
\vareps_{0}' = \Gamma_{\underline{012r\theta\phi}} \vareps_0~.
\end{equation}
The conditions \eqref{eta0con2} and \eqref{vareps0con2} are compatible with \eqref{conMWrel1}.  Together they impose the following projection conditions on the Majorana--Weyl spinors $\eta_0,\vareps_0$:
\begin{equation}\label{conMWproj}
\vareps_0 = \Gamma^{\underline{r\theta\phi y}} \vareps_0 ~, \qquad \eta_0 = \Gamma^{\underline{r\theta\phi y}} \eta_0 ~.
\end{equation}

In summary, when $\vec{z}_0 = 0$ the D5-brane embedding preserves sixteen supersymmetries, parameterized by the Majorana--Weyl spinors $\vareps_0,\eta_0$ of positive and negative chirality, respectively, and additionally satisfying the projections \eqref{conMWproj}.  When $\vec{z}_0 \neq 0$, the $\eta_0$ must be set to zero and the embedding preserves eight supersymmetries.  In the following we will use the Majorana--Weyl spinor $\vareps(x^a;x_{0}^m)$ as our generator of supersymmetry transformations in the Yang--Mills theory.  ($\vareps'$ is determined in terms of it.)  We will simply write $\vareps$ henceforth, with the understanding that we are alway evaluating at $x^m = x_{0}^m$.  With respect to a frame in which $\{ \Gamma^{\umu}, \Gamma^{\ur},\Gamma^{\utheta},\Gamma^{\uphi} \}$ are constant, it has components
\begin{equation}\label{braneKS}
(\vareps)_{S^2} = \left\{ \begin{array}{l l} (\mu r)^{-1/2} \Gamma^{\ur} h_{S^2}(\theta,\phi) \eta_0 + (\mu r)^{1/2} h_{S^2}(\theta,\phi) \left[ \vareps_0 - (\mu x^\mu) \Gamma_{\umu} \eta_0 \right]~, & \quad \vec{z}_0 = 0~, \\[1ex] \sqrt{\mu} (r^2 + z_{0}^2)^{1/4} h_{S^2}(\theta,\phi) \vareps_0~, & \quad \vec{z}_0 \neq 0~. \end{array} \right.
\end{equation}
(Note that $v \to r$ when $\vec{z}_0 \to 0$.)  The same analysis can be carried out in the Cartesian-like frame, where $\{ \Gamma^{\underline{\mu}}, \Gamma^{\underline{r_i}} \}$ are constant and the bulk Killing spinor has the form \eqref{epscart}.  The analogous result is
\begin{equation}\label{braneKScart}
(\vareps)_{\rm cart} =  \left\{ \begin{array}{l l} (\mu r)^{-1/2} \, \hat{r}_i \Gamma^{\underline{r_i}} \eta_0 + (\mu r)^{1/2} \left[ \vareps_0 - (\mu x^\mu) \Gamma_{\umu} \eta_0 \right]~, & \quad \vec{z}_0 = 0~, \\[1ex] \sqrt{\mu} (r^2 + z_{0}^2)^{1/4} \vareps_0~, & \quad \vec{z}_0 \neq 0~, \end{array} \right.
\end{equation}
where we used the shorthand $\hat{r}_i = r_i/r$.  The conditions on $\vareps_0,\eta_0$ are the same; in particular the projection conditions \eqref{conMWproj} can equivalently be written as 
\begin{equation}\label{conMWproj2}
\vareps_0 = \Gamma^{\underline{r_1 r_2 r_3 y}} \vareps_0 ~, \qquad \eta_0 = \Gamma^{\underline{r_1 r_2 r_3 y}} \eta_0 ~.
\end{equation}

Finally, we would like to derive a `Killing spinor equation' for $\vareps$ alone, the solutions of which can be equivalently represented by \eqref{braneKS} or \eqref{braneKScart}.  This can be done by looking at the real and imaginary parts of \eqref{bulkKSeqn}, restricting to $x^m = x_{0}^m$, and imposing \eqref{kappaconYM}.  We relegate the details to appendix \ref{app:bkseqn} and state the final result here:
\begin{equation}\label{braneKSeqn}
\left[ D_a + \frac{M_{\Psi}}{2} \Gamma^{\underline{\theta\phi y}} \Gamma_a \right] \vareps = 0~,
\end{equation}
where $D_a = \pd_a + \frac{1}{4} \omegabar_{bc,a} \Gamma^{bc}$ is the spinor covariant derivative with respect to the 6D metric (but still utilizing the 10D gamma matrices).  

\subsection{Supersymmetry of the worldvolume theory}\label{sec:ftsusy}

We now turn the supersymmetry of the worldvolume theory. Recall that the six-dimensional Yang--Mills action is the sum of \eqref{SYMb} and \eqref{SYMf} and can be written 
\begin{equation}\label{SYM}
S_{\rm ym} := \frac{1}{g_{{\rm ym}_6}^2} \int \ed^6 x \sqrt{-g_6} \, \LL + S^{\rm bndry} ~,
\end{equation}
where the `bulk' Lagrangian density is
\begin{align}\label{bulkL}
\mathcal{L} :=&~ -  \Tr \bigg\{ \frac{1}{4} F_{ab} F^{ab} + \half D_a \Phi_{\um} D^a \Phi^{\um} + \frac{i}{2} \Psibar \left( \Gamma^a D_a + M_\Psi \Gamma^{\underline{\theta\phi y}} \right) \Psi  + 2 M_\Psi \epsilon^{\alpha\beta} F_{\alpha\beta} \Phi^{\uy} +  \cr
&~ \qquad \quad +  \half M_{y}^2 (\Phi^{\uy})^2 + \half M_{z}^2 (\vec{\Phi}^{\uz})^2  + \frac{i}{2} \Psibar \Gamma^{\um} [\Phi_{\um}, \Psi] + \frac{1}{4} [\Phi^{\um},\Phi^{\un}] [\Phi_{\um},\Phi_{\un}] \bigg\}~, \qquad \qquad \raisetag{20pt}
\end{align}
where the background metric is given by \eqref{6Dmet}, and the ($r$-dependent) masses by \eqref{SYMmasses}.  The boundary action, which we discuss in the next subsection, is the sum of \eqref{Sbbndry} and \eqref{Sfbndry}:
\begin{align}\label{Sbndry}
S^{\rm bndry}  :=&~ \frac{1}{g_{{\rm ym}_6}^2} \int_{~~\,\mathclap{\pd M_6}} \ed^5 x \sqrt{-g_{(\pd)} }\Tr \left\{ \frac{M_{\Psi}}{2} \left( (\Phi^{\uy})^2 - (\vec{\Phi}^{\uz})^2 \right) + \half \Phi^{\uy} \epsilon^{\alpha\beta} F_{\alpha\beta} - \frac{i}{4} \Psibar \Gamma^{\underline{\theta\phi y}} \Psi  \right\}~.
\end{align}
Here we focus on supersymmetry invariance of the bulk action modulo boundary terms. Note again that while we derived this action from a D-brane system where the relevant Lie algebra is $\mathfrak{su}(N_f)$, it describes super Yang-Mills theory on $AdS_4\times S^2$ for \emph{any} simple Lie group.\footnote{with the appropriately defined $\Tr$.  See the comments in the first paragraph of section \ref{sec:Myers}.}

Motivated by the form of the fermion mass term, the Killing spinor equation \eqref{braneKSeqn}, and the philosophy espoused in \cite{Blau:2000xg},  we make the following ansatz for the supersymmetry variations of the fields:
\begin{align}\label{SUSY}
& \delta_{\varepsilon} A_a = -i \, \varepsbar \Gamma_a \Psi ~, \qquad  \delta_{\varepsilon} \Phi_{\um} = -i \, \varepsbar \Gamma_{\um} \Psi~, \cr
& \delta_{\vareps} \Psi = \left[ \half F_{ab} \Gamma^{ab} + D_a \Phi_{\um} \Gamma^{a\um} + \half [\Phi_{\um},\Phi_{\un}]\Gamma^{\um\un} + \alpha \Gamma^{\underline{\theta\phi y}} \Gamma_{\um} \Phi^{\um} \right] \vareps~,
\end{align}
where $\alpha$ is a parameter---possibly a function of $r$---to be determined.

Standard manipulations, \emph{without} making use of the Killing spinor equation, lead to
\begin{align}\label{LLvar}
i \delta_{\vareps} \LL =&~ \nabla_a \BB^a + \cr
&~ + \overline{D_a\vareps} \Tr \bigg\{ \left( \half F_{bc} \Gamma^{bc} + (D_b \Phi_{\um}) \Gamma^{b\um} + \half [\Phi_{\um},\Phi_{\un}] \Gamma^{\um\un} + \alpha \Phi^{\um} \Gamma_{\um} \Gamma^{\underline{\theta\phi y}} \right) \Gamma^a \Psi \bigg\} + \cr
&~ + \varepsbar \Tr \bigg\{ - \half M_\Psi F_{ab} \Gamma^{ab} \Gamma^{\underline{\theta\phi y}} \Psi - 2 M_\Psi \epsilon^{\alpha\beta} F_{\alpha\beta} \Gamma^{\uy} \Psi + \cr
&~ \qquad \quad + (D_a \Phi_{\um}) \left( \alpha \Gamma^{\um} \Gamma^{\underline{\theta\phi y}} \Gamma^a - M_\Psi \Gamma^{a\um} \Gamma^{\underline{\theta\phi y}} \right) \Psi + 4 M_\Psi (D_\alpha \Phi^{\uy}) \epsilon^{\alpha\beta} \Gamma_\beta \Psi + \cr
&~ \qquad \quad - [\Phi_{\um},\Phi_{\un}] \left( \half M_\Psi \Gamma^{\um\un} \Gamma^{\underline{\theta\phi y}} + \alpha \Gamma^{\um} \Gamma^{\underline{\theta\phi y}} \Gamma^{\un} \right) \Psi +  \cr
&~ \qquad \quad +  \Phi_{\um} \Gamma^{\um} \left[ - \left( \pd_r \alpha + \frac{2 z_{0}^2}{r (r^2 + z_{0}^2)} \alpha \right) \Gamma^r \Gamma^{\underline{\theta\phi  y}}  +  \alpha M_\Psi - M_{m}^2 \right] \Psi \bigg\} ~. \qquad
\end{align}
The terms appearing here are naturally divided into three sets.  First there are the total derivative terms of the first line; the \emph{boundary current} $\BB^a$ is
\begin{align}\label{Bcurrent}
\BB^a :=&~  \varepsbar \Tr \bigg\{ -\half F_{bc} \Gamma^{bca} \Psi + (D_b \Phi_{\um}) \Gamma^{\um} \Gamma^{ba} \Psi + \cr
&~  \qquad \quad - \half [\Phi_{\um},\Phi_{\un}] \Gamma^{\um\un} \Gamma^a \Psi - \alpha \Phi^{\um} \Gamma_{\um} \Gamma^{\underline{\theta\phi y}} \Gamma^a \Psi \bigg\} + \half \Tr \left\{ \Psibar \Gamma^a \delta_{\vareps} \Psi\right\}  ~, \qquad
\end{align}
The contribution of these terms to $\delta_{\vareps} S_{\rm ym}$ will have to be canceled by the variation of the boundary action; we analyze this in subsection \ref{sec:bndrySUSY} below.  Then there are the terms proportional to $\overline{D_a \vareps}$ in the second line, and those proportional to $\varepsbar$ in the remaining lines.\footnote{The appearance of $\pd_r \alpha$ term and especially the term that accompanies it inside the round brackets in the last line of \eqref{LLvar} is somewhat subtle.  It arises in the process of moving the $D_a$ in $-\alpha \varepsbar \Tr\left(\Phi^{\um} \Gamma_{\um} \Gamma^{\underline{\theta\phi y}} \Gamma^a D_a \Psi\right)$ off of $\Psi$.  When $z_0 \neq 0$, the covariant derivative does not commute with $\Gamma^{\underline{\theta\phi y}}$ due to nonzero mixed components of the spin connection of the type $\omega_{\alpha r, \beta}$.  The commutator of $D_a$ with $\Gamma^{\underline{\theta\phi y}}$ is what gives rise to the term with the $2 z_{0}^2/(r (r^2 + z_{0}^2))$ factor.}

Some cancellations have already occurred to arrive a \eqref{LLvar}---namely those  that would have occurred in the flat space limit, $\mu r \to 0$.  The remaining terms are present precisely because we are working on a nontrivial background.  They involve either the derivative of the supersymmetry parameter, or the mass-type couplings.  From the D-brane point of view the latter are induced from the non-flat normal bundle to the brane worldvolume and the background RR five-form flux.

The next step is to make use of the Killing spinor equation, or rather its conjugate:
\begin{equation}
\overline{D_a \varepsilon} = \frac{M_\Psi}{2} \varepsbar \Gamma_a \Gamma^{\underline{\theta\phi y}}~.
\end{equation}
Supersymmetry invariance then requires that all of the resulting terms from the second line of \eqref{LLvar} cancel with the terms in the remaining lines.  This must be checked by explicitly working out the coefficients for each of the possible index structures of the worldvolume fields, \eg\ $F_{\mu\nu}$ $F_{\mu r}$, $F_{\mu\alpha}$, $F_{\alpha\beta}$, $\DD_\mu \Phi^{\uy}$, \etc.  One indeed finds that supersymmetry is preserved, modulo boundary terms, provided the following three conditions on $\alpha$, the masses, and the Killing spinor hold:
\begin{align}\label{susycondition}
i \delta_{\vareps} \LL =&~ \nabla_a \BB^a  \cr
\iff \quad & 
\left\{ \begin{array}{l} 0 = M_\Psi - \alpha ~, \\[2ex]
0 =  \varepsbar \left[ \left( \pd_r \alpha + \frac{2 z_{0}^2}{r (r^2 + z_{0}^2)} \alpha \right) \Gamma^r\Gamma^{\underline{\theta\phi  y}}  +  4 \alpha M_\Psi - M_{y}^2\right]~, \\[2ex]
0 = \varepsbar \left[ - \left( \pd_r \alpha + \frac{2 z_{0}^2}{r (r^2 + z_{0}^2)} \alpha \right) \Gamma^r \Gamma^{\underline{\theta\phi  y}}  -2 \alpha M_\Psi - M_{z}^2\right]~.  \end{array} \right.
\end{align}
All three conditions are met by taking
\begin{equation}
\alpha = M_\Psi = \frac{\mu r}{\sqrt{r^2 + z_{0}^2}} ~.
\end{equation}
When $z_0 =0$ this follows directly from the expressions for the masses \eqref{SYMmasses}.  When $z_0 \neq 0$ one can show that with $\alpha = M_{\Psi}$ the latter two equations are proportional to the projector $\half (\mathbbm{1} - \Gamma^{\underline{r\theta\phi y}})$.  But when $z_0 \neq 0$ we must set the $\eta_0$ to zero in $\vareps$, and then $\varepsbar$ is indeed annihilated by this projector acting to the left.  (See \eqref{varepseigen}.)  

\section{Classical vacua, boundary conditions, and asymptotic analysis}\label{boundarystuff}

In this section we describe the (classical) vacuum structure of the Yang--Mills theory and formulate appropriate boundary conditions on the fields.  The latter is based on consistency of the variational principle and a large-$r$ asymptotic analysis of field modes with an eye towards holographic applications.

It will be convenient to start with the action  \eqref{SYMb1} given in terms of the $\Phi^m$ (rather than the $\Phi^{\um}$ we have been using so far).  We have
\begin{equation}\label{SYM2}
S_{\rm ym} = \frac{1}{g_{{\rm ym}_6}^2} \int \ed^6 x \sqrt{-g_6} \LL' + S_{f}^{\rm bndry}~,
\end{equation}
with
\begin{align}\label{primeL}
\LL' =&~ - \Tr \bigg\{ \frac{1}{4} F_{ab} F^{ab} + \half \Gbar_{mn} D_a \Phi^m D^a \Phi^m + \frac{1}{4} \Gbar_{mk} \Gbar_{nl} [\Phi^m, \Phi^n] [\Phi^k, \Phi^l]  +  \cr
&~  \qquad \qquad - \half \tilde{\epsilon}^{r_i r_j r_k} \mu^4 (r^2 + z_{0}^2)^2 (D_{r_i} \Phi^y) F_{r_j r_k} + \cr
&~  \qquad \qquad  + \frac{i}{2} \Psibar \left( \Gamma^a D_a + M_{\Psi} \Gamma^{\underline{\theta\phi y}} \right) \Psi +  \frac{i}{2} \Psibar \Gamma_m [\Phi^m, \Psi] \bigg\}~. 
\end{align}
Only the fermion has an explicit boundary action, \eqref{Sfbndry}.  

A first variation of \eqref{SYM2} leads to the equations of motion,
\begin{align}\label{eoms}
0 =&~ D_a F^{ab} - \Gbar_{mn} [\Phi^m, D^b \Phi^n] + \frac{i}{2} [\Psibar,\Gamma^b \Psi] + \tilde{\epsilon}^{r\alpha\beta} \mu^4 \left[\pd_r (r^2 + z_{0}^2)^2\right] (D_\alpha \Phi^y) \delta_{\beta}^{\phantom{\beta}b} ~, \cr
0 =&~ D^a (\Gbar_{yy} D_a \Phi^y) + \Gbar_{mn} \Gbar_{yy} [\Phi^m, [\Phi^n,\Phi^y]] + \frac{i}{2} [\Psibar, \Gamma_y \Psi] - \half \tilde{\epsilon}^{r\alpha\beta} \mu^4 \left[\pd_r (r^2 + z_{0}^2)^2\right] F_{\alpha\beta} ~, \cr
0 =&~ D^a ( \Gbar_{z_i z_j} D_a \Phi^{z_j}) + \Gbar_{mn} \Gbar_{z_i z_j} [\Phi^m, [\Phi^n, \Phi^{z_j}]] + \frac{i}{2} [\Psibar, \Gamma_{z_i} \Psi] ~, \cr
0 =&~ \left( \slashed{D} + M_\Psi \Gamma^{\underline{\theta\phi y}} \right) \Psi + \Gamma_m [\Phi^m, \Psi ]~.
\end{align}
On a solution to these equations the variation reduces to a set of boundary terms:
\begin{align}\label{Svaronshell}
(\delta S_{\rm ym})^{\textrm{o-s}} =&~  -\lim_{r\to\infty} \frac{1}{\mu^2g_{{\rm ym}_6}^2} \int \ed^3x\ed\Omega \Tr \bigg\{ \PP^{\mu} \delta A_\mu + \PP^\beta \delta A_\beta  + \cr
&~ \qquad \qquad \qquad \qquad \qquad \qquad \quad  + \delta_{ij} \PP^{z_i} \delta \Phi^{z_i} +  \PP^{y} \delta \Phi^{y} + \PP^{\Psi} \delta \Psi \bigg\}~, \qquad
\end{align} 
where
\begin{align}\label{boundaryPs}
\PP^\nu :=&~ \mu^2 r^2 F_{r\mu} \eta^{\mu\nu} ~, \qquad \PP^\beta := \mu^6 (r^2 + z_{0}^2)^2 \left[ F_{r\alpha} \tilde{g}^{\alpha\beta} + (D_\alpha \Phi^y) \tilde{\epsilon}^{\alpha\beta} \right]~, \cr
\PP^{z_i} :=&~ \mu^2 r^2 (D_r \Phi^{z_i})~, \qquad  \PP^y := \mu^6 (r^2 + z_{0}^2)^2 \left[ r^2 D_r \Phi^y - \half \tilde{\epsilon}^{\alpha\beta} F_{\alpha\beta} \right] ~, \cr
\PP^{\Psi} :=&~ -\frac{i\mu^3}{2} r^2 (r^2 + z_{0}^2)^{1/2} \Psibar \left( \Gamma^{\ur} + \Gamma^{\underline{\theta\phi y}} \right)~.
\end{align}
Here we have made all powers of $r$ explicit.  In particular, $\tilde{g}^{\alpha\beta}$ and $\tilde{\epsilon}^{\alpha\beta}$ are the inverse metric and Levi--Civita tensor on the round $S^2$ of unit radius.  We also note that there is no $\delta A_r$ term and that the $\Psi$-term receives contributions from the variation of both bulk and boundary actions.  

The on-shell variation, \eqref{Svaronshell}, must vanish in order to ensure a consistent variational principle. This, in turn, restricts the asymptotic behavior of bulk field configurations.  In taking this approach we are excluding the additional boundary action that arises in the context of holographic renormalization.  Holographic renormalization is a procedure that introduces a cutoff surface at large $r$ and determines a set of boundary counterterms that are to be added to the action \cite{Henningson:1998gx,Balasubramanian:1999re,Emparan:1999pm,deHaro:2000vlm,Skenderis:2002wp}.  Originally these terms were determined from the condition of having a finite on-shell action, but later it was understood that the same terms are required to render the variational principle well-defined when one allows for field configurations with divergent behavior, (i.e. non-normalizable modes), as $r \to \infty$ \cite{Papadimitriou:2005ii,Papadimitriou:2010as}.  This procedure corresponds to the standard renormalization of UV divergences in the holographic dual, and it is appropriate for constructing the generator of correlation functions.  In this paper, however, we are interested in classical finite-energy BPS field configurations of the 6D Yang--Mills theory which, roughly speaking, should be the appropriate leading-order description of BPS states in the holographic dual in the limit $N_c \gg g_s N_c \gg 1$.  In this case, notions of finite energy, consistency of the variational principle, and the like should descend from the classical Yang--Mills action, without the cut-off boundary and associated boundary terms.    

Before examining this in detail it is useful to first consider the vacuum structure and perturbative spectrum of the theory.  

\subsection{Classical (flux) vacua}\label{sec:fluxvacua}

The classical vacua are the absolute minima of the energy functional.  We construct the Yang--Mills energy functional, or Hamiltonian, by performing a Legendre transform of the Lagrangian in \eqref{SYM2} with respect to the natural time coordinate $t = x^0$ of the Minkowski foliation of AdS.  Let $x^p$, $p=1,2$, denote the spatial coordinates in $x^\mu$ so that $x^a = (t,x^p,r_i)$, and let $(E_p,E_{r_i}) = (F_{p0},F_{r_i0})$ be the components of the non-abelian electric field.  Then the bosonic part of the Yang--Mills Hamiltonian takes the form
\begin{align}\label{Hbos1}
H_{\rm ym}^{\rm bos} =&~ \frac{1}{g_{{\rm ym}_6}^2} \int \ed^5 x \sqrt{-g_6} \left( \KK + \VV -  g^{00} \Tr \left\{E^p D_p A_0 + E^{r_i} D_{r_i} A_0 + \Gbar_{mn} D_0 \Phi^m [A_0,\Phi^n] \right\} \right)~,
\end{align}
where the kinetic and potential energy densities are
\begin{align}
\KK :=&~ - \frac{1}{2} g^{00} \Tr \left\{ E^p E_p + E^{r_i} E_{r_i} + \Gbar_{mn} D_0 \Phi^m D_0 \Phi^n \right\}~, \label{kinenergy} \\
\VV :=&~ \Tr \bigg\{ \frac{1}{4} F^{r_i r_j} F_{r_i r_j} + \half (F^{12} F_{12} + F^{p r_i} F_{p r_i}) +  \half \Gbar_{mn} (D_p \Phi^m D^p \Phi^n + D_{r_i} \Phi^m D^{r_i} \Phi^n)  + \qquad \cr
&~ \qquad + \frac{1}{4} \Gbar_{mn} \Gbar_{m'n'} [\Phi^m, \Phi^{m'}] [\Phi^{n}, \Phi^{n'}] - \half \mu^4 (r^2 + z_{0}^2)^2 \tilde{\epsilon}^{r_i r_j r_k} (D_{r_i} \Phi^y) F_{r_j r_k}  \bigg\}~. \label{potenergy} \raisetag{20pt}
\end{align}

Together, the last terms of \eqref{Hbos1}, proportional to $g^{00}$, are a total derivative when we restrict the space of field configurations to the constraint surface defined by the Gauss Law,
\begin{equation}\label{localGauss}
D^{r_i} (g^{00} E_{r_i}) + g^{00}\left( D^p E_p - \Gbar_{mn} [\Phi^m, D_0 \Phi^{n}] \right) = 0~.
\end{equation}
We refer to this as the local Gauss Law constraint.  It is equivalent to the $A_0$ equation of motion in \eqref{eoms} when the fermi field is set to zero.  There is also a boundary Gauss Law constraint that comes from demanding that the total derivative term vanishes:
\begin{equation}\label{globalGauss}
\int \ed^5 x \left[ \pd_r \left( \sqrt{-g_6} \Tr\{E^{r} A^0\} \right) \right] = 0~.
\end{equation}
We will return to this condition after we have analyzed the field asymptotics, but for now we simply assume it is satisfied.  It will merely amount to imposing appropriate fall-off conditions on $A_0$.  

Restricting to field configurations that satisfy both of these constraints, our energy functional for the bosonic fields is
\begin{equation}\label{HKV}
H_{\rm ym}^{\rm bos} = \frac{1}{g_{{\rm ym}_6}^2} \int \ed^5 x \sqrt{-g_6}\left( \KK + \VV\right)~.
\end{equation}

On general grounds, supersymmetry implies that this functional is positive semi-definite.  The presence of the last term in the potential \eqref{potenergy}, however, makes this property slightly non-obvious.  Recall that this term originates from the Chern--Simons part of the D5-brane action and is present because of the RR-flux of the string background.  Positivity is established by noting that it can be combined with two other terms to make a complete square:
\begin{align}\label{monopoleVacua}
& \frac{1}{4} F^{r_i r_j} F_{r_i r_j} + \half \Gbar_{mn} D_{r_i} \Phi^y D^{r_i} \Phi^y - \half \mu^4 (r^2 + z_{0}^2)^2 \tilde{\epsilon}^{r_i r_j r_k} (D_{r_i} \Phi^y) F_{r_j r_k} \cr
& \qquad = \frac{1}{4} \mu^4 (r^2 + z_{0}^2)^2 \left( F_{r_i r_j} - \tilde{\epsilon}_{r_i r_j r_k} D_{r_k} \Phi^y \right)^2~.
\end{align}
Here, repeated downstairs indices are contracted with a flat Euclidean metric $\delta^{r_i r_j}$, and we have taken advantage of the fact that $g^{r_i r_j} = \mu^2 (r^2 + z_{0}^2) \delta^{r_i r_j} = \Gbar_{yy} \delta^{r_i r_j}$.  With this observation it is then manifest that $\KK + \VV$ is a positive sum of squares.

Hence the space of classical vacua of the Yang--Mills theory is the space of gauge-inequivalent zero energy configurations: 
\begin{equation}\label{Mvac}
M_{\rm vac} := \{ [(A_a,\Phi^m)] ~|~ \KK = \VV =0 \}~.
\end{equation}
This is an extremely interesting space, mostly because of the observation \eqref{monopoleVacua}.  One of its components is the usual sort of field theory Coulomb branch that we expect to have in a supersymmetric Yang--Mills theory---mutually commuting, constant vevs for the Higgs fields, $\Phi_{\infty}^m$.  Since the vevs are mutually commuting they can be simultaneously diagonalized to a Cartan subalgebra $\mathfrak{t} \subset \mathfrak{g}$.  Residual gauge transformations act by Weyl conjugation.  Hence this component of the vacuum has the form
\begin{equation}\label{Cbranch}
M_{{\rm vac}}^0 = \mathfrak{t}^{\otimes 4}/W~,
\end{equation}
where $W$ is the Weyl group.

However, the vacuum space has an infinite number of additional components associated with nontrivial zeros of \eqref{monopoleVacua}.  These consist of field configurations $(A_{r_i}, \Phi^y)$ that solve
\begin{equation}\label{monopoleVacua2}
({\rm B}): \qquad F_{r_i r_j} - \tilde{\epsilon}_{r_i r_j r_k} D_{r_k} \Phi^y = 0~,
\end{equation}
together with $(A_0, A_p, \Phi^{z_i})$ that ensure the remaining terms in $H_{\rm ym}^{\rm bos}$ vanish.  The condition \eqref{monopoleVacua2} is none other than the Bogomolny equation for Euclidean monopoles on the $\mathbb{R}^3$ parameterized by $r_i$!

It might sound strange that monopole moduli spaces are part of the \emph{vacuum} manifold of the theory.  Ordinarily, they parameterize local, but not global, minima of the energy functional, which are associated with soliton masses.  The reason the story is different here is again due to the `extra' term in the action arising from the background RR flux.  Normally, one has to complete the square by hand, adding and subtracting such a term.  Also, this cross term is usually a total derivative, so the term that is added in this process is topological and provides the mass of the soliton.  Here, this is not the case.  The cross term is dynamical, and already present in the action from the beginning.

Monopole moduli spaces are defined as spaces of gauge-inequivalent solutions to \eqref{monopoleVacua2}, and therefore they play an essential role in defining the vacuum manifold of \eqref{HKV}.  The data that goes into specifying an ordinary monopole moduli space are the asymptotic Higgs vev, $\Phi_{\infty}^y$, and a magnetic charge $P$.  They determine asymptotic boundary conditions on the solutions such that
\begin{equation}\label{monoasbc}
({\rm bc}_{\infty}): \qquad \Phi^y = \Phi_{\infty}^y - \frac{P}{2r} + \cdots ~, \qquad F = \frac{P}{2} \omega_{S^2} + \cdots ~, \qquad r \to \infty~,
\end{equation}
where $\omega_{S^2} = \sin{\theta} \ed\theta \ed\phi$ is the standard volume form on the two-sphere and the ellipses represent subleading terms.

For reasons to be explained shortly, we should actually consider a more general notion of monopole moduli space that allows for magnetic singularities at specified points $\vec{r} = \vec{v}_{\sigma}$, corresponding to the insertion of 't Hooft defects \cite{'tHooft:1977hy}.  These singularities are defined by imposing boundary conditions on the fields as $\vec{r} \to \vec{v}_{\sigma}$ of the form \cite{Kapustin:2005py,Moore:2014jfa}
\begin{equation}\label{monotHooftbc}
({\rm bc}_{\sigma}): \qquad \Phi^y = - \frac{{\rm P}_{\sigma}}{2 |\vec{r} - \vec{v}_\sigma|} + \cdots ~, \qquad F = \frac{{\rm P}_\sigma}{2} \omega_{S^2}^{(\sigma)} + \cdots ~, \qquad |\vec{r} - \vec{v}_\sigma| \to 0~,
\end{equation}
where $\omega_{S^2}^{(\sigma)}$ is the standard volume form on a two-sphere centered on $\vec{v}_\sigma$ and ${\rm P}_\sigma$ is the 't Hooft charge of the defect.  The associated \emph{moduli space of singular monopoles} is defined as the space of gauge-inequivalent solutions to \eqref{monopoleVacua2} obeying the asymptotic boundary conditions \eqref{monoasbc} as $r \to \infty$ and the 't Hooft defect boundary conditions \eqref{monotHooftbc} as $\vec{r} \to \vec{v}_\sigma$:
\begin{equation}
\MM( \{{\rm P}_\sigma,\vec{v}_\sigma\}; P, \Phi_{\infty}^y) := \left\{ (A_{r_i}, \Phi_{\infty}^y)~ |~ ({\rm B}) ~\&~ ({\rm bc}_{\infty}) ~\&~ ({\rm bc}_\sigma) \right\} \bigg/ \GG_{ \{{\rm P}_\sigma\} }^0 ~.
\end{equation}
Here $\GG_{ \{ {\rm P}_\sigma \}}^0$ is the group of gauge transformations that approach the identity at infinity and leave the 't Hooft charges invariant.  These spaces have been studied intensely since the initial work of Kronheimer \cite{Kronheimer}, with important contributions in \cite{MR1624279,Cherkis:1997aa,Kapustin:2006pk}, to name a few.  See \cite{Moore:2014jfa} and references therein for a more complete discussion.

Note that the 't Hooft and asymptotic charges are quantized.  Factoring out the center of mass $U(1)$ results in the $\mathfrak{su}(N_f)$ charges taking values in the co-character lattice of $U(N_f)/U(1) \cong PSU(N_f)$.  This lattice consists of integer linear combinations of the fundamental magnetic weights.  For $\mathfrak{su}(2)$ the fundamental magnetic weight is half of the simple co-root. 

A detailed description of how the space of vacua, $M_{\rm vac}$, is defined in terms of the moduli spaces $\MM(\{{\rm P}_\sigma,\vec{v}_\sigma\},P,\Phi_{\infty}^y)$ is beyond the scope of the present paper.  We will limit ourselves, instead, to using the intersecting D3/D5 system to indicate what the various data defining $M_{\rm vac}$ correspond to in terms of branes.  This will also lead to a natural description of the corresponding vacua in the holographic dual, though we leave a detailed matching to future work.

\begin{figure}
\begin{center}
\includegraphics[scale=0.4]{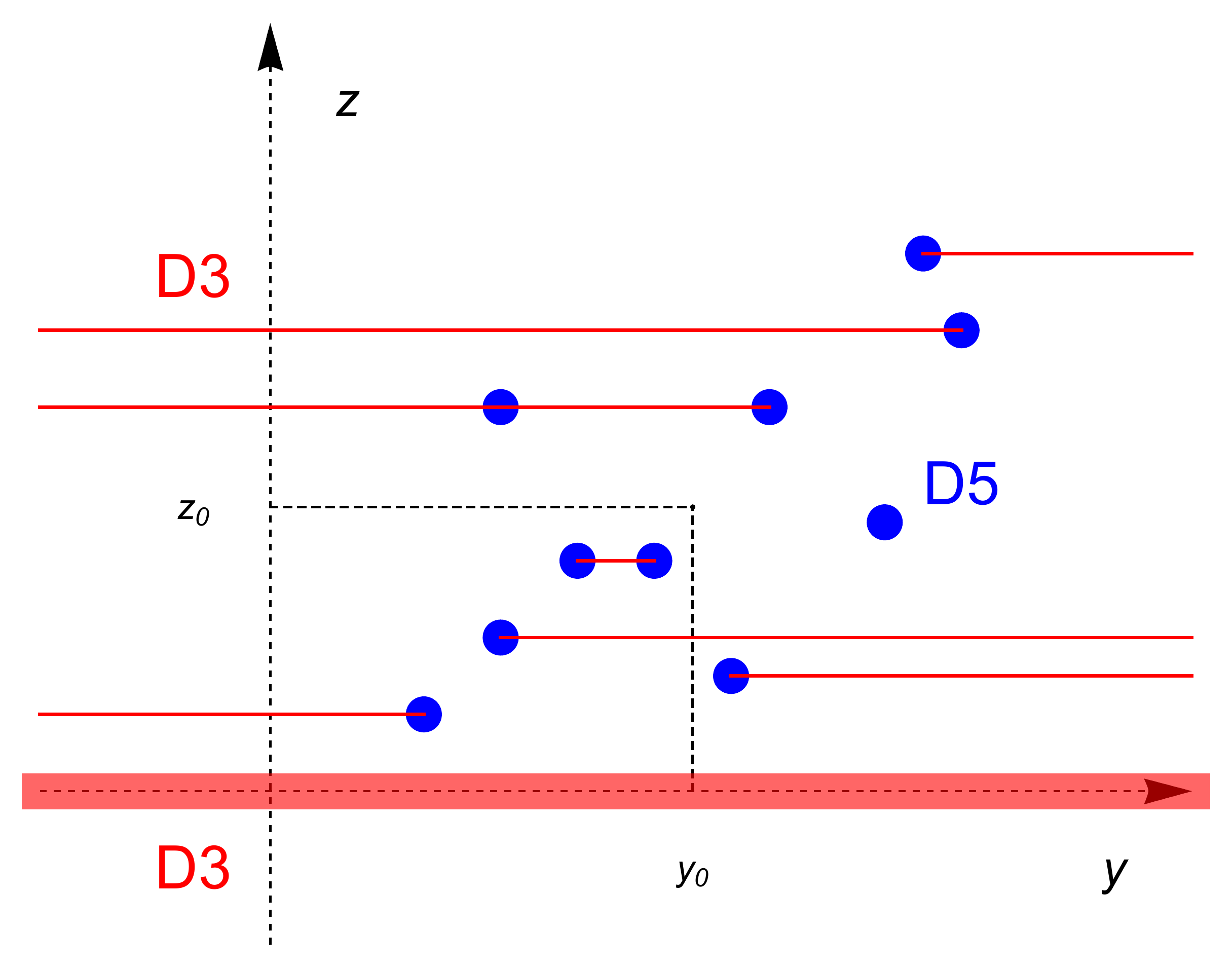}
\caption{The brane configuration corresponding to a generic point of the vacuum $M_{\rm vac}$.  The thick red line along the $y$-axis represents the original color D3-branes.  The blue dots are the D5-branes.  Their relative separations in the $y$ and $z_i$ directions are dictated by the vevs $\Phi_{\infty}^y,\Phi_{\infty}^{z_i}$.  Additional finite and/or semi-infinite D3-branes can begin and end on the D5-branes.  The precise numbers of these will be controlled by the asymptotic and 't Hooft charges.}
\label{fig4}
\end{center}
\end{figure}

A generic vacuum configuration is depicted in Figure \ref{fig4}.  The vevs $(\Phi_{\infty}^{z_i},\Phi_{\infty}^y)$ characterize the relative separation of the D5-branes from each other in the $z_i$ and $y$ directions respectively.\footnote{Giving a precise transcription from the vevs to the separations requires a little more Lie algebra notation than we need in the rest of this paper, so we refer the reader to \cite{Moore:2014gua} for details.}  The center-of-mass position of the D5-branes is instead parameterized by $(y_0,\vec{z}_0)$.  We are free to set $y_0 = 0$ since translations in $y$ are an isometry of the background.

It is well known that finite-length D3-branes stretched between D5-branes appear as smooth monopoles in the worldvolume theory of the D5-branes \cite{Diaconescu:1996rk}, while semi-infinite D3-branes ending on D5-branes appear as 't Hooft defects \cite{Hanany:1996ie,Cherkis:1997aa,Moore:2014gua}.  Thus, nontrivial configurations $(A_{r_i}, \Phi^y)$ carrying asymptotic magnetic and 't Hooft charges, correspond to additional finite-length and semi-infinite D3-branes, stretching between and ending on the D5-branes.  The additional D3-branes should run parallel to the color D3-branes so as to preserve the same supersymmetries as the vacua without flux, \eqref{Cbranch}.

To gain intuition for the properties of these vacua, let us consider some special cases.  Suppose first that all of the vevs $\Phi_{\infty}^{z_i}$ are vanishing, so the D5-branes are separated from each other in the $y$ direction only.  A detailed description of this moduli space $\MM( \{{\rm P}_\sigma,\vec{v}_\sigma\}; P, \Phi_{\infty}^y)$ in terms of brane configurations was given in \cite{Moore:2014gua}.  This includes an accounting for the dimension of the moduli space, which was derived in \cite{Moore:2014jfa}, in terms of mobile D3-brane segments that can slide along parallel D5-branes.\footnote{The description used in \cite{Moore:2014gua} employed a D1/D3 system that is T-dual to the D3/D5 system here.  The relationship between monopoles and brane configurations is identical for the two.}  It also includes formulae for the 't Hooft charges in terms of the numbers of semi-infinite D3-branes ending on each D5-brane, and for the decomposition of these charges into $\mathfrak{u}(1)$ and $\mathfrak{su}(N_f)$ components.   

Second, consider the case that $\Phi_{\infty}^y$ is vanishing, meaning that the D5-branes are separated from each other in the $z_i$ directions only.  Then these brane configurations are those studied in \cite{Gaiotto:2008sa}.  The authors of \cite{Gaiotto:2008sa} focused on the description of the vacua from the point of view of the D3-branes, with the  goal of classifying all half-BPS boundary conditions for $\NN = 4$ SYM on the half-space.  Note that $\NN = 4$ SYM on the half space can be used to describe $\NN =4$ SYM with a defect at $y_0 = 0$ using the folding trick, as they discussed.

From the perspective of the D3-brane theory, semi-infinite D3-branes ending on a stack of D5-branes at $y_0 =0$ are described by solutions to Nahm's equation with a pole at $y_0 = 0$.  The six adjoint-valued scalars of the $\NN = 4$ SYM are divided into two triplets, $\vec{R}_{\rm D3}$ and $\vec{Z}_{\rm D3}$, which encode transverse fluctuations in the $r_i$ and $z_i$ directions respectively.\footnote{These scalars were denoted $\vec{X}$ and $\vec{Y}$ respectively in \cite{Gaiotto:2008sa}.}  The $\vec{R}_{\rm D3}$ solve Nahm's equation on the semi-infinite interval $y \in \phantom{[}(-\infty,y_0]\phantom{)}$ with the pole at $y_0 =0$ specified in terms of an $\mathfrak{su}(2)$ representation.  The components of the 't Hooft charge, which dictate the number of D3-branes ending on each D5-brane, are encoded in the dimensions of the irreducible components of the $\mathfrak{su}(2)$ representation.  The scalars $\vec{Z}_{\rm D3}$ are required to take constant values that commute with the $\vec{R}_{\rm D3}$, and will be related to the vevs $\Phi_{\infty}^{z_i}$.  Finally the position of the 't Hooft charges, $\vec{v}_\sigma$, will be related to the asymptotic values of the $\vec{R}_{\rm D3}$ as $y \to -\infty$.

Then, based on \cite{Moore:2014gua,Gaiotto:2008sa}, and the known relation between singular monopoles and solutions to Nahm's equations on a semi-infinite interval \cite{Cherkis:1997aa}, we expect that the generic configuration depicted in Figure \ref{fig4} corresponds to a solution to the $(\vec{R}_{\rm D3},\vec{Z}_{\rm D3})$-system analogous to those described in \cite{Gaiotto:2008sa}, but with multiple parallel defects at different values of $y$, as dictated by the vev $\Phi_{\infty}^y$.  The solution will involve a solution to Nahm's equation on a union of semi-infinite and finite intervals, with appropriate boundary conditions at each of the defects.  This will provide the holographically dual description of the space of vacua, \eqref{Mvac}.\footnote{All of these vacua preserve eight supersymmetries, so we do not expect the space of classical vacua to be lifted by quantum effects.  See \eqref{vacsusy} below.}  It would be highly desirable to have a complete description of this space from both points of view, and we will return to this issue in future work.

Let us note that in the abelian case, a 't Hooft defect is also known as a BIon spike \cite{Gibbons:1997xz,Callan:1997kz}.  The case of a single BIon spike of charge $p$ at $r = 0$ has been studied extensively in the AdS/dCFT literature.  See \eg\ \cite{deLeeuw:2017dkd} and references therein.  On the gravity side of the correspondence, a $U(1)$ magnetic flux through the $S^2$ is accompanied by a modification of the D5-brane embedding \cite{Karch:2000gx,Skenderis:2002vf,Arean:2006vg}.  This is described by the nontrivial profile for the abelian $\Phi^y$ in \eqref{monoasbc} with $P \to p$, where now the displayed terms are the full solution.  As $r \to 0$ the D5-brane curves and begins to run parallel with the D3-branes.  The induced metric on the embedding turns out to be another $AdS_4 \times S^2$, but the $AdS_4$ slice is different.  Its radius depends on the flux and is different than the AdS radius of the ambient $AdS_5$.  In the holographic dual, nontrivial $U(1)$ flux of charge $p$ corresponds to a defect that implements a jump in the rank of the 4D $\NN = 4$ SYM gauge group from $SU(N_c)$ to $SU(N_c - p)$.\footnote{Since we work at leading order in the large $N_c$ limit, we can't distinguish the difference between a jump from $SU(N_c)$ to $SU(N_c - p)$ versus a jump from $SU(N_c+p)$ to $SU(N_c)$.  All that matters for this discussion is that the rank changes by $p$.}  The triplet $\vec{R}_{\rm D3}$ obeys Nahm's equations with a pole at the defect given in terms of the $p$-dimensional irreducible $\mathfrak{su}(2)$ representation, as pointed out in \cite{Arean:2006vg}.  A generic configuration of branes like Figure \ref{fig4} will involve an abelian defect of the type just described together with a singular $\mathfrak{su}(N_f)$ monopole configuration.

A simple non-abelian generalization of the BIon spike discussed above is a Cartan-valued flux vacuum, in which the fields take the form
\begin{equation}\label{Pvacua}
\tilde{F} = \frac{P}{2} \sin{\theta} \ed\theta\ed\phi~, \qquad \tilde{\Phi}^y = \Phi_{\infty}^y - \frac{P}{2r} ~, \qquad \tilde{\Phi}^{z_i} = \Phi_{\infty}^{z_i}~,
\end{equation}
where all of the Higgs vevs and the charge $P$ are constant and mutually commuting.  This can be viewed as a Dirac monopole embedded in the non-abelian gauge group via the homomorphism $U(1) \to PSU(N_f)$ specified by $P$.  See \cite{Moore:2014jfa} for a detailed discussion of these solutions in the context of singular monopole moduli spaces.  It is a convenient background to use for the analysis of the perturbative spectrum below since the linearized equations of motion around this vacuum are tractable and the fields of any vacuum configuration will take this form asymptotically.  Note that for these solutions $P$ plays the role of both an asymptotic magnetic charge and a 't Hooft defect charge at $r = 0$.  For this reason, and because there will be other magnetic charges that make an appearance below, we will often refer to $P$ in this context as a 't Hooft charge.

Finally, let us consider the supersymmetry of the vacua \eqref{Mvac}, starting with the example of the flux vacua \eqref{Pvacua}.  Making use of the relations \eqref{canonscalar}, one finds that the variation of the Fermi field evaluated on \eqref{Pvacua} can be expressed in the form
\begin{align}\label{vacsusy}
\delta_{\vareps} \Psi \bigg|_{(\tilde{A},\tilde{\Phi})} =&~ \left\{ \frac{\mu^2 z_{0}^2}{2 r^2}  P \Gamma^{\underline{r y}} + \mu^2 r \Phi_{\infty}^y \Gamma^{\underline{r y}} - \frac{r}{r^2 + z_{0}^2} \Phi_{\infty}^{z_i} \Gamma^{\underline{r z_i}} \right\} \left( \mathbbm{1} - \Gamma^{\underline{r\theta\phi y}} \right) \vareps \cr
=&~ \left\{ \mu^2 r \Phi_{\infty}^y \Gamma^{\underline{r y}} - \frac{r}{r^2 + z_{0}^2} \Phi_{\infty}^{z_i} \Gamma^{\underline{r z_i}} \right\} \left( \mathbbm{1} - \Gamma^{\underline{r\theta\phi y}} \right) \vareps ~.
\end{align}
The second step followed because if $z_0 \neq 0$ then $\vareps$ satisifes $\Gamma^{\underline{r\theta\phi y}} \vareps = \vareps$, while if $z_0 =0$ the $P$ term vanishes trivially.  Hence we learn the following.  On the one hand, if $z_0 \neq 0$, then $\delta_{\vareps} \Psi = 0$.  In other words no further supersymmetry is broken when we turn on $\Phi_{\infty}^{m}$ beyond that broken by $z_0$ already.  On the other, if $z_0 = 0$ then turning on any nonzero $\Phi_{\infty}^m$ breaks the supersymmetries generated by $\eta_0$.  This is expected since they generate supersymmetries associated with the superconformal symmetries in the holographic dual, and separating the D5-branes breaks scale invariance.  Notice that $P$ completely drops out.  Hence the flux vacua \eqref{Pvacua} with vanishing Higgs vevs preserve all sixteen supersymmetries for any $P$.  The same conclusion was previously shown to hold for the $U(1)$ magnetic flux vacua in  \cite{Skenderis:2002vf}.

More general monopole vacua preserve all eight supersymmetries generated by $\vareps_0$.  To show this we proceed as follows.  Assuming $\eta_0$ vanishes, the supersymmetry variation of the fermion can be simplified.  In this case $\vareps$ is an eigenspinor of $\Gamma^{\underline{r\theta\phi y}} = \Gamma^{\underline{r_1 r_2 r_3 y}}$, and this allows us to absorb the $M_\Psi$ term into the $D_a \Phi$ terms by switching to scalars with coordinate indices rather than tangent space indices.  We work in the Cartesian frame where
\begin{equation}
(\vareps)_{\rm cart} = \sqrt{\mu} (r^2 + z_{0}^2)^{1/4} \vareps_0 ~,
\end{equation}
and it is the $\Gamma^{\underline{r_i}}$ that are constant.  Then the relations we need, following from \eqref{canonscalar} and $\Gamma^{\underline{r_1 r_2 r_3 y}} \vareps_0 = \vareps_0$, are
\begin{align}\label{Phirescale}
& \left[ D_{r_i} \Phi_{\uy} \Gamma^{r_i \uy} + M_\Psi \Gamma^{\underline{\theta\phi y}} \Gamma^{\underline{y}} \Phi_{\uy} \right] \vareps_0 = \mu^2 (r^2 + z_{0}^2) D_{r_i} \Phi^y \Gamma^{\underline{r_i y}} \, \vareps_0~ \cr
& \left[ D_{r_i} \Phi_{\underline{z_j}} \Gamma^{r_i \underline{z_j}} + M_\Psi \Gamma^{\underline{\theta\phi y}} \Gamma^{\underline{z_j}} \Phi_{\underline{z_j}} \right] \vareps_0 = D_{r_i} \Phi^{z_j} \Gamma^{\underline{r_i z_j}} \, \vareps_0~,
\end{align}
where we have also used that $\Gamma^{\underline{\theta\phi y}} = \Gamma^{\ur} \Gamma^{\underline{r\theta\phi y}} = (\hat{r}_i \Gamma^{\underline{r_i}}) \Gamma^{\underline{r_1 r_2 r_3 y}}$ .

Since $\Phi^{z_i}$ must be covariantly constant on a global minimum of the energy functional, the $D_{r_i} \Phi^{z_j}$ terms drop out.  The $D_\mu \Phi^y$ terms as well as the terms involving the components of $F_{ab}$ that do not participate in the Bogomolny equation \eqref{monopoleVacua} drop out as well.  Then the supersymmetry variation reduces to
\begin{align}
\delta \Psi \bigg|_{\rm vac} =&~ \mu^2 (r^2 + z_{0}^2) \left[ \half F_{r_i r_j} \Gamma^{\underline{r_i r_j} }+ D_{r_k} \Phi^y \Gamma^{\underline{r_k y}} \right] \vareps \cr
=&~ \half \mu^{5/2} (r^2 + z_{0}^2)^{5/4} \left( F_{r_i r_j} - \tilde{\epsilon}_{r_i r_j r_k} D_{r_k} \Phi^y \right) \Gamma^{\underline{r_i r_j}} \vareps_0 = 0~,
\end{align}
where in the second step we used $\Gamma^{\underline{r_1 r_2 r_3 y}} \vareps_0 = \vareps_0$, and in the last step we used the Bogomolny equation.  Hence all of the vacua in $M_{\rm vac}$ preserve (at least) eight supersymmetries.  In fact they preserve the full 3D $\NN = 4$ super-Poincar\'e algebra.

Next we turn to the analysis of the perturbative spectrum around these vacua.

\subsection{Perturbative spectrum}\label{sec:pertspec}

The spectrum of linearized fluctuations on the D5-brane has been computed in the abelian theory.  Each field gives rise to a Kaluza--Klein tower of modes on the (asymptotically) $AdS_4$ space after expanding in an orthogonal basis on the $S^2$.  The modes organize into short multiplets of the superconformal algebra and are holographically dual to a tower of operators in the dCFT localized on the defect, identified in \cite{DeWolfe:2001pq}.  This analysis was originally carried out for the $AdS_4$ background with $z_0 = 0$, and extended to the $z_0 \neq 0$ case in \cite{Arean:2006pk,Myers:2006qr}.\footnote{This was further generalized in \cite{Arean:2006vg} to the case of the abelian D5-brane embedding that includes both nonzero $z_0$ and a $U(1)$ magnetic flux on the worldvolume.  There, unlike in our case, the presence of $U(1)$ flux together with nonzero $z_0$ leads to a continuous 3D spectrum.}  In the latter case $z_0$ provides a scale, breaking conformal invariance in the IR, and leading to a discrete set of normalizable radial modes for each KK mode.  

These analyses can be further extended to the flux vacua, \eqref{Pvacua}, of the non-abelian theory, and the results pertaining to the large $r$ behavior of the modes will be valid around any of the monopole vacua of the previous subsection.  In order to describe the results we first introduce some notation.  We use $(a,\phi)$ to denote field fluctuations around \eqref{Pvacua}, so that
\begin{equation}\label{flucexpand}
A_a = \tilde{A}_a + a_a~, \qquad \Phi^m = \tilde{\Phi}^m + \phi^m~.
\end{equation}
Then we expand the fluctuations in components along a basis of the Lie algebra $\{T^s\}$, writing \eg\ $a_a = a_{a}^{s} T^s$.  

Basis elements of the real Lie algebra $\mathfrak{g}$ are represented by anti-Hermitian matrices in our conventions, but it is more convenient to employ a basis of the complexified Lie algebra $\mathfrak{g}_{\mathbb{C}}$ that utilizes raising and lowering operators associated with a root decomposition of $\mathfrak{g}_{\mathbb{C}}$.  Since the background data $\Phi_{\infty}^m,P$ are mutually commuting, they can be taken to lie in a Cartan subalgebra such that their adjoint action is diagonal.  Thus we introduce masses $(\vec{m}_{z,s},m_{y,s})$ and charges $p_s$ such that
\begin{equation}\label{adaction}
[\Phi_{\infty}^{z_i},T^s] = -i m_{z_i,s} T^s~, \qquad [\Phi_{\infty}^{y},T^s] = -i m_{y,s} T^s~, \qquad [P, T^s] = -i p_s T^s~.
\end{equation}
The quantization of $P$ implies that the $p_s$ are integers.  The index $s$ runs over values of the form $\{ \pm \upalpha,\mathfrak{i}\}$, where the $\{\upalpha\}$ are a set of positive roots and $\mathfrak{i}$ in an index labeling a basis of generators for the Cartan subalgebra.  We can choose the basis such that $\Tr(T^{-\upalpha} T^{\upbeta}) = \delta^{\upalpha\upbeta}$, $\Tr(T^{\mathfrak{i}} T^{\mathfrak{j}}) = \delta^{\mathfrak{i}\mathfrak{j}}$, with the rest vanishing.  If $\Phi_{\infty}^y$ is a regular element of $\mathfrak{g}$ we fix the Cartan subalgebra uniquely by requiring that it be in the fundamental Weyl chamber of $\mathfrak{t}$.  This means that all $m_{y,s} > 0$ for every $s$ corresponding to a positive root, and all D5-branes have distinct $y$-positions in Figure \ref{fig4}.  In any case we partially fix the choice of Cartan by requiring $m_{y,s} \geq 0$ for all $s$.  The masses and charges vanish for $s = \mathfrak{i}$, and satisfy $p_{-\upalpha} = - p_{\upalpha}$ \etc.\ for the nonzero roots.  The components of a  real adjoint-valued field $\Phi = \Phi^s T^s$ satisfy $\Phi^{-\upalpha} = (\Phi^{\upalpha})^\ast$ and $(\Phi^{\mathfrak{i}})^\ast = \Phi^{\mathfrak{i}}$.  In an effort to keep the notation manageable, we will avoid making the decomposition into root and Cartan directions explicit, and instead just write $p_{-s} = - p_{s}$ and $(\Phi^{s})^\ast = \Phi^{-s}$. 

We plug these expansions back into the equations of motion \eqref{eoms} and linearize in the fluctuations.  Some details of this procedure are given in appendix \ref{app:modes}.  Here we summarize the key points.  First, the equations for $a_{\mu}$ and $\phi^{z_i}$ can be decoupled from the rest and take an identical form after choosing a convenient gauge-fixing condition (described in the appendix), so it is useful to start with them:
\begin{align}\label{amufluc}
& \bigg\{ \pd_{r}^2 + \frac{2}{r} \pd_r + \frac{1}{r^2} \tilde{D}_{S^2}^2 - \left(m_{y,s} - \frac{p_s}{2r} \right)^2 + \frac{( \eta^{\mu\nu} \pd_\mu \pd_\nu - \vec{m}_{z,s}^2)}{\mu^4 (r^2 + z_{0}^2)^2} \bigg\} (a_{\mu}^s,\phi^{z_i,s}) = 0~.
\end{align}
Here $\tilde{D}_{S^2}^2 \equiv \tilde{g}^{\alpha\beta} \tilde{D}_{\alpha} \tilde{D}_{\beta}$ is the covariant background Laplacian on the two-sphere constructed from the background gauge field, $\tilde{D}_\alpha = \pd_\alpha + [\tilde{A}_\alpha, \, \cdot \,]$.  When $\vec{m}_{z,s},m_{y,s}$, and $p_s =0$ this equation coincides with the analogous one given in \cite{Arean:2006pk,Myers:2006qr}, as do the equations for the remaining fluctuations.  The effects of the nontrivial background \eqref{Pvacua} are qualitatively the same for all the fluctuations, so we describe them in the context of \eqref{amufluc}.

Regarding the large $r$ behavior of the solution, the most important question is whether $m_{y,s}$ is zero or positive---\ie\ whether the fluctuation commutes with $\Phi_{\infty}^y$ or not.  When $m_{y,s}$ is positive then $r = \infty$ is an essential singularity of the ODE.  Solutions either blow up or decay exponentially, $(a_\mu, \phi^{z_i}) \sim e^{\pm m_{y,s} r}$, and the boundary conditions we impose below will allow for the decaying behavior only.  Due to their exponential rather than power-law fall-off, these modes are not dual to local operators in the holographic dual.  Henceforth we restrict attention to those Lie algebra components such that $m_{y,s} = 0$.  (If $\Phi_{\infty}^y$ is generic---\ie\ a regular element of $\mathfrak{g}$---then these will be the components along the Cartan subalgebra that is uniquely determined by it.)

When $m_{y,s}$ is zero, the effects of nonzero $\vec{m}_{z,s},p_s$ are easily accommodated by making slight modifications to the analysis of \cite{DeWolfe:2001pq,Arean:2006pk,Myers:2006qr}.  First, $\vec{m}_{z,s}$ always appears as a shift of the $\mathbbm{R}^{1,2}$ wave operator: $\eta^{\mu\nu} \pd_\mu\pd_\nu \to \eta^{\mu\nu} \pd_\mu \pd_\nu - \vec{m}_{z,s}^2$.  This just leads to a constant shift for the 3D spectrum of each mode.  It does not affect the leading order large $r$ behavior of solutions.  

The background flux on the two-sphere is dealt with by making a mode expansion that diagonalizes $\tilde{D}_{S^2}^2$.  The background gauge field $\tilde{A}_\alpha$ is that of a Dirac monopole, and this is a well-known problem with a complete and explicit solution.  When acting on a scalar, the eigenfunctions are spin-weighted spherical harmonics, with eigenvalues
\begin{equation}\label{swharmonic}
\tilde{D}_{S^2}^2 \bigg|_{T^s} {}_{m'}Y_{jm}(\theta,\phi) = \left( -j(j+1) + \frac{p_{s}^2}{4} \right) {}_{m'}Y_{jm}(\theta,\phi) ~, \qquad \textrm{where} \quad m' = - \frac{p_s}{2}~.
\end{equation}
Here $(j,m)$ are the usual angular momentum quantum numbers, but $j$ is restricted to start at the minimum value $|p_s|/2$ and increase in integer steps.  Thus the set of $j$'s are all integers or all half-integers depending on whether $p_s$ is even or odd respectively.  $m$ runs from $-j$ to $j$ in integer steps as usual.  If $p_s = 0$ then these reduce to the ordinary spherical harmonics.  The combination $\tilde{D}_{S^2}^2 - \frac{p_{s}^2}{4}$  in \eqref{amufluc} almost always appears together: the second term comes from the background $\tilde{\Phi}^y$.  Hence, after expanding, say, $a_{\mu}^s$ in spin-weighted spherical harmonics the resulting equation for each KK mode is
\begin{equation}\label{amumodeeqn}
\bigg\{ \pd_{r}^2 + \frac{2}{r} \pd_r - \frac{j(j+1)}{r^2} + \frac{( \eta^{\mu\nu} \pd_\mu \pd_\nu - \vec{m}_{z,s}^2)}{\mu^4 (r^2 + z_{0}^2)^2} \bigg\} a_{\mu,(j,m)}^s = 0~,
\end{equation}
and similarly for $\phi_{(j,m)}^{z_i,s}$.  (Here we are assuming $m_{y,s}= 0$.)  One can further make a Fourier expansion along $\mathbb{R}^{1,2}$ so that $\eta^{\mu\nu} \pd_{\mu} \pd_\nu \to -k^2$, and then this equation becomes identical in form to the corresponding ones in \cite{Arean:2006pk,Myers:2006qr}.

Exact solutions to \eqref{amumodeeqn} for the radial dependence are available in terms on hypergeometric functions when $z_0 \neq 0$.  Demanding that the series solution truncate (in order to have normalizability) leads to the introduction of a radial quantum number $n$ and a discrete spectrum of masses, $k^2 = M_{n,j}^2$ for `meson' states in the holographic dual.  See \cite{Arean:2006pk,Myers:2006qr} for further details.  (These masses will now all be shifted from the values given in \cite{Arean:2006pk,Myers:2006qr} by $\vec{m}_{z,s}^2$.)  If $z_0 = 0$ we instead get the usual continuous spectrum indicative of a conformal dual, and the radial wavefunctions can be given in terms of Bessel functions.

Our focus is on the large $r$ behavior of these modes, where $z_0$ can be neglected and the analysis reduces to the $AdS_4$ case studied in \cite{DeWolfe:2001pq}.  At large $r$ one has from \eqref{amumodeeqn} that\footnote{This agrees with \cite{DeWolfe:2001pq} after taking into account that $a_{\mu} \sim r a_{\umu}$ and $\vec{\phi}^z \sim r \vec{\phi}^{\underline{z}}$.}
\begin{equation}\label{amuas}
(a_{\mu,(j,m)}^s , \phi_{(j,m)}^{z_i,s}) \sim \left\{ \begin{array}{l} r^{-(j+1)} = r^{1 -\Delta_j} ~, \\ r^j = r^{1 -3+ \Delta_j}~, \end{array} \right.
\end{equation}
where $\Delta_j$, the conformal dimension of the dual operator, is given by $\Delta_j = 2 + j$.  The first behavior corresponds to the `normalizable mode' solution for the given KK mode, and the second behavior corresponds to the `non-normalizable mode'.  In the dual theory, the ($x^\mu$-dependent) coefficient of the non-normalizable mode is the source for the dual operator, while the coefficient of the normalizable mode is its vev.  This holographic interpretation of the field asymptotics is an important guide to the type of boundary conditions that one should consider. We will have more to say about this in the next subsection.  First, however, let us summarize the rest of the perturbative spectrum.

Like $a_{\mu}$ and $\phi^{z_i}$, the Lie algebra components of the other field fluctuations along directions with nonzero $m_{y,s}$ have exponential decay at large $r$.  Thus we focus on the case $m_{y,s} = 0$.  Then the equation for the radial component of the gauge field fluctuations, $a_{r,(j,m)}^s$, decouples from the rest.  The asymptotic behavior of solutions is
\begin{equation}
a_{r,(j,m)}^s \sim  \left\{ \begin{array}{l} r^{-(j+2)} = r^{-\Delta_j} ~, \\ r^{j-1} = r^{-3 + \Delta_j} ~, \end{array} \right. 
\end{equation}
for the normalizable and non-normalizable mode respectively.

The remaining bosonic degrees of freedom are contained in $a_\alpha$ and $\phi^y$.  These are usefully repackaged into three adjoint-valued scalars $\lambda,f,\sy$ defined by
\begin{equation}\label{AalphaY}
a_\alpha = \tilde{D}_\alpha \lambda + \tilde{\epsilon}_{\alpha\beta} \tilde{g}^{\beta\gamma} \tilde{D}_\gamma f~, \qquad \phi^y = -\frac{1}{2r} [P,\lambda] + \sy ~,
\end{equation}
which can then be expanded in Lie algebra components and in spin-weighted spherical harmonics on the $S^2$.  The $\lambda$ modes are nondynamical and determined by a gauge-fixing condition.  The $j=0$ modes (which are only possible when $p_s = 0$) can be set to zero while the higher $j$ modes are given by
\begin{equation}\label{lambdasolve}
\lambda_{(j,m)}^s = \frac{1}{j(j+1)} \left[ \pd_r \left( r^2 a_{r}^s \right) + \frac{i p_s}{2} \left( r \sy_{(j,m)}^s - f_{(j,m)}^s \right) \right]~.
\end{equation}
Notice that, if $a_r =0$, then $\lambda^s$ is only present when there is a nontrivial flux, $p_s$.  In the absence of flux, $\lambda$ is pure gauge and can be set to zero.  This is consistent with the assumptions made in \cite{DeWolfe:2001pq}, which did not consider turning on flux.\footnote{Reference \cite{DeWolfe:2001pq} also works in a different gauge, related to ours by a shift involving $a_r$ that removes the $a_r$ dependence from $\lambda$ at the price of introducing it back into \eg\ the $a_\mu$ equation.  The gauge choice we use is consistent with the analysis of \cite{Myers:2006qr}.  See appendix \ref{app:modes} for further details.}  

The $\sy$ and $f$ modes form a coupled system that must be diagonalized, as in \cite{DeWolfe:2001pq,Arean:2006pk,Myers:2006qr}.  The diagonal combinations are $\phi^{\pm}$ defined by
\begin{equation}\label{diagonalizeyf}
\phi_{(j,m)}^{+} = \frac{1}{\sqrt{2j+1}} \left( j f_{(j,m)} + r \sy_{(j,m)} \right)~, \qquad \phi_{(j,m)}^{-} = \frac{1}{\sqrt{2j+1}} \left( (j+1) f_{(j,m)} - r \sy_{(j,m)} \right) ~.
\end{equation}
The respective asymptotics at large $r$ are
\begin{equation}
\phi_{(j,m)}^{\pm,s} \sim \left\{ \begin{array}{l} r^{-\Delta_{j}^{\pm}} ~, \\[1ex]  r^{-3 + \Delta_{j}^{\pm}} ~, \end{array} \right.
\end{equation}
where
\begin{equation}
\Delta_{j}^+ = j+4 ~, \qquad \Delta_{j}^- = j ~.
\end{equation}
Here it is important to note that only the $+$ mode is physical for the lowest, $j = |p_s|/2$, rung of the KK tower: one can show that the $j= |p_s|/2$ mode of $\phi^-$ drops out of the expressions \eqref{AalphaY} for $a_\alpha$ and $\phi^y$.  (This includes the $j=0$ modes along directions with $p_s = 0$.)

Finally there are two KK towers of four-dimensional fermions $\eta_{(j,m)}^s, \chi_{(j,m)}^s$, coming from $\Psi$.  The spectrum of these modes has not been previously discussed in the literature.  We provide the details in appendix \ref{app:modesf}.  Let us denote by $\Psi_{(j,m)}^{(\chi)s}$ the restriction of $\Psi$ to the case in which only the $\chi_{(j,m)}^s$ mode is turned on, and analogously for $\Psi_{(j,m)}^{(\eta)s}$.  For each mode, the two possible behaviors---normalizable and non-normalizable---are correlated with a specific chirality of $\Psi$ with respect to $\Gamma^{\underline{r\theta\phi y}}$.  (This is true for the leading order behavior of the mode; the other chirality will be turned on at subleading order.)  Defining
\begin{equation}
\Psi^{\pm} := \half (\mathbbm{1} \pm \Gamma^{\underline{r\theta\phi y}}) \Psi ~,
\end{equation}
the leading asymptotics are
\begin{equation}
\Psi_{(j,m)}^{(\chi)s,\pm} \sim r^{-\frac{3}{2} \pm m_{j}^{(\chi)}} ~, \qquad \Psi_{(j,m)}^{(\eta)s,\pm} \sim r^{-\frac{3}{2} \pm m_{j}^{(\eta)}}~,
\end{equation}
where the $AdS_4$ masses are given by
\begin{equation}\label{fspectrum}
m_{j}^{(\chi)} = j - \half ~, \qquad m_{j}^{(\eta)} = - \left( j + \frac{3}{2} \right)~.
\end{equation}
For the $\chi$-type modes, $j$ starts at $\half (|p_s| + 1)$ and increases in integers steps.  For the $\eta$-type modes, $j$ starts at $\half (|p_s|-1)$ if $p_s \neq 0$, or $\half$ if $p_s = 0$, and increases in integer steps.  The quantum number $m$ runs from -$j$ to $j$ in integer steps as usual.  These asymptotics are valid when $m_{y,s} = 0$, otherwise the normalizable (non-normalizable) modes are exponentially decaying (blowing up).  The conformal dimensions of the dual operators are $\Delta_{j}^{(\chi,\eta)} = \frac{3}{2} + |m_{j}^{(\chi,\eta)}|$. 

Observe that the masses of the $\eta$-type modes are always $\leq - \frac{3}{2}$, while the masses of the $\chi$-type modes are always $\geq 0$.  For each $s$ such that $p_s = 0$ (and $m_{y,s} = 0$), the $j=1/2$ $\chi$-type modes provide a doublet of massless fermions on the asymptotically $AdS_4$ space.  Also, since the masses have opposite sign for the $\chi$- and $\eta$-type modes, the leading behavior of $\Psi$ sits in opposite $\Gamma^{\underline{r\theta\phi y}}$ eigenspaces.  For example, the normalizable $\chi$-type modes correspond to $\Psi^-$ while the normalizable $\eta$-type modes correspond to $\Psi^+$.

In fact, for the massless $\chi$ modes it is not obvious from this analysis whether  $\Psi^{(\chi),-}$ or $\Psi^{(\chi),+}$ should be identified with the normalizable mode, as both have the same $O(r^{-3/2})$ behavior.  We will see in subsection \ref{sec:bndrySUSY} below that supersymmetry dictates that $\Psi^{(\chi),-}$ is the normalizable mode while $\Psi^{(\chi),+}$ is the non-normalizable mode.

The $S^2$ singlet modes of the fields will appear often in the following.  We will use a simplified notation for them,
\begin{align}
 (a,\phi)_{(0,0)}(x^\mu,r) Y_{00} \equiv &~ (a,\phi)(x^\mu,r)~, 
\end{align}
absorbing the constant factor $Y_{00} = (4\pi)^{-1/2}$ into the definition of the modes.  It will be clear from the context whether we are using lowercase $(a,\phi)$ to refer to the $S^2$ singlet mode only, or to the sum over all modes as in \eqref{flucexpand}.
  
\subsection{Boundary conditions and consistency of the variational principle}\label{sec:cvp}

With the asymptotic behavior of the linearized perturbations in hand, we can return to the questions of boundary conditions and consistency of the variational principle.  As we saw above, each KK mode has an associated  normalizable and non-normalizable mode.  Correspondingly, one can consider two types of boundary condition.  For the first type one turns on the non-normalizable mode and holds it fixed.  Then, according to the AdS/CFT correspondence, the on-shell action as a function of these sources gives (a leading saddle point approximation to) the generator of correlation functions for the dual operators.  This boundary condition generally requires the addition of boundary terms in order to maintain consistency with the variational principle, and the systematic procedure for doing this is known as holographic renormalization.  (See the discussion following \eqref{boundaryPs}.)  The second type of boundary condition sets the coefficient of the non-normalizable mode to zero and lets the coefficient of the normalizable mode fluctuate.  This is appropriate for the construction of the Hilbert space of states in the bulk theory, which is equivalent to the Hilbert space of states in the dual CFT.  Here we follow the Lorentzian-signature version of the correspondence as described in \cite{Balasubramanian:1998sn}.

In this paper we generally want to consider the latter type of boundary condition.  However for some special modes at the bottom of the KK towers it is possible (and useful) to turn on the non-normalizable mode without spoiling consistency of the variational principle.  In fact we have already done so for two types of mode.  The vevs $\Phi_{\infty}^{z_i}$ and $\Phi_{\infty}^y$ can be viewed as finite, constant coefficients for the $j=0$ non-normalizable modes of $\phi^{z_i}$ and $\sy = r^{-1} \phi^{+}$ respectively.  In contrast, the flux $P$ is not associated with any non-normalizable (or normalizable) mode; this is consistent with the fact that it is quantized and cannot be adiabatically tuned.  Rather, $P$ determines a superselection sector of the theory under consideration.

We will not turn on any further non-normalizable modes for the $\phi^{z_i}$.  However, we will allow for the $S^2$ singlet modes corresponding to the vevs to be spacetime varying.  We write
\begin{align}\label{fieldasPhiz}
\Phi^{z_i} =&~ \phi^{z_i}(x^\mu,r) + \sum_{j\geq \half ,m,s} \phi_{(j,m)}^{z_i,s}(x^\mu,r) \left( {}_{-\frac{p_s}{2}} Y_{jm}(\theta,\phi)\right) T^s~, \qquad \textrm{with} \cr
 \phi^{z_i}(x^\mu,r) =&~  \phi_{({\rm nn})}^{z_i}(x^\mu) + \frac{1}{\mu r} \phi_{({\rm n})}^{z_i}(x^\mu) + O(r^{-2})~, \qquad \phi_{(j,m)}^{z_i,s} = O(r^{-(j+1)})~,
\end{align}
The $S^2$ singlets $\phi^{z_i}$ are Lie algebra-valued and commute with $P,\Phi_{\infty}^y$.  The more general asymptotics are encoded by the $\phi_{({\rm nn})}^{z_i}(x^\mu)$, which can be mutually non-commuting.  This will still give rise to finite energy field configurations provided they have sufficient decay properties at large $x^\mu$.  Specifically, we  require them to approach the (mutually commuting) constant vevs $\Phi_{\infty}^{z_i}$ at spatial infinity and have vanishing time derivatives as $t \to \pm\infty$.  Since the non-normalizable mode is held fixed in the variational principle, $\delta \Phi^{z_i} = O(1/r)$, $D_r \Phi^{z_i} = O(1/r^2)$, and it follows that the $\delta_{ij} \PP^{z_i} \delta \Phi^{z_i}$ term drops out of \eqref{Svaronshell}, as required for consistency.  

There are additional types of non-normalizable modes we wish to consider.  The first type is the $j=0$ non-normalizable mode of the gauge field components $A_\mu = a_{\mu}$.  According to \eqref{amuas}, these modes have a finite limit as $r \to \infty$.  We set all higher $j$ non-normalizable modes of $A_{\mu}$ to zero.  We also set all non-normalizable modes of $A_r$ to zero, so that the leading behavior of this field at large $r$ is $O(1/r^2)$, corresponding to the normalizable $S^2$ singlet.  Thus our boundary conditions for the $AdS_4$ part of the gauge field are 
\begin{align}
A_{\mu} =&~ a_{\mu}(x^\nu,r) + \sum_{j\geq \half,m,s} a_{\mu,(j,m)}(x^\nu,r) \left({}_{-\frac{p_s}{2}}Y_{jm}(\theta,\phi)\right) T^s~, \qquad \textrm{with} \cr
a_{\mu}(x^\nu,r) =&~ a_{\mu}^{({\rm nn})}(x^\nu) + \frac{1}{\mu r} a_{\mu}^{({\rm n})}(x^\nu) + O(1/r^2)~, \qquad a_{\mu,(j,m)} = O(r^{-(j+1)})~, \qquad \textrm{and} \nonumber \\[1ex]  \label{fieldasAdSA}
A_r =&~ a_r(x^\nu,r) +  \sum_{j\geq \half,m,s} a_{r,(j,m)}(x^\nu,r) \left({}_{-\frac{p_s}{2}}Y_{jm}(\theta,\phi)\right) T^s~, \qquad \textrm{with} \cr
a_r(x^\nu,r) =&~ \frac{1}{\mu^2 r^2} a_{r}^{({\rm n})}(x^\nu) + O(r^{-3}) ~, \qquad a_{r,(j,m)} = O(r^{-(j+2)}) ~.
\end{align}
The singlets $a_{\mu,r}$ commute with $P,\Phi_{\infty}^y$ but need not commute with one another.  The data $a_{\mu}^{({\rm nn})}$ is held fixed, and we assume it approaches a pure gauge configuration as $x^\mu \to \infty$.  These boundary conditions imply that $F_{r\mu} = O(1/r^2)$ and the variation $\delta A_\mu = O(1/r)$, ensuring that the $\PP^\mu \delta A_\mu$ term drops out of \eqref{Svaronshell}.  

Next, consider the $j=1$ angular momentum triplet of modes at the bottom of the KK tower for $\phi^-$.  In general the low $j$ modes of $\phi^-$ are subtle, due to the small value of the  conformal dimension, $\Delta_{j}^- = j$.  We can immediately assume that all non-normalizable modes of $\phi^+$ are set to zero (besides the singlet giving the vev $\Phi_{\infty}^y$) and all non-normalizable modes of $\phi^-$ for $j > \frac{5}{2}$ are set to zero---these conditions ensure that the fluctuations are subleading to background $(\tilde{\Phi}^y, \tilde{A}_\alpha)$.  The $j \leq \frac{5}{2}$ modes of $\phi^-$ are all decaying, and a closer look at consistency of the variational principle is required.

Restricting consideration to the $\phi^-$ type modes and utilizing \eqref{flucexpand} and \eqref{AalphaY} through \eqref{diagonalizeyf}, one obtains the following results for the KK modes of the conjugate variables $\PP^y$ and $\PP^\alpha$, \eqref{boundaryPs}:
\begin{align}\label{PPy}
\PP_{(j,m)}^{y,s} =&~ \mu^6 r^4 \bigg\{ - \left(j - \frac{p_{s}^2}{4j}\right)\frac{ (r \pd_r + j) \phi_{(j,m)}^{s,-}}{(2j+1)^{1/2}} + r^2 [a_r, \phi^y]_{(j,m)}^s -\frac{1}{\sin{\theta}}[a_\theta,a_\phi]_{(j,m)}^s + O(r^{-3}) \bigg\} ~,
\end{align}
and
\begin{align}\label{PPalpha}
\PP_{(j,m)}^{\theta,s} =&~ \mu^6 r^4 \bigg\{ \frac{1}{r} \left[ \frac{1}{\sin{\theta}} \tilde{D}_\phi - \frac{i p_s}{2j} \pd_\theta \right] \frac{(r \pd_r + j) \phi_{(j,m)}^{s,-}}{(2j+1)^{1/2}} + [a_r, a_\theta]_{(j,m)}^{s} + \cr
&~ \qquad \qquad - \frac{1}{\sin{\theta}} [a_\theta, \phi^y]_{(j,m)}^{s} + O(r^{-4}) \bigg\} \cr
\PP_{(j,m)}^{\phi,s} =&~ \frac{\mu^6 r^4}{\sin{\theta}} \bigg\{ - \frac{1}{r} \left[ \pd_\theta + \frac{i p_s}{2j \sin{\theta}} \tilde{D}_\phi \right] \frac{(r\pd_r + j) \phi_{(j,m)}^{s,-}}{(2j+1)^{1/2}} + [a_r, a_\phi]_{(j,m)}^s + \cr
&~ \qquad \qquad + [a_\theta,\phi^y]_{(j,m)}^s + O(r^{-4}) \bigg\}~.
\end{align}
We have computed these to the order necessary for taking the $r\to \infty$ limit in \eqref{Svaronshell}, taking into account that $\delta \Phi^y = O(r^{-2})$ and $\delta A_\alpha = O(1/r)$.  In obtaining these results we dropped terms proportional to $\pd_{r}^2 (r^2 a_{r,(j,m)}) - j(j+1) a_{r,(j,m)}$.  One can use the linearized equation of motion for $a_{r,(j,m)}$ and the allowed asymptotics \eqref{fieldasAdSA} to conclude that this combination of terms is $O(r^{-4})$.

We can now argue that the non-normalizable modes for $j = 5/2,2,3/2$ should be set to zero.  If they are not, then they give the dominant asymptotics of $\phi_{(j,m)}^-$, which leads to $(r\pd_r + j) \phi_{(j,m)}^- = O(r^{j-3})$, implying contributions to the $\PP$ of $\PP^y \sim O(r^{j+1})$, $\PP^\alpha \sim O(r^j)$.  However $\delta \phi^-$, which is the order of the normalizable mode, since the non-normalizable mode is held fixed, contributes to the variations according to $\delta  \Phi^y = O(r^{-1-j})$ and $\delta A_\alpha = O(r^{-j})$.  Hence, $\PP^y \delta \Phi^y + \PP^\alpha \delta A_\alpha$ in \eqref{Svaronshell} would have a finite limit as $r \to \infty$.  Thus for these values of $j$, as with all higher values, the non-normalizable mode of $\phi_{(j,m)}^-$ should be set to zero, or else one must resort to holographic renormalization.

In the case $j=1$ the dominant asymptotics correspond to the normalizable mode, $\phi^- \sim r^{-j} = r^{-1}$.  The operator $(r\pd_r + 1)$ annihilates this, however.  Hence it is still the non-normalizable mode that gives the leading contribution to $(r\pd_r + 1) \phi^-$.  It is useful to introduce a real basis for the $j=1$ triplet of scalars, $\vec{\XX} = \vec{\XX}(x^\mu,r)$, such that
\begin{equation}\label{XXtripdef}
\sum_{m=-1}^1 \phi^{s,-}_{(1,m)}(x^\mu,r) Y_{1m}(\theta,\phi) =: -\frac{\sqrt{3}}{\mu^2 r} \hat{r} \cdot \vec{\XX}^s(x^\mu,r)~,
\end{equation}
where $\hat{r} = (\sin{\theta}\cos{\phi},\sin{\theta}\sin{\phi},\cos{\theta})$.  Note that these modes only exist for those $s$ such that $p_s = 0$.  In other words the adjoint-valued $\vec{\XX} = \vec{\XX}^s T^s$ satisfies $[\vec{\XX}, P] = 0 = [\vec{\XX}, \Phi_{\infty}^y]$.  We denote the coefficients of the two modes by $\vec{\XX}_{({\rm n})}$ and $\vec{\XX}_{\rm nn}'$  such that
\begin{equation}\label{Ytriplet}
\vec{\XX}(x^\mu,r) =  \vec{\XX}_{({\rm n})}(x^\nu) + \frac{1}{\mu r} \vec{\XX}_{({\rm nn})}'(x^\nu) + O(1/r^2)~.
\end{equation}
As we will see, $\vec{\XX}_{({\rm nn})}'$ is not quite the conjugate of $\vec{\XX}_{({\rm n})}$.  One then has the following expansions:
\begin{align}\label{fieldasAtpY}
\Phi^y =&~ \Phi_{\infty}^y - \frac{P}{2r} +  \frac{1}{\mu^2 r^2} \hat{r} \cdot \vec{\XX}(x^\nu,r) + O(r^{-5/2})~, \cr
A_\theta =&~ - \frac{1}{\mu^2 r} \hat{\phi} \cdot \vec{\XX}(x^\nu,r) + O(r^{-3/2})~, \cr
A_\phi =&~ \frac{P}{2}(\pm1 - \cos{\theta}) + \frac{\sin{\theta}}{\mu^2 r} \hat{\theta} \cdot \vec{\XX}(x^\nu,r) + O(r^{-3/2})~,
\end{align}
where $\hat{\theta} = \pd_\theta \hat{r}$ and $\sin{\theta} \hat{\phi} = \pd_\phi \hat{r}$.  We do not specify the mode expansion of the subleading terms in detail since they receive contributions from both $\phi_{(j,m)}^{\pm}$ and $\lambda_{(j,m)}$ in a rather nontrivial way.  In particular the $\phi_{(2,m)}^-$ modes will contribute at the same order in $1/r$ as the $\vec{\XX}_{\rm nn}'$ piece of $\vec{\XX}$.  This does not lead to any ambiguities below, however, as they correspond to orthogonal harmonics.

Using these one finds that the conjugate momenta \eqref{PPy}, \eqref{PPalpha} take the form
\begin{align}\label{Pylarger}
\PP^y =&~ \mu^6 r^4 \left\{ - \frac{1}{\mu^3 r^2} \hat{r} \cdot \vec{\XX}_{({\rm nn})}' + \frac{1}{\mu^4 r^2} [a_{r}^{({\rm n})}, \hat{r} \cdot \vec{\XX}_{({\rm n})}] + \frac{1}{\mu^4 r^2} [\hat{\phi} \cdot \vec{\XX}_{({\rm n})} , \hat{\theta} \cdot \vec{\XX}_{({\rm n})}] + O(r^{-5/2}) \right\} \cr
\equiv&~ \mu^2 r^2 \, \hat{r} \cdot  \left( \vec{\XX}_{({\rm nn})} + O(r^{-1/2}) \right)~,  \raisetag{18pt}
\end{align}
and similarly
\begin{equation}\label{Palphalarger}
\PP^{\theta} = -\mu^2 r \, \hat{\phi} \cdot \left( \vec{\XX}_{({\rm nn})} + O(r^{-1/2}) \right)~, \qquad \PP^\phi = \frac{\mu^2 r}{\sin{\theta}} \, \hat{\theta} \cdot \left( \vec{\XX}_{({\rm nn})} + O(r^{-1/2}) \right)~,
\end{equation}
where
\begin{align}\label{Xmom}
\vec{\XX}_{({\rm nn})} :=&~ \lim_{r\to \infty} \left( \mu^2 r^2 D_r \vec{\XX} - \half [\vec{\XX}, \times \vec{\XX}] \right) = - \mu \vec{\XX}_{({\rm nn})}' + [a_{r}^{({\rm n})},\vec{\XX}_{({\rm n})}] - \half [\vec{X}_{({\rm n})}, \times \vec{\XX}_{{(\rm n)}}]~.
\end{align}
The $\times$ notation refers to the Cartesian cross product on Euclidean $\mathbb{R}^3$ such that $( [\vec{\XX}, \times \vec{\XX}] )^i = \tilde{\epsilon}^{i}_{\phantom{i}jk} [\XX^j, \XX^k]$.  We have that $\XX_{({\rm nn})} = - \mu \vec{\XX}_{({\rm nn})}'$ plus non-abelian terms.  $\XX_{({\rm nn})}$ is indeed the momentum conjugate to $\vec{\XX}_{({\rm n})}$ in the sense that
\begin{equation}\label{PXdX}
\left\{ \PP^y \delta \Phi^y + \PP^\theta \delta A_\theta + \PP^\phi \delta A_\phi \right\} = \vec{\XX}_{({\rm nn})} \cdot \delta \vec{\XX}_{({\rm n})} + O(r^{-1/2}) ~.
\end{equation}

We see from \eqref{PXdX} that in order for the variational principle to be well-defined, we must either hold $\vec{\XX}_{({\rm n})}$ fixed, or set $\vec{\XX}_{({\rm nn})} = 0$.  This however is not consistent with the holographic interpretation.  The modes $\vec{\XX}$ were identified with a triplet of conformal dimension one operators in the dCFT in \cite{DeWolfe:2001pq}, and $\vec{\XX}_{({\rm nn})}$ is the source dual to these operators.  The action should be extremized when $\vec{\XX}_{({\rm nn})}$ is held fixed.  Hence, following \cite{Klebanov:1999tb}, it is not the original on-shell Yang--Mills action that provides a holographic description of the dCFT, but rather its Legendre transform with respect to the pair $(\vec{\XX}_{({\rm n})},\vec{\XX}_{({\rm nn})})$:
\begin{align}\label{Shol}
S_{\rm hol}[\vec{\XX}_{({\rm nn})}, \ldots] := \left[ (S_{\rm ym})^{\textrm{o-s}}[\vec{\XX}_{({\rm n})}, \ldots] + \frac{4\pi}{g_{{\rm ym}_6}^2 \mu^2} \int \ed^3 x \Tr \left\{ \vec{\XX}_{({\rm n})} \cdot \vec{\XX}_{({\rm nn})} \right\} \right]_{\vec{\XX}_{({\rm n})} = \vec{\XX}_{({\rm n})}[\vec{\XX}_{({\rm nn})}]}~,
\end{align}
where $\vec{\XX}_{({\rm n})}[\vec{\XX}_{({\rm nn})}]$ is the $\vec{\XX}_{({\rm n})}$ that extremizes the quantity in square brackets.  The addition of this term ensures that the variation of $S^{\rm hol}$ is proportional to $\delta \vec{\XX}_{({\rm nn})}$ and vanishes when we hold $\vec{\XX}_{({\rm nn})}$ fixed.  In the next subsection we will show that the additional term is also necessary for the cancelation of boundary contributions to the supersymmetry variation.  A very similar situation is nicely analyzed in the recent works \cite{Freedman:2016yue,Mezei:2017kmw}.

Finally, the normalizable modes of the fermion can be isolated by requiring $\Psi^- = O(r^{-3/2})$ and $\Psi^+ = O(r^{-5/2})$.  These conditions ensure that all normalizable modes of $\chi$ and $\eta$ type are admissible while none of the non-normalizable ones are.  In particular, we are identifying the normalizable mode of the $j=1/2$ $\chi$-type modes with the $O(r^{-3/2})$ behavior of $\Psi^-$ asymptotically.  These modes have a vanishing 4D mass, and it would also be consistent with the variational principle to identify the normalizable mode with an $O(r^{-3/2})$ $\Psi^+$ component instead.  We will see below that supersymmetry requires the identifications we have made.  Nonetheless it is useful to turn on the non-normalizable modes for the massless fermions since, as we will see, they sit in a supermultiplet with some of the other non-normalizable modes.  

Let
\begin{equation}
\Psi_{j=1/2}^{(\chi)} := \Psi_{(\half,\half)}^{(\chi)} + \Psi_{(\half,-\half)}^{(\chi)}~,
\end{equation}
be the restriction of $\Psi$ to the $j=1/2$ $\chi$-type modes.  Then
\begin{equation}
\Psi^+ = \Psi_{j=1/2}^{(\chi),+} + O(r^{-2})~, \qquad  \Psi^- = \Psi_{j=1/2}^{(\chi),-} + O(r^{-5/2})~,
\end{equation}
and we show appendix \ref{app:fermsol} that $\Psi_{j=1/2}^{(\chi)}$ takes the form
\begin{equation}\label{Psi2uppsiS2}
(\Psi_{j=1/2}^{(\chi)})_{S^2} = h_{S^2}(\theta,\phi) \uppsi^+(x^\mu,r) + h_{S^2}(-\theta,\phi) \uppsi^-(x^\mu,r)~,
\end{equation}
with respect to a basis in which $\Gamma^{\ur},\Gamma^{\utheta},\Gamma^{\uphi}$ are constant, and $h_{S^2}$ is given by \eqref{hS2def}.  Here $\uppsi^{\pm} = \half (\mathbbm{1} \pm \Gamma^{\underline{r\theta\phi  y}}) \uppsi$, where $\uppsi(x^\mu,r)$ is Majorana--Weyl.  If we instead work in a natural basis with respect to the Cartesian frame in which the $\Gamma^{\underline{r_i}}$ are constant, we have
\begin{equation}\label{Psi2uppsi}
(\Psi_{j=1/2}^{(\chi)})_{\rm cart} = \uppsi^+(x^\mu,r) + (\hat{r} \cdot \vec{\Gamma}_{(r)}) \Gamma^{\underline{r_3}} \uppsi^-(x^\mu,r)~,
\end{equation}
where we've introduced the notation $\vec{\Gamma}_{(r)} := (\Gamma^{\underline{r_1}},\Gamma^{\underline{r_2}},\Gamma^{\underline{r_3}})$.  The asymptotics of $\uppsi^{\pm}$ are
\begin{align}\label{fieldasPsi}
& \uppsi^+ = \frac{1}{(\mu r)^{3/2}} \uppsi_{0}^{({\rm nn})}(x^\mu) + O(r^{-5/2})~, \qquad \uppsi^- = \frac{1}{(\mu r)^{3/2}} \Gamma^{\underline{r_3}} \, \uppsi_{0}^{({\rm n})}(x^\mu) + O(r^{-5/2})~,
\end{align}
where $(\uppsi_{0}^{({\rm nn})} ,\uppsi_{0}^{({\rm n})})$ encode the non-normalizable and normalizable modes of the $j=1/2$ $\chi$-type doublet.  They are 10D Majorana--Weyl spinors satisfying the same chirality and projection conditions as $(\vareps_0,\eta_0)$.

The non-normalizable mode is to be held fixed so $(\delta \Psi)^+ = O(r^{-5/2})$.  Since $\Gamma^{\ur}$ anticommutes with $\Gamma^{\underline{r\theta\phi y}}$ we then have
\begin{equation}\label{PPsidPsi}
 \half \Psibar \Gamma^{\ur} \left(1 + \Gamma^{\underline{r\theta\phi y}}\right) \delta \Psi = \Psibar^- (\delta \Psi)^+ = O(r^{-7/2})~,
 \end{equation}
which implies that the $\PP^{\Psi} \delta \Psi$ term drops out of \eqref{Svaronshell}.  The addition of the fermion boundary action \eqref{Sfbndry} was crucial for this to work.

Summarizing, the field asymptotics are given by \eqref{fieldasPhiz}, \eqref{fieldasAdSA}, \eqref{fieldasAtpY}, and \eqref{fieldasPsi}, which we collect here:  
\begin{equation}\label{fieldas1}
\begin{gathered}
A_\mu = a_\mu(x^\nu,r) + O(r^{-3/2})~, \qquad A_r = a_r(x^\nu,r) +O(r^{-5/2})~, \\
\Phi^{z_i} = \phi^{z_i}(x^\nu,r) + O(r^{-3/2})~, \\
A_\theta = -\frac{1}{\mu^2 r} \hat{\phi} \cdot \vec{\XX}(x^\nu, r) + O(r^{-3/2})~, \\
A_\phi = \frac{P}{2} (\pm 1 - \cos{\theta}) + \frac{\sin{\theta}}{\mu^2 r} \hat{\theta} \cdot \vec{\XX}(x^\nu ,r) + O(r^{-3/2})~, \\
\Phi^y = \Phi_{\infty}^y - \frac{P}{2r} + \frac{1}{\mu^2 r^2} \hat{r} \cdot \vec{\XX}(x^\nu,r) + O(r^{-5/2})~, \\
\Psi^+ = \Psi_{j=1/2}^{(\chi),+} + O(r^{-2})~, \qquad  \Psi^- = \Psi_{j=1/2}^{(\chi),-} + O(r^{-5/2})~,
\end{gathered}
\end{equation}
with
\begin{equation}\label{fieldas2}
\begin{gathered}
a_\mu = a_{\mu}^{({\rm nn})} + \frac{1}{\mu r} a_{\mu}^{({\rm n})} + O(r^{-2}) ~, \qquad a_r = \frac{1}{\mu^2 r^2} a_{r}^{({\rm n})} + O(r^{-3})~, \\
\phi^{z_i} = \phi_{({\rm nn})}^{z_i} + \frac{1}{\mu r} \phi_{({\rm n})}^{z_i} + O(1/r^2) ~, \\
\vec{\XX}(x^\nu,r) = \vec{\XX}_{({\rm n})} + \frac{1}{\mu r} \vec{\XX}_{({\rm nn})}' + O(r^{-2})~, \\
\Psi_{j=1/2}^{(\chi)} = \frac{1}{(\mu r)^{3/2}} \left( \uppsi_{0}^{({\rm nn})} + (\hat{r} \cdot \vec{\Gamma}_{(r)}) \uppsi_{0}^{({\rm n})} \right) + O(r^{-5/2})~.
\end{gathered}
\end{equation}
The non-normalizable data $(a_{\mu}^{({\rm nn})}, \phi^{z_i}_{({\rm nn})}, \vec{\XX}_{({\rm nn})}, \uppsi_{0}^{({\rm nn})}; \Phi_{\infty}^y)$ and the 't Hooft flux $P$ are to be held fixed while all remaining modes vary.  This is consistent with the variational principle for the Legendre transformed action, \eqref{Shol}.  Here $\vec{\XX}_{({\rm nn})}$ is related to $\vec{\XX}$ through \eqref{Xmom}.

\subsection{Conservation of supersymmetry on the boundary}\label{sec:bndrySUSY}

With the aid of the field asymptotics \eqref{fieldas1}, \eqref{fieldas2} we can now complete the supersymmetry analysis.  In section \ref{sec:ftsusy} we established that the variation of $S_{\rm ym}$ with respect to the supersymmetry transformations, \eqref{SUSY}, reduces to a set of boundary terms:
\begin{align}
\delta_{\vareps} S_{\rm ym} =&~ -\frac{i}{g_{{\rm ym}_6}^2} \int \ed^6 x \sqrt{-g_6} \nabla_a \BB^a + \delta_{\vareps} S^{\rm bndry}~,
\end{align}
where the boundary current $\BB^a$ is given in \eqref{Bcurrent} and the boundary action in \eqref{Sbndry}.  We will assume the fields are sufficiently regular at $r \to 0$ so that there is no contribution for this component of the boundary.  Then the first term above reduces to the contribution from the $r \to \infty$ boundary.  Defining $\BB^{\rm bndry}$ such that
\begin{equation}\label{Bbndrydef}
\delta_{\vareps} S^{\rm bndry} = \frac{-i}{g_{{\rm ym}_6}^2} \int_{~~\,\mathclap{\pd M_6}} \ed^5 x \sqrt{-g_{(\pd)} } \BB^{\rm bndry}~,
\end{equation}
we have
\begin{equation}\label{dSBB}
\delta_{\vareps} S_{\rm ym} = \frac{-i}{g_{{\rm ym}_6}^2} \int_{~~\,\mathclap{\pd M_6}} \ed^5 x \sqrt{-g_{(\pd)} } \left( \BB^{\underline{r}} + \BB^{\rm bndry} \right)~.
\end{equation}
Recall that the boundary measure is given in \eqref{bmeasure} and goes as $r^3$ as $r \to \infty$.

The large $r$ behavior of $\BB^{\underline{r}} + \BB^{\rm bndry}$ following from the field asymptotics \eqref{fieldas1} is analyzed in appendix \ref{app:bsusy}.  It is useful to separate the contribution from the variation of the fermion from the rest.  We eventually obtain the following expression:
\begin{align}\label{BBsum}
& \BB^{\ur} + \BB^{\rm bndry} = \cr
& =   -\frac{1}{(\mu r)^3} \varepsbar_0(x^\nu) \Tr \left\{ \left( \half f_{\mu\nu}^{({\rm nn})} \Gamma^{\underline{\mu\nu}} + D_{\mu}^{({\rm nn})} \phi_{({\rm nn})}^{z_i} \Gamma^{\underline{\mu z_i}} + \half [\phi_{({\rm nn})}^{z_i},\phi_{({\rm nn})}^{z_j}] \Gamma^{\underline{z_i z_j}} \right) \uppsi_{0}^{({\rm n})} \right\} + \cr
& \quad + \frac{1}{(\mu r)^3} \Tr \left\{ \left[ \left(2\mu \etabar_0 \Gamma^{\uy} + \varepsbar_{0}(x^\nu) \Gamma^{\uy} (\Gamma^{\umu} D_{\mu}^{({\rm nn})} + \Gamma^{\underline{z_i}} \ad(\phi_{({\rm nn})}^{z_i}) ) \right) (\vec{\Gamma}_{(r)} \cdot \vec{\XX}_{({\rm n})}) \right] \uppsi_{0}^{({\rm nn})} \right\}  \cr
& \quad +  \half \Tr \left\{ \Psibar \Gamma^{\underline{r}} \left(\mathbbm{1} + \Gamma^{\underline{r\theta\phi y}} \right) \delta_{\vareps} \Psi \right\} + O(r^{-7/2})~,
\end{align}
where $f_{\mu\nu}^{({\rm nn})} = 2 \pd_{[\mu},a_{\nu]}^{({\rm nn})} + [a_{\mu}^{({\rm nn})}, a_{\nu}^{({\rm nn})}]$ is the fieldstrength of the boundary gauge field, $D_{\mu}^{({\rm nn})} = \pd_\mu + \ad(a_{\mu}^{({\rm nn})})$ is the corresponding covariant derivative, and the terms we neglected decay sufficiently fast so as not to contribute to \eqref{dSBB}.  We also introduced the shorthand 
\begin{equation}
\varepsbar_0(x^\mu) := \varepsbar_0 + \mu x^\nu \etabar_0 \Gamma_{\unu} ~.
\end{equation}

The leading $O(r^{-3/2})$ asymptotics of $(\delta_{\vareps} \Psi)^+$, which encode the supersymmetry variation of $\uppsi_{0}^{({\rm nn})}$, turn out to be
\begin{align}\label{Psivaras}
(\delta_{\vareps} \Psi)^+ =&~ \frac{1}{(\mu r)^{3/2}} \delta_{\vareps} \uppsi_{0}^{({\rm nn})} + O(r^{-2}) \cr
=& \frac{1}{(\mu r)^{3/2}} \bigg[ \half f_{\mu\nu}^{({\rm nn})} \Gamma^{\underline{\mu\nu}} + D_{\mu}^{({\rm nn})} \phi_{({\rm nn})}^{z_i} \Gamma^{\underline{\mu z_i}} + \cr
&~ \qquad \qquad  + \half [\phi_{({\rm nn})}^{z_i}, \phi_{({\rm nn})}^{z_j}] \Gamma^{\underline{z_i z_j}} - \Gamma^{\uy} \vec{\Gamma}_{(r)} \cdot  \vec{\XX}_{({\rm nn})} \bigg] \vareps_0(x^\nu) +  O(r^{-2})~.
\end{align}
Three terms cancel when $(\delta_\vareps \Psi)^+$ is plugged back into \eqref{BBsum}, but the $\vec{\XX}_{({\rm nn})}$ term remains and gives an additional finite contribution to $\delta_{\vareps} S_{\rm ym}$, \eqref{dSBB}.  All terms that contribute in the $r \to \infty$ limit are independent of $\theta,\phi$ so we can trivially integrate over the $S^2$, leading to
\begin{align}\label{dSgivesP}
\delta_{\vareps} S_{\rm ym} =&~ \frac{4\pi i}{g_{{\rm ym}_6}^2\mu^2 } \int_{\mathbb{R}^{1,2}} \ed^3 x \Tr \bigg\{ \varepsbar_0(x^\nu) \Gamma^{\uy} (\vec{\Gamma}_{(r)} \cdot \vec{\XX}_{({\rm nn})}) \uppsi_{0}^{({\rm n})} + \cr
&~ - \left[ \left(2\mu \etabar_0 \Gamma^{\uy} + \varepsbar_{0}(x^\nu) \Gamma^{\uy} (\Gamma^{\umu} D_{\mu}^{({\rm nn})} + \Gamma^{\underline{z_i}} \ad(\phi_{({\rm nn})}^{z_i}) ) \right) (\vec{\Gamma}_{(r)} \cdot \vec{\XX}_{({\rm n})}) \right] \uppsi_{0}^{({\rm nn})}  \bigg\} ~. \qquad
\end{align}
This is our final result for the supersymmetry variation of $S_{\rm ym}$.  We see that supersymmetry invariance of $S_{\rm ym}$ can be achieved by taking, for example, $\vec{\XX}_{({\rm nn})} = \uppsi_{0}^{({\rm nn})} = 0$.

However it is the Legendre transformed action, \eqref{Shol}, that is relevant for the holographic dual.  We wish to show, therefore, that the supersymmetry variation of the $\vec{\XX}_{({\rm n})} \cdot \vec{\XX}_{({\rm nn})}$ term cancels \eqref{dSgivesP}.  The variation of the boundary data is determined from the large $r$ asymptotics of the variations of the bosons in \eqref{SUSY}.  In order to extract the necessary information, one must describe the asymptotic behavior of the fermi field $\Psi$ in some detail.  Specifically, in order to extract the supersymmetry variation of $\vec{\XX}_{({\rm nn})}'$ in $\vec{\XX}_{({\rm nn})}$, we will need to determine the first subleading corrections in \eqref{fieldasPsi}.  This is done by solving the fermion equation of motion asymptotically, in terms of the boundary data.  Note it is to be expected that the equations of motion must be used, as the Legendre transform in \eqref{Shol} takes place at the level of the on-shell action, viewed as a functional of boundary data.  The solution is obtained in appendix \ref{app:fermsol} and takes the form
\begin{align}\label{Psisolcart}
\uppsi^+ =&~ \frac{1}{(\mu r)^{3/2}} \bigg\{ \left[ \mathbbm{1} + \frac{1}{\mu^2 r} \left( \ad(a_{r}^{({\rm n})}) -\Gamma^{\uy} \vec{\Gamma}_{(r)} \cdot \ad(\vec{\XX}_{({\rm n})}) \right) \right] \uppsi_{0}^{({\rm nn})}(x^\mu) + \cr
&~ \qquad \qquad \qquad \qquad - \frac{1}{\mu^2 r} \left[ \Gamma^{\umu} D_{\mu}^{({\rm nn})} + \Gamma^{\underline{z_i}} \ad(\phi_{\rm nn}^{z_i}) \right] \uppsi^{({\rm n})}_0(x^\mu) \bigg\} + O(r^{-7/2})~, \cr
\uppsi^- =&~ \frac{1}{(\mu r)^{3/2}} \Gamma^{\underline{r_3}} \bigg\{ \left[ \mathbbm{1} + \frac{1}{\mu^2 r} \left( \ad(a_{r}^{({\rm n})}) + \Gamma^{\uy} \vec{\Gamma}_{(r)} \cdot \ad(\vec{\XX}_{({\rm n})}) \right) \right] \uppsi^{({\rm n})}_0(x^\mu) + \cr
&~ \qquad \qquad \qquad \qquad + \frac{1}{\mu^2 r} \left[ \Gamma^{\umu} D_{\mu}^{({\rm nn})} + \Gamma^{\underline{z_i}} \ad(\phi_{\rm nn}^{z_i}) \right] \uppsi^{({\rm nn})}_0(x^\mu) \bigg\} + O(r^{-7/2})~. \qquad 
\end{align}

Equations \eqref{Psisolcart} and \eqref{Psi2uppsi} in conjunction with \eqref{braneKScart} can be straightforwardly used to obtain the supersymmetry variations of the bosonic boundary data.  One simply compares the asymptotic expansions of the left- and right-hand sides of \eqref{SUSY} order by order.  The results are\footnote{One simply finds $\delta_{\vareps} \Phi_{\infty}^y = 0$ under our assumptions.  It would be sourced by the non-normalizable mode of a massive fermi field on $AdS_4$.  This is consistent with findings in \cite{DeWolfe:2001pq}, which identified the dual operator as the lowest component in a different supermultiplet associated with a higher KK mode of the $S^2$ expansion.}
\begin{align}\label{boundarydatavar}
\delta_{\vareps} a_{\mu}^{({\rm nn})} =&~  -i \varepsbar_0(x^\nu) \Gamma_{\umu} \uppsi^{({\rm nn})}_0 ~, \cr
\delta_{\vareps} \phi_{({\rm nn})}^{z_i} =&~ -i \varepsbar_0(x^\nu) \Gamma^{\underline{z_i}} \uppsi^{({\rm nn})}_0~, 
\end{align}
and
\begin{align}
\delta_{\vareps} a_{r}^{({\rm n})} =&~ - i \varepsbar_0(x^\nu)  \uppsi^{({\rm n})}_0~, \cr
\delta_{\vareps} \vec{\XX}_{({\rm n})} =&~ - i \varepsbar_0(x^\nu) \Gamma^{\uy} \vec{\Gamma}_{(r)} \,  \uppsi^{({\rm n})}_0~, \cr
\delta_{\vareps} \vec{\XX}_{({\rm nn})}' =&~ i \left[ \etabar_0 \Gamma^{\uy} \vec{\Gamma}_{(r)} - \mu^{-1} \varepsbar_0(x^\nu) \Gamma^{\uy} \vec{\Gamma}_{(r)} \left( \Gamma^{\umu} D_{\mu}^{({\rm nn})} + \Gamma^{\underline{z_i}} \ad(\phi_{({\rm nn})}^{z_i}) \right) \right]  \uppsi^{({\rm nn})}_0 + \cr
&~ -  i\mu^{-1} \varepsbar_0(x^\nu) \left[ \Gamma^{\uy} \vec{\Gamma}_{(r)} \ad(a_{r}^{({\rm n})}) + \Gamma^{\uy} \vec{\Gamma}_{(r)} \times \ad(\vec{\XX}_{({\rm n})}) - \ad(\vec{\XX}_{({\rm n})}) \right]  \uppsi^{({\rm n})}_0 ~. \qquad
\end{align}
With the aid of the last three and \eqref{Xmom}, one can show that the variation of $\vec{\XX}_{({\rm nn})}$ is sourced by $ \uppsi^{({\rm nn})}_0$ only:
\begin{align}\label{XXnnvar}
\delta_{\vareps} \vec{\XX}_{({\rm nn})} =&~ - i \left[\mu \etabar_0 \Gamma^{\uy} \vec{\Gamma}_{(r)} - \varepsbar_0(x^\nu) \Gamma^{\uy} \vec{\Gamma}_{(r)} \left( \Gamma^{\umu} D_{\mu}^{({\rm nn})} + \Gamma^{\underline{z_i}} \ad(\phi_{({\rm nn})}^{z_i}) \right) \right]  \uppsi^{({\rm nn})}_0 ~.
\end{align}

Notice that $\{a_{\mu}^{({\rm nn})},\phi_{({\rm nn})}^{z_i},\Psi_{({\rm nn})}^0,\vec{\XX}_{({\rm nn})}\}$ is a closed system under the supersymmetry transformations, \eqref{Psivaras}, \eqref{boundarydatavar}, and \eqref{XXnnvar}.  This cements our identification of the non-normalizable modes of the massless fermions.  In fact these are the transformations of an off-shell 3D $\NN = 4$ vector-multiplet, with $\vec{\XX}_{({\rm nn})}$ playing the role of the triplet of auxiliary fields.\footnote{ABR thanks Dan Butter for an enlightening discussion on this point.}  (See \eg\ \cite{Banerjee:2015uee}.)  This is consistent with the supersymmetry discussion in \cite{DeWolfe:2001pq} which identifies $SU(2)_z$ as the $SU(2)_V$ under which the triplet of scalars in an $\NN = 4$ vector-multiplet is charged.  The auxiliary fields of the vector-multiplet transform as a triplet of the other $SU(2)_r \equiv SU(2)_H$, under which the scalars in an $\NN = 4$ hypermultiplet are charged.  The non-normalizable data $\{a_{\mu}^{({\rm nn})},\phi_{({\rm nn})}^{z_i},\Psi_{({\rm nn})}^0,\vec{\XX}_{({\rm nn})}\}$ is a vector-multiplet of sources for the bottom KK multiplet of relevant operators in the dCFT.\footnote{See the last column in the table on page 32 of \cite{DeWolfe:2001pq}.  The map of notation for the modes is $a_\mu \to b_\mu$, $\phi^{z_i} \to \psi$, and $\vec{\XX} \to (b+z)^{(-)}$.}

Using these results, the variation of the $\vec{\XX}_{({\rm n})} \cdot \vec{\XX}_{({\rm nn})}$ term in \eqref{Shol} takes the form
\begin{align}
& \delta_{\vareps} \left( \frac{4\pi}{g_{{\rm ym}_6}^2\mu^2} \int_{\mathbbm{R}^{1,2}} \ed^3 x \Tr \left\{ \vec{\XX}_{({\rm nn})} \cdot \vec{\XX}_{({\rm n})} \right\}  \right) = \cr
& = - \frac{4\pi i}{g_{{\rm ym}_6}^2\mu^2} \int_{\mathbbm{R}^{1,2}} \ed^3 x \Tr \bigg\{ \varepsbar_0(x^\nu) \Gamma^{\uy} (\vec{\Gamma}_{(r)} \cdot \vec{\XX}_{({\rm nn})}) \uppsi_{0}^{({\rm n})} + \cr
&~~~  +  \left[\mu \etabar_0 \Gamma^{\uy} (\vec{\Gamma}_{(r)} \cdot \vec{\XX}_{({\rm n})}) - \varepsbar_0(x^\nu) \Gamma^{\uy} (\vec{\Gamma}_{(r)} \cdot \vec{\XX}_{({\rm n})}) \left( \Gamma^{\umu} D_{\mu}^{({\rm nn})} + \Gamma^{\underline{z_i}} \ad(\phi_{({\rm nn})}^{z_i}) \right) \right]  \uppsi^{({\rm nn})}_0 \bigg\}~. \qquad
\end{align}
Adding this to \eqref{dSgivesP}, one sees that the $\uppsi_{0}^{({\rm n})}$ term cancels.  Remarkably, the $\uppsi_{0}^{({\rm nn})}$ terms combine into a total derivative, so that
\begin{equation}
\delta_{\vareps} S_{\rm hol} = \frac{4\pi i}{g_{{\rm ym}_6}^2\mu^2} \int_{\mathbbm{R}^{1,2}} \ed^3 x \, \pd_\mu \Tr \left\{ \varepsbar_0(x^\nu) \Gamma^{\umu} \Gamma^{\uy} (\vec{\Gamma}_{(r)} \cdot \vec{\XX}_{({\rm n})}) \uppsi_{0}^{({\rm nn})} \right\}~.
\end{equation}
In particular we used $\pd_\mu \varepsbar_{0}(x^\nu) \Gamma^{\umu} = 3 \mu \etabar_0$.  Hence

\begin{equation}\label{Sholinvariance}
\delta_{\vareps} S_{\rm hol} = 0~,
\end{equation}
provided we assume sufficient fall-off conditions on $\uppsi_{0}^{({\rm nn})}$ as we go to infinity in the Minkowski space on the boundary.

\section{A consistent truncation to $\NN = 4$ SYM on $AdS_4$}\label{sec:ct}

In this section we show that the six-dimensional Yang--Mills theory can be consistently truncated to a four-dimensional theory by keeping the modes $(a_\mu, a_r, \phi^{z_i},\vec{\XX}, \uppsi)$.  When $z_0 = 0$ this gives a consistent truncation of the six-dimensional theory on $AdS_4 \times S^2$ to maximally supersymmetric $\NN = 4$ SYM on $AdS_4$.  Turning on $z_0$ yields a one-parameter family of consistent truncations to Yang--Mills on asymptotically $AdS_4$ spaces preserving half of the supersymmetry.  The gauge group of the reduced theory is generated by the centralizer $C(P,\Phi_{\infty}^y) \subseteq \mathfrak{g}$.\footnote{or more generally, the commutant of the vacuum monopole configuration we are expanding around.  See section \ref{sec:fluxvacua}.}  In the extreme cases this will be the full group if $\Phi_{\infty}^y,P$ are vanishing, or a Cartan torus if either is generic.

Having a consistent truncation means that every solution to the equations of motion of the lower-dimensional theory can be uplifted to a solution of the parent theory.  In particular, the fields we want to keep must not source the modes we want to discard in the full nonlinear equations of motion.  

Consistent truncations of gauge theories on coset spaces are implicit in ansatze  for instanton and monopole configurations that are based on spherical symmetry. See \cite{tHooft:1974kcl,Polyakov:1974ek,Belavin:1975fg,Witten:1976ck}. These ideas were formalized and generalized in \cite{Forgacs:1979zs}.  In this approach one identifies the action of the isometry group of the internal space with the action of gauge transformations on the fields, and as a result the gauge group of the truncated theory is reduced.  

The consistent truncation described here is different in that the gauge group need not be reduced.  We can start with any simple gauge group in the parent theory and it need not be reduced at all in the truncated theory.  This is despite the fact that some of the modes we keep have nontrivial dependence on the two-sphere, namely the fermions and the triplet of scalars parameterized by $\vec{\XX}$.  We do not have a clear conceptual understanding of why this truncation works, but we observe that the Chern--Simons-like term in the 6D theory plays a crucial role.  The 6D Lagrangian evaluated on the reduction ansatz would not be an $S^2$ singlet without it, and the $S^2$ dependence of the 6D equations of motion would not factor out.  Recall this term originates from \eqref{MyersCS} and is present thanks to the Ramond-Ramond flux of the string background.

The ansatz for the 6D degrees of freedom in terms of the 4D degrees of freedom is
\begin{equation}\label{truncation}
\begin{gathered}
A_{\mu,r} = a_{\mu,r}(x^\nu,r)~, \qquad \Phi^{z_i} = \phi^{z_i}(x^\nu,r)~, \\
A_\theta = - \frac{1}{\mu^2 r} \hat{\phi} \cdot \vec{\XX}(x^\nu,r)~, \qquad A_\phi = \frac{P}{2}(\pm 1 - \cos{\theta}) + \frac{\sin{\theta}}{\mu^2 r} \hat{\theta} \cdot \vec{\XX}(x^\nu,r)~, \\
\Phi^y = \Phi_{\infty}^y - \frac{P}{2 r} + \frac{1}{\mu^2 r^2} \hat{r} \cdot \vec{\XX}(x^\nu,r) ~, \\
\Psi = h_{S^2}(\theta,\phi) \uppsi^+ + h_{S^2}(-\theta,\phi) \uppsi^- ~,
\end{gathered}
\end{equation}
where $(a_{\mu,r},\phi^{z_i},\vec{\XX},\uppsi)$ are taken to commute with $\Phi_{\infty}^y$ and $P$.\footnote{The $\Phi_{\infty}^y$ and $P$ terms in $(A_{r_i}, \Phi^y)$ could be generalized to any of the monopole vacua described in section \ref{sec:fluxvacua} provided we restrict $(a_{\mu,r},\phi^{z_i},\vec{\XX},\uppsi)$ to the commutant of the vacuum configuration.}  Note we are using \eqref{Psi2uppsiS2} for the fermion ansatz.  It will be much more convenient in this section to work in a natural basis with respect to the $S^2$ frame in which $\Gamma^{\ur},\Gamma^{\utheta},\Gamma^{\uphi}$ are constant.

In the following we combine the $AdS_4$ directions into a single notation, 
\begin{equation}
x^{\hat{\mu}} = (x^\mu,r) ~,
\end{equation}
with indices $\hat{\mu}$ running over $0,1,2,r$, and lowered with the metric
\begin{equation}\label{4Dmet}
g_{\hat{\mu}\hat{\nu}} \ed x^{\hat{\mu}} \ed x^{\hat{\nu}} = \mu^2 (r^2 + z_{0}^2) \eta_{\mu\nu} \ed x^\mu \ed x^\nu + \frac{ \ed r^2}{\mu^2 (r^2 + z_{0}^2)} ~.
\end{equation}
We also introduce a new transverse metric that will be used to contract the $j=1$ triplet indices of $\XX^i$:
\begin{equation}\label{hypermet}
\Gbar_{h_ih_j} := \frac{(r^2 + z_{0}^2)}{\mu^2 r^4} \delta_{ij} ~,
\end{equation}
and collect the transverse metric and scalars as follows:
\begin{equation}
\phi^{I} = (\phi^{z_1},\phi^{z_2},\phi^{z_3}, \XX^1, \XX^2, \XX^3)~, \qquad \Gbar_{IJ} = \diag ( \Gbar_{z_i z_j}, \Gbar_{h_ih_j})~,
\end{equation}
where $I,J = 1,\ldots 6$.  The letter $h$ is for `hypermultiplet' in the new triplet of indices.

Let us denote the right-hand sides of the 6D equations of motion \eqref{eoms} by ${\rm EOM}^a$, ${\rm EOM}_y$, ${\rm EOM}_{z_i}$, and ${\rm EOM}_\Psi$ respectively.  We insert $A = a + \delta A$, \etc.\ into these equations, where $\delta A$ collectively represents all remaining degrees of freedom that we wish to discard.  As an intermediate step, in appendix \ref{app:trunc} one can find expressions for the components of the fieldstrength and covariant derivatives on the ansatz \ref{truncation}.  Some tedious but straightforward computations lead to
\begin{align}\label{Etoe}
& {\rm EOM}^{\hat{\mu}} = {\rm eom}^{\hat{\mu}} + O(\delta A)~,  \quad {\rm EOM}_{z_i} = {\rm eom}_{z_i} + O(\delta A)~, \cr
& {\rm EOM}^{\theta} = - \mu^2 r \, \hat{\phi} \cdot \overrightarrow{{\rm eom}}_{(h)} + O(\delta A)~, \cr  
& {\rm EOM}^{\phi} = \frac{\mu^2 r}{\sin{\theta}} \, \hat{\theta} \cdot \overrightarrow{{\rm eom}}_{(h)} + O(\delta A)~, \cr 
& {\rm EOM}_{y} = \mu^2 r^2 \, \hat{r} \cdot \overrightarrow{{\rm eom}}_{(h)} + O(\delta A)~, \cr
& {\rm EOM}_{\Psi} = h_{S^2}(\theta,\phi) {\rm eom}_{\uppsi}^+ + h_{S^2}(-\theta,\phi) {\rm eom}_{\uppsi}^- + O(\delta A) ~,
\end{align}
where $\overrightarrow{{\rm eom}}_{(h)} = ({\rm eom}_{h_1},{\rm eom}_{h_2},{\rm eom}_{h_3})$, and all of these ${\rm eom}$'s are $S^2$-independent quantities given by
\begin{align}\label{4Deoms}
{\rm eom}^{\hat{\nu}} :=&~ \frac{1}{r^2} D_{\hat{\mu}} (r^2 F^{\hat{\mu}\hat{\nu}})   - \Gbar_{IJ} [\phi^I, D^{\hat{\nu}}\phi^J] + \frac{i}{2} [\overbar{\uppsi}, \Gamma^{\hat{\nu}} \uppsi] ~, \cr
{\rm eom}_{z_i} :=&~ \frac{1}{r^2} D_{\hat{\mu}} (r^2 \Gbar_{z_i z_j} D^{\hat{\mu}} \phi^{z_j}) + \Gbar_{z_i z_j} \Gbar_{IJ} [\phi^{I}, [\phi^{J}, \phi^{z_j}]] + \frac{i}{2} [\overbar{\uppsi}, \Gamma_{z_i} \uppsi]~, \cr
{\rm eom}_{h_i} :=&~ \frac{1}{r^2} D_{\hat{\mu}} ( r^2 \Gbar_{h_ih_j} D^{\hat{\mu}} \XX^{j}) + \Gbar_{h_ih_j} \Gbar_{IJ} [\phi^{I}, [\phi^{J}, \XX^j]] -2 m_{\uppsi} \epsilon_{h_ih_jh_k} [\XX^j, \XX^k]  + \frac{i}{2} [\overbar{\uppsi},\Gamma_{h_i} \uppsi]~, \cr
{\rm eom}_{\uppsi} :=&~ \left( \Gamma^{\hat{\mu}} D_{\hat{\mu}} + m_{\uppsi} \Gamma^{\underline{h_1h_2h_3}} \right) \uppsi + \Gamma_I [\phi^I, \uppsi] ~.
\end{align}
Some details of the derivation of the fermion equation can be found in appendix \ref{app:fermsol}.  $F$ and $D$ are to be understood as the fieldstrength and covariant derivative associated with $a$.  The ($r$-dependent) fermion mass is
\begin{equation}\label{4Dfermimass}
m_{\uppsi} := - \frac{\mu z_{0}^2}{r (r^2 + z_{0}^2)^{1/2}}~,
\end{equation}
and an orthonormal frame is employed along the new $h_i$ directions such that
\begin{equation}
\epsilon_{h_ih_j h_k} = \frac{(r^2 + z_{0}^2)^{3/2}}{\mu^3 r^6} \, \tilde{\epsilon}_{ijk} ~, \qquad \Gamma_{h_i} = \frac{ (r^2 + z_{0}^2)^{1/2}}{\mu r^2} \, \Gamma^{\underline{h_i}}~,
\end{equation}
with $\tilde{\epsilon}_{123} = 1$ as usual.  The new triplet of gamma matrices is related to the old one by
\begin{equation}\label{hyperGtrip}
\vec{\Gamma}_{(h)} := (\Gamma^{\underline{h_1}},\Gamma^{\underline{h_2}},\Gamma^{\underline{h_3}}) := (\Gamma^{\underline{\phi}}, -\Gamma^{\underline{\theta}}, \Gamma^{\underline{y}})~.
\end{equation}

When acting on a spinor of definite $\Gamma^{\underline{r\theta\phi y}}$-chirality, this triplet is related to the $\vec{\Gamma}_{(r)}$ that have appeared before:
\begin{align}
\vec{\Gamma}_{(h)} \uppsi^{\pm} =&~  (\pm \Gamma^{\underline{r\theta y}}, \pm \Gamma^{\underline{r\phi  y}}, \Gamma^{\underline{r}} \Gamma^{\underline{r y}}) \uppsi^{\pm} = \Gamma^{\uy} \left\{ \begin{array}{l}  \Gamma^{\underline{r}} (\Gamma^{\utheta}, \Gamma^{\uphi}, \Gamma^{\ur}) \uppsi^+ \\[1ex] (\Gamma^{\utheta},\Gamma^{\uphi},\Gamma^{\ur}) \Gamma^{\underline{r}} \uppsi^-  \end{array} \right. \cr
=&~ \Gamma^{\uy} \left\{ \begin{array}{l} ( U \Gamma^{\underline{r_3}} \vec{\Gamma}_{(r)} U^{-1} ) \uppsi^+ \\[1ex]  ( U \vec{\Gamma}_{(r)} \Gamma^{\underline{r_3}} U^{-1} ) \uppsi^- \end{array} \right. ~,
\end{align}
where $U = h_{S^2}(\theta,\phi)^{-1}$ is the unitary transformation sending $(\Gamma^{\utheta},\Gamma^{\uphi},\Gamma^{\ur})$ to $\vec{\Gamma}_{(r)}$.  (See \eqref{CtoS}.)  For the consistent truncation, it is more natural, however, to use $\vec{\Gamma}_{(h)}$ and the $S^2$-based frame because this makes it clear that the directions associated with $\XX^i$ can be viewed as internal and independent of the radial direction $r$ of the four-dimensional spacetime.  Note also that $\Gamma^{\underline{h_1 h_2 h_3}} = \Gamma^{\underline{\theta\phi y}}$ is the same combination that appears in the 6D fermion mass term, \eqref{eoms}.

Since the ${\rm eom}$ in \eqref{4Deoms} are $S^2$-independent, the 6D equations of motion restricted to the truncation ansatz, \eqref{truncation}, we have
\begin{equation}
{\rm EOM} \bigg|_{\delta A = 0} = 0 \quad \iff \quad {\rm eom} = 0~.
\end{equation}
The truncation will be consistent iff the equations of motion of the reduced action are equivalent to ${\rm eom} = 0$.

We insert the truncation ansatz into the 6D Yang--Mills action in the form \eqref{SYM2}, with \eqref{primeL} and \eqref{Sfbndry}.  After some effort one finds that the density \eqref{primeL} can be put in the form
\begin{align}\label{Lprimetrunc}
\LL' \bigg|_{\delta A = 0} =&~ -\Tr \bigg\{ \frac{1}{4} F_{\hat{\mu}\hat{\nu}} F^{\hat{\mu}\hat{\nu}} + \half \Gbar_{IJ} D_{\hat{\mu}} \phi^I D^{\hat{\mu}} \phi^J + \frac{1}{4} \Gbar_{IJ} \Gbar_{KL} [\phi^I, \phi^K] [\phi^J,\phi^L] + \cr
& \qquad \qquad - \frac{(r^2 + z_{0}^2)^2}{2\mu^2 r^6} \tilde{\epsilon}_{ijk} (D_r \XX^i) [\XX^j, \XX^k] + \cr
& \qquad \qquad + \frac{i}{2} \overbar{\uppsi} \left( \Gamma^{\hat{\mu}} D_{\hat{\mu}} + m_{\uppsi} \Gamma^{\underline{h_1 h_2 h_3}} \right) \uppsi + \frac{i}{2} \overbar{\uppsi} \Gamma_I [\phi^I,\uppsi] \bigg\}~.
\end{align}
The 6D measure can be expressed in terms of the 4D measure associated with \eqref{4Dmet}, at the price of introducing an $r$-dependent 4D Yang--Mills coupling.  Since \eqref{Lprimetrunc} is an $S^2$-invariant we can carry out the integral over $S^2$ as well:
\begin{align}\label{6to4measure}
\frac{1}{g_{{\rm ym}_6}^2} \int \ed^6 x \sqrt{-g_6}  =&~ \frac{1}{g_{{\rm ym}_6}^2} \int \ed^3 x \ed r \ed \Omega \, r^2 \to  \int \ed^4 x \sqrt{-g_4} \, \frac{1}{g_{{\rm ym}_4}(r)^2}~,
\end{align}
where $g_4 = \det(g_{\hat{\mu}\hat{\nu}})$ is the determinant of \eqref{4Dmet} and
\begin{equation}\label{gYM4}
g_{{\rm ym}_4}(r)^2 = g_{{\rm ym}_6}^2 \frac{\mu^2 (r^2 + z_{0}^2)}{4\pi r^2} = \frac{\pi^{3/2} \sqrt{g_s N_c}}{N_c} \cdot \frac{(r^2 + z_{0}^2)}{r^2}~.
\end{equation}
Meanwhile the boundary action \eqref{Sfbndry} at $r = r_b$ reduces to
\begin{align}\label{Sfbndrytrunc}
S_{f}^{\rm bndry} \bigg|_{\delta A = 0} =&~ -\frac{i}{4} \int \ed^3 x \sqrt{-g_{(\pd 4)}} \frac{1}{g_{{\rm ym}_4}(r_b)^2} \Tr \left\{ \overbar{\uppsi} \Gamma^{\underline{h_1 h_2 h_3}} \uppsi \right\}~,
\end{align}
where $\sqrt{-g_{(\pd 4)}} = \mu^3 (r_{b}^2 + z_{0}^2)^{3/2}$ is the induced measure.  However the asymptotics of $\uppsi$ imply that the leading $r_b \to \infty$ behavior is finite and hence the $z_0$'s in the measure and coupling can be dropped in the limit.

The $\tilde{\epsilon}_{ijk} (D_r \XX^i) [\XX^j, \XX^j]$ term in \eqref{Lprimetrunc} descends from the Chern--Simons-like term in the 6D action.  The $\ad(a_r)$ part of $D_r$ actually drops out of this term by the Jacobi identity.  We then integrate by parts.  Keeping in mind the factor of $r^2$ out in front, \eqref{6to4measure}, the overall prefactor is asymptotically constant, or exactly constant if $z_0 = 0$.  Hence we get a boundary term and a bulk term that vanishes when $z_0 \to 0$ or $r \to \infty$.  The bulk term can be expressed in terms of $m_{\uppsi}$.  The boundary term is finite in the $r_b \to \infty$ limit and adds to the boundary term we already have, \eqref{Sfbndrytrunc}.  We write the result for the truncated action as follows:
\begin{equation}\label{Strunc}
S_{\rm trnc} := S_{\rm ym} \bigg|_{\delta A =0}  = \int \ed^4 x \sqrt{-g_4} \, \frac{1}{g_{{\rm ym}_4}(r)^2} \LL_{\rm trnc}' +S^{\prime\, {\rm bndry}}_{\rm trnc} ~,
\end{equation}
with
\begin{align}
\LL_{\rm trnc}' :=& -\Tr \bigg\{ \frac{1}{4} F_{\hat{\mu}\hat{\nu}} F^{\hat{\mu}\hat{\nu}} + \half \Gbar_{IJ} D_{\hat{\mu}} \phi^I D^{\hat{\mu}} \phi^J + \frac{1}{4} \Gbar_{IJ} \Gbar_{KL} [\phi^I, \phi^K] [\phi^J,\phi^L] + \cr
& \qquad \quad + 2 m_{\uppsi} \epsilon_{h_ih_j h_k} \XX^i [\XX^j, \XX^k] + \frac{i}{2} \overbar{\uppsi} \left( \Gamma^{\hat{\mu}} D_{\hat{\mu}} + m_{\uppsi} \Gamma^{\underline{h_1 h_2 h_3}} \right) \uppsi + \frac{i}{2} \overbar{\uppsi} \Gamma_I [\phi^I,\uppsi] \bigg\}~, \cr \quad
\end{align}
and
\begin{equation}
S^{\prime\, {\rm bndry}}_{\rm trnc} = \frac{1}{g_{{\rm ym}_4}^2} \int \ed^3 x \sqrt{-g_{(\pd 4)}} \Tr \left\{ \epsilon_{h_i h_j h_k} \XX^i [\XX^j, \XX^k] - \frac{i}{4} \overbar{\uppsi} \Gamma^{\underline{h_1 h_2 h_3}} \uppsi \right\} ~,
\end{equation}
where $g_{{\rm ym}_4}^2$ denotes the $r \to \infty$ limit of \eqref{gYM4}.  The equations of motion one derives from \eqref{Strunc} are precisely the vanishing of the ${\rm eom}$'s, \eqref{4Deoms}:
\begin{equation}
\delta S_{\rm trnc} = 0 \quad \iff \quad {\rm eom} = 0~.
\end{equation}
Hence the truncation \eqref{truncation} is consistent.

We can also express the truncated action in terms of scalars with canonical kinetic terms by setting
\begin{align}\label{canonicalXX}
& \phi^{z_i} = \mu (r^2 + z_{0}^2)^{1/2} \phi^{\underline{z_i}} ~, \qquad \XX^{i} = \frac{\mu r^2}{(r^2 + z_{0}^2)^{1/2}} \XX^{\underline{i}} ~.
\end{align}
This introduces both mass terms and additional boundary terms.  The latter are linearly divergent and necessary to cancel divergences in the variational principle, energy, \etc, if we use this form of the action.  The result is
\begin{equation}\label{Strunc2}
S_{\rm trnc} = \int \ed^4 x \sqrt{-g_4} \, \frac{1}{g_{{\rm ym}_4}(r)^2} \LL_{\rm trnc} +S^{{\rm bndry}}_{\rm trnc} ~,
\end{equation}
with
\begin{align}
\LL_{\rm trnc} :=& - \Tr \bigg\{ \frac{1}{4} F_{\hat{\mu}\hat{\nu}} F^{\hat{\mu}\hat{\nu}} + \half D_{\hat{\mu}} \phi^{\underline{I}} D^{\hat{\mu}} \phi^{\underline{I}} + \half \left( m_{z}^2 (\phi^{\underline{z_i}})^2 + m_{\XX}^2 (\XX^{\underline{i}})^2 \right)+  \frac{1}{4} [\phi^{\underline{I}}, \phi^{\underline{J}}]^2 +  \cr
&~ \qquad \quad  + 4 m_{\uppsi} \XX^{\underline{1}} [\XX^{\underline{2}}, \XX^{\underline{3}}]  + \frac{i}{2}\overbar{\uppsi} \left( \Gamma^{\hat{\mu}} D_{\hat{\mu}} + m_{\uppsi} \Gamma^{\underline{h_1 h_2 h_3}} \right)\uppsi + \frac{i}{2} \overbar{\uppsi} \Gamma^{\underline{I}} [\phi^{\uI}, \uppsi]   \bigg\}~, \qquad
\end{align}
and
\begin{equation}
S^{\rm bndry}_{\rm trnc} = \frac{1}{g_{{\rm ym}_4}^2} \int \ed^3 x \sqrt{-g_{(\pd 4)}} \Tr \left\{ 2 \XX^{\underline{1}} [\XX^{\underline{2}},\XX^{\underline{3}}] - \frac{\mu}{2} (\phi^{\underline{I}})^2 - \frac{i}{4} \overbar{\uppsi} \Gamma^{\underline{h_1 h_2 h_3}} \uppsi \right\} ~,
\end{equation}
where the new $r$-dependent masses are
\begin{equation}
m_{z}^2(r) := - \mu^2 \left(\frac{2 r^2 + 3 z_{0}^2}{r^2 + z_{0}^2}\right) ~, \qquad m_{\XX}^2(r) := - \mu^2 \left( \frac{ 2 r^4 + r^2 z_{0}^2 - 2 z_{0}^4}{r^2 (r^2 + z_{0}^2)} \right)~. \qquad
\end{equation}

Finally we must take into account the Legendre transform, \eqref{Shol}, which involves only the degrees of freedom we are keeping in the truncation.  Thus we define
\begin{equation}
S_{\rm trnc}^{\rm hol} := S_{\rm trnc} + \frac{1}{g_{{\rm ym}_4}^2} \int \ed^3 x \Tr \left\{ \vec{\XX}_{({\rm nn})} \cdot \vec{\XX}_{({\rm n})} \right\}~.
\end{equation}
Now recall that $\XX_{({\rm n})}^i$ is the boundary value of $\XX^i$, while $\XX_{({\rm nn})}^i$ can be expressed in terms of the boundary values of $\XX^i$ and $D_r \XX^i$ via \eqref{Xmom}.  Taking into account the rescaling \eqref{canonicalXX} we can write
\begin{align}
\vec{\XX}_{({\rm nn})} \cdot \vec{\XX}_{({\rm n})} =&~ \lim_{r \to \infty} \left( \sqrt{-g_{(\pd 4)}} \Tr \left\{ \mu r \XX^{\underline{i}} D_r \XX^{\underline{i}} + \mu (\XX^{\underline{i}})^2 - 3 \XX^{\underline{1}} [\XX^{\underline{2}},\XX^{\underline{3}}] \right\} \right)~,
\end{align}
and hence
\begin{equation}
S^{\rm hol}_{\rm trnc} = \int \ed^4 x \sqrt{-g_4} \, \frac{1}{g_{{\rm ym}_4}(r)^2} \LL_{\rm trnc} +S^{\textrm{hol-bndry}}_{\rm trnc} ~,
\end{equation}
where the boundary action is
\begin{align}\label{Sholbndry}
S^{\textrm{hol-bndry}}_{\rm trnc} =&~ \frac{1}{g_{{\rm ym}_4}^2} \int \ed^3 x \sqrt{-g_{(\pd 4)}} \Tr \bigg\{ \mu r \XX^{\underline{i}} D_r \XX^{\underline{i}} + \frac{\mu}{2} (\XX^{\underline{i}})^2 - \frac{\mu}{2} (\phi^{\underline{z_i}})^2 + \cr
&~ \qquad \qquad \qquad \qquad \qquad  \quad - 2 \XX^{\underline{1}} [\XX^{\underline{2}},\XX^{\underline{3}}]  - \frac{i}{4} \overbar{\uppsi} \Gamma^{\underline{h_1 h_2 h_3}} \uppsi \bigg\} ~.
\end{align}
We have expressed the result in this fashion because \eqref{Sholbndry} is precisely the set of boundary terms obtained recently in \cite{Mezei:2017kmw}.  These authors carried out a supersymmetric localization computation in maximally supersymmetric Yang--Mills on $AdS_4$, and found that this set of terms is necessary and sufficient for the preservation of supersymmetry in the presence of non-normalizable mode boundary data.  We have arrived at the same set of terms via an independent analysis.

One can check that the supersymmetry variations close on the truncation \eqref{truncation}, so the reduced theory enjoys the same amount of supersymmetry as the parent theory.  Indeed, when $z_0 =0$ we see that the background is $AdS_4$, the Yang--Mills coupling is constant, $m_{\uppsi}$ vanishes, the bosonic masses go to the conformally coupled value, $m^2 = -2\mu^2$, and the bulk Lagrangian takes the $\NN = 4$ maximally supersymmetric form
\begin{align}\label{N4MSYM}
\lim_{z_0 \to 0} \LL_{\rm trnc} =&~ - \Tr \bigg\{ \frac{1}{4} F_{\hat{\mu}\hat{\nu}} F^{\hat{\mu}\hat{\nu}} + \half D_{\hat{\mu}} \phi^{\uI} D^{\hat{\mu}} \phi^{\uI} - \mu^2 (\phi^{\uI})^2 + \frac{1}{4} [\phi^{\uI}, \phi^{\uJ}]^2 + \cr
&~ \qquad \quad +  \frac{i}{2} \overbar{\uppsi} \left( \Gamma^{\umu} D_{\umu} + \Gamma^{\uI} \ad(\phi_{\uI}) \right) \uppsi \bigg\}~.
\end{align}
Turning on $z_0$ gives a deformation that preserves half of the supersymmetries---namely those satisfying the 3D $\NN = 4$ super-Poincar\'e algebra.

We make two further comments on this result before concluding the section.  First, the discussion in section \ref{sec:YMleeft} shows that the on-shell six-dimensional action, $S^{\rm hol}$, viewed as a functional of the non-normalizable modes $(a_{\mu}^{({\rm nn})}, \phi^{z_i}_{({\rm nn})}, \uppsi_{0}^{({\rm nn})}, \vec{\XX}_{({\rm nn})})$, is the leading approximation to the generating functional of correlators amongst the lowest KK multiplet defect operators in the dual, in the regime $g_s N_c \gg N_c \gg 1$ and $N_f \ll N_c/\sqrt{g_s N_c}$.  In principle it is the generating functional for the full tower of KK multiplets of defect operators, but one would have to include the additional boundary terms required by holographic renormalization for the higher multiplets.  And more precisely, it is the generating functional for those operators transforming in the $\mathfrak{su}(N_f)$ subalgebra of the $\mathfrak{u}(N_f)$ flavor symmetry of the dual.  The results of this section imply that we can just as well use $S_{\rm trnc}^{\rm hol}$ as the leading-order generating functional in this regime.  The fact that the truncation is consistent implies that the effects of the higher KK multiplets on the lowest multiplet can only enter through loops.

Second, recall that, on the one hand, that ten-dimensional type IIB supergravity has a consistent truncation on $S^5$ to the five-dimensional maximally supersymmetric gauged supergravity of \cite{Gunaydin:1984qu,Pernici:1985ju}.\footnote{This had long been suspected since the work of \cite{deWit:1986oxb} on eleven-dimensional supergravity on $S^7$, and after a series of partial results in this direction it was recently demonstrated for type IIB on $S^5$ in full generality.  See \cite{Baguet:2015sma} for the complete nonlinear reduction ansatze, and for further discussion and references.}  On the other hand, we have just shown that the low-energy effective theory on the D5-branes has a consistent truncation in the regime where the coupling to gravity is suppressed.  Although we will not provide any evidence here, it seems natural to conjecture that the coupled IIB supergravity plus D5-brane system has a consistent truncation to maximal 5D gauged supergravity with a half-BPS codimension one defect hosting a $U(N_f)$ super-Yang--Mills theory.  Specifically, we suggest that the bosonic couplings of the defect theory to gravity should be obtained by applying the combined reduction ansatze for the supergravity and D5-brane modes to the non-abelian Myers action.  One should expand the Myers action to second order in the open string field variables, \eqref{osev}, but keep all orders in closed string fluctuations.\footnote{In the language of appendix \ref{app:fluc}, one makes the open string expansion as in \eqref{SDBIopenex} through \eqref{VDBI2open} and \eqref{SCSopenex} through \eqref{VCS2open}, and evaluates the result on the reduction ansatze.}

\section{Bogomolny equations and monopoles}\label{sec:monopoles}

We now turn to a preliminary study of BPS field configurations in both the 6D Yang--Mills theory and its 4D truncation.  These are solutions to the equations of motion that preserve some supersymmetry.  We are especially interested in finite energy configurations that can be interpreted as solitons in $AdS_4$, and that can serve as the starting point for a description of BPS particle states in the corresponding quantum (string) theory. 

\subsection{BPS equations as generalized self-duality equations}

We set the fermion to zero and derive a system of first order equations for the bosonic fields by demanding that the fermion's supersymmetry variation (parameterized by $\vareps$) vanish as well:
\begin{equation}\label{BPSPsicon}
 \delta_{\vareps} \Psi = \left[ \half F_{ab} \Gamma^{ab} + D_a \Phi_{\um} \Gamma^{a\um} + \half [\Phi_{\um},\Phi_{\un}]\Gamma^{\um\un} + M_{\Psi} \Gamma^{\underline{\theta\phi y}} \Gamma_{\um} \Phi^{\um} \right] \vareps \stackrel{!}{=} 0~.
\end{equation}
What conditions should we impose on $\varepsilon$ in order to satisfy this requirement?  One certainly expects that superconformal symmetries are broken in the presence of BPS states, and we can argue for this as follows.  If we expand $\delta_{\vareps} \Psi$ near $r \to \infty$, then we already know the leading order equation---namely that $\delta_\vareps \Psi_{({\rm nn})}^0$, \eqref{Psivaras}, should vanish:
\begin{equation}\label{leadingScon}
\left[  \half f_{\mu\nu}^{({\rm nn})} \Gamma^{\underline{\mu\nu}} + D_{\mu}^{({\rm nn})} \phi_{({\rm nn})}^{z_i} \Gamma^{\underline{\mu z_i}} + \half [\phi_{({\rm nn})}^{z_i}, \phi_{({\rm nn})}^{z_j}] \Gamma^{\underline{z_i z_j}} - \Gamma^{\uy} \vec{\Gamma}_{(r)} \cdot \vec{\XX}_{({\rm nn})} \right] \left(\vareps_0 - (\mu x^\nu) \Gamma_{\unu} \eta_0 \right) \stackrel{!}{=} 0~.
\end{equation}
We know from \eqref{vacsusy} that having a covariantly constant $\phi_{({\rm nn})}^{z_i} = \Phi_{\infty}^{z_i}$ breaks superconformal symmetry.  Then we see from \eqref{leadingScon} that having any other nontrivial boundary data will require a condition on $\vareps(x^\nu) = \vareps_0 - (\mu x^\nu) \Gamma_{\unu} \eta_0$ of the form $\Gamma^{\underline{M_1 M_2}} \vareps^+ = \pm \Gamma^{\underline{M_3 M_4}}\vareps^+$ where the $M$'s are different and at least one of them is $0,1$, or $2$.  This condition should hold for all values of $x^\mu$ and leads to incompatible projection conditions on $\eta_0$.  Therefore we set $\eta_0 =0$.

With $\eta_0$ set to zero, $\vareps$ becomes an eigenspinor of $\Gamma^{\underline{r\theta\phi y}} = \Gamma^{\underline{r_1 r_2 r_3 y}}$.  Then one can use the previous observation, \eqref{Phirescale}, to simplify \eqref{BPSPsicon}.  

Now, $\vareps_0$ has eight real independent parameters corresponding to the 3D $\NN = 4$ Poincar\'e supersymmetry of the boundary theory.  The most general BPS (particle) states in this theory are $1/4$-BPS, preserving two supersymmetries.  Therefore we should be able to impose two additional projections on $\vareps_0$ beyond $\Gamma^{\underline{r_1 r_2 r_3 y}}\, \vareps_0 = \vareps_0$, such that all three projections are mutually compatible and such that the $SO(2)$ little group of the 3D Lorentz group is preserved.  There is an $S^2 \times S^2$ family of choices parameterized by two fixed unit vectors $\hat{n}_{(r)} \in \mathbb{R}_{(r)}^3$ and $\hat{n}_{(z)} \in \mathbb{R}_{(z)}^3$, given by
\begin{equation}\label{susychoices}
\Gamma^{\underline{12}}\, \vareps_0 = \half \epsilon_{ijk} \hat{n}_{(r)}^i \Gamma^{\underline{r_j r_k}}\,  \vareps_0 = \half \epsilon_{ijk} \hat{n}_{(z)}^i \Gamma^{\underline{z_j z_k}}\, \vareps_0 ~.
\end{equation}
We need to keep track of the parameters $\hat{n}_{(r)},\hat{n}_{(z)}$.  The Bogomolny bound we derive below will depend on them, and they must be allowed to vary to achieve the strongest possible bound.  However there is no need to carry them around explicitly: we can always choose our $\{ r_i \}$ and $\{z_i\}$ axes so that the $\hat{n}$ point in the respective `3' directions, and then restore the dependence on them at the end using covariance.  Working in such a basis for now, the full set of projections satisfied by $\vareps_0$ are
\begin{align}\label{7proj}
& \Gamma^{\underline{r_1 r_2 r_3 y}} \vareps_0 = \vareps_0 ~, \qquad \Gamma^{\underline{1 2 r_3 y}} \vareps_0 =  \vareps_0 ~, \qquad \Gamma^{\underline{12 r_1 r_2}} \vareps_0 = -  \vareps_0~, \cr
& \Gamma^{\underline{12 z_1 z_2}} \vareps_0 = -  \vareps_0~, \qquad \Gamma^{\underline{r_1 r_2 z_1 z_2}} = -  \vareps_0~, \qquad \Gamma^{\underline{r_3 z_1 z_2 y}} \vareps_0 =  \vareps_0~, \qquad \Gamma^{\underline{0 z_3}} \vareps_0 =  \vareps_0 ~. \qquad
\end{align}

With the aid of \eqref{7proj} and \eqref{Phirescale} we collect terms in $\delta_\vareps \Psi$ that are proportional to equivalent $\Gamma^{\underline{M_1 M_2}} \, \vareps_0$ structures.  Setting the coefficient of each linearly independent structure to zero yields the following system of 22 BPS equations:
\begin{equation}\label{primaryBPS}
\begin{gathered}
 F_{12} + [\Phi^{z_1}, \Phi^{z_2}] + \mu^4 (r^2 + z_{0}^2)^2 \left( F_{r_1 r_2} - D_{r_3} \Phi^y\right) = 0 ~, \\
F_{1r_3} + D_2 \Phi^y  = 0~, \qquad F_{2 r_3} -  D_1 \Phi^y = 0~, \\
F_{r_1 r_3} + D_{r_2} \Phi^y = 0~, \qquad F_{r_2 r_3} - D_{r_1} \Phi^y = 0~, \\
F_{1 r_1} -  F_{2 r_2} = 0~, \qquad F_{2 r_1} +  F_{1 r_2} = 0~, \\
D_1 \Phi^{z_1} -  D_2 \Phi^{z_2} = 0~, \qquad D_2 \Phi^{z_1} +  D_1 \Phi^{z_2} = 0~, \\
D_{r_1} \Phi^{z_1} -  D_{r_2} \Phi^{z_2} = 0~, \qquad D_{r_1} \Phi^{z_2} +  D_{r_2} \Phi^{z_1} = 0~, \\
D_{r_3} \Phi^{z_1} +  [\Phi^y, \Phi^{z_2}] = 0~, \qquad D_{r_3} \Phi^{z_2} -  [\Phi^y, \Phi^{z_1}] = 0~,
\end{gathered}
\end{equation}
and
\begin{equation}\label{secondaryBPS}
\begin{gathered}
F_{p0} - D_p \Phi^{z_3} = 0~, \qquad F_{r_i0} -  D_{r_i} \Phi^{z_3} = 0~, \qquad D_0 \Phi^{z_p} -  [\Phi^{z_3}, \Phi^{z_p}] = 0~, \\
D_0 \Phi^y -  [\Phi^{z_3}, \Phi^y] = 0~, \qquad D_0 \Phi^{z_3} = 0~,
\end{gathered}
\end{equation}
where we recall that the indices $p,q = 1,2$.  We refer to the first set of equations, \eqref{primaryBPS}, as the \emph{primary} or \emph{magnetic} system of BPS equations and the second set, \eqref{secondaryBPS}, as the \emph{secondary} or \emph{electric} system of BPS equations.  This is due to their close analogy with BPS equations for 4D $\NN = 4,2$ theories that split in a similar fashion; see \eg\ \cite{Lee:1998nv,Gauntlett:1999vc}.  

The primary equations do not involve $A_0,\Phi^{z_3}$ and can be solved independently of these fields.  By working in \emph{generalized temporal gauge}, defined by
\begin{equation}\label{gtgauge}
A_0 = \Phi^{z_3}~,
\end{equation}
one sees that the secondary equations reduce to time-independence for the fields participating in the primary equations:
\begin{equation}
\pd_0 (A_p,A_{r_i},\Phi^{z_i}) = 0~.
\end{equation}
This in particular applies to the boundary data, $(a_{p}^{({\rm nn})},\vec{\XX}_{({\rm nn})},\phi^{z_p}_{({\rm nn})},\Phi_{\infty}^y;P)$.  

Given a solution to the primary system, the secondary BPS equations are satisfied with \eqref{gtgauge}.  We clearly need one further equation to specify an independent solution for $A_0$ or $\Phi^{z_3}$.  This can be taken to be the Gauss Law constraint, \eqref{localGauss}.  Using \eqref{secondaryBPS}, this equation takes the form of a gauge-covariant Laplacian, constructed from the solution $(A_p,A_{r_i},\Phi^{z_p},\Phi^y)$ of \eqref{primaryBPS}, annihilating $\Phi^{z_3}$.  

One nice way to repackage the primary BPS system introduces complex covariant derivatives,
\begin{align}\label{cmplxconnection}
& \DD_1 := \pd_1 + i \pd_2 + \ad(A_1 + i A_2) ~,  & \DD_3 := \pd_{r_3} + \ad(A_{r_3} - i \Phi^y) ~, \cr
& \DD_2 := \pd_{r_1} + i \pd_{r_2} + \ad(A_{r_1} + i A_{r_2}) ~,  & \DD_4 := \ad(\Phi^{z_1} + i \Phi^{z_2})~.
\end{align}
Then the last twelve of \eqref{primaryBPS} are equivalent to the six complex equations
\begin{equation}\label{cmplxHYM4}
[ \DD_{\bf p} , \DD_{\bf q} ] = 0~,
\end{equation}
for ${\bf p},{\bf q} = 1,\ldots ,4$.  The first equation can be written in the form
\begin{equation}
[\DD_1, \DD_{1}^\dag] + [\DD_4, \DD_{4}^\dag] + \mu^4 (r^2 + z_{0}^2)^2 \left( [\DD_2, \DD_{2}^\dag] + [\DD_3, \DD_{3}^\dag] \right) = 0~.
\end{equation}
The advantage of this approach, which applies to many of the standard self-duality type equations, is that a subset of the equations---\eqref{cmplxHYM4} in this case---have an extended gauge invariance.  They are invariant under gauge transformations of the complexified gauge group, $G_{\mathbb{C}}$.  This can be a powerful tool, both for studying the space of solutions and for constructing model solutions---\eg\ Donaldson's approach to monopoles via rational maps \cite{MR769355}.  The results of such analyses, of course, depend heavily on the boundary conditions, and we have described a class of boundary conditions that are natural from the holographic perspective in \eqref{fieldas1}.  However we prefer to leave a detailed analysis of these issues to future work.  

The primary system \eqref{primaryBPS} also has the structure of a generalized self-duality equation of the type introduced in \cite{Corrigan:1982th,Ward:1983zm}.  Indeed these equations were studied from a supersymmetry point of view in \cite{Bak:2002aq}, and we recognize our system as a curved space version of the eight-dimensional `$\tfrac{2}{16}$-BPS' case given in their equation (53).  Let us view
\begin{equation}\label{auxAtrip}
(\hat{A}_{z_1}, \hat{A}_{z_2}, \hat{A}_y) := (\Phi^{z_1}, \Phi^{z_2}, \Phi^y)~,
\end{equation}
as the remaining three components of a gauge field $\hat{A}_A = (A_p,A_{r_i},\hat{A}_{z_p},\hat{A}_y)$ on the Riemannian eight-manifold with metric
\begin{align}\label{8metric}
\ed \hat{s}^2 :=&~ \hat{g}_{AB} \ed \hat{x}^A \ed \hat{x}^B \cr
:=&~ \mu^2 (r^2 + z_{0}^2) \delta_{pq} (\ed x^p \ed x^p + \ed \hat{z}^p \ed \hat{z}^q) + \frac{1}{\mu^2 (r^2+ z_{0}^2)} (\delta_{ij} \ed r^i \ed r^j + \ed \hat{y}^2 )~.
\end{align}
Here $A,B = 1,\ldots,8$, and we use hatted coordinates $(\hat{z}^p,\hat{y})$ to emphasize that these are not coordinates of the original 10D spacetime.  Note that $\hat{g}_{\hat{z}^p \hat{z}^q} = \Gbar^{z_p z_q}$ and $\hat{g}_{\hat{y}\hat{y}} = \Gbar^{yy}$.\footnote{\label{foot:T}These relations suggest an interpretation in terms of T-duality, but we will not explore that possibility here.}  

Now introduce the four-form
\begin{align}\label{asdomega}
\omega_4 :=&~ \mu^4 (r^2 + z_{0}^2)^2 \ed x^1 \ed x^2 \ed \hat{z}_1 \ed \hat{z}_2 + \frac{1}{\mu^4 (r^2 + z_{0}^2)^2}  \ed \hat{y} \ed r_1 \ed r_2 \ed r_3 + \cr
&~ + \left( \ed x^1 \ed x^2 +  \ed \hat{z}^1 \ed \hat{z}^2\right) \wedge \left(  \ed \hat{y} \ed r_3 + \ed r_1 \ed r_2  \right)~,
\end{align}
and observe that it is anti-self-dual with respect to \eqref{8metric}:
\begin{equation}
\hat{\star} \, \omega_4 = -\omega_4 ~.
\end{equation}
Then one can check that the primary BPS system \eqref{primaryBPS} is equivalent to
\begin{equation}\label{gsd}
\hat{\star} \, \hat{F} = \omega_4 \wedge \hat{F}~,
\end{equation}
restricted to configurations that are translation-invariant with respect to $\hat{z}^1,\hat{z}^2$ and $\hat{y}$.  Taking the dual of \eqref{gsd} we find $F = {\rm i}_F (\hat{\star} \omega_4)= - {\rm i}_F \omega_4$.  The first step is valid generally, and in the second step we used that $\omega_4$ and anti-self-dual.  In components, this result takes the form $F_{AB} = -\half (\omega_4)_{ABCD} F^{CD}$, which is the form given in \cite{Corrigan:1982th,Bak:2002aq}.  This equation is consistent thanks to the identity
\begin{equation}\label{omegasdid}
(\omega_4)^{A_1 A_2 AB} (\omega_4)_{A_3 A_4 AB}  = 4 (\omega_4)^{A_1 A_2}_{\phantom{A_1 A_2}A_3 A_4} + 6 (\delta^{A_1}_{\phantom{A_1}A_3} \delta^{A_2}_{\phantom{A_2}A_4} - \delta^{A_1}_{\phantom{A_1}A_4} \delta^{A_2}_{\phantom{A_2}A_3}) ~.
\end{equation}

The secondary BPS equations can also be written in terms of \eqref{auxAtrip} and \eqref{8metric}.  Let $\hat{E}^A := \hat{F}^{A}_{\phantom{A}0}$ be the electric field vector associated with the gauge field $(A_0,\hat{A})$ on the 9D spacetime with metric $g_{00} \ed t^2 + \ed \hat{s}^2$.  Then the equations \eqref{secondaryBPS} are equivalent to
\begin{equation}\label{8Dsecondary}
\hat{E}_A - D_A \Phi^{z_3} = 0~, \qquad D_0 \Phi^{z_3} = 0~.
\end{equation}

We will discuss some of the mathematical background of these generalized self-duality equations further in subsection \ref{sec:octonions} below, after introducing one additional generalization.  In the next subsection we will see how the repackaging \eqref{gsd} is useful for obtaining a Bogomolny bound on the energy functional.

A generic solution to \eqref{primaryBPS} and \eqref{secondaryBPS} will preserve two supersymmetries, but special types of solutions can preserve additional supersymmetry.  For example, one can repeat the supersymmetry analysis imposing only the first projection in \eqref{susychoices}.  This leads to the system of equations
\begin{equation}\label{primaryhalfBPS}
\begin{gathered}
 F_{12} + \mu^4 (r^2 + z_{0}^2)^2 \left( F_{r_1 r_2} - D_{r_3} \Phi^y\right) = 0 ~, \\
F_{1r_3} + D_2 \Phi^y  = 0~, \qquad F_{2 r_3} -  D_1 \Phi^y = 0~, \\
F_{r_1 r_3} + D_{r_2} \Phi^y = 0~, \qquad F_{r_2 r_3} - D_{r_1} \Phi^y = 0~, \\
F_{1 r_1} -  F_{2 r_2} = 0~, \qquad F_{2 r_1} +  F_{1 r_2} = 0~, \\
D_{p} \Phi^{z_q} = D_{r_i} \Phi^{z_q} = [\Phi^{y}, \Phi^{z_q}] = [\Phi^{z_1}, \Phi^{z_2}] = 0~,
\end{gathered}
\end{equation}
together with the conditions that 
\begin{equation}\label{secondaryhalfBPS}
F_{p0} = F_{r_i 0} = D_0 \Phi^{m} = 0~, \qquad D_{p} \Phi^{z_3} = D_{r_i} \Phi^{z_3} = [\Phi^m, \Phi^{z_3}] = 0~.
\end{equation}
In other words the electric field vanishes and the $\Phi^{z_i}$ are covariantly constant, mutually commuting, and commuting with $\Phi^y$.  Field configurations satisfying \eqref{primaryhalfBPS} and \eqref{secondaryhalfBPS} also solve \eqref{primaryBPS} and \eqref{secondaryBPS}, but preserve twice as many supersymmetries---\ie\ four supercharges.

We can also write \eqref{primaryhalfBPS} in the form \eqref{gsd} with $\omega_{4} \to \tilde{\omega}_4$, given by
\begin{align}\label{tomega4}
\tilde{\omega}_{4} :=&~  \mu^4 (r^2 + z_{0}^2)^2 \ed x^1 \ed x^2 \ed \hat{z}_1 \ed \hat{z}_2  +  \ed \hat{z}_1 \ed \hat{z}_2  \left( \ed\hat{y} \ed r_3 +  \ed r_1 \ed r_2 \right)  ~.
\end{align}
This four-form is neither self-dual nor anti-self-dual.  This, however, is not required.  All that is required for consistency of \eqref{gsd} is that $- \hat{\star} \, \tilde{\omega}_4$ satisfies \eqref{omegasdid}, and one can check that it does.  We note that $\omega_4$ and $\tilde{\omega}_4$ are related by $\omega_4 = (1 - \hat{\star}) \tilde{\omega}_4$.

\subsection{Bogomolny bound on the energy}\label{sec:Bbound}

Recall that the Yang--Mills energy functional takes the form \eqref{HKV}, provided the Gauss Law constraints \eqref{localGauss} and \eqref{globalGauss} hold.  Now that we have a handle on the field asympototics, let us take a closer look at the boundary constraint, \eqref{globalGauss}.

First we explain why the $r =0$ term can always be dropped in such boundary integrals.  Any constant time slice of the asymptotically $AdS_4 \times S^2$ spacetime is asymptotically $H_3 \times S^2$, where $H_3$ is hyperbolic three-space.  Thus the boundary we are integrating over is $\pd H_3 \times S^2$.   The boundary of (the conformal compactification of) $H_3$ is a two-sphere, viewed as the one-point compactification $\mathbb{R}^2 \cup \{ \infty \}$.  The $\mathbb{R}^2 \times S^2$ part of $\pd H_3 \times S^2$ is reached by sending $r \to \infty$ for any fixed $x^p$.  When $z_0 = 0$, the $\{\infty\} \times S^2$ part of $\pd H_3 \times S^2$ can be reached by either sending $x^p \to \infty$ for any fixed $r$ (including $r =\infty$), or by sending $r \to 0$ for any fixed $x^p$.  (See, for example, Figure 2.8 in the review \cite{Aharony:1999ti}.)  In contrast, when $z_0 \neq 0$, the points at $r =0$ with any finite $x^p$ are regular points in the interior, and $\{\infty\} \times S^2$ is only reached by sending $x^p \to \infty$ (for any fixed $r$).  The reason is that, when $z_0 \neq 0$, the two-sphere shrinks to zero size and the space smoothly caps off as $r \to 0$.  

In either case, the intersection of the locus $r =0$ with the boundary $\pd H_3 \times S^2$ is merely providing the set of points $\{\infty\} \times S^2$ that compactifies $\mathbb{R}^2 \times S^2$.  Therefore $r=0$ makes no contribution to boundary integrals provided the integral over $\mathbb{R}^2 \times S^2$ at $r \to \infty$ is finite.  We will always impose asymptotic conditions on boundary data, as we go to infinity in the $\mathbb{R}^2$ parameterized by $x^p$, such that this integral is finite.

Therefore in analyzing \eqref{globalGauss} we can set
\begin{equation}
\int \ed^5 x \left[ \pd_r \left( \sqrt{-g_6} \Tr\{E^{r} A^0\} \right) \right] = \lim_{r \to \infty} \int_{\mathbb{R}^2 \times S^2} \ed^2 x \ed \Omega \, r^2 \Tr\{ F_{r0} A_0 \}~.
\end{equation}
The field asymptotics in \eqref{fieldas1}, \eqref{fieldas2} allow for $F_{r0} = O(r^{-2})$, and thus the boundary Gauss constraint requires us to set the non-normalizable $S^2$ singlet mode of $A_0$ to zero.  This condition is not compatible with generalized temporal gauge, \eqref{gtgauge}, when $\phi_{({\rm nn})}^{z_3} \neq 0$, but it can be easily accommodated by making a time-dependent gauge transformation that eliminates $a_{0}^{({\rm nn})}$.  This is completely analogous to going from the Julia--Zee form of the dyon solution in 4D Yang--Mills--Higgs theory \cite{Julia:1975ff} to the Gibbons--Manton form \cite{Gibbons:1986df}, in which the dyon field configuration takes the form of the monopole field configuration dressed with a simple time dependence that generates the requisite electric field.    

Hence we can assume that the Gauss constraints hold, and therefore the energy functional is given by \eqref{HKV}.  We make use of the notation \eqref{auxAtrip}, \eqref{8metric}, to write the latter as
\begin{align}\label{H8D}
H_{\rm ym}^{\rm bos} =&~ \frac{1}{g_{{\rm ym}_6}^2} \int \ed^8 \hat{x} \sqrt{\hat{g}} \Tr \bigg\{ - \half g^{00} (\hat{E}_A \hat{E}^A + D_{A} \Phi^{z_3} D^A \Phi^{z_3}) + \half (g^{00})^2 (D_0 \Phi^{z_3})^2 + \cr
&~ \qquad \qquad \qquad \qquad \quad + \frac{1}{4} \hat{F}_{AB} \hat{F}^{AB} - \half \mu^4 (r^2 + z_{0}^2)^2 (\hat{F} \wedge \hat{F})_{r_1 r_2 r_3 y} \bigg\}~.  \raisetag{20pt}
\end{align}
Since nothing depends on $(\hat{z}_p,\hat{y})$ the integral over these directions is trivial.  We take them to be periodic with periodicity one so that \eqref{H8D} reproduces \eqref{HKV}.  The first line gives the contribution from $\KK$ and the second line gives the contribution from $\VV$.  We've also used that $\Gbar_{z_3 z_3} = -g^{00}$.

Now consider the quantity
\begin{align}
\left| \hat{\star} \hat{F} - \omega_4 \wedge \hat{F} \right|^2 :=&~ (\hat{\star} \hat{F} - \omega_4 \wedge \hat{F}) \wedge \hat{\star} (\hat{\star} \hat{F} - \omega_4 \wedge \hat{F})  \cr
=&~ \left| \hat{F} \right|^2 - 2 \omega_4 \wedge \hat{F} \wedge \hat{F} + \left| {\rm i}_{\hat{F}} \omega_4 \right|^2 \cr
=&~ 2 \hat{F}_{AB} \hat{F}^{AB} \ed^8 \hat{x} - 4 \omega_4 \wedge \hat{F} \wedge \hat{F} ~,
\end{align}
where we used \eqref{omegasdid}.  Similarly, with $\hat{E} = \hat{E}_A \ed \hat{x}^A$ and $D \Phi^{z_3} = (D_A \Phi^{z_3}) \ed \hat{x}^A$, we have
\begin{align}
\left| \hat{E} - D \Phi^{z_3} \right|^2 =&~ \left( \hat{E}_A \hat{E}^A + D_A \Phi^{z_3} D^A \Phi^{z_3} \right) \ed^8\hat{x} - 2 D\Phi^{z_3} \wedge \hat{\star} \hat{E} ~.
\end{align}
Therefore
\begin{align}\label{Hbog1}
H_{\rm ym}^{\rm bos} =&~ \frac{1}{g_{{\rm ym}_6}^2} \int \Tr \bigg\{ \frac{1}{8} \left| \hat{\star} \hat{F} - \omega_4 \wedge \hat{F} \right|^2 - \half g^{00} \left| \hat{E} - D \Phi^{z_3} \right|^2 + \half (g^{00})^2 \left| D_0 \Phi^{z_3} \right|^2 + \cr
&~ \qquad ~~+ \half \left( \omega_4 - \mu^4 (r^2 + z_{0}^2)^{2} \ed x^1 \ed x^2 \ed\hat{z}^1 \ed\hat{z}^2 \right) \wedge \hat{F} \wedge \hat{F} - g^{00} D \Phi^{z_3} \wedge \hat{\star} \hat{E} \bigg\}~.  \qquad
\end{align}

The last line of \eqref{Hbog1} is in fact a boundary term.  While $\omega_4$ is not closed, the shifted four-form
\begin{align}
\omega_4' :=&~ \omega_4 - \mu^4 (r^2 + z_{0}^2)^{2} \ed x^1 \ed x^2 \ed\hat{z}_1 \ed\hat{z}_2 \cr
=&~ \frac{1}{\mu^4 (r^2 + z_{0}^2)^2} \ed \hat{y} \ed r_1 \ed r_2 \ed r_3 + \left( \ed x^1 \ed x^2 +  \ed \hat{z}_1 \ed \hat{z}_2\right) \wedge \left( \ed r_1 \ed r_2 + \ed y \ed r_3 \right)~, \qquad
\end{align}
clearly is.  Since $\Tr (\hat{F} \wedge \hat{F}) = \ed \omega_{\rm CS}(\hat{A})$, we have $\omega_{4}' \wedge \Tr(\hat{F} \wedge \hat{F}) = \ed ( \omega_4' \wedge \omega_{\rm CS}(\hat{A}) )$, where the Chern--Simons three-form is
\begin{equation}\label{CS3form}
\omega_{\rm CS}(\hat{A}) := \Tr \left( \hat{F} \wedge \hat{A} - \tfrac{1}{3} \hat{A} \wedge \hat{A} \wedge \hat{A}\right)~.
\end{equation}
Furthermore the last term is a total derivative by the local Gauss constraint \eqref{localGauss}.

Hence we have brought the Hamiltonian to Bogomolny form:
\begin{align}
H_{\rm ym}^{\rm bos} =&~ \frac{1}{g_{{\rm ym}_6}^2} \int \Tr \left\{ \frac{1}{8} \left| \hat{\star} \hat{F} - \omega_4 \wedge \hat{F} \right|^2 + \half (-g^{00}) \left| \hat{E} - D \Phi^{z_3} \right|^2 + \half (g^{00})^2 \left| D_0 \Phi^{z_3} \right|^2 \right\} + \cr
&~ + \frac{1}{g_{{\rm ym}_6}^2} \int \ed \left( \half \omega_{4}' \wedge \omega_{\rm CS}(\hat{A}) - g^{00}  \Tr \{\Phi^{z_3} \wedge \hat{\star} \hat{E} \} \right)~.  \raisetag{20pt}
\end{align}
The first line is a sum of squares with positive coefficients, so we can immediately infer the bound
\begin{equation}\label{BPSbound1}
H_{\rm ym}^{\rm bos} \geq \frac{1}{g_{{\rm ym}_6}^2} \int_{\pd \hat{M}_8} \left( \half \omega_{4}' \wedge \omega_{\rm CS}(\hat{A}) - g^{00}  \Tr \{\Phi^{z_3} \wedge \hat{\star} \hat{E} \} \right)~,
\end{equation}
which is saturated on field configurations satisfying the first order equations \eqref{gsd}, \eqref{8Dsecondary}:
\begin{equation}\label{BPSeqns}
\hat{\star} \hat{F} = \omega_4 \wedge \hat{F} ~, \qquad \hat{E}_A - D_A\Phi^{z_3} = 0~, \qquad D_0 \Phi^{z_3} = 0~.
\end{equation}
Finally, on a solution to these equations the local Gauss constraint takes the form
\begin{equation}\label{8DLaplace}
D^A \left( \Gbar_{z_3 z_3} D_A \Phi^{z_3} \right) = 0~.
\end{equation}

In appendix \ref{app:BPSEnergy} we evaluate the BPS energy \eqref{BPSbound1} in terms of the field asymptotics \eqref{fieldas1}, \eqref{fieldas2}.  The results are summarized here.  The magnetic energy (the $\omega_{4}' \wedge \omega_{\rm CS}$ term) receives to types of contribution in general.  First, there is a contribution proportional to the vev $\Phi_{\infty}^y$. It has the form of a standard monopole mass term, where the relevant magnetic charge is expressed in terms the magnetic flux through the $\mathbbm{R}^2$ boundary associated to one of the $j=1$ triplet modes of gauge fields.  Let us introduce a vector notation for this triplet, analogous to \eqref{XXtripdef}, such that
\begin{equation}
\hat{r} \cdot \vec{a}_{\mu}(x^\nu, r) := - \frac{1}{\sqrt{3}} \sum_{m=-1}^m a_{\mu,(1,m)}(x^\nu,r) Y_{1m}(\theta,\phi) ~.
\end{equation}
The mode analysis determined that the leading behavior of this triplet is
\begin{equation}
\vec{a}_{\mu}(x^\nu, r) = \frac{1}{\mu^2 r^2} \vec{a}_{\mu}^{\, ({\rm n})}(x^\nu) + O(r^{-3})~,
\end{equation}
where $\vec{a}_{\mu}^{\, ({\rm n})}$ commutes with the vev $\Phi_{\infty}^y$ and the 't Hooft charge $P$.  Then we define a triplet of magnetic fluxes in terms of these:
\begin{equation}\label{fluxtrip}
\vec{\gamma}_{\rm m} := \frac{1}{2\pi} \int_{\mathbbm{R}^2} \vec{f}^{\, ({\rm n})} ~, \qquad \vec{f}_{\mu\nu}^{\,({\rm n})} := \pd_{\mu} \vec{a}_{\nu}^{\, ({\rm n})} - \pd_{\nu} \vec{a}_{\mu}^{\, ({\rm n})} ~.
\end{equation}
The first contribution to the magnetic energy is proportional to $\Tr \{ \Phi_{\infty}^y \gamma_{\rm m}^3 \}$.  The fact that the third component of the flux vector is picked out can be traced back to our choice for the unit vector $\hat{n}_{(r)}$ in the supersymmetry projection \eqref{susychoices} to be along the three-direction in $\vec{r}$ space.  

The second type of magnetic contribution is proportional to $\XX_{(\rm n)}^3$, the third component of the triplet $\vec{\XX}_{(\rm n)}$.  In fact there are two types of terms---one that depends on the local value of $\XX_{({\rm n})}^3$ and is integrated over $\mathbbm{R}^2$, and a line integral around the circle at infinity, \ie\ the boundary of the boundary, that hence only depends on the asymptotic value of $\XX_{({\rm n})}^3$.  In both of these terms, $\XX_{({\rm n})}^3$ is traced against quantities constructed from the non-normalizable $S^2$ singlet modes $\{a_{\mu}^{({\rm nn})}, \phi_{({\rm nn})}^{z_i}\}$.

Then there is the contribution from the electric energy term.  This can be expressed in terms of the non-normalizable $S^2$ singlet $\phi_{({\rm nn})}^{z_3}$, and the radial component of the normalizable $S^2$ singlet electric field.  Note that the $z_3$ component is picked out in \eqref{BPSbound1} because of our choice of unit vector $\hat{n}_{(z)}$.  Only the coefficient of the leading $O(1/r^2)$ component of the electric field contributes.  We define this coefficient by
\begin{equation}\label{asEflux}
f_{r0}^{({\rm n})}(x^\mu) := \lim_{r \to \infty} (\mu^2 r^2 F_{r0})~.
\end{equation}
On a solution to the BPS equations one can use $F_{r0} = D_r \Phi^{z_3}$, and so this quantity depends on the first subleading, $O(r^{-1})$ behavior of $\Phi^{z_3}$, where $\Phi^{z_3}$ is required to solve \eqref{8DLaplace} subject to the boundary condition $\Phi^{z_3} \to \phi_{({\rm nn})}^{z_3}$ as $r \to \infty$.

In terms of these quantities one then finds the following expression for the Bogomolny bound:
\begin{align}\label{BPSbound2}
H_{\rm ym}^{\rm bos} \geq &~  \frac{4\pi}{\mu^2 g_{{\rm ym}_6}^2} \int_{\mathbbm{R}^2} \ed^2 x \Tr \left\{ \XX_{({\rm n})}^3 \left( f_{12}^{({\rm nn})} + [\phi_{({\rm nn})}^{z_1}, \phi_{({\rm nn})}^{z_2}] \right) + \phi_{({\rm nn})}^{z_3} f_{r0}^{({\rm n})} \right\} +    \cr
&~ - \frac{4\pi^2}{\sqrt{3} \mu^2 g_{{\rm ym}_6}^2} \Tr \left\{ \Phi_{\infty}^y \gamma_{\rm m}^3 \right\} - \frac{2\pi}{\mu^2 g_{{\rm ym}_6}^2} \oint_{S_{\infty}^1} \Tr \left\{ \XX_{({\rm n})}^3 a^{({\rm nn})} \right\} ~.
\end{align}

This gives the bound on the Hamiltonian associated with $S_{\rm ym}$.  Recall, however, that it is $S_{\rm hol}$, \eqref{Shol}, rather than the on-shell value of $S_{\rm ym}$, that is the relevant functional for the holographic correspondence.  This means that the holographic energy for static solutions to the equations of motion is given by the Legendre transform
\begin{equation}\label{Hhol}
H_{\rm hol} = \left[ (H_{\rm ym})^{\textrm{o-s}} - \frac{4\pi}{g_{{\rm ym}_6}^2 \mu^2} \int_{\pd H_3} \ed^2 x \Tr \left\{ \vec{\XX}_{({\rm nn})} \cdot \vec{\XX}_{({\rm n})} \right\} \right]_{\vec{\XX}_{({\rm n})} = \vec{\XX}_{({\rm n})}[\vec{\XX}_{({\rm nn})}]} ~.
\end{equation}
Extremization with respect to $\vec{\XX}_{({\rm n})}$ leads to\footnote{Write $(H_{\rm ym})^{\textrm{o-s}} = H_{\rm pos} + H_{\rm ym}^{\rm BPS}$ where $H_{\rm pos}$ is the on-shell value of the positive-definite sum-of-squares term in $H_{\rm ym}$ and $H_{\rm ym}^{\rm BPS}$ is the right-hand side of \eqref{BPSbound2}.  Any static solution to the equations of motion for fixed boundary data $\vec{\XX}_{({\rm n})}$ will be a local minimum of $H_{\rm pos}$, and hence the first variation of the on-shell value with respect to $\vec{\XX}_{({\rm n})}$ will vanish: $\frac{\delta H_{\rm pos}}{\delta \XX_{({\rm n})}} = \frac{\delta H_{\rm pos}}{\delta \phi} \cdot \frac{\delta \phi}{\delta \XX_{({\rm n})}} = 0$.}
\begin{align}
\begin{array}{r c l} 0 &=& \XX_{({\rm nn})}^{1,2}~, \\[1ex]  0 &=& f_{12}^{({\rm nn})} + [\phi_{({\rm nn})}^{z_1}, \phi_{({\rm nn})}^{z_2} ]  - \XX_{({\rm nn})}^3~. \end{array}
\end{align}
In particular, these relations are consistent with the asymptotics of the BPS equations.  There are some cancellations in $H_{\rm hol}$ upon using them.

Note that in the extremization with respect to $\vec{\XX}_{({\rm n})}$, we hold the asymptotics of $\vec{\XX}_{({\rm n})}$ fixed, as we go to infinity on the two-plane.  Therefore the last term of \eqref{BPSbound2} does not vary.  Finiteness of the energy suggests that the appropriate boundary conditions at $S_{\infty}^1$ should be of vortex type.  Letting $(\varrho,\varphi)$ be plane-polar coordinates, we impose the following asymptotic behavior:
\begin{equation}\label{vortexbc}
\vec{\XX}_{({\rm n})} = \vec{\mathrm{v}} + O(\varrho^{-1}) ~, \qquad  a^{({\rm nn})} = \gm \ed\varphi + O(\varrho^{-2})~, \qquad \textrm{as} ~ \varrho \to \infty~,
\end{equation}
where the triplet of vevs $\vec{\mathrm{v}}$ is constant and mutually commuting with the magnetic charge, $\gm$.  Note that the asymptotics of $a^{({\rm nn})}$ are consistent with the definition
\begin{equation}
\gm := \frac{1}{2\pi} \int_{\mathbbm{R}^2} f^{({\rm nn})} = \frac{1}{2\pi} \oint_{S_{\infty}^1} a^{({\rm nn})} ~.
\end{equation}
We then see that the last term of \eqref{BPSbound2} is proportional to $\Tr\{ \mathrm{v}^3 \gm \}$.  The bound on the holographic energy takes the form
\begin{equation}
H_{\rm hol} \geq  \frac{4\pi}{\mu^2 g_{{\rm ym}_6}^2} \bigg\{ \int_{\mathbbm{R}^2} \Tr \left\{ \phi_{({\rm nn})}^{z_3} f_{r0}^{({\rm n})} \right\} - \frac{\pi}{\sqrt{3}} \Tr \left\{ \Phi_{\infty}^y \gamma_{\rm m}^3 \right\} - \pi \Tr \left\{ \mathrm{v}^3 \gm \right\} \bigg\} ~,
\end{equation}
which is saturated on solutions to \eqref{BPSeqns}.

Finally we must restore the dependence on the unit vectors $\hat{n}_{(r,z)}$, as discussed under \eqref{susychoices}, and vary to achieve the strongest bound.  The `$3$' component of the triplets $\vec{\gamma}_{\rm m}$ and $\vec{v}$ refers to their component along $\hat{n}_{(r)}$, while $\phi_{({\rm nn})}^{z_3}$ is the component of the triplet $\vec{\phi}_{({\rm nn})}^{z}$ along $\hat{n}_{(z)}$.  Thus we have the bound
\begin{equation}
H_{\rm hol} \geq \hat{n}_{(r)} \cdot \vec{M}_{\rm m} + \hat{n}_{(z)} \cdot \vec{M}_{\rm e}~,
\end{equation}
where we've introduced the triplets of magnetic and electric masses
\begin{align}\label{solitonmasses}
\vec{M}_{\rm m} :=&~ \frac{4\pi^2}{\mu^2 g_{{\rm ym}_6}^2} \Tr \left\{ \frac{1}{\sqrt{3}} \Phi_{\infty}^y \vec{\gamma}_{\rm m} + \vec{\mathrm{v}} \, \gm \right\} ~, \cr
\vec{M}_{\rm e} :=&~  -\frac{4\pi}{\mu^2 g_{{\rm ym}_6}^2} \int_{\mathbbm{R}^2} \ed^2x \Tr \left\{ \vec{\phi}_{({\rm nn})}^{z} f_{r0}^{({\rm n})} \right\}~.
\end{align}
The strongest bound is achieved by taking
\begin{equation}
\hat{n}_{(r)} = \frac{\vec{M}_{\rm m}}{|\vec{M}_{\rm m}|} ~, \qquad \hat{n}_{(z)} = \frac{\vec{M}_{\rm e}}{|\vec{M}_{\rm e}|}~,
\end{equation}
which gives
\begin{equation}\label{BPSsolenergy}
H_{\rm hol} \geq H_{\rm BPS} := | \vec{M}_{\rm m}| + |\vec{M}_{\rm e}| ~, \qquad \left( \textrm{$\tfrac{1}{4}$-BPS}\right) ~.
\end{equation}
Here we've emphasized that this bound is saturated on solutions preserving two supersymmetries, \ie\ $1/4$ of the supersymmetries of the 3D $\NN = 4$ Poincare superalgebra.

The masses \eqref{solitonmasses} transform in the $({\bf 3},{\bf 1})$ and $({\bf 1},{\bf 3})$ of $SU(2)_r \times SU(2)_z = SU(2)_V \times SU(2)_H$ respectively.  This is consistent with the central charges of the 3D $\NN = 4$ superalgebra.  It would be nice to derive these charges independently, a la \cite{Witten:1978mh}, by computing the commutator of Noether charges associated with the supersymmetry transformations.

The story can be repeated for the $1/2$-BPS system \eqref{primaryhalfBPS}, by using the $\tilde{\omega}_4$ defined in \eqref{tomega4}, in place of $\omega_4$ in the energy bound.  The result is the same, except that the electric contribution vanishes:
\begin{equation}\label{halfBPSbound}
H_{\rm hol} \geq H_{\rm BPS} := | \vec{M}_{\rm m}|  ~, \qquad \left( \textrm{$\tfrac{1}{2}$-BPS}\right) ~.
\end{equation}
This is analogous to 4D $\NN = 4$ supersymmetry: monopoles are 1/2-BPS while dyons are $1/4$-BPS.

In the Conclusion (Section \ref{sec:future}) we will comment a bit more on the nature of solutions to \eqref{BPSeqns}, and possible descriptions in terms of the holographic dual and in terms of D-brane systems. We leave a complete analysis of these issues to future work.  

\subsection{Domain walls and dyonic octonionic instantons}\label{sec:octonions}

The BPS equations discussed above are the most general ones giving rise to configurations that have a soliton-particle interpretation.  If we consider extended objects, however, we can impose one further projection condition on $\vareps$, bringing us all the way down to a single preserved supersymmetry.  There is a $U(1)^3$ family of choices, corresponding to choosing directions in the $x^1$-$x^2$, $r_1$-$r_2$, and $z_1$-$z_2$ planes.  The latter two are the planes orthogonal to $\hat{n}_{(r)} \in \mathbb{R}_{(r)}^3$ and $\hat{n}_{(z)} \in \mathbb{R}_{(z)}^3$ respectively.  For now we will take these directions to be along the respective $1$-axes and then restore the dependence on this choice at the end.  Hence our final projection condition is
\begin{equation}\label{lastproj}
\Gamma^{\underline{1 r_1 z_1 y}} \vareps_0 = \vareps_0~,
\end{equation}
which is mutually compatible with all previous ones, \eqref{7proj}.  We will see that the corresponding field configurations give a holographic description of codimension-one domain walls \emph{within} the defect CFT---that is, $(1+1)$-dimensional strings inside the $(1+2)$-dimensional defect CFT---and more generally soliton-domain wall junctions.

By combining the new projection with the previous ones we find that the 28 $\Gamma^{\underline{AB}}$ break into seven sets of four, where each member of a given set is equivalent when acting on $\vareps_0$:
\begin{equation}\label{Gquads}
\begin{gathered}
\left\{ \Gamma^{\underline{12}}, \Gamma^{\underline{r_1 r_2}}, \Gamma^{\underline{z_1 z_2}}, - \Gamma^{\underline{r_3 y}} \right\} ~, \cr\left\{ \Gamma^{\underline{z_1 y}}, \Gamma^{\underline{2 r_2}}, - \Gamma^{\underline{1 r_1}} , - \Gamma^{\underline{z_2 r_3}} \right\}~, \\
\left\{ \Gamma^{\underline{r_1 y}}, -\Gamma^{\underline{2 z_2}}, \Gamma^{\underline{1 z_1}} , - \Gamma^{\underline{r_2 r_3}} \right\}~, \\
\left\{ \Gamma^{\underline{2 r_1}}, \Gamma^{\underline{z_2 y}}, \Gamma^{\underline{z_1 r_3}}, \Gamma^{\underline{1 r_2}} \right\}~, \\
\left\{ \Gamma^{\underline{r_1 z_1}}, \Gamma^{\underline{2 r_3}}, - \Gamma^{\underline{1 y}}, - \Gamma^{\underline{r_2 z_2}} \right\} ~, \\
\left\{ \Gamma^{\underline{1 z_2}}, - \Gamma^{\underline{r_2 y}}, - \Gamma^{\underline{r_1 r_3}}, \Gamma^{\underline{2 z_1}} \right\} ~, \\
\left\{ \Gamma^{\underline{1 r_3}}, \Gamma^{\underline{r_2 z_1}}, \Gamma^{\underline{r_1 z_2}}, \Gamma^{\underline{2 y}} \right\} ~.
\end{gathered}
\end{equation}
We also still have the electric-type projection, $\Gamma^{\underline{0 z_3}} \vareps_0 = \vareps_0$, which is unaffected by the above.  Setting the supersymmetry variation of the fermion to zero, we get the same set of electric BPS equations as before, but the magnetic equations are modified.  The four-term equation we had in \eqref{primaryBPS} remains, as it corresponds to the first quadruplet in \eqref{Gquads}.  The twelve two-term equations combine into six four-term equations, so that the new magnetic system is
\begin{equation}\label{newmagBPS}
\begin{gathered}
F_{12} +  [\Phi^{z_1}, \Phi^{z_2}] + \mu^4 (r^2 + z_{0}^2)^2 \left( F_{r_1 r_2} - D_{r_3} \Phi^y \right) = 0~, \\
D_2 \Phi^{z_2} - D_{1} \Phi^{z_1} +  \mu^4 (r^2 + z_{0}^2)^2 \left( F_{r_2 r_3} - D_{r_1} \Phi^y \right) = 0~, \\
D_{2} \Phi^{z_1} + D_1 \Phi^{z_2} +  \mu^4 (r^2 + z_{0}^2)^2 \left( F_{r_3 r_1} - D_{r_2} \Phi^y \right) = 0 ~, \\
F_{1 r_3} + D_2 \Phi^y  + D_{r_1} \Phi^{z_2} + D_{r_2} \Phi^{z_1} = 0 ~, \\
F_{2 r_3} - D_1 \Phi^y  + D_{r_1} \Phi^{z_1} - D_{r_2} \Phi^{z_2} = 0~, \\
F_{1 r_1} - F_{2 r_2} - \left( D_{r_3} \Phi^{z_2} - [\Phi^y, \Phi^{z_1}] \right) = 0~, \\
F_{1 r_2} + F_{2 r_1} - \left( D_{r_3} \Phi^{z_1}  + [\Phi^y, \Phi^{z_2}] \right) = 0~.
\end{gathered}
\end{equation}

This is another example of a generalized self-duality equation in eight dimensions, \cite{Corrigan:1982th,Ward:1983zm,Bak:2002aq}---the ``$\tfrac{1}{16}$-BPS'' case in the latter reference.  Utilizing \eqref{auxAtrip} and \eqref{8metric}, one can show that \eqref{newmagBPS} is equivalent to
\begin{equation}\label{octinstantons}
\hat{\star} \hat{F} = \Omega_4 \wedge \hat{F}~,
\end{equation}
where the new anti-self-dual four-form, $\Omega_4 = - \star_8 \Omega_4$, has some additional terms relative to \eqref{asdomega}:
\begin{align}\label{asdOmega}
\Omega_4 =&~ \omega_4 + (\ed\hat{y} \ed r_1 + \ed r_2 \ed r_3) \wedge (\ed x^2 \ed \hat{z}_2 - \ed x^1 \ed \hat{z}_1) + \cr
&~ + (\ed \hat{y} \ed r_2 + \ed r_3 \ed r_1) \wedge (\ed x^2 \ed \hat{z}_1 + \ed x^1 \ed \hat{z}_2)~.
\end{align}
More precisely, \eqref{newmagBPS} is equivalent to \eqref{octinstantons} when the latter is restricted to configurations with $\mathbb{R}^3$ invariance corresponding to translations of $\hat{z}^1,\hat{z}^2,\hat{y}$.

Equation \eqref{octinstantons} with \eqref{asdOmega} is also known as the octonionic instanton equation, or the $Spin(7)$ instanton equation, \cite{Corrigan:1982th,Ward:1983zm,ReyesCarrion:1998si,Lewis:1998,MR1634503,MR2893675,Cherkis:2014xua}.  Let us briefly review the connection to octonions, following \cite{Corrigan:1982th}.  An arbitrary element $q \in \mathbb{O}$ can be written
\begin{equation}
q = \sum_{A' = 1}^8 q_{\uA'} {\bf e}_{\uA'}~,
\end{equation}
where ${\bf e}_{\underline{8}} = 1$ and ${\bf e}_{\ua'}$ $a' =1, \ldots, 7$ are the unit octonions.  The reason for the index notation is that we will soon identify this $\mathbb{R}^8$ with the tangent space of our eight-manifold, \eqref{8metric}, where the coordinates $x^{A'}$ are a simple reshuffling of the $x^A$.  The unit octonions satisfy
\begin{equation}
{\bf e}_{\ua'} {\bf e}_{\ub'} = -\delta_{\ua'\ub'} + C_{\ua' \ub' \uc'} {\bf e}_{\uc'}~,
\end{equation}
where the structure constants are totally antisymmetric and satisfy
\begin{equation}
\{ C^{\uc'}, C^{\ud'} \}_{\ua'\ub'} = {\delta^{\uc'}}_{\ua'} {\delta^{\ud'}}_{\ub'} + {\delta^{\uc'}}_{\ub'} {\delta^{\ud'}}_{\ua'} -2 \delta^{\uc' \ud'} \delta_{\ua' \ub'}~,
\end{equation}
with $(C^{\uc'})_{\ua'\ub'} = C_{\uc'\ua'\ub'}$.  Then one way to write the octonionic instanton equation is
\begin{equation}
\hat{F}_{\underline{8}\ua'} = \half C_{\ua'\ub'\uc'} \hat{F}_{\ub'\uc'}~,
\end{equation}
where $(C^{\uc'})_{\uA'\uB'} = 0$ when $A'$ or $B' = 8$.  Taking the basis for the structure constants to be
\begin{equation}
1 = C_{\underline{127}} = C_{\underline{163}} = C_{\underline{154}} = C_{\underline{253}} = C_{\underline{246}} = C_{\underline{347}} = C_{\underline{567}}~,
\end{equation}
one finds the equations\footnote{Another interesting occurrence of these equations in everyday physics is observed in \cite{Harvey:1990eg}, where they describe heterotic string solitons.}
\begin{equation}
\begin{gathered}
\hat{F}_{\underline{12}} + \hat{F}_{\underline{34}} + \hat{F}_{\underline{56}} + \hat{F}_{\underline{78}} = 0~, \\
\hat{F}_{\underline{13}} + \hat{F}_{\underline{42}} + \hat{F}_{\underline{57}} + \hat{F}_{\underline{86}} = 0~, \\
\hat{F}_{\underline{14}} + \hat{F}_{\underline{23}} + \hat{F}_{\underline{76}} + \hat{F}_{\underline{85}} =0~, \\
\hat{F}_{\underline{15}} + \hat{F}_{\underline{62}} + \hat{F}_{\underline{73}} + \hat{F}_{\underline{48}} = 0~, \\
\hat{F}_{\underline{16}} + \hat{F}_{\underline{25}} + \hat{F}_{\underline{38}} + \hat{F}_{\underline{47}} = 0~, \\
\hat{F}_{\underline{17}} + \hat{F}_{\underline{82}} + \hat{F}_{\underline{35}} + \hat{F}_{\underline{64}} = 0~, \\
\hat{F}_{\underline{18}} + \hat{F}_{\underline{27}} + \hat{F}_{\underline{63}} + \hat{F}_{\underline{54}} = 0~.
\end{gathered}
\end{equation}
The system \eqref{newmagBPS} is equivalent to this upon using \eqref{auxAtrip}, going to an orthonormal frame associated with \eqref{8metric}, and relabeling indices according to
\begin{equation}\label{translation}
\{1,2,r_1,r_2,\hat{y},r_3,\hat{z}^1,\hat{z}^2\} \leftrightarrow \{1,2,3,4,5,6,7,8\}~.
\end{equation}

Eight-manifolds with $Spin(7)$ structure provide the natural geometric setting for these equations.  The structure constants for the octonions yield a canonical self-dual\footnote{The map \eqref{translation} sends the canonical orientation on $\mathbb{R}^8$ to the negative of the orientation we chose on $\hat{M}_8$, hence the anti-self-duality of $\Omega_4$ on $\hat{M}_8$.} four-form on $\mathbb{R}^8 = \mathbb{R}^7 \oplus \mathbb{R}$, known as the Cayley form:
\begin{align}\label{Spin7G2}
& \Omega_4 = \star_7 \varphi_3 + \varphi_3 \wedge \ed x^{\underline{8}}~, \qquad \textrm{with} \qquad \varphi_3 := \frac{1}{3!} C_{\ua' \ub' \uc'} \ed x^{\ua'} \ed x^{\ub'} \ed x^{\uc'} ~.
\end{align}
$\Omega_4$ is a $Spin(7)$ structure on $\mathbb{R}^8$:  It is invariant under an irreducible $Spin(7)$ subgroup of the $SO(8)$ rotation group.  This subgroup is the little group of a constant unit spinor in the positive chirality Weyl spinor representation of $Spin(8)$.  A general self-dual four-form transforms in a ${\bf 35}$ of $SO(8)$, which decomposes into ${\bf 1} \oplus {\bf 7} \oplus {\bf 27}$ under $Spin(7)$.  The Cayley form sits in the singlet.  Note that $\varphi_3$ in \eqref{Spin7G2} is a $G_2$-structure on $\mathbb{R}^7$ and that $G_2$ is the automorphism group of the octonion algebra.

We used the same notation for the Cayley form in \eqref{Spin7G2} as for the anti-self-dual four-form, \eqref{asdOmega}.  They are indeed the same upon identifying the canonical frame $\{ \ed x^{\uA'} \}$ on $\mathbb{R}^8$ with the natural orthonormal frame $\{ e^A \}$ associated to \eqref{8metric} via \eqref{translation}.  Hence $\Omega_4$ is a $Spin(7)$ structure on $(\hat{M}_8,\hat{g}_{AB})$.  More precisely, it is a \emph{non-integrable} $Spin(7)$, or \emph{almost}-$Spin(7)$ structure, \cite{MR2893675}, because it is not closed.  If it would have been closed then $(\hat{M}_8,\hat{g}_{AB},\Omega_4)$ would have been a $Spin(7)$-holonomy manifold.  See \eg\ \cite{MR2292510}.  A non-integrable $Spin(7)$ structure, however, is already enough to define the octonionic, or $Spin(7)$, instanton equation, and is sufficient to guarantee certain nice properties of this equation.  For example, on a closed eight-manifold, the linearized equations determining gauge-inequivalent deformations of this equation form an elliptic complex whose index has been computed in \cite{ReyesCarrion:1998si,Lewis:1998}.

A new Bogomolny bound on the energy functional, which is saturated on solutions to \eqref{octinstantons}, can be derived by repeating an identical sequence of steps as before, but with $\{ \omega_4, \omega_{4}^{\prime} \} \to \{ \Omega_4 , \Omega_{4}^{\prime} \}$, where
\begin{align}
\Omega_{4}^{\prime} :=&~ \Omega_4 - \mu^4 (r^2 +  z_{0}^2)^2 \ed x^1 \ed x^2 \ed \hat{z}_1 \ed \hat{z}_2~. 
\end{align}
The shift is exactly what is needed to guarantee that $\Omega_{4}'$ is closed.  The bound on the Yang--Mills functional is
\begin{equation}\label{dwBPSbound1}
H_{\rm ym}^{\rm bos} \geq \frac{1}{g_{{\rm ym}_6}^2} \int_{\pd \hat{M}_8} \left( \half \Omega_{4}' \wedge \omega_{\rm CS}(\hat{A}) - g^{00}  \Tr \{\Phi^{z_3} \wedge \hat{\star} \hat{E} \} \right)~.
\end{equation}
The shift of $\Omega_4$ to $\Omega_{4}'$ in the boundary term is again ultimately due to the presence of the background RR flux.  

Let us comment on this further.  If the $Spin(7)$ structure $\Omega_4$ had been closed, then the $Spin(7)$ instanton equations would have implied the standard Yang--Mills equations: 
\begin{equation}
D \star F = \Omega_4 \wedge D F = 0~, \qquad \textrm{if $\ed\Omega_4 = 0$}~.
\end{equation}
In the first step we used that $\Omega_4$ is closed and in the second step we used the Bianchi identity.  Our $\Omega_4$ is not closed, but $\Omega_{4}'$ is.\footnote{This situation, in which one has a pair $(\Omega_{4}, \Omega_{4}')$ consisting of a non-integrable $Spin(7)$ structure and a closed four-form such that the Yang--Mills energy functional is given by a boundary integral in terms of it, was actually considered in \cite{MR2893675}, where it was referred to as having a \emph{tamed} (non-integrable) $Spin(7)$ structure.}  Correspondingly, our second-order equations of motion are not the standard Yang--Mills equations.  They have an extra piece,
\begin{equation}
D \star F = \ed \Omega_4 \wedge F = 4 \mu^4 (r^2 + z_{0}^2) r \ed r \ed x^1 \ed x^2 \ed \hat{z}_1 \ed \hat{z}_2 \wedge F~.
\end{equation}
This `extra' term in the equations of motion comes precisely from the coupling to the background RR flux.  One can view this as another instance of background fluxes in string theory naturally leading to modified or generalized geometric structures.  Here, this occurs in the context of field theories on curved backgrounds, and dovetails nicely with the philosophy recently presented in \cite{Maxfield:2015evr}.

The bound \eqref{dwBPSbound1} is evaluated on the field asymptotics \eqref{fieldas1} in appendix \ref{app:BPSEnergy}.  In order to describe the result in a relatively compact fashion we introduce a little notation.  Let $x^{\tp} = \{ x^1, x^2, \hat{z}_1, \hat{z}_2 \}$ parameterize a Euclidean $\mathbbm{R}^4$ with standard orientation, and let $(\eta^i)_{\tp\tq}$, or equivalently $\vec{\eta}_{\tp\tq}$, denote the triplet of self-dual 't Hooft matrices.  (See appendix \ref{app:BPSEnergy} for our conventions.)  We collect the gauge field $A_p$ and the scalars $\Phi^{z_p}$, $p=1,2,$ into a 4D gauge field $A_{\tp} = \{ A_1, A_2, \Phi^{z_1}, \Phi^{z_2} \}$.  Since the $A_p$ and $\Phi^{z_p}$ have the same asymptotics, we can consistently define all of the corresponding normalizable and non-normalizable modes for this 4D gauge field.  Then, with these definitions, the bound \eqref{dwBPSbound1} takes the form
\begin{align}\label{dwBPSbound2}
H_{\rm ym}^{\rm bos} \geq &~ \frac{2\pi}{\mu^2 g_{{\rm ym}_6}^2} \int_{\mathbbm{R}^2} \ed^2 x \, \vec{\eta}\,{}^{\tp\tq} \cdot \bigg\{ \Tr \{ \vec{\XX}_{({\rm n})} f_{\tp \tq}^{({\rm nn})} \} - \frac{1}{2\sqrt{3}} \Tr \{ \Phi_{\infty}^y \vec{f}_{\tp\tq}^{\,({\rm n})} \}  - \pd_{\tp} \left[ \Tr \{ \vec{\XX}_{({\rm n})} a_{\tq}^{({\rm nn})} \}\right] \bigg\} +  \cr
&~ + \frac{4\pi}{\mu^2 g_{{\rm ym}_6}^2} \int_{\mathbbm{R}^2} \ed^2 x  \Tr \{ \phi_{({\rm nn})}^{z_3} f_{r0}^{({\rm n})} \}~.   \raisetag{20pt}
\end{align}
In this expression it should be understood that nothing depends on the coordinates $\hat{z}_p$, so for example the total derivative term vanishes those values of the $\tp$ index.

The previous bound, \eqref{BPSbound2}, can be recovered by dropping the terms proportional to the first two 't Hooft matrices.  This corresponds to restricting $\Omega_{4}'$ to $\omega_{4}'$, as discussed in appendix \ref{app:BPSEnergy}.

The bound \eqref{dwBPSbound2} can be Legendre transformed to a bound on the energy functional $H_{\rm hol}$, \eqref{Hhol}.  Variation with respect to $\vec{\XX}_{({\rm n})}$ now results in the triplet of equations
\begin{equation}\label{asBPStripeqn}
\half \vec{\eta}\,{}^{\tp\tq} f_{\tp \tq}^{({\rm nn})} - \vec{\XX}_{({\rm nn})} = 0~,
\end{equation}
if we assume that the total derivative term in the first line of \eqref{dwBPSbound2} does not vary.  It can be shown that \eqref{asBPStripeqn} is consistent with the asymptotics of \eqref{newmagBPS}.  (This will be demonstrated in the next subsection.  See the paragraph below \eqref{magBPSred1}.)  Plugging \eqref{asBPStripeqn} back in, one finds an expression for $H_{\rm hol}$ that looks identical to \eqref{dwBPSbound2} except that the first term is absent.

However it is likely that the total derivative term of \eqref{dwBPSbound2} does in fact contribute to the variation with respect to $\vec{\XX}_{({\rm n})}$.  This quantity evaluates to
\begin{align}\label{dwenergy}
\vec{\eta}\,{}^{\tp\tq} \cdot \pd_{\tp} \left[ \Tr \{ \vec{\XX}_{({\rm n})} a_{\tq}^{({\rm nn})} \}\right]  =&~ \pd_1 \left[ \Tr  \left\{ \XX_{({\rm n})}^2 \phi_{({\rm nn})}^{z_2} - \XX_{({\rm n})}^1 \phi_{({\rm nn})}^{z_1} \right\} \right] + \cr
&~ + \pd_2 \left[ \Tr \left\{ \XX_{({\rm n})}^2 \phi_{({\rm nn})}^{z_1} + \XX_{({\rm n})}^1 \phi_{({\rm nn})}^{z_2} \right\} \right] + \cr
&~ + \pd_1 \left[ \Tr \{ \XX_{({\rm n})}^3 a_{2}^{({\rm nn})} \} \right] - \pd_2 \left[ \Tr \{ \XX_{({\rm n})}^3 a_{1}^{({\rm nn})} \} \right] ~.
\end{align}
The last line comes from the $\eta^3$ terms, and produces the line integral we found previously in \eqref{BPSbound2}.  The first two lines are new, relative to \eqref{BPSbound2}, and are clearly related to the possible presence of a domain wall in the plane.  Integrating the first term over $\mathbbm{R}^2$ reduces it to a line integral of a tension over the $x^2$ direction, for example, where the tension is given by a discontinuity in $x^1$.  Which discontinuity conditions are consistent with the BPS equations requires further study.    

One must also restore the dependence on the parameters determining the supersymmetry projections---the unit vectors $\hat{n}_{(r)},\hat{n}_{(z)}$ and the three $U(1)$ rotations in the $x^1$-$x^2$ plane and the planes orthogonal to these vectors.  This can be done by defining appropriately rotated versions of the 't Hooft symbols.  One should then vary with respect to all of these parameters to achieve the strongest bound.  We postpone both of these analyses to a future publication.

 \subsection{Dimensional reduction and the Haydys--Witten equations}\label{sec:reducedBPS}

Given the consistent truncation of section \ref{sec:ct}, it is natural to ask how the BPS equations reduce when restricted to the ansatz \eqref{truncation}.  We apply the ansatz directly to the system \eqref{newmagBPS}.  With the help of the formulae collected in appendix \ref{app:trunc}, the first three equations reduce to
\begin{align}\label{magBPSred1}
F_{12} +  [\phi^{z_1}, \phi^{z_2}] - \frac{\mu^2 (r^2 + z_{0}^2)^2}{r^2} \left( D_r \XX^3 - \frac{1}{\mu^2 r^2} [\XX^1,\XX^2] \right) = 0~, \cr
D_2 \phi^{z_2} - D_{1} \phi^{z_1} - \frac{\mu^2 (r^2 + z_{0}^2)^2}{r^2} \left( D_r \XX^1 - \frac{1}{\mu^2 r^2} [\XX^2,\XX^3] \right) = 0~, \cr
D_{2} \phi^{z_1} + D_1 \phi^{z_2} - \frac{\mu^2 (r^2 + z_{0}^2)^2}{r^2} \left( D_r \XX^2 - \frac{1}{\mu^2 r^2} [\XX^3,\XX^1] \right) = 0 ~. 
\end{align}

Let $x^{\tp} = (x^1,x^2,\hat{z}_1,\hat{z}_2)$ and $a_{\tp} = (a_1,a_2,\phi^{z_1},\phi^{z_2})$, and let $\vec{\eta}_{\tp\tq}$ denote the 't Hooft matrices as in the previous subsection.  Then the first two terms in each of these equations can be written as $\half (\eta^i)^{\tp\tq} F_{\tp\tq}$ for $i=3,1,2$ respectively.  Recalling the definition \eqref{Xmom} for $\vec{\XX}_{({\rm nn})}$ as well, one sees that the leading terms in the $r \to \infty$ limit of these equations reproduce the constraint \eqref{asBPStripeqn} on the non-normalizable boundary data.  Since the leading asymptotic behavior in the full theory is controlled by precisely the degrees of freedom we are keeping in the dimensional reduction, \eqref{asBPStripeqn} gives the leading asymptotics of the first three BPS equations in the full theory too.

Each of the last four equations in \eqref{newmagBPS} implies a triplet of equations on the truncation ansatz, due to dependence on the three $j=1$ modes of the two-sphere.  However there are redundancies in these twelve equations such that only four are independent.  Hence the remaining BPS equations in the reduced theory are
\begin{align}\label{magBPSred2}
& D_2 \XX^1 + D_1 \XX^2 + \mu^2 r^2 D_r \phi^{z_2} - [\XX^3, \phi^{z_1}] = 0~, \cr
& D_2 \XX^2 - D_1 \XX^1 + \mu^2 r^2 D_r \phi^{z_1} + [\XX^3, \phi^{z_2}] = 0~, \cr
& \mu^2 r^2 F_{1r} + D_2 \XX^3 + [\XX^1, \phi^{z_1}] - [\XX^2, \phi^{z_2}] = 0~, \cr
& \mu^2 r^2 F_{2r} - D_1 \XX^3 - [\XX^1, \phi^{z_2}] - [\XX^2, \phi^{z_1}] = 0~.
\end{align}

We can work in the generalized temporal gauge $a_0 = \phi^{z_3}$ to solve the electric equations as before.  The Gauss law constraint for $\phi^{z_3}$ is then found to be
\begin{align}
0 =&~ \frac{1}{\mu^2 (r^2 +  z_{0}^2)} \left(D_p D_p + \ad(\phi^{z_p}) \ad(\phi^{z_p})\right) \phi^{z_3}  + \cr
&~ +  \mu^2 (r^2 + z_{0}^2) \left( D_{r}^2 + \frac{2}{r} D_r + \frac{1}{\mu^4 r^4} \ad(\vec{\XX}) \cdot \ad(\vec{\XX}) \right) \phi^{z_3} ~.
\end{align}

The form of these equations simplifies a bit if we work with the inverse radial coordinate
\begin{equation}
s = \frac{1}{\mu^2 r} ~,
\end{equation}
such that $s \in \phantom{(}[0,\infty)\phantom{]}$ with the holographic boundary at $s=0$.  One finds that the first three equations are
\begin{equation}\label{HWeqns1}
\half \vec{\eta}\,{}^{\tp\tq} F_{\tp\tq} + (1 + \mu^4 z_{0}^2 s^2)^2 \left( D_s \vec{\XX} + \half [\vec{\XX}, \times \vec{\XX}] \right) = 0~,
\end{equation}
while the remaining four equations can be put in the form
\begin{align}\label{HWeqns2}
& F_{s1} + D_2 \XX^3 -D_{z_1} \XX^1 + D_{z_2} \XX^2 = 0~, \qquad F_{s2} - D_1 \XX^3 + D_{z_1} \XX^2 + D_{z_2} \XX^1 = 0~, \cr
& F_{s z_1} + D_1 \XX^1 - D_2 \XX^2 + D_{z_2} \XX^3 = 0~, \qquad F_{s z_3} - D_1 \XX^2 - D_2 \XX^1 - D_{z_1} \XX^3 = 0~.
\end{align}
Here the indices $\tp,\tq = 1,2,z_1,z_2$, are raised and lowered with the flat Euclidean metric on $\mathbbm{R}^4$.

These equations are closely related to ones written down recently by Haydys in \cite{MR3338833}, and by Witten in \cite{Witten:2011zz}.  In particular, the latter reference used them to develop a gauge theory formulation of Khovanov homology for knot invariants.  In order to put the equations in the form given in \cite{Witten:2011zz}, we first set
\begin{equation}
B_{\tp\tq} := - \vec{\XX} \cdot \vec{\eta}_{\tp\tq} ~.
\end{equation}
$B$ is then an adjoint-valued self-dual two-form on the $\mathbb{R}^4$ parameterized by $x^{\tp}$.  We define the cross product
\begin{equation}
(B \times B)_{\tp\tq} := [B_{\tp\tilde{r}} , {B_{\tq}}^{\tilde{r}} ]~.
\end{equation}
This gives another adjoint-valued self-dual two-form, a fact that can be seen from the product formula for the 't Hooft matrices, $\eta^i \eta^j = -\delta^{ij} + \epsilon^{ij}_{\phantom{ij}k} \eta^k$.  Explicitly, 
\begin{equation}
(B \times B)_{\tp\tq} = - [\vec{\XX}, \times \vec{\XX}] \cdot \vec{\eta}_{\tp\tq}~.
\end{equation}

Then, using also that 
\begin{equation}
\frac{1}{4} \vec{\eta}_{\tp\tq} \cdot \vec{\eta}\,{}^{\tilde{r}\tilde{s}} = {(\Pi^+)_{\tp\tq}}^{\tilde{r}\tilde{s}} := \frac{1}{4} \left( {\delta_{\tp}}^{\tilde{r}} {\delta_{\tq}}^{\tilde{s}} - {\delta_{\tp}}^{\tilde{s}} {\delta_{\tq}}^{\tilde{r}} + {\epsilon_{\tp\tq}}^{\tilde{r}\tilde{s}} \right)~,
\end{equation}
the projector onto the self-dual forms, one finds that \eqref{HWeqns1} and \eqref{HWeqns2} can be written in the form
\begin{align}\label{HWeqns}
& F^+ - \half (1 + \mu^4 z_{0}^2 s^2)^2 \left( D_s B + \frac{1}{2} B \times B \right) = 0~, \cr
& F_{s\tq} + D^{\tp} B_{\tp\tq} = 0~,
\end{align}
where $F^+ := \Pi^+(F)$.  When $z_0 = 0$ these are the Haydys--Witten (HW) equations.  More precisely, when $z_0 = 0$ they are the HW equations on $M_4 \times \mathbb{R}_+$ where the four-manifold is $M_4 = \mathbb{R}^4$.\,\footnote{Reference \cite{Witten:2011zz} uses $y$ for the coordinate parameterizing $\mathbb{R}_+$, so $s_{\rm here} = y_{\rm there}$.}  Furthermore we only obtain these equations when they are restricted to translationally invariant configurations along two of the directions in $\mathbb{R}^4$, such that $D_{z_p} \to \ad(\phi^{z_p})$ for $p=1,2$.

Nonzero $z_0$ (which, we recall, is the separation between the D3-branes and D5-branes \`a la Figure \ref{D3D5system}),  apparently leads to a rather interesting deformation of the HW equations.  This deformation modifies the form of the equations in the interior, but the equations quickly approach their standard form as we approach the boundary at $s = 0$.  The case of nonzero $z_0$ deserves further study, but we leave it to the future and henceforth set
\begin{equation}
z_0 = 0~,
\end{equation}
for the rest of this subsection.

Translationally invariant forms of the HW equations of the type appearing here are in fact closely related to another set of equations introduced earlier by Kapustin and Witten \cite{Kapustin:2006pk}.  These equations also play an important role in the study of Khovanov homology and knot invariants \cite{Witten:2011zz,Gaiotto:2011nm}.  We follow the discussion in \cite{Witten:2011zz}.  

Suppose we start with the HW equations on $M_4 \times \mathbb{R}_+$.  Now suppose that $M_4 = \mathbb{R} \times M_3$ and we look for solutions that are translationally invariant along the first factor.  Then \eqref{HWeqns} reduces to the KW equations on $M_3 \times \mathbb{R}_+$.  These are equations for a gauge field $a$ and an adjoint-valued one-form $b$ given by\footnote{These are the KW equations with the parameter $\mathrm{t} = 1$.  This parameter can be restored by restoring dependence of the supersymmetry projection \eqref{lastproj} on the choice of $U(1)$ phase in the $z_1$-$z_2$ plane.}
\begin{equation}\label{KWeqns}
F - b \wedge b + \star D b = 0 = D \star b~.
\end{equation}
The one-form is constructed from the components of $B$ and the component of the gauge field along the first $\mathbb{R}$ factor.  The components of $B$ provide the legs along $M_3$ while the component of the gauge field along the direction associated with translation invariance is reinterpreted as the component of $b$ along $\mathbb{R}_+$.   (The precise details can be found in \cite{Witten:2011zz}; they will not be important here.)

Now suppose we have a second translation invariance.  We look for solutions to \eqref{KWeqns} on $M_3 \times \mathbb{R}_+$ where $M_3$ is of the form $M_3 = \mathbb{R} \times M_2$ and we assume translation invariance along the first factor.  This corresponds to solutions of the HW equations on $\mathbb{R}^2 \times M_2 \times \mathbb{R}_+$ that are translationally invariant in the two-plane associated with the first factor.  This is precisely the situation we have here, where the two-plane is parameterized by $(\hat{z}_1,\hat{z}_2)$ and the $M_2$ factor is $M_2 = \mathbb{R}^2$ parameterized by $(x^1,x^2)$.

The extended Bogomolny equations, also introduced in \cite{Kapustin:2006pk}, arise from this system, \ie\ the system we have in \eqref{HWeqns1}, \eqref{HWeqns2}, upon specializing to $\Phi^{z_1} = \Phi^{z_2} = 0$.  Explicitly, they are
\begin{equation}\label{exBogo}
\begin{gathered}
F_{12} + D_{s} \XX^3 + [\XX^1, \XX^2] = 0~, \\
D_s \XX^1 + [\XX^2, \XX^3] = 0~, \qquad D_s \XX^2 + [\XX^3,\XX^1] = 0~, \\
F_{s1} + D_2 \XX^3 = 0~, \qquad F_{2s} + D_1 \XX^3 = 0~, \\  
D_1 \XX^1 - \DD_2 \XX^2 = 0~, \qquad D_1 \XX^2 - D_2 \XX^1 = 0~.
\end{gathered}
\end{equation}
We have them on the half-space $\mathbb{R}^2 \times \mathbb{R}^+$, parameterized by $(x^1,x^2,s)$, with flat Euclidean metric.  They are an interesting mishmash of the (ordinary) Bogomolny equations for monopoles, the Hitchin equations, and the Nahm equations, and can be reduced to all of these upon further specializations.  The Bogomolny equations arise by setting $\XX^1 = \XX^2 = 0$, the Hitchin equations arise by setting $\XX^3 = a_s = 0$ and restricting to $s$-independent field configurations, and the Nahm equations arise by setting $a_1 = a_2 = 0$ and restricting to $x^1$- and $x^2$-independent configurations.

Although \eqref{exBogo} arises from \eqref{HWeqns1} and \eqref{HWeqns2} upon setting $\phi^{z_p} = 0$, this is not as significant of a restriction as it sounds.  In fact it is sufficient to set the boundary values $\phi_{({\rm nn})}^{z_p} = 0$.  A vanishing theorem then implies that $\phi^{z_p} = 0$ identically \cite{Witten:2011zz}.\footnote{See also the paragraph containing equation (4.11) in \cite{Witten:2010cx}.}  

The vanishing theorem can be seen from the Bogomolny bounds we have derived in this paper.  When we set $\phi_{({\rm nn})}^{z_p} = 0$, the terms that depend on $\phi_{({\rm nn})}^{z_p}$ do not contribute to the BPS energy on the right-hand side of \eqref{dwBPSbound2}.  We can also set the electric contribution to the energy to zero, since $\phi^{z_3}$ does not participate in the extended Bogomolny equations.  In this situation, the bound \eqref{dwBPSbound2} is the same as the bound we derived earlier for configurations preserving four supercharges, \eqref{halfBPSbound}.  (The $\Phi_{\infty}^y$ term drops out of both because the triplet of fieldstrengths, $\vec{f}_{12}$, has been set to zero by the truncation ansatz, \eqref{truncation}.)  But if the field configuration saturates \eqref{halfBPSbound}, then it must satisfy the stronger system of BPS equations that were used in deriving that bound---namely \eqref{primaryhalfBPS}.

Indeed, if we evaluate \eqref{primaryhalfBPS} on the truncation ansatz and set $z_0 = 0$, we recover the extended Bogomolny equations, \eqref{exBogo}, together with the conditions
\begin{equation}
D_{p} \phi^{z_q} = D_s \phi^{z_q} = [\vec{\XX},\phi^{z_q}] = [\phi^{z_1}, \phi^{z_2}] = 0~.
\end{equation}
However if $\phi^{z_p}$ is covariantly constant in $s$ and vanishing as $s = 0$, then it vanishes everywhere.

In summary, we have shown that the extended Bogomolny equations on $\mathbb{R}^2 \times \mathbb{R}_+$ arise as equations for finite-energy BPS field configurations, preserving four supercharges in maximally supersymmetric Yang--Mills on $AdS_4$.  Furthermore we have shown how maximally supersymmetric Yang--Mills on $AdS_4$ is obtained from a consistent truncation of SYM on $AdS_4 \times S^2$.  The latter is the low energy effective description of D5-branes on the bulk side of the defect AdS/CFT correspondence, as depicted in Figure \ref{fig2}.  This opens the door to the possibility of using holography to study knot invariants.  However in order to do so we will need to generalize the class of boundary conditions we have considered so far in this paper.  We sketch this idea a little further in the conclusions.

We make one final comment in closing.  After setting $z_0 = 0$, the three-dimensional equations we obtained in this section are defined on the half-space with \emph{Euclidean} metric.  Since these are BPS equations for solitons in supersymmetric Yang--Mills on $AdS_4$, one might have expected to find equations on the half-space with \emph{hyperbolic} metric.  This was the initial expection of the authors, at least.  Indeed, one of the initial motivations for this project was to embed hyberbolic monopoles into a string theory brane system.  With hindsight, the reason we get Euclidean self-duality equations seems clear.  In order to have a supersymmetric theory on $AdS_4$, the Higgs fields have to be conformally coupled.  (See \eqref{N4MSYM}.)  This means that their second-order equations of motion can be mapped to flat space equations by a conformal transformation.  In light of this, it is not surprising that the BPS equations also appear as flat space equations.  

Nevertheless the equations are defined on a manifold with boundary, and the boundary is the holographic boundary of the AdS/dCFT set-up.  There are many exciting directions to pursue, a few of which are sketched next.

\section{Conclusions and future directions}\label{sec:future}

In this paper we constructed a six-dimensional supersymmetric Yang--Mills theory on $AdS_4 \times S^2$, with $\mathfrak{osp}(4|4)$ symmetry.  We showed that, for $\mathfrak{g} = \mathfrak{su}(N_f)$ and in the regime $N_c \gg g_s N_c \gg 1$ and $N_f \ll N_c/\sqrt{g_s N_c}$, this is a good low-energy effective description of $N_f$ D5-branes probing the near-horizon geometry of $N_c$ D3-branes.  The probe D5-branes are defects on the bulk side of an AdS/dCFT correspondence that  generalizes the original, single probe set-up of \cite{Karch:2001cw,Karch:2000gx,DeWolfe:2001pq}.  The primary motivation driving this work is the application of holography to the study of curved space Yang--Mills solitons (described in greater detail below), within the context of a controlled, top-down string theory framework.

With that goal in mind, we analyzed the vacuum structure and  perturbative spectrum of the 6D SYM theory.  We also derived systems of first order equations for finite-energy BPS solitons with various fractions of supersymmetry.  Solutions to these equations saturate bounds on the energy functional, and we evaluated these bounds in terms of asymptotic data at the holographic boundary.  We left questions about the existence of solutions, and the structure of the space of solutions, to future work.  We believe that the holographic perspective will be useful here.  We describe some ongoing work along these lines below.

We also showed that the 6D theory has a nonlinear consistent truncation on the two-sphere to maximally supersymmetric Yang--Mills on $AdS_4$.  This means in particular that every solution of the lower-dimensional theory can be uplifted to a solution of the higher-dimensional one --- though of course the higher-dimensional theory contains many more solutions.

We now sketch three avenues for future work:

\paragraph{The holographic dual.}  In this paper we alluded to general features of the holographic dual on several occasions, but mostly focused on the bulk side of the correspondence.  In forthcoming work \cite{DR2}, we will construct the holographic dual in detail.  We study its vacuum structure and compare with the picture described in section \ref{sec:fluxvacua}.  For those vacua that preserve superconformal symmetry, the dCFT is a simple extension of the one in \cite{DeWolfe:2001pq}, in which the global $U(1)$ ``baryon'' symmetry is enhanced to a global $U(N_f)$.  The basic structure of the defect theory consists of a 3D $\NN = 4$ hypermultiplet, which contains a doublet of complex scalars $(q_1,q_2)$. These transform in the bi-fundamental of $SU(N_c) \times U(N_f)$, where the first factor is gauged by including couplings to (the restriction to the defect of) the 4D $\NN = 4$ vector-multiplet on the D3-branes.

We will also use the dual theory to elaborate on the structure of BPS states.  The non-normalizable modes we identified in the field asymptotics \eqref{fieldas1}, \eqref{fieldas2}, play a double role.  On the one hand they provide boundary values for the D5-brane fields participating in the various generalized self-duality equations of section \ref{sec:monopoles}.  On the other, they appear as sources for a class of dual operators constructed out of the scalars ($q_{1,2}$) in the boundary defect theory.  The correspondence suggests that supersymmetric solutions for the $q$'s in the presence of these sources will exist if and only if the same sources serve as boundary values for a supersymmetric bulk solution.  In this way we obtain a characterization of boundary values that lead to SYM solitons in the bulk.  This characterization will be given in terms of the integrability of a different system of equations for the $q$'s.

In carrying out this analysis, we will be guided by two key points.  The first is supersymmetry:  the action of supersymmetry on the defect theory will determine the relevant system of first-order BPS equations.  We are also using supersymmetry to motivate the comparison between supersymmetric solutions in the two systems.  Holography is of course a strong/weak duality, so one does not expect a priori that semiclassical techniques will be useful on both sides of the correspondence.  Our working assumption is that supersymmetry supplies the necessary rigidity  to justify the comparison.

The second point that guides the analysis is the decoupling of ambient modes.  The decoupling of closed string fluctuations in the bulk should be mirrored by the decoupling of D3-brane fluctuations in the boundary theory.  We can thus look for solutions to the boundary equations in which the D3-brane fields are restricted to their vacuum configuration.  Note that the latter can still involve a nontrivial solution to Nahm's equations, as described in section \ref{sec:fluxvacua}.

\paragraph{D-brane interpretation of solitons.}  One advantage of embedding Yang--Mills--Higgs theory into a D-brane system is that the D-brane system provides a geometric interpretation for solitons, in terms of branes extending in extra dimensions.  The basic example is the one we already mentioned in section \ref{sec:fluxvacua}:  the D3/D5 system makes manifest the equivalent descriptions of monopoles as solutions to the Bogomolny equations or as solutions to the Nahm equations \cite{Diaconescu:1996rk}.  In our case, the picture of \cite{Diaconescu:1996rk} is merely describing the \emph{vacua} of the 6D theory. But what about the soliton configurations of section \ref{sec:monopoles}?

The supersymmetry projections \eqref{susychoices} and \eqref{lastproj}, as well as the charges that appear in the Bogomolny bounds, suggest the identifications in Table \ref{table2}.  As we work our way down the list we decrease the supersymmetry by half at each stage.  Starting at the top we have the vacuum configurations of D5-branes, the original color D3-branes, and additional semi-infinite or finite-length `vacuum' D3-branes, as depicted in Figure \ref{fig4}.  These configurations preserve eight supercharges, or sixteen in special cases.

\begin{table}
\begin{center}
\caption{Solitonic D-branes} \vspace{4pt}
\begin{tabular}{c | c c c ;{2pt/2pt} c c c ;{2pt/2pt} c c c ;{2pt/2pt} c}
\hline\hline
  & 0 & 1 & 2 & $r_1$ & $r_2$ & $r_3$ & $z_1$ & $z_2$ & $z_3$ & $y$ \T\T \B\B \\
 \hline
 D3$_{c,v}$ & X & X & X &  & & & & & & X \T\T \\
 D5$_{\rm ym}$ & X & X & X & X & X & X & & & & \B\B \\
 \hdashline[2pt/2pt]
 D3$_{\rm m}$ & X & & & X & X & & & & &  X \T\T\B\B \\
 \hdashline[2pt/2pt]
 D3$_{\rm e}$ & X & & & &  & & X & X & & X \T\T \\
 F1$_{\rm e}$ & X & & & & & & & & X &  \B\B \\
 \hdashline[2pt/2pt]
 D3$_{\rm d}$ &X  & X  & & X &  &  & X &  &  & \T\T
 \end{tabular}
 \label{table2}
\end{center}
\end{table}
%

The `magnetic' D3-branes denoted by D3$_{\rm m}$ correspond to solutions to the system \eqref{primaryhalfBPS} from the D5-brane worldvolume point of view, and preserve four supercharges.  They provide the purely magnetic contribution to the BPS energy \eqref{BPSsolenergy}.  This contribution has two pieces, one proportional to the Higgs vev $\Phi_{\infty}^y$ and one proportional to the vev $|\vec{{\rm v}}|$ that gives the asymptotic value of $\XX_{({\rm n})}$ as we send $|x|$ to infinity on the two-plane boundary.  Recall that $\vec{\XX}_{({\rm n})}$ is in the commutant of $\Phi_{\infty}^y$.  (See \eqref{vortexbc}.)  These observations suggest that the $\Phi_{\infty}^y$-contribution to the energy is associated with finite length D3$_{\rm m}$-branes stretched between D5-branes at different $y$-positions, while naively the $|\vec{{\rm v}}|$ contribution is associated with infinitesimal D3$_{\rm m}$-branes stretched between coincident D5-branes.

In the case of the finite-length D3$_{\rm m}$-branes, solutions will necessarily depend on both $x^p$ and $(r,\theta,\phi)$ directions.  The general solution will describe a combination of D3$_{\rm m}$'s and D3$_{v}$'s.  In contrast, solutions describing the infinitesimal D3$_{\rm m}$'s can be obtained in the truncated theory, where the relevant BPS equations are the extended Bogomolny equations, \eqref{exBogo}.\footnote{The corresponding uplifted solutions in the D5-brane theory will also depend on $\theta,\phi$, but the dependence is specified by the truncation ansatz \eqref{truncation} and is relatively simple.}

Configurations with both D3$_{\rm m}$'s and `electric' D3-branes, D3$_{\rm e}$, preserve two supersymmetries.  We expect that they correspond to solutions to the system \eqref{primaryBPS} in the D5-brane worldvolume theory.  We've included macroscopic fundamental strings as part of this system.  They are responsible for depositing electric charge on the D5-brane worldvolume.  They should stretch along the $z_3$ direction with one end on the D3$_{\rm e}$-branes and the other end on the D5-branes.  An abelian version of this D5/D3$_{\rm 3}$/F1$_{\rm e}$ system is described in detail in \cite{Callan:1998iq}.  

Further evidence for the identification of this set of branes is provided by the following observation.  Any pair of distinct members from the set $\{ {\rm D3}_{v}, {\rm D3}_{\rm m}, {\rm D3}_{\rm e} \}$ have a $1+1$-dimensional common worldvolume.  For a given pair, let the two sets of orthogonal directions be parameterized by two complex coordinates.  Abelian intersecting D3-brane systems of this type can be deformed into a `diamond,' described by a holomorphic profile in the corresponding $\mathbb{C}^2$ \cite{Gauntlett:1998vk,Aganagic:1999fe}.  This should be an abelian version of the observation we made in \eqref{cmplxHYM4} regarding the complex form of the BPS equations.

Finally we come to the `defect' D3-branes.  When these are present with all of the other types of branes the supersymmetry is reduced to one preserved supercharge.  We expect that general solutions to the system \eqref{newmagBPS} are described by such configurations.  In particular, we showed how the BPS bound for this system, \eqref{dwBPSbound1} with \eqref{dwenergy}, receives contributions associated with domain walls in the boundary defect.  The D3$_{\rm d}$'s are mutually BPS with all other branes listed in the table, and we expect that they can be interpreted as these domain walls.

One interesting direction going forward would be to use the various D-brane identifications suggested here to provide dual descriptions of BPS solitons -- analogous to the Nahm description of monopoles.  In other words, we can analyze the supersymmetry conditions on the worldvolume theories of these various probe D3-branes, as they stretch between D5-branes and sit in the background geometry of the color D3's.  In particular, an analysis of the D3$_{\rm d}$'s should provide insight into the appropriate implementation of singularity/jumping conditions discussed under \eqref{dwenergy}.  The bouquet of branes appearing in Table \ref{table2} is reminiscent of Nekrasov's brane origami for the construction of spiked and crossed instantons \cite{Nekrasov:2016qym,Nekrasov:2016gud,Nekrasov:2016ydq}.  It would be interesting to investigate possible connections between them.

\paragraph{A holographic construction for knot invariants.}  As we discussed in section \ref{sec:reducedBPS}, the extended Bogomolny equations \eqref{exBogo} on $\mathbb{R}^2 \times \mathbb{R}_+$ play a prominent role in  Witten's gauge theory approach to Khovanov homology \cite{Witten:2011zz}.  A critical part of the construction of \cite{Witten:2011zz} is the Nahm pole boundary condition on the triplet of scalar fields, and its generalization representing the insertion of a knot.  

We are pursuing the implementation of this boundary condition in the holographic set-up developed in this paper.  Note that a Nahm pole in $\vec{\XX}$ as $s \to 0$, corresponds to an $O(1/r)$ term in the asymptotic behavior of $\Phi^y$.  This has the same fall-off as the term involving the magnetic charge, $P$, but it would be in the commutant of $P$ and described by a triplet of $\mathfrak{su}(2)$ matrices.  Given this, as well as the D-brane identifications of Table \ref{table2}, we suspect that the Nahm pole and knot boundary conditions are related to placing D3$_{\rm m}$-branes at $r = 0$. If so, these would be probe D3-branes of the type considered in another case of the defect AdS/CFT correspondence \cite{Constable:2002xt} --- and results from that analysis could be used to understand the holographic dual of the Nahm pole boundary condition!  This would be a first step towards constructing a holographically dual description of knot invariants.\footnote{Ultimately, one would like to promote the Higgs fields, $\phi^{z_p}$, in the translationally-invariant form of the HW equations, \eqref{HWeqns1}, \eqref{HWeqns2}, to honest covariant derivatives---\ie\ incorporate dependence on the corresponding $\hat{z}_p$ directions.  As we also remarked in footnote \ref{foot:T}, T-duality might be of relevance.}

\section*{Acknowledgements} We thank Ofer Aharony, Nima Arkani-Hamed, Ibrahima Bah, Katrin Becker, Melanie Becker, Dan Butter, Luca Delacr\'etaz, Hadi Godazgar, Mahdi Godazgar, Aki Hashimoto,  William Linch, Nick Manton, Greg Moore, Chris Pope, Michael Singer, and Paul Sutcliffe for helpful discussions.  SKD thanks the Mitchell Institute at Texas A\&M University for hospitality during the preparation of this work, as well as the Center for Cosmology and Particle Physics at NYU, where she was a employed when this work began. ABR thanks the Center for Cosmology and Particle Physics at NYU for hospitality.  He is supported by the Mitchell Family Foundation.

\appendix
\section{Fluctuation expansion of the Myers action}\label{app:fluc}

In this appendix we fill in some of the details in going from \eqref{Myers} to \eqref{D5expanded} and \eqref{Vocs}.

\subsection{The DBI action}

We want to make a double expansion of the DBI action \eqref{MyersDBI} in the open string variables $\mathfrak{O} \in (F_{ab}, D_a \Phi^m, [\Phi^m,\Phi^n], \mu \Phi^m)$ and the closed string fluctuations $\mathfrak{C}  \in (h_{MN},b_{MN},\varphi, c^{(n)})$.  This is carried out in three steps.  The first step is to write the NS sector closed string fields $G_{MN},B_{MN}$ in terms of their ``near-horizon'' analogs: $G_{MN} = (L\mu)^2 \tilde{G}_{MN}$ and $B_{MN} = (L\mu)^2 \tilde{B}_{MN}$.  Equivalently, $E_{MN} = (L \mu)^2 \tilde{E}_{MN}$.  The dilaton, $\Delta\upphi$, does not require rescaling.  This will enable us to write the open string fields in terms of the $\mathfrak{O}$, and we will see that factors of $\mathfrak{O}$ are accompanied by factors of $\epsilon_{\rm op}$.  

The matrix $Q^{m}_{\phantom{m}n}$, \eqref{MyersQ}, becomes
\begin{align}
Q^{m}_{\phantom{m}n} =&~ \delta^{m}_{\phantom{m}n} + (L\mu)^2 \lambda^{-1} [X^m, X^k] \tilde{E}_{kn}(X)  \cr
=&~ \delta^{m}_{\phantom{m}n} + \epsilon_{\rm op}  [\Phi^m, \Phi^k] \tilde{E}_{kn}(X) ~.
\end{align}
In the second step we introduced the scalars $\Phi^m$ according to \eqref{XtoPhi}, and then the open string expansion parameter $\epsilon_{\rm op}$ according to \eqref{epsop}.  We also emphasize that the closed string fields still depend on the transverse fluctuation scalars; $E(X)$ is shorthand for $E(x^a; -iX^m)$.  We will also need the inverse,
\begin{align}
(Q^{-1})^{m}_{\phantom{m}n} =&~ \delta^{m}_{\phantom{m}n} -  \epsilon_{\rm op} [\Phi^m, \Phi^k] \tilde{E}_{kn}(X) + \epsilon_{\rm op}^2 [\Phi^m, \Phi^k] \tilde{E}_{kl}(X) [\Phi^l, \Phi^{m'}] \tilde{E}_{m' n}(X) + O(\epsilon_{\rm op}^3)~.
\end{align}
The second index of $(Q^{-1} - \delta)^{mn}$ is raised using $E^{mn}$ which is, by definition, the inverse of $E_{km}$ \cite{Myers:1999ps}.  Hence
\begin{equation}
(Q^{-1} - \delta)^{mn} = (L\mu)^{-2} \left\{ - \epsilon_{\rm op} [\Phi^m,\Phi^n] + \epsilon_{\rm op}^2 [\Phi^m, \Phi^k] \tilde{E}_{kl}(X) [\Phi^l, \Phi^n] + O(\epsilon_{\rm op}^3) \right\}~.
\end{equation}

Applying the pullback operation \eqref{PofE}, the quantity appearing in the first determinant is
\begin{align}
\GG_{ab} :=&~ P[E_{ab}(X)] + P\left[ E_{am}(X) (Q^{-1} - \delta)^{mn} E_{nb}(X) \right] -i \lambda F_{ab} \cr
=&~ (L\mu)^2 \bigg\{ \tilde{E}_{ab}(X) + \cr
&~ -i \epsilon_{\rm op} \left[ (D_a \Phi^m) \tilde{E}_{mb}(X) + \tilde{E}_{am}(X) (D_b \Phi^m)  -i \tilde{E}_{am}(X) [\Phi^m, \Phi^n] \tilde{E}_{nb}(X) + F_{ab} \right] + \cr
&~ - \epsilon_{\rm op}^2 \bigg[ (D_a \Phi^m) \tilde{E}_{mn}(X) (D_b \Phi^n) - i (D_a \Phi^k) \tilde{E}_{km}(X) [\Phi^m,\Phi^n] \tilde{E}_{nb}(X) + \cr
&~ \quad - i \tilde{E}_{am}(X) [\Phi^m,\Phi^n] \tilde{E}_{nk}(X) (D_b \Phi^k) - \tilde{E}_{am}(X) [\Phi^m, \Phi^k] \tilde{E}_{kl}(X) [\Phi^l, \Phi^n] \tilde{E}_{nb}(X) \bigg] + \cr
&~ + O(\epsilon_{\rm op}^3) \bigg\}~.
\end{align}
Before we can take the determinant, however, we must\footnote{The reason is as follows.  The standard manipulation, $\det{(E_{ab} + M_{ab})} = \det{(E_{ab})} \det{(\mathbbm{1} + E^{-1} M)}$, is \emph{not valid} under the $\STr$ if $E_{ab}$ is a functional of the matrix-valued $X^m$, due to the fact that $\STr(A,A^{-1},B,C) \neq \STr(B,C)$.  In fact this is not an issue for us since we only work to second order in the open string $\mathfrak{O}$'s, but we prefer to present a systematic approach that could be carried out to higher orders.} extract the transverse fluctuation scalars from the closed string functionals $\tilde{E}_{MN}(X)$ according to \eqref{Etaylor}.  This finally brings us to
\begin{equation}\label{GGopenexp}
\GG_{ab} = (L\mu)^2 \left\{ \tilde{E}_{ab} -i \epsilon_{\rm op} \GG_{ab}^{(1)} - \epsilon_{\rm op}^2 \GG_{ab}^{(2)} + O(\epsilon_{\rm op}^3)  \right\}~,
\end{equation}
where
\begin{align}
\GG_{ab}^{(1)} :=&~ \Phi^m (\pd_m \tilde{E}_{ab}) |_{x_{0}^m} + (D_a \Phi^m) \tilde{E}_{mb} + \tilde{E}_{am} (D_b \Phi^m) - i \tilde{E}_{am} [\Phi^m,\Phi^n] \tilde{E}_{nb} + F_{ab}~,
\end{align}
and
\begin{align}
\GG_{ab}^{(2)} :=&~ \half \Phi^m \Phi^n (\pd_m \pd_n  \tilde{E}_{ab}) |_{x_{0}^m} + (D_a \Phi^m) \Phi^k (\pd_k \tilde{E}_{mb}) |_{x_{0}^m} + \Phi^k (\pd_k \tilde{E}_{am}) |_{x_{0}^m} (D_b \Phi^m) + \cr
& + (D_a \Phi^m) \tilde{E}_{mn} (D_b \Phi^n) - i \left( \Phi^k (\pd_k \tilde{E}_{am}) |_{x_{0}^m} + (D_a \Phi^k) \tilde{E}_{km} \right) [\Phi^m,\Phi^n] \tilde{E}_{nb} + \cr
& - i \tilde{E}_{am} [\Phi^m, \Phi^n] \left( \Phi^k (\pd_k \tilde{E}_{nb}) |_{x_{0}^m} + \tilde{E}_{nk} (D_b \Phi^k) \right) - \tilde{E}_{am} [\Phi^m, \Phi^k] \tilde{E}_{kl} [\Phi^l, \Phi^n] \tilde{E}_{nb} . \qquad
\end{align}
In these expressions, factors of $\tilde{E}_{MN}$ without explicit $|_{x_{0}^m}$ are to be understood as evaluated at $x^m = x_{0}^m$.  In particular the first term, $\tilde{E}_{ab}$, in \eqref{GGopenexp} is a scalar with respect to $U(N_f)$.  Similarly, for the second determinant in the DBI action we need the expansion
\begin{equation}
Q^{m}_{\phantom{m}n} = \delta^{m}_{\phantom{m}n} -i \epsilon_{\rm op} (Q^{(1)})^{m}_{\phantom{m}n} - \epsilon_{\rm op}^2 (Q^{(2)})^{m}_{\phantom{m}n} + O(\epsilon_{\rm op}^3)~,
\end{equation}
with
\begin{align}
& (Q^{(1)})^{m}_{\phantom{m}n} := i [\Phi^m, \Phi^k] \tilde{E}_{kn} ~, \qquad (Q^{(2)})^{m}_{\phantom{m}n} = i [\Phi^m,\Phi^k] \Phi^l (\pd_l \tilde{E}_{kn}) |_{x_{0}^m}~.
\end{align}

The next step is to evaluate the determinants perturbatively in $\epsilon_{\rm op}$:
\begin{align}
\sqrt{-\det(\GG_{ab})} =&~ (L\mu)^6 \sqrt{-\det(\tilde{E}_{ab})} \bigg\{ 1 - \frac{i \epsilon_{\rm op}}{2} \tilde{E}^{ab} (\GG^{(1)})_{ba} + \cr
&~ \quad - \epsilon_{\rm op}^2 \left[ \half \tilde{E}^{ab} (\GG^{(2)})_{ba} - \frac{1}{4} \tilde{E}^{ab} (\GG^{(1)})_{bc} \tilde{E}^{cd} (\GG^{(1)})_{da} + \frac{1}{8} \left( \tilde{E}^{ab} (\GG^{(1)})_{ba} \right)^2 \right] + \cr
&~ \quad + O(\epsilon_{\rm op}^3) \bigg\}~,
\end{align}
where $\tilde{E}^{ab}$ is defined to be the inverse of $\tilde{E}_{bc}$.  A similar formula applies for $\sqrt{\det(Q^{m}_{\phantom{m}n})}$.  The integrand of the DBI action is a product of these two determinants and the dilaton factor, which must also be expanded in open string fluctuations:
\begin{align}
e^{-\Delta\upphi(X)} =&~ e^{-\Delta\upphi} \bigg\{ 1 +i \epsilon_{\rm op} \Phi^m (\pd_m \Delta\upphi) |_{x_{0}^m} + \cr
&~ \qquad \qquad -  \frac{\epsilon_{\rm op}^2}{2} \Phi^m \Phi^n \left. \left( \pd_m \Delta\upphi \pd_n \Delta\upphi - \pd_m \pd_n \Delta\upphi \right) \right|_{x_{0}^m} + O(\epsilon_{\rm op}^3) \bigg\}~.
\end{align}

This results in the open string expansion of the DBI action,
\begin{equation}\label{SDBIopenex}
S_{\rm DBI} = - \tau_{\rm D5} (L\mu)^6 \int \ed^6 x \left\{ V^{\rm DBI}_{(0)} + \epsilon_{\rm op} V^{\rm DBI}_{(1)} + \epsilon_{\rm op}^2 V^{\rm DBI}_{(2)} + O(\epsilon_{\rm op}^3) \right\}~,
\end{equation}
with
\begin{align}
V^{\rm DBI}_{(0)} =&~ -\sqrt{- \det(\tilde{E}_{ab})} e^{-\Delta\upphi} \STr(1)~, \cr
V^{\rm DBI}_{(1)} =&~ i \sqrt{- \det(\tilde{E}_{ab})} e^{-\Delta\upphi} \STr \left\{ \half \left( \tilde{E}^{ab} (\GG^{(1)})_{ba} +  (Q^{(1)})^{m}_{\phantom{m}m} \right) - \Phi^m (\pd_m \Delta\upphi) |_{x_{0}^m} \right\}~,  \qquad
\end{align}
and
\begin{align}\label{VDBI2open}
V^{\rm DBI}_{(2)} =&~ \sqrt{- \det(\tilde{E}_{ab})} e^{-\Delta\upphi} \STr \bigg\{ \half \left( \tilde{E}^{ab} (\GG^{(2)})_{ba} +  (Q^{(2)})^{m}_{\phantom{m}m} - \Phi^m \Phi^n (\pd_m \pd_n \Delta\upphi) |_{x_{0}^m} \right) + \cr
&~ \qquad \qquad \qquad \qquad - \frac{1}{4} \left( \tilde{E}^{ab} (\GG^{(1)})_{bc} \tilde{E}^{cd} (\GG^{(1)})_{da} + (Q^{(1)})^{m}_{\phantom{m}n} (Q^{(1)})^{n}_{\phantom{n}m} \right) + \cr
&~ \qquad \qquad \qquad \qquad + \frac{1}{8} \left(  \tilde{E}^{ab} (\GG^{(1)})_{ba} +  (Q^{(1)})^{m}_{\phantom{m}m} - 2\Phi^m (\pd_m \Delta\upphi) |_{x_{0}^m}  \right)^2 \bigg\}~. \qquad \qquad \raisetag{22pt}
\end{align}

The final step is to expand in closed string fluctuations.  This is straightforward using
\begin{align}
\tilde{E}_{MN} =&~ e^{\kappabar \varphi/2} \left(\Gbar_{MN} + \kappabar (h_{MN} + b_{MN})\right) ~,
\end{align}
and gives expansions
\begin{equation}
V_{(n_o)}^{\rm DBI} = \sqrt{- g_6} \sum_{n_c} \kappabar^{n_c} \, V_{n_o,n_c}^{\rm DBI}~,
\end{equation}
for each of the $V_{(n_o)}^{\rm DBI}$ above, where $g_6 \equiv \det(\Gbar_{ab}(x^a,x_{0}^m))$.  This results in the contributions from the DBI action to the $V_{n_o,n_c}$'s appearing in \eqref{Vocs}.

\subsection{The CS action}

Now we carry out the analogous steps for the CS action, \eqref{MyersCS}.  First consider the near-horizon rescaling.  Given the form of $C^{(n)}$ and $B_{MN}$ in \eqref{csfluc}, it is natural to define $\tilde{\CC}^{(n)}$ such that\footnote{The sum over $n$ starts at $n=0$, runs over even values, and truncates at $n=10$, the top degree for a form on spacetime.}
\begin{equation}
\mathcal{C} = \sum_n (L\mu)^n \tilde{\CC}^{(n)} ~.
\end{equation}
Then, for example,
\begin{align}
\tilde{\CC}^{(4)} = \tilde{C}^{(4)} + \tilde{C}^{(2)} \wedge e^{\Delta\upphi/2} \tilde{B} + \half \tilde{C}^{(0)} \wedge e^{\Delta\upphi} \tilde{B}^2~,
\end{align}
where $\tilde{C}^{(4)} = \Cbar^{(4)} + \kappa c^{(4)}$, $\tilde{C}^{(n)} = \kappabar c^{(n)}$ otherwise, and $\tilde{B} = \kappabar b$.  

Now observe that
\begin{equation}
\exp \left( \lambda^{-1} {\rm i}_X {\rm i}_X \right) = \exp \left(  (L\mu)^{-2} \epsilon_{\rm op} \, {\rm i}_{\Phi} {\rm i}_{\Phi} \right)~,
\end{equation}
when acting on any $n$-form, and that
\begin{equation}
\exp(-i \lambda F) = \exp \left(-i (L \mu)^2 \epsilon_{\rm op} \, F \right)~.
\end{equation}
Hence each power of ${\rm i}_{\Phi}^2$ comes with a factor of $(L\mu)^{-2} \epsilon_{\rm op}$ and reduces the degree of the form it acts on by two.  Meanwhile each power of $F$ comes with a factor of $(L\mu)^2 \epsilon_{\rm op}$ and increases the degree of the form by two.  Since the integral picks out the six-form part, it follows that every term scales as $(L\mu)^6$ and we have the equality
\begin{equation}
\tau_{\rm D5} \int \STr \bigg\{ P \left[ e^{\lambda^{-1} {\rm i}_X {\rm i}_X} \CC \right] \wedge e^{-i\lambda F} \bigg\} = \tau_{\rm D5} (L\mu)^6 \int \STr \bigg\{ P \left[ e^{\epsilon_{\rm op} \, {\rm i}_{\Phi} {\rm i}_{\Phi}} \tilde{\CC} \right] \wedge e^{-i\epsilon_{\rm op} F} \bigg\} ~,
\end{equation}
where $\tilde{\CC} := \sum_{n} \tilde{\CC}^{(n)}$.  Since each power of ${\rm i}_{\Phi}^2$ will produce a factor of $[\Phi^m,\Phi^n]$, and each power of $(D_a \Phi^m)$ from the pullback operation comes with an $\epsilon_{\rm op}$, it is clear that the expansion in open string variables $\mathfrak{O}$ is organized by $\epsilon_{\rm op}$.

The next step is to carry out the open string expansion, which is simply a matter of pealing away the various operations on $\tilde{\CC}$.  Working to $O(\epsilon_{\rm op}^2)$ we first have that
\begin{align}\label{chFexp}
\bigg\{ P\bigg[ e^{\epsilon_o {\rm i}_{\Phi}^2} \tilde{\CC}(X) \bigg] e^{-i \epsilon_o F} \bigg\}^{(6)} =&~ \bigg\{ P\bigg[ e^{i \epsilon_{\rm op} \,{\rm i}_{\Phi}^2} \tilde{\CC}(X) \bigg] \bigg\}^{(6)} -i \epsilon_{\rm op} \bigg\{ P\bigg[ e^{\epsilon_{\rm op} \, {\rm i}_{\Phi}^2} \tilde{\CC}(X) \bigg] \bigg\}^{(4)} \wedge F + \cr
&~ - \half \epsilon_{\rm op}^2 \bigg\{ P\bigg[ e^{\epsilon_{\rm op} \, {\rm i}_{\Phi}^2} \tilde{\CC}(X) \bigg] \bigg\}^{(2)} \wedge F^2 + O(\epsilon_{\rm op}^3)~. \raisetag{20pt}
\end{align}
Next we expand $e^{\epsilon_{\rm op} \, {\rm i}_{\Phi} {\rm i}_{\Phi}}$ as far as necessary in each term:
\begin{align}\label{interiorexp}
\bigg[ e^{\epsilon_{\rm op} \, {\rm i}_{\Phi}^2} \tilde{\CC}(X) \bigg]^{(6)} =&~ \tilde{\CC}^{(6)}(X) + \epsilon_{\rm op} \, {\rm i}_{\Phi}^2 \tilde{\CC}^{(8)}(X) + \half \epsilon_{\rm op}^2 \, {\rm i}_{\Phi}^2 {\rm i}_{\Phi}^2 \tilde{\CC}^{(10)}(X)~, \cr
\bigg[ e^{\epsilon_{\rm op} \, {\rm i}_{\Phi}^2} \tilde{\CC}(X) \bigg]^{(4)} =&~ \tilde{\CC}^{(4)}(X) + \epsilon_{\rm op} \, {\rm i}_{\Phi}^2 \tilde{\CC}^{(6)}(X) + O(\epsilon_{\rm op}^2)~, \cr
\bigg[ e^{\epsilon_{\rm op} {\rm i}_{\Phi}^2} \tilde{\CC}(X) \bigg]^{(2)} =&~ \tilde{\CC}^{(2)}(X) + O(\epsilon_{\rm op})~.
\end{align}
Then the pullbacks that need to be computed are
\begin{align}\label{PCexp}
P[\tilde{\CC}^{(6)}]_{abcdef} =&~ \tilde{\CC}^{(6)}_{abcdef} - 6i \epsilon_{\rm op} (D_{[a} \Phi^m) \tilde{\CC}^{(6)}_{|m|bcdef]} + \cr
&~ -  6 \cdot 5 \epsilon_{\rm op}^2 (D_{[a} \Phi^m) (D_b \Phi^n) \tilde{\CC}^{(6)}_{|mn|cdef]} + O(\epsilon_{\rm op}^3) ~, \nonumber \\[1ex]
\left[ P \left( {\rm i}_{\Phi}^2 \tilde{\CC}^{(8)} \right) \right]_{abcdef} =&~  \left( {\rm i}_{\Phi}^2 \tilde{\CC}^{(8)} \right)_{abcdef} - 6i \epsilon_{\rm op} (D_{[a} \Phi^m)  \left( {\rm i}_{\Phi}^2 \tilde{\CC}^{(8)} \right)_{|m|bcdef]} + O(\epsilon_{\rm op}^2) ~, \nonumber \\[1ex]
P[\tilde{\CC}^{(4)}]_{abcd} =&~ \tilde{\CC}^{(4)}_{abcd} - 4i \epsilon_{\rm op} (D_{[a} \Phi^m) \tilde{\CC}^{(4)}_{|m|bcd]} + O(\epsilon_{\rm op}^2)~,
\end{align}
while the remaining ones can be evaluated at leading order.

Assembling the pieces brings us to the following expression for the CS integrand:
\begin{align}\label{openCSexp}
& \bigg\{ P\bigg[ e^{\epsilon_{\rm op} \, {\rm i}_{\Phi}^2} \tilde{\CC} \bigg] e^{\epsilon_{\rm op} F} \bigg\}^{(6)} = \cr
& =  \epsilon^{abcdef} \bigg\{ \frac{1}{6!} \tilde{\CC}^{(6)}_{abcdef} + \cr
& \qquad \qquad -i \epsilon_{\rm op} \bigg[ \frac{1}{5!} (D_{a} \Phi^m) \tilde{\CC}^{(6)}_{mbcdef} + \frac{i}{6!} \left( {\rm i}_{\Phi}^2 \tilde{\CC}^{(8)} \right)_{abcdef} + \frac{1}{4!2!} \tilde{\CC}_{abcd}^{(4)} F_{ef} \bigg] + \cr
& \qquad \qquad - \epsilon_{\rm op}^2 \bigg[ \frac{1}{4!} (D_{a} \Phi^m) (D_b \Phi^n) \tilde{\CC}^{(6)}_{mncdef} + \frac{1}{5!} (D_{a} \Phi^m)  \left( {\rm i}_{\Phi}^2 \tilde{\CC}^{(8)} \right)_{mbcdef} + \cr
& \qquad \qquad \qquad - \frac{1}{6! 2} \left( {\rm i}_{\Phi}^2 {\rm i}_{\Phi}^2 \tilde{\CC}^{(10)} \right)_{abcdef} + \left[ \frac{1}{3! 2!} (D_a \Phi^m) \tilde{\CC}_{mbcd}^{(4)} + \frac{i}{4! 2!} \left({\rm i}_{\Phi}^2 \tilde{\CC}^{(6)}\right)_{abcd} \right] F_{ef} + \cr
& \qquad \qquad \qquad + \frac{1}{16} \tilde{\CC}_{ab}^{(2)} F_{cd} F_{ef} \bigg] \bigg\} \sqrt{-g_6} \ed^6 x+ O(\epsilon_{\rm op}^3)~.
\end{align}
We have suppressed the arguments of the $\tilde{\CC}^{(n)}$ in \eqref{interiorexp}, \eqref{PCexp}, and \eqref{openCSexp}, but it should be understood that they are still functionals of the transverse scalars $X^m$ at this point.  The final step in the open string expansion is to expand them according to \eqref{Etaylor}.  This brings us to the result
\begin{equation}\label{SCSopenex}
S_{\rm CS} = \tau_{\rm D5} (L \mu)^6 \int \ed^6 x \sqrt{-g_6} \left\{ V_{(0)}^{\rm CS} + \epsilon_{\rm op} V_{(1)}^{\rm CS} + \epsilon_{\rm op}^2 V_{(2)}^{\rm CS} +O(\epsilon_{\rm op}^3) \right\}~,
\end{equation}
where
\begin{align}
V_{(0)}^{\rm CS} =&~ -\frac{1}{6!} \epsilon^{abcdef} \tilde{\CC}^{(6)}_{abcdef} \STr (1 )~, \qquad \textrm{and} \cr
V_{(1)}^{\rm CS} =&~ i\epsilon^{abcdef} \STr \bigg\{  \frac{1}{6!} \Phi^m (\pd_m \tilde{\CC}_{abcdef}) |_{x_{0}^m} + \frac{1}{5!} (D_{a} \Phi^m) \tilde{\CC}^{(6)}_{mbcdef} + \frac{i}{6!} \left( {\rm i}_{\Phi}^2 \tilde{\CC}^{(8)} \right)_{abcdef} + \cr
&~ \qquad \qquad \qquad +  \frac{1}{4!2!} \tilde{\CC}_{abcd}^{(4)} F_{ef} \bigg\}~.
\end{align}
Rather than writing out the full $V_{(2)}^{\rm CS}$ we just give the single term that will contribute at leading order in the closed string expansion:
\begin{equation}\label{VCS2open}
V_{(2)}^{\rm CS} = \epsilon^{abcdef} \STr \left\{ \frac{1}{3!2} (D_a \Phi^m) \tilde{\CC}_{mbcd}^{(4)} F_{ef} + \cdots \right\}~.
\end{equation}

The final step is the expansion of the above in closed string fluctuations,
\begin{equation}
V_{(n_o)}^{\rm CS} = \sum_{n_c} \kappabar^{n_c} \, V_{n_o,n_c}^{\rm CS}~.
\end{equation}
Some relevant observations are
\begin{align}
& \tilde{\CC}^{(6)}_{abcdef} = \kappabar \, c_{abcdef}^{(6)} + \kappabar^2 \frac{6!}{4! 2} c_{[abcd}^{(4)} b_{ef]} + O(\kappabar^3)~, \cr
& \tilde{\CC}_{mbcdef}^{(6)} = \kappabar \left( c_{mbcdef}^{(6)} + \frac{5!}{3! 2} \Cbar_{m [bcd}^{(4)} b_{ef]} \right) + O(\kappabar^2)~, \cr
& \tilde{\CC}^{(4)}_{mbcd} = \Cbar^{(4)}_{mbcd} + \kappabar \, c^{(4)}_{mbcd} + O(\kappabar^2)~, \qquad \tilde{\CC}_{abcd}^{(4)} = \kappabar \, c_{abcd}^{(4)} + O(\kappabar^2)~.
\end{align}
The rest is straightforward and these results together with the $V_{n_o,n_c}^{\rm DBI}$ give the $V_{n_0,n_c}$ quoted in \eqref{Vocs} through
\begin{equation}
V_{n_o,n_c} = V_{n_o,n_c}^{\rm DBI} - V_{n_o,n_c}^{\rm CS}~.
\end{equation}
%

\section{Background geometry and Killing spinors}\label{app:spinors}

Let $x^{\tilde{\mu}}$, $\tilde{\mu} = 0,\ldots, 3$ denote the collection of coordinates $x^{\tilde{\mu}} = (x^\mu,y)$, let $v^{I}$, $I = 1,\ldots, 6$ denote the collection of coordinates $v^{I} = (r_i, z_j)$, and set $v = \sqrt{v^I v^I}$.  The 10D background metric is
\begin{equation}
\ed \sbar_{10}^2 = (\mu v)^2 \eta_{\tilde{\mu}\tilde{\nu}} \ed x^{\tilde{\mu}} \ed x^{\tilde{\nu}} + \frac{\delta_{IJ} \ed v^I \ed v^J}{(\mu v)^2}~,
\end{equation}
and we take the frame to be
\begin{equation}
e^{\utmu} = (\mu v) \ed x^\tmu ~, \qquad e^{\uI} = \frac{\ed v^I}{(\mu v)}  ~.
\end{equation}
This is the `Cartesian-like' frame introduced in \eqref{Cartframe}.  The components of the spin connection are
\begin{equation}
\omegabar_{\underline{\tmu I},\utnu} = (\mu \hat{v}_{\uI}) \eta_{\underline{\tmu\tnu}} ~, \qquad \omegabar_{\uI\uJ,\uK} = \mu (\hat{v}_{\uI} \delta_{\underline{JK}} - \hat{v}_{\uJ} \delta_{\underline{IK}} )~,
\end{equation}
and so the covariant derivatives, $\hat{D}_{P}^{(0)} := \pd_M + \frac{1}{4} \omegabar_{\underline{MN},P} \Gamma^{\underline{MN}}$, are
\begin{equation}
\hat{D}_{\tmu}^{(0)} = \pd_{\tmu} + \frac{\mu}{2} \Gamma_{\tmu} (\hat{v}_{\uI} \Gamma^{\uI}) ~, \qquad \hat{D}_{I}^{(0)} = \pd_I + \frac{\mu}{4} \left( (\hat{v}_{\uJ} \Gamma^{\uJ}) \Gamma_{I} - \Gamma_I (\hat{v}_{\uJ} \Gamma^{\uJ}) \right)~,
\end{equation}
where we use the shorthand $\hat{v}_{\uI} := v_I/v$.

Now $\Fbar^{(5)} = 4\mu (1 + \star) \vol_{AdS_5}$, and in these coordinates
\begin{equation}
\vol_{AdS_5} = (\mu v)^3 \ed^4 x \wedge \ed v = e^{\underline{0}} \wedge e^{\underline{1}} \wedge e^{\underline{2}} \wedge e^{\underline{y}} \wedge \hat{v}_{\uI} e^{\uI}~.
\end{equation}
It follows that
\begin{align}
\frac{1}{5!} \Gamma^{\underline{M_1 \cdots M_5}} \Fbar_{\underline{M_1 \cdots M_5}}^{(5)}  =&~ 4 \mu \hat{v}_{\uI} \left( \Gamma^{\underline{012y I}} + \frac{1}{5!} \epsilon_{\underline{I_1 \cdots I_5}}^{\phantom{I_1 \cdots I_5} \underline{I 012 y}} \Gamma^{\underline{I_1 \cdots I_5}} \right) \cr
=&~ 4 \mu \hat{v}_I \Gamma^{\underline{I 012y}} \left( 1 - \Gammabar \right)~,
\end{align}
where we are using that $\epsilon^{\underline{012r_1r_2r_3 z_1 z_2 z_3 y}} = 1$ and $\Gammabar := \Gamma^{\underline{012r_1r_2r_3 z_1 z_2 z_3 y}}$ is the 10D chirality operator.

Hence the $M = \tmu$ components of the Killing spinor equation \eqref{bulkKSeqn} take the form
\begin{equation}
\left[ \pd_{\tmu} + \frac{\mu}{2} \Gamma_{\tmu} (\hat{v}_{\uI} \Gamma^{\uI}) \left( \mathbbm{1} - i \Gamma^{\underline{012 y}} \right) \right] \epsilon = 0~.
\end{equation}
We write $\epsilon = \epsilon_+ + \epsilon_-$ with $\epsilon_{\pm} = \pm i \Gamma^{\underline{012 y}} \epsilon_{\pm}$ and project the equation onto $i \Gamma^{\underline{012 y}}$ eigenspaces:
\begin{equation}
\pd_{\tmu} \epsilon_- = 0~, \qquad \pd_{\tmu} \epsilon_+ = - \mu \Gamma_{\tmu} (\hat{v}_{\uI} \Gamma^{\uI}) \epsilon_- ~.
\end{equation}
The solutions can be parameterized as
\begin{equation}\label{Mmusol}
\epsilon_- = (\hat{v}_{\uJ} \Gamma^{\uJ}) \tilde{\epsilon}_-(v) ~, \qquad \epsilon_+ = \tilde{\epsilon}_+(v) - (\mu v) (\mu x^{\mu} \Gamma_{\umu}) \tilde{\epsilon}_-(v)~,
\end{equation}
where $\tilde{\epsilon}_{\pm}$ are functions of the $v^I$ only and satisfy $i \Gamma^{\underline{012 y}} \tilde{\epsilon}_{\pm} = \pm \tilde{\epsilon}_{\pm}$.  Note that $\tilde{\epsilon}_{+}$ has the same 10D chirality as $\epsilon$ itself, but that $\tilde{\epsilon}_-$ has the opposite.  In our conventions, $\Gammabar \tilde{\epsilon}_{\pm} = \pm \tilde{\epsilon}_{\pm}$.

Turning to the $M = I$ components of \eqref{bulkKSeqn} and projecting the equation onto $i \Gamma^{\underline{012y}}$ eigenspaces gives
\begin{align}
\left[ \pd_I + \mu (\hat{v}_{\uJ} \Gamma^{\uJ}) \Gamma_{I} - \frac{v_I}{2v^2} \right] \epsilon_- = 0~, \qquad   \left[ \pd_I  - \frac{v_I}{2v^2} \right] \epsilon_+ = 0 ~,
\end{align}
and one finds that these equations are solved by taking
\begin{equation}
\tilde{\epsilon}_+(v) = \frac{1}{\sqrt{\mu v}} \epsilon_{+}^0 ~, \qquad \tilde{\epsilon}_-(v) = \sqrt{\mu v} \, \epsilon_{-}^0~,
\end{equation}
where $\epsilon_{\pm}^0$ are constant spinors satisfying the same projection conditions as the $\tilde{\epsilon}_{\pm}$.  Plugging these back into \eqref{Mmusol}, one finds that $\epsilon = \epsilon_+ + \epsilon_-$ takes the form given in \eqref{epscart}.

\subsection{Frame rotations}

The relationship between the triplets $\{ \Gamma^{\underline{r_i}} \}$ and $\{\Gamma^{\ur}, \Gamma^{\utheta}, \Gamma^{\uphi} \}$ is expressed in terms of the usual rotation sending the $\{\hat{r}_3, \hat{r}_1, \hat{r}_2 \}$ frame to the $\{\hat{r},\hat{\theta},\hat{\phi}\}$ frame in $\mathbb{R}^3$:
\begin{align}\label{CtoS}
\Gamma^{\underline{r}} =&~ \Gamma^{\underline{r_3}} \cos{\theta} + (\Gamma^{\underline{r_1}} \cos{\phi} + \Gamma^{\underline{r_2}} \sin{\phi}) \sin{\theta} = U(\theta,\phi) \Gamma^{\underline{r_3}} U(\theta,\phi)^{-1} ~, \cr
\Gamma^{\underline{\theta}} =&~ - \Gamma^{\underline{r_3}} \sin{\theta} + (\Gamma^{\underline{r_1}} \cos{\phi} + \Gamma^{\underline{r_2}} \sin{\phi}) \cos{\theta} = U(\theta,\phi) \Gamma^{\underline{r_1}} U(\theta,\phi)^{-1} ~, \cr
\Gamma^{\underline{\phi}} =&~ - \Gamma^{\underline{r_1}} \sin{\phi} + \Gamma^{\underline{r_2}} \cos{\phi} = U(\theta,\phi) \Gamma^{\underline{r_2}} U(\theta,\phi)^{-1}~,
\end{align}
where
\begin{equation}
U(\theta,\phi)^{-1} = \exp \left( \frac{\theta}{2} \Gamma^{\underline{r_3 r_1}} \right) \exp \left( \frac{\phi}{2} \Gamma^{\underline{r_1 r_2}} \right) ~.
\end{equation}
These relationships are basis independent and hold for any matrix representations of the $\Gamma$.  If we work in a basis (of sections of the Dirac spinor bundle) where the matrix elements of the $\Gamma^{\underline{r_i}}$ are constant then, as is clear from \eqref{CtoS}, the matrix elements of $\Gamma^{\ur},\Gamma^{\utheta},\Gamma^{\uphi}$ will not be.  Conversely, if we work in a basis where the matrix elements of $\Gamma^{\ur},\Gamma^{\utheta},\Gamma^{\uphi}$ are constant, then those of the $\Gamma^{\underline{r_i}}$ will not be.  Typically, one assumes a basis with respect to which gamma matrices with tangent frame indices are constant.  Such an assumption was implicit in \eg\ writing the solutions \eqref{Mmusol} above---when we said the spinors $\epsilon_{\pm}^0$ are constant, this meant constant with respect to such a basis.  A basis in which gamma matrices carrying the same tangent space indices as the frame have constant matrix elements will be referred to as a \emph{natural} basis associated with the given frame.  

In expressions like \eqref{Mmusol} containing `constant' spinors a natural basis, with respect to which the $\Gamma^{\underline{M}}$ that appear in the expression have constant matrix elements, will always be assumed unless explicitly stated otherwise.  When we want to emphasize the choice of basis, we will write brackets, $(\cdot)$ with a  subscript label, around the quantity in question.  We write $(\cdot)_{\rm cart}$, for `Cartesian,' for a natural basis with respect to the frame $\{ e^{\underline{r_i}} \}$, and we write $(\cdot)_{S^2}$ for a natural basis with respect to the frame $\{ e^{\ur}, e^{\utheta}, e^{\uphi} \}$.

The transformation \eqref{CtoS} takes an active point of view: we are rotating the $\Gamma$ themselves, rather than any basis we may choose to express their matrix elements with respect to.  However, when we wish to understand how the presentation of a solution such as \eqref{Mmusol} changes when we change our choice of frame, then we must take a passive point of view.  The change of basis transformation, $h_{S^2}(\theta,\phi)$, that maps components with respect to the Cartesian basis to components with respect to the $S^2$ frame is precisely the (lift to the Dirac bundle of the) inverse of the frame rotation:
\begin{equation}\label{GammaCOB}
(\Gamma^{\ur,\utheta,\uphi})_{S^2} = h_{S^2}(\theta,\phi) (\Gamma^{\ur,\utheta,\uphi})_{\rm cart} h_{S^2}(\theta,\phi)^{-1} ~, \qquad \textrm{with} \quad h_{S^2}(\theta,\phi) = U^{-1}(\theta,\phi)~.
\end{equation}
Note that the relationship $h_{S^2} = U^{-1}$ allows us to express $h_{S^2}$ in terms of $\Gamma^{\ur},\Gamma^{\utheta},\Gamma^{\uphi}$ instead of $\Gamma^{\underline{r_i}}$:  On the one hand, $U h_{S^2} U^{-1} = U U^{-1} U^{-1} = U^{-1} = h_{S^2}$, while on the other hand from \eqref{CtoS} we have
\begin{equation}
U h_{S^2} U^{-1} = \exp\left( \frac{\theta}{2} \Gamma^{\underline{r\theta}} \right) \exp \left(\frac{\phi}{2} \Gamma^{\underline{\theta\phi}}\right)~.
\end{equation}
This gives $h_{S^2}$ as we defined it in \eqref{hS2def}.  Note also that $h_{S^2} = U^{-1}$ implies the relations
\begin{equation}\label{sillyGrel}
(\Gamma^{\utheta}, \Gamma^{\uphi}, \Gamma^{\ur})_{S^2} = (\Gamma^{\underline{r_1}}, \Gamma^{\underline{r_2}}, \Gamma^{\underline{r_3}})_{\rm cart}~,
\end{equation}
among the matrix elements of different $\Gamma$'s referred to different bases.

Thus, if $(\epsilon)_{\rm cart}$ and $(\epsilon)_{S^2}$ denote the components of the Killing spinor with respect to a natural basis associated with the $\{ e^{\underline{r_i}} \}$-frame and $\{ e^{\ur},e^{\utheta},e^{\uphi} \}$-frame respectively, then they are related by $h_{S^2}$.  To see how this gives \eqref{bulkKSsol} from \eqref{epscart}, we first observe from \eqref{CtoS} that
\begin{equation}
(\Gamma^{\ur})_{\rm cart} = \hat{r}_i (\Gamma^{\underline{r_i}})_{\rm cart}~,
\end{equation}
so we can write $(\epsilon)_{\rm cart}$ as
\begin{equation}
(\epsilon)_{\rm cart} = \frac{1}{\sqrt{\mu v}} \left( \frac{r (\Gamma^{\ur})_{\rm cart} + z_i \Gamma^{\underline{z_i}} }{v} \right) \epsilon_{-}^0 + \sqrt{\mu v} \left[ \epsilon_{+}^0 - \mu (x^p \Gamma_{\up} + y \Gamma_{\uy}) \epsilon_{-}^0 \right]~,
\end{equation}
Then we set
\begin{equation}
(\epsilon)_{S^2} = h_{S^2}(\theta,\phi) (\epsilon)_{\rm cart} ~,
\end{equation}
and use \eqref{GammaCOB}.  This results in the expression \eqref{bulkKSsol}.  Note that the remaining gamma matrices, $\{ \Gamma^{\up}, \Gamma^{\underline{z_i}},\Gamma^{\underline{y}}\}$, are the same with respect to both bases.

Additional frame rotations can be made to bring the bulk Killing spinors into a form found more commonly in the literature.  First consider introducing spherical coordinates $(z,\zeta,\chi)$ for the $\vec{z}$ directions such that $(z_1,z_2,z_3) = z(\sin{\zeta}\cos{\chi},\sin{\zeta}\sin{\chi},\cos{\zeta})$.  Then $\hat{z}_i (\Gamma^{\underline{z_i}})_{\rm cart} = (\Gamma^{\uz})_{\rm cart}$ and the Cartesian basis is transformed to an $S^2$ basis by a completely analogous set of formulae.  Referring to this 10D frame and its associated natural bases by the subscript $S^2 \times S^2$ one finds that the components of the Killing spinor are
\begin{align}\label{KSS2S2}
(\epsilon)_{S^2 \times S^2} =&~ h_{S^2}(\zeta,\chi) h_{S^2}(\theta,\phi) (\epsilon)_{\rm cart} \cr
=&~ \frac{1}{\sqrt{\mu v}} \left( \frac{r \Gamma^{\ur} + z \Gamma^{\uz}}{v} \right) h_{S^2}(\zeta,\chi) h_{S^2}(\theta,\phi) \epsilon_{-}^0 + \cr
&~ + \sqrt{\mu v} h_{S^2}(\zeta,\chi) h_{S^2}(\theta,\phi) \left[ \epsilon_{+}^0 - \mu (x^p \Gamma_{\up} + y \Gamma_{\uy}) \epsilon_{-}^0 \right]~,
\end{align}
where
\begin{equation}
h_{S^2}(\zeta,\chi) = \exp\left( \frac{\zeta}{2} \Gamma^{\underline{z\zeta}} \right) \exp \left( \frac{\chi}{2} \Gamma^{\underline{\zeta\chi}} \right)~.
\end{equation}

Then one can exchange coordinates $(r,z)$ for $(v,\psi)$ via
\begin{equation}
r = v \cos{\psi}~, \qquad z = v \sin{\psi}~,
\end{equation}
for $\psi \in [0,\pi/2]$, which brings the metric to the form used in \cite{DeWolfe:2001pq}:
\begin{align}
\ed \sbar_{10}^2 =&~ (\mu v)^2 \eta_{\mu\nu} \ed x^\mu \ed x^\nu + \frac{\ed v^2}{(\mu v)^2} + \mu^{-2} \ed \Omega_{5}^2~, \qquad \textrm{with} \cr
\ed \Omega_{5}^2 =&~ \ed \psi^2 + \cos^2{\psi} \left( \ed \theta^2 + \sin^2{\theta} \ed\phi^2 \right) + \sin^2{\psi} \left( \ed \zeta^2 + \sin^2{\zeta} \ed \chi^2 \right)~.
\end{align}
Taking $e^{\underline{v}} = \ed v/(\mu v)$ and $e^{\underline{\psi}} = \mu^{-1} \ed\psi$, the frames (and hence gamma matrices) are related by
\begin{align}\label{rz2vpsi}
\Gamma^{\underline{v}} =&~ \cos{\psi} \Gamma^{\underline{r}} + \sin{\psi} \Gamma^{\underline{z}} =   \exp\left( - \frac{\psi}{2} \Gamma^{\underline{r z}} \right) \Gamma^{\underline{\ur}} \exp\left( \frac{\psi}{2} \Gamma^{\underline{r z}} \right)  \cr
\Gamma^{\underline{\psi}} =&~ -\sin{\psi} \Gamma^{\underline{r}} + \cos{\psi} \Gamma^{\underline{z}} = \exp\left( - \frac{\psi}{2} \Gamma^{\underline{r z}} \right)  \Gamma^{\underline{\uz}} \exp\left( \frac{\psi}{2} \Gamma^{\underline{r z}} \right)  ~.
\end{align}
We will use the subscript $S^5$ to refer to a natural basis associated with the ten-dimensional frame $\{ e^{\umu}, e^{\underline{v}}, e^{\upsi},e^{\utheta},e^{\uphi}, e^{\underline{\zeta}}, e^{\underline{\chi}} \}$.  The change of basis transformation is the inverse of the active one appearing in \eqref{rz2vpsi}:
\begin{align}
& \{ (\Gamma^{\underline{v}})_{S^5}, (\Gamma^{\underline{\psi}})_{S^5} \} = h_{\psi} \{ (\Gamma^{\underline{v}})_{S^2 \times S^2} , (\Gamma^{\underline{\psi}})_{S^2 \times S^2} \} h_{\psi}^{-1} ~, \qquad \textrm{with} \cr
& h_{\psi} := \exp\left(\frac{\psi}{2} \Gamma^{\underline{v \psi}} \right) = \exp \left( \frac{\psi}{2} \Gamma^{\underline{r z}} \right)~.
\end{align}

Then noting first that \eqref{KSS2S2} can be written
\begin{align}
(\epsilon)_{S^2 \times S^2} =&~ \frac{1}{\sqrt{\mu v}} (\Gamma^{\underline{v}})_{S^2 \times S^2} h_{S^2}(\zeta,\chi) h_{S^2}(\theta,\phi) \epsilon_{-}^0  + \cr
&~ +  \sqrt{\mu v} h_{S^2}(\zeta,\chi) h_{S^2}(\theta,\phi) \left[ \epsilon_{+}^0 - \mu (x^p \Gamma_{\up} + y \Gamma_{\uy}) \epsilon_{-}^0 \right]~,
\end{align}
we find that
\begin{align}
(\epsilon)_{S^5} =&~ h_{\psi} (\epsilon)_{S^2 \times S^2} \cr
=&~ \frac{1}{\sqrt{\mu v}} \Gamma^{\underline{v}} h_{S^5}(\theta^A) \epsilon_{-}^0 + \sqrt{\mu v} h_{S^5}(\theta^A) \left[ \epsilon_{+}^0 - \mu (x^p \Gamma_{\up} + y \Gamma_{\uy}) \epsilon_{-}^0 \right]~,
\end{align}
where we use the shorthand $\theta^A = (\psi,\theta,\phi,\zeta,\chi)$ and with
\begin{align}
h_{S^5}(\theta^A) :=&~ \exp\left( \frac{\psi}{2} \Gamma^{\underline{v \psi}} \right) \exp\left( \frac{\theta}{2} \Gamma^{\underline{v \theta}} \right) \exp\left( \frac{\phi}{2} \Gamma^{\underline{\theta\phi}} \right) \exp\left( \frac{\zeta}{2} \Gamma^{\underline{\psi \zeta}} \right) \exp\left( \frac{\chi}{2} \Gamma^{\underline{\zeta\chi}} \right)~.
\end{align}
This has the form of the Killing spinors found in \eg\ \cite{Lu:1998nu,Skenderis:2002vf}.

\subsection{D5-brane Killing spinor equation}\label{app:bkseqn}

Taking the real part of \eqref{bulkKSeqn} and utilizing \eqref{kappaconYM} for the $\Fbar^{(5)}$ term gives
\begin{equation}\label{braneKSeqn1}
\left[ \pd_a + \frac{1}{4} \omegabar_{\underline{MN},a} \Gamma^{\underline{MN}} \right] \varepsilon + \frac{\sigma}{16 \cdot 5!} (\Gamma^{M_1 \cdots M_5} \Fbar_{M_1 \cdots M_5}^{(5)})  \Gamma_a \Gamma_{\underline{012r_1r_2r_3}} \varepsilon = 0~,
\end{equation}
where we are working in the Cartesian-like frame.  One computes that
\begin{equation}\label{FdotGamma}
\frac{1}{16 \cdot 5!} \Gamma^{M_1 \cdots M_5} \Fbar_{M_1 \cdots M_5}^{(5)} = \frac{\mu}{4} \left( \frac{ r_i \Gamma^{\underline{r_i}} + z_{0,i} \Gamma^{\underline{z_i}} }{\sqrt{r^2 + z_{0}^2}} \right) \Gamma^{\underline{012y}} (\mathbbm{1} - \Gammabar)~,
\end{equation}
on the brane worldvolume.  There are also contributions to the spin connection from directions transverse to the brane.  The nonzero components are
\begin{equation}\label{normalspin}
\omegabar_{\underline{\nu z_i},\mu} = \mu \, e^{\underline{\mu}}_{\phantom{\up}\mu} \, \eta_{\underline{\nu\mu}} \frac{z_{0,i}}{\sqrt{r^2 + z_{0}^2}} ~, \qquad \omegabar_{\underline{r_j z_k}, r_i} = - \mu \, e^{\underline{r_i}}_{\phantom{\underline{r_i}} r_i} \delta_{\underline{r_i r_j}} \frac{z_{0,k}}{\sqrt{r^2 + z_{0}^2}}~,
\end{equation}
in addition to the $\omegabar_{\underline{bc},a}$.  The $z_{0,i}$ terms of \eqref{FdotGamma} and \eqref{normalspin} can be combined in \eqref{braneKSeqn1}, such that that equation becomes
\begin{equation}\label{braneKSeqn2}
\left[ D_a \mp \frac{\mu}{2} \frac{ z_{0,i} \Gamma^{\underline{z_i}}}{\sqrt{r^2 + z_{0}^2}} \Gamma_a \left( \mathbbm{1} - \Gamma^{\underline{r_1 r_2 r_3 y}} \right) + \frac{\mu}{2} \frac{r_i \Gamma^{\underline{r_i}}}{\sqrt{r^2 + z_{0}^2}} \Gamma^{\underline{r_1 r_2 r_3 y}} \Gamma_a \right] \vareps = 0~,
\end{equation}
where we recall that $D_a := \pd_a + \frac{1}{4} \omegabar_{bc,a} \Gamma^{bc}$, and the top (bottom) sign is for the case $a = p$ ($a = r_i$).

In fact the middle term of \eqref{braneKSeqn2} annihilates $\varepsilon$, as we can see from \eg\ \eqref{braneKScart}.  On the one hand, if $\vec{z}_0 = 0$ then this term is simply not present.  On the other hand, if $\vec{z}_0 \neq 0$ then $\vareps$ itself is actually an eigenspinor\footnote{Equivalently, $\Gamma^{\underline{r\theta\phi y}} \vareps = \vareps$ when $\vec{z}_0 \neq 0$, which is evident from \eqref{braneKS} upon noting that $\Gamma^{\underline{r\theta\phi y}}$ commutes with $h_{S^2}$.} of $\Gamma^{\underline{r_1r_2r_3 y}}$:
\begin{equation}\label{varepseigen}
\Gamma^{\underline{r_1 r_2 r_3 y}} \vareps = \vareps ~~ \textrm{when} ~~ \vec{z}_0 = 0~.
\end{equation}
Hence the Killing spinor equation satisfied by $\vareps$ is
\begin{equation}\label{braneKSeqncart}
\left[ D_a + \frac{\mu}{2} \frac{r_i \Gamma^{\underline{r_i}}}{\sqrt{r^2 + z_{0}^2}} \Gamma^{\underline{r_1 r_2 r_3 y}} \Gamma_a \right] \vareps = 0 ~.
\end{equation}
Working in terms of $\{ \Gamma^{\ur},\Gamma^{\utheta},\Gamma^{\uphi} \}$ instead, using that $\hat{r}_i \Gamma^{\underline{r}_i} = \Gamma^{\ur}$, and recalling the definition of $M_{\Psi}$ in \eqref{SYMmasses}, gives the desired result, \eqref{braneKSeqn}.  One can directly check that \eqref{braneKScart} and \eqref{braneKS} are the solutions of this equation.

\section{Mode analysis}\label{app:modes}

\subsection{Bosons}

Our starting point is the equations of motion \eqref{eoms}.  For the gauge field equations we take advantage of the identity $\nabla_a T^{ab} = \frac{1}{\sqrt{|g|}} \pd_a \left( \sqrt{|g|} \, T^{ab} \right)$, which holds for any antisymmetric tensor $T^{ab} = T^{[ab]}$ on a (pseudo-)Riemannian manifold with metric $g_{ab}$ with determinant $g$.  Then plugging in \eqref{flucexpand} with the background \eqref{Pvacua}, using \eqref{adaction}, and working to linear order in $(a,\phi)$, we eventually find the following results for the $a_{\mu}, \phi^{z_i}$, and $a_r$ equations:
\begin{align}\label{amuarphiz}
0 =&~ \bigg\{ \pd_{r}^2 + \frac{2}{r} \pd_r + \frac{1}{r^2} \tilde{D}_{S^2}^2 - \left(m_{y,s} - \frac{p_s}{2r} \right)^2 + \frac{( \eta^{\mu\nu} \pd_\mu \pd_\nu - \vec{m}_{z,s}^2)}{\mu^4 (r^2 + z_{0}^2)^2} \bigg\} a_{\mu}^s + \cr
&~  - \frac{1}{\mu^2 (r^2 + z_{0}^2)} \pd_\mu \left( \tilde{D}_M a^{M,s} \right)~, \cr
0 =&~ \bigg\{ \pd_{r}^2 + \frac{2}{r} \pd_r + \frac{1}{r^2} \tilde{D}_{S^2}^2 - \left(m_{y,s} - \frac{p_s}{2r} \right)^2 + \frac{( \eta^{\mu\nu} \pd_\mu \pd_\nu - \vec{m}_{z,s}^2)}{\mu^4 (r^2 + z_{0}^2)^2} \bigg\} \phi^{z_i , s} + \cr
&~ - \frac{1}{\mu^2 (r^2 + z_{0}^2)} \left[ \tilde{\Phi}^{z_i}, \tilde{D}_M a^{M,s} \right]~, \cr
0 =&~  \frac{1}{r^2} \pd_{r}^2 \left(r^2 a_{r}^s \right) + \frac{1}{r^2} \tilde{D}_{S^2}^2 a_{r}^s - \left(m_{y,s} - \frac{p_s}{2r} \right)^2 a_{r}^s + \frac{( \eta^{\mu\nu} \pd_\mu \pd_\nu - \vec{m}_{z,s}^2)}{\mu^4 (r^2 + z_{0}^2)^2} a_{r}^s - \frac{2 i m_{y,s}}{r} \phi^{y,s} + \cr
&~ +\frac{1}{(r^2 + z_{0}^2)^2} \left[ \pd_r \frac{(r^2 + z_{0}^2)^2}{r^2} \right] \left( \pd_r (r^2 a_{r}^s) + \frac{1}{\sqrt{\tilde{g}}} \tilde{D}_\alpha (\sqrt{\tilde{g}} \tilde{g}^{\alpha\beta} a_{\beta}^s) + r^2 [\tilde{\Phi}^y, \phi^y ]^s \right) + \cr
&~ - \frac{1}{\mu^4 (r^2 + z_{0}^2)^2} \pd_r \left( g^{rr} \tilde{D}_M a^{M,s} \right)~.
\end{align}
The new object in these expressions is the ten-dimensional covariant divergence, based on the metric $\Gbar_{MN} = \diag(g_{ab}, \Gbar_{mn})$ evaluated at $x^m = x_{0}^m$, with determinant satisfying $(-\Gbar)^{1/2} = r^2/(\mu^2(r^2 + z_{0}^2))$:
\begin{align}
\tilde{D}_M a^M =&~ (-\Gbar)^{-1/2} \pd_M \left( (-\Gbar)^{1/2} \Gbar^{MN} a_{N} \right) + \Gbar^{MN} [\tilde{A}_M, a_N] \cr
=&~ \mu^2 (r^2 + z_{0}^2) \left( \frac{1}{r^2} \pd_r \left( r^2 a_r \right) + \frac{1}{r^2 \sqrt{\tilde{g}}} \tilde{D}_\alpha \left( \sqrt{\tilde{g}} \,\tilde{g}^{\alpha\beta} a_\beta \right) + [\tilde{\Phi}^y, \phi^y] \right) + \cr
&~  +  \frac{1}{\mu^2 (r^2 + z_{0}^2)} \left( \eta^{\mu\nu} \pd_\mu a_{\nu} +  \delta_{ij} [\tilde{\Phi}^{z_i}, \phi^{z_j} ] \right)~.
\end{align}
Here we have collected the bosonic fluctuations into a ten-dimensional gauge field, $a_M = (a_a,\phi_m)$, which is translation invariant along the $x^m$ directions.  

It is useful to fix a gauge before proceeding further.  A natural gauge-fixing condition is
\begin{equation}\label{gaugefix1}
\tilde{D}_M a^M = 0~,
\end{equation}
as it decouples the equations for $a_\mu$ and $\phi^{z_i}$ from the rest of the fluctuations.  This still leaves us the freedom to make gauge transformations $a_M \to a_M - D_M \epsilon$ that preserve this condition.  The gauge-fixing condition will be preserved if the parameter $\epsilon$ is annihilated by the ten-dimensional background covariant Laplacian:
\begin{align}
0 =&~ (-\Gbar)^{-1/2} D_M \left( (-\Gbar)^{1/2} \Gbar^{MN} D_N \epsilon \right) \cr
=&~ \bigg\{ \pd_{r}^2 + \frac{2}{r} \pd_r + \frac{1}{r^2} \tilde{\DD}^2 - \left(m_{y,s} - \frac{p_s}{2r} \right)^2 + \frac{( \eta^{\mu\nu} \pd_\mu \pd_\nu - \vec{m}_{z,s}^2)}{\mu^4 (r^2 + z_{0}^2)^2} \bigg\} \epsilon^s ~.
\end{align}
Notice that this the same operator appearing in the $a_{\mu}$ and $\phi^{z_i}$ equations.  Hence we could use residual gauge freedom to set any one component of these fields to zero.  However a better choice is the following.  Observe that the combination $\eta^{\mu\nu} \pd_\mu a_\nu + \delta_{ij} [\tilde{\Phi}^{z_i}, \phi^{z_j}]$ will also be annihilated by $\tilde{\Delta}_{10}$, using the condition \eqref{gaugefix1} and the equations of motion, since both $\pd_\nu$ and $\ad(\Phi_{\infty}^{z_i})$ commute with it.  Thus we can use the residual gauge freedom to additionally set
\begin{equation}\label{gaugefix2}
\eta^{\mu\nu} \pd_\mu a_\nu + \delta_{ij} [\tilde{\Phi}^{z_i}, \phi^{z_j}] = 0~.
\end{equation}
This condition together with \eqref{gaugefix1} also imply
\begin{equation}\label{gaugefix3}
\pd_r \left( r^2 a_r \right) + \frac{1}{\sqrt{\tilde{g}}} \tilde{D}_\alpha \left( \sqrt{\tilde{g}} \,\tilde{g}^{\alpha\beta} a_\beta \right) + r^2 [\tilde{\Phi}^y, \phi^y] = 0~.
\end{equation}
Note that this quantity is exactly what appears in the second line of the $a_r$ equation of motion.  These conditions define the gauge that we work in.  In this gauge the equations \eqref{amuarphiz} simplify to
\begin{align}\label{amuarphiz2}
0 =&~ \bigg\{ \pd_{r}^2 + \frac{2}{r} \pd_r + \frac{1}{r^2} \tilde{D}_{S^2}^2 - \left(m_{y,s} - \frac{p_s}{2r} \right)^2 + \frac{( \eta^{\mu\nu} \pd_\mu \pd_\nu - \vec{m}_{z,s}^2)}{\mu^4 (r^2 + z_{0}^2)^2} \bigg\} (a_{\mu}^s ~,~ \phi^{z_i, s}) ~, \cr
0 =&~  \frac{1}{r^2} \pd_{r}^2 \left(r^2 a_{r}^s \right) + \frac{1}{r^2} \tilde{D}_{S^2}^2 a_{r}^s - \left(m_{y,s} - \frac{p_s}{2r} \right)^2 a_{r}^s + \frac{( \eta^{\mu\nu} \pd_\mu \pd_\nu - \vec{m}_{z,s}^2)}{\mu^4 (r^2 + z_{0}^2)^2} a_{r}^s - \frac{2 i m_{y,s}}{r} \phi^{y,s} ~. \qquad 
\end{align}

Next there are the equations for $\phi^y$ and $a_\alpha$.  We immediately plug in the parameterization of $a_\alpha$ in terms of adjoint valued scalars $\lambda,f$ given in \eqref{AalphaY}.  Using \eqref{gaugefix1}, we eventually obtain the $\phi^y$ equation,
\begin{align}
0 =&~ \bigg\{ \frac{1}{r^2 (r^2 + z_{0}^2)^2} \pd_r \left[ r^2 (r^2 + z_{0}^2)^2 \pd_r \right] + \frac{1}{r^2} \tilde{D}_{S^2}^2  - \left(y_\sigma - \frac{p_\sigma}{2r}\right)^2 + \frac{ \eta^{\mu\nu} \pd_\mu \pd_\nu - \vec{m}_{z,s}^2 }{\mu^4 (r^2 + z_{0}^2)^2} \bigg\} \phi^{y,s} + \cr
&~ -2  \left[ \pd_r \tilde{\Phi}^y , a_r \right]^s - \frac{4 r}{r^2 + z_{0}^2} \left( [\tilde{\Phi}^y, a_r]^s  + [\pd_r \tilde{\Phi}^y, \lambda]^s - \frac{1}{r^2} \tilde{\DD}^2 f^s \right)~, \raisetag{20pt}
\end{align}
and the $a_\alpha$ equation,
\begin{align}\label{aalphaeqn}
0 =&~ \tilde{g}^{\alpha\beta} \tilde{D}_\beta \bigg\{ \frac{ (\eta^{\mu\nu} \pd_\mu \pd_\nu - \vec{m}_{z,s}^2)}{r^2} \lambda^s + \frac{\mu^4}{r^2} \pd_r \left[ (r^2 + z_{0}^2)^2 \pd_r \lambda^s \right] - \frac{\mu^4 (r^2 + z_{0}^2)^2}{r^2} \left(y_\sigma - \frac{p_\sigma}{2r} \right)^2 \lambda^s + \cr
&~ \qquad \qquad + \frac{\mu^4 (r^2 + z_{0}^2)^2}{r^4} \left(\tilde{D}_{S^2}^2 \lambda^s + \lambda^s +2 r^2 [\pd_r \tilde{\Phi}^y, f]^s \right) - \mu^4 \left[ \pd_r \frac{(r^2 + z_{0}^2)^2}{r^2}\right] a_{r}^s  + \cr
&~ \qquad \qquad - \frac{\mu^4 (r^2 + z_{0}^2)^2}{r^4} \lambda^s + \left[ f, ~\frac{2 \mu^4 (r^2 + z_{0}^2)^2}{r^2} \pd_r \tilde{\Phi}^y + \frac{\mu^4}{r^2} \left[ \pd_r (r^2 + z_{0}^2)^2 \right] \tilde{\Phi}^y \right]^s \bigg\} + \cr
& + \tilde{\epsilon}^{\alpha\beta} \tilde{D}_\beta \bigg\{  \frac{(\eta^{\mu\nu} \pd_\mu \pd_\nu - \vec{m}_{z,s}^2)}{r^2} f^s + \frac{\mu^4}{r^2} \pd_r \left[ (r^2 + z_{0}^2)^2 \pd_r f^s \right] - \frac{\mu^4 (r^2 + z_{0}^2)^2}{r^2} \left(y_\sigma - \frac{p_\sigma}{2r} \right)^2 f^s + \cr
&~~ \qquad \qquad + \frac{\mu^4 (r^2 +  z_{0}^2)^2}{r^4} \left( \tilde{D}_{S^2}^2 f^s + f^s - 2r^2 [\pd_r \tilde{\Phi}^y, \lambda]^s \right) - \frac{\mu^4}{r^2} \left[\pd_r (r^2 + z_{0}^2)^2 \right] \phi^{y,s} + \cr
&~~ \qquad \qquad - \frac{\mu^4 (r^2 + z_{0}^2)^2}{r^4} f^s - \left[ \lambda,~ \frac{2 \mu^4 (r^2 + z_{0}^2)^2}{r^2} \pd_r \tilde{\Phi}^y + \frac{\mu^4}{r^2} \left[ \pd_r (r^2 + z_{0}^2)^2 \right] \tilde{\Phi}^y \right]^s \bigg\}~. 
\end{align}
In order to obtain these results we used, for example,
\begin{align}
\tilde{\epsilon}^{\alpha\beta} \tilde{D}_\alpha a_\beta =&~ \tilde{\epsilon}^{\alpha\beta} \tilde{D}_\alpha \tilde{D}_\beta \lambda - \frac{1}{\sqrt{\tilde{g}}} \tilde{D}_\alpha (\sqrt{\tilde{g}} \, \tilde{g}^{\alpha\beta} \tilde{D}_\beta f)  \cr
=&~ \half \tilde{\epsilon}^{\alpha\beta} [\tilde{F}_{\alpha\beta}, \lambda] - \tilde{D}_{S^2}^2 f \cr
=&~ r^2 [\pd_r \tilde{\Phi}^y, \lambda] - \tilde{D}_{S^2}^2 f~,
\end{align}
and
\begin{align}
\tilde{D}_{S^2}^2 \left(\tilde{g}^{\alpha\beta} a_\beta \right) =&~ \tilde{g}^{\alpha\beta} \tilde{D}_\beta \left( \tilde{D}_{S^2}^2 \lambda + \lambda + 2 r^2 [\pd_r \tilde{\Phi}^y, f] \right) + \tilde{\epsilon}^{\alpha\beta} \tilde{D}_\beta \left( \tilde{\DD}^2 f + f - 2 r^2 [\pd_r \tilde{\Phi}^y, \lambda ] \right)~.
\end{align}
Terms in the latter arise from the commutator of covariant derivatives, which involves both a Riemann curvature term for the two-sphere and a fieldstrength term for the background gauge field.

The $a_\alpha$ equation has the form $0 = \tilde{g}^{\alpha\beta} \tilde{D}_\beta \Lambda + \tilde{\epsilon}^{\alpha\beta} \tilde{D}_\beta F$.  This leads to two separate equations, $\Lambda = 0$ and $F = 0$, which can be viewed as the equations associated with $\lambda$ and $f$ respectively.  One can show, however, that the $\lambda$ equation follows from the gauge-fixing condition \eqref{gaugefix3} together with the equations of motion for $a_r,\phi^y$, and $f$.\footnote{Apply the operator $\pd_r \left[ r^2 (r^2 + z_{0}^2)^2 \cdot \right]$ to the $a_r$ equation and go from there.}  Hence we can drop the $\lambda$ equation and instead use the constraint \eqref{gaugefix3} to solve for $\lambda$.  Hence the remaining second order equations in addition to \eqref{amuarphiz2} are
\begin{align}\label{phiyfpair}
0 =&~ \bigg\{\frac{1}{r^2 (r^2 + z_{0}^2)^2} \pd_r \left[ r^2 (r^2 + z_{0}^2)^2 \pd_r \right] + \frac{1}{r^2} \tilde{D}_{S^2}^2 - \left(y_\sigma - \frac{p_\sigma}{2r}\right)^2 +  \frac{\eta^{\mu\nu} \pd_\mu \pd_\nu - \vec{m}_{z,s}^2 }{\mu^4 (r^2 + z_{0}^2)^2}    \bigg\} \phi^{y,s} + \cr
&~ - 2 [\pd_r \tilde{\Phi}^y, a_r ]^s - \frac{4 r}{r^2 + z_{0}^2} \left( [\tilde{\Phi}^y, a_r]^s + [\pd_r \tilde{\Phi}^y, \lambda]^s - \frac{1}{r^2} \tilde{D}_{S^2}^2 f^s \right)~, \cr
0 =&~  \bigg\{\frac{1}{(r^2 + z_{0}^2)^2} \pd_r \left[ (r^2 + z_{0}^2)^2 \pd_r \right] + \frac{1}{r^2} \tilde{D}_{S^2}^2  - \left(y_\sigma - \frac{p_\sigma}{2r}\right)^2 + \frac{\eta^{\mu\nu} \pd_\mu \pd_\nu - \vec{m}_{z,s}^2 }{\mu^4 (r^2 + z_{0}^2)^2}  \bigg\} f^s + \cr
&~ - \frac{4 r}{r^2 + z_{0}^2} \left( \phi^{y,s} - [\tilde{\Phi}^y, \lambda]^s \right)~,  \raisetag{20pt}
\end{align}
and the gauge constraint expressed in terms of these variables is
\begin{align}\label{gaugefix3b}
0 =&~  \pd_r (r^2 \delta a_r) +  \tilde{D}_{S^2}^2 \lambda +  r^2 [\pd_r \tilde{\Phi}^y, f] + r^2 [\tilde{\Phi}^y, \phi^y] ~.
\end{align}

It appears that after using \eqref{gaugefix3b} to eliminate $\lambda$, the equations for $a_r, \phi^y$, and $f$ form a coupled system.  However it is actually possible to remove the $a_r$ dependence from the $\phi^y$-$f$ system by one further shift of variables.  We introduce new fluctuations $\sy$ and $\sf$ by setting
\begin{equation}
\phi^{y,s} = \sy^s + \frac{i p_s}{2r} \lambda^s ~, \qquad f_{(j,m)}^s = \sf_{(j,m)}^s + \frac{i m_{y,s}  r^2}{j(j+1) - \frac{p_s m_{y,s} r}{2}} \,  a_{r}^s ~.
\end{equation}
Plugging these back into the pair \eqref{phiyfpair}, all dependence on $\lambda$ and $a_r$ remarkably drops out, and the two equations can be cast into the form
\begin{align}\label{yfsystem}
0 =&~ \bigg\{ \left[ \frac{1}{(r^2 + z_{0}^2)^2} \pd_r \left[ (r^2 + z_{0}^2)^2 \pd_r \right] - m_{y,s}^2 + \frac{p_s m_{y,s}}{r} + \frac{ \eta^{\mu\nu} \pd_\mu \pd_\nu - \vec{m}_{z,s}^2}{\mu^4 (r^2 + z_{0}^2)^2} \right] \mathbbm{1} + \cr
&~ \quad - {\bf J}(r) - {\bf Y}(r) \bigg\} \left( \begin{array}{c} r \sy_{(j,m)}^s \\ \sf_{(j,m)}^s \end{array} \right)~,
\end{align}
where the two-by-two matrices ${\bf J}, {\bf Y}$ are
\begin{align}
{\bf J} = &~\left( \begin{array}{c c} \frac{j(j+1)}{r^2} + \frac{4}{r^2 + z_{0}^2} ~&~ \frac{4 j (j+1)}{r^2 + z_{0}^2}  \\[1ex] \frac{4}{r^2 + z_{0}^2} ~&~ \frac{j(j+1)}{r^2} \end{array}\right)~,
\end{align}
and
\begin{align}\label{Ymatrix}
{\bf Y} =&~ \left( \begin{array}{c c} 0 ~&~ - \frac{2 p_\sigma m_{y,s} r}{r^2 + z_{0}^2} \\[1ex]   m_{y,s}^2 \gamma + m_{y,s} r \left( m_{y,s} - \frac{p_\sigma}{2r} \right) \left(2 \gamma' + \frac{4 r \gamma}{r^2 + z_{0}^2}\right) ~&~  \frac{p_\sigma m_{y,s}}{2}  \left(2 \gamma' + \frac{4 r \gamma}{r^2 + z_{0}^2}\right) \end{array} \right)~,
\end{align}
with $\gamma(r) \equiv \left( j(j+1) - \frac{p_s m_{y,s} r}{2} \right)^{-1}$.  Hence we can in principle solve the system \eqref{yfsystem}.  Then the solutions will appear as inhomogeneous sources in the equation for $a_r$, \eqref{amuarphiz2}.  Finally from \eqref{gaugefix3b} (and remembering \eqref{swharmonic}) we find that $\lambda$ is given by
\begin{equation}\label{lambdageneral}
\lambda_{(j,m)}^s = \pd_r \left[ \frac{ r^2 a_{r,(j,m)}^s }{j(j+1) - \frac{p_s m_{y,s} r}{2}} \right] + \frac{ \frac{i p_s}{2} \left( r \sy_{(j,m)}^s - \sf_{(j,m)}^s \right) - i r^2 m_{y,s} \sy_{(j,m)}^s }{j(j+1) - \frac{p_s m_{y,s} r}{2}} ~.
\end{equation}

The appearance of the quantity $j(j+1) - \frac{p_s m_{y,s} r}{2}$ in denominators might seem disturbing: If the sign of $p_s m_{y,s}$ is positive then such expressions become singular at a physical value of the radial coordinate.  We believe these singularities are indicative of an instability in the system that occurs when $p_s m_{y,s} > 0$ for any $s$.  Recall that positive $m_{y,s}$ indicates a separation between consecutive D5-branes in the $y$-direction, while $p_s$ indicates the presence of vacuum D3-branes, parallel to the color D3-branes, stretched between these D5-branes or ending on them.  The sign of $p_s$ is what determines whether these are D3-branes or $\overbar{{\rm D3}}$-branes, and supersymmetry requires a specific choice.  Hence we expect that supersymmetry leads to the condition $p_s m_{y,s} < 0$.

An important fact about ${\bf Y}$, which is not obvious from \eqref{Ymatrix}, is that it vanishes as $r \to \infty$.  More precisely, the off-diagonal entries are $O(1/r)$ and the lower right entry is $O(1/r^2)$.  Hence the dominant term in \eqref{yfsystem} at large $r$ is the $- m_{y,s}^2 \mathbbm{1}$ term, when $m_{y,s} \neq 0$.  The same is true of the other fluctuation equations, \eqref{amuarphiz2}.  It follows that the asymptotic behavior of all fluctuations is $e^{\pm m_{y,s} r}$ when $m_{y,s}$ is nonzero.

The matrix ${\bf J}$ can be diagonalized by an $r$-independent similarity transformation:
\begin{equation}
{\bf S} = \left( \begin{array}{c c} 1 ~&~ j \\ -1 ~&~ j+1 \end{array} \right)~, \qquad {\bf S} {\bf J} {\bf S}^{-1} = \left( \begin{array}{c c} \frac{j(j+1)}{r^2} + \frac{4(j+1)}{r^2 + z_{0}^2} ~&~ 0 \\[1ex] 0 ~&~ \frac{j(j+1)}{r^2} - \frac{4 j}{r^2 + z_{0}^2} \end{array} \right)~.
\end{equation}
In general this is not useful since ${\bf S}$ does not diagonalize ${\bf Y}$.  However in the case $m_{y,s} = 0$, ${\bf Y}$ vanishes and then ${\bf S}$ can be used to diagonalize the system \eqref{yfsystem}.  ${\bf S}$ acting on $(r \sy, \sf)^T$ defines the scalar fluctuations $\phi^{\pm}$ introduced in \eqref{diagonalizeyf}.  Also in the $m_{y,s} = 0$ case we have $\sf = f$ and \eqref{lambdageneral} reduces to \eqref{lambdasolve}.

\subsection{Fermions}\label{app:modesf}

We now turn to the last of \eqref{eoms}.  The fermion is already first order in fluctuations, so we evaluate the gauge field and the scalars on their background values \eqref{Pvacua}.  The first step is to write the equation in terms of a six-dimensional Dirac spinor $\psi$.  In general we follow the conventions in appendix B of Polchinski, \cite{Polchinski:1998rr}.  We decompose the ten-dimensional gamma matrices according to 
\begin{equation}\label{6plus4split}
\Gamma^a = \Sigma^a \otimes \mathbbm{1}_4 ~, \qquad \Gamma^{5 +m} = \Sigmabar \otimes \rho^m ~,
\end{equation}
where $\Sigma^{a}$ are $Spin(1,5)$ gamma matrices,
\begin{equation}
\Sigmabar := - \Sigma^{\underline{012345}} = -\Sigma^{\underline{012r\theta\phi}}
\end{equation}
is the corresponding chirality operator, and the $\rho^m$ are $Spin(4)$ gamma matrices.  Specifically, for the $\rho^m$ we take
\begin{equation}
\rho^{\um} := \left( \begin{array}{c c} 0 & \taubar^{\um} \\ \tau^{\um} & 0 \end{array} \right)~, \qquad \tau^{\um} := (\vec{\sigma},-i \mathbbm{1}_2)~, \qquad \taubar^{\um} := (\vec{\sigma},i \mathbbm{1}_2)~,
\end{equation}
where the $\vec{\sigma}$ are the Pauli matrices.  Then with $\rhobar := \rho^{\underline{1234}} = \diag(\mathbbm{1}_2, - \mathbbm{1}_2)$ we have
\begin{equation}
\Gammabar = - \Sigmabar \otimes \rhobar~.
\end{equation}

The Majorana condition 
\begin{equation}\label{Mcondition}
\Psi^\ast = B_{10} \Psi
\end{equation}
is implemented with the intertwiner $B_{10}$ defined such that $(\Gamma^{M})^\ast = B_{10} \Gamma^M B_{10}^{-1}$.  It can be taken as
\begin{equation}
B_{10} := \prod_{ \textrm{imag}} \Gamma^{\uM}~,
\end{equation}
and we can always choose our basis such that $B_{10} = B_{10}^\ast = B_{10}^T = B_{10}^{-1}$.  (This choice means in particular that we take $\Gamma^{\uzero}$ to be real, and hence antisymmetric.)    The six-dimensional counterpart,
\begin{equation}
B_{6} := \prod_{\textrm{imag}} \Sigma^{\ua} ~,
\end{equation}
satisfying $(\Sigma^a)^\ast = B_{6} \Sigma^a B_{6}^{-1}$, will then have $B_6 = B_{6}^\ast = - B_{6}^T = - B_{6}^{-1}$.  

With these conventions a MW spinor $\Psi$, satisfying \eqref{Mcondition} and $\Psi = \Gammabar\Psi$, takes the form
\begin{equation}\label{psiNPsi}
\Psi = \left( \begin{array}{c} \psi_{-} \\ B_6 \psi_{-}^\ast \\ \psi_{+} \\ - B_6 \psi_{+}^\ast \end{array} \right)~,
\end{equation}
where $\psi_{\pm} = \pm \Sigmabar \psi_{\pm}$ are the positive and negative chirality components of a six-dimensional (complex Dirac) spinor $\psi = \psi_+ + \psi_-$.  One finds that the fermionic action \eqref{SYMf} expressed in terms of $\psi$ takes the form
\begin{align}
S_{{\rm ym},f} =&~ -\frac{i}{2g_{{\rm ym}_6}^2} \int \ed^6 x \sqrt{-g_6} \Tr \bigg\{ \bar{\psi} \slashed{D}_6 \psi - \bar{\psi} \overleftarrow{\slashed{D}}_6 \psi + 2i M_{\Psi} \bar{\psi} \Sigma^{\underline{\theta\phi}} \psi + 2i \psibar [\Phi_{\uy}, \psi]  + \cr
&~ + 2 \psibar [\Phi_{\underline{z_3}},  \Sigmabar \psi] - \psi^T \Sigma^{\uzero} [ \Phi_{\underline{z_1}} + i \Phi_{\underline{z_2}} , B_6 \psi]  - \psibar [ \Phi_{\underline{z_1}} - i \Phi_{\underline{z_2}}, B_{6} \psi^\ast ] \bigg\} + S_{f}^{\rm bndry} , \qquad
\end{align}
where $\slashed{D}_{6} := \Sigma^a D_a$, $\psibar := \psi^\dag \Sigma^{\uzero}$, and $i \psibar\overleftarrow{\slashed{D}} \psi$ is the conjugate of $-i \psibar \slashed{D}_6 \psi$.  They are equal up to a total derivative, (but the total derivative can be nonvanishing).  Varying with respect to $\psibar$ gives the equation of motion, 
\begin{align}
0 =&~ \left( \slashed{D}_6 + i M_{\Psi} \Sigma^{\underline{\theta\phi}} \right) \psi + i [\Phi_{\uy}, \psi] +  [\Phi_{\underline{z_3}}, \Sigmabar \psi] - [\Phi_{\underline{z_1}} - i \Phi_{\underline{z_2}} , B_6 \psi^\ast] ~, 
\end{align}
which is equivalent to the last of \eqref{eoms}.  Evaluating the bosonic fields on their background values gives the linearized equation
\begin{align}\label{6Dfeqn}
0 =&~ \left\{ \Sigma^\mu D_\mu + \Sigma^r D_r + \frac{\mu (r^2 + z_{0}^2)^{1/2}}{r} \tilde{\Sigma}^\alpha \tilde{D}_\alpha + i M_{\Psi} \Sigma^{\underline{\theta\phi}} + \mu (r^2 + z_{0}^2)^{1/2} \left( y_\sigma - \frac{p_\sigma}{2r} \right) \right\} \psi + \cr
&~ - \frac{i}{\mu (r^2 + z_{0}^2)^{1/2}} \left\{ z_{3,\sigma} \Sigmabar \psi - (z_{1,\sigma} - i z_{2,\sigma}) B_6 \psi^\ast \right\}~, \qquad  \raisetag{22pt}
\end{align}
where $\tilde{\Sigma}^\alpha$ is constructed using zweibein on the unit-radius $S^2$.

To analyze the spectrum of modes on the asymptotically $AdS_4$ space we choose an adapted basis for the six-dimensional gamma matrices:
\begin{equation}\label{4plus2split}
\Sigma^{\mu,r} = \gamma^{\mu,r} \otimes \sigma^3~, \qquad \Sigma^{\utheta,\uphi} = \mathbbm{1}_4 \otimes \sigma^{1,2}~.
\end{equation}
The next step is then to diagonalize the operator $\sigma^\alpha \tilde{D}_\alpha := \sigma^1 \tilde{D}_\theta + \frac{1}{\sin{\theta}} \sigma^2 \tilde{D}_\phi$ over a complete set of eigenspinors on the two-sphere.  This is an $S^2$ Dirac operator coupled to a Dirac monopole background.  The eigenvalue equation
\begin{equation}
\sigma^\alpha \tilde{D}_\alpha \xi = i M \xi~,
\end{equation}
is equivalent to the $\dim{\mathfrak{g}}$ equations
\begin{equation}
\left[ \sigma^\alpha D_\alpha - \frac{i p_s}{2\sin{\theta}} (\epsilon 1 - \cos{\theta}) \sigma^2 \right] \xi^{(\epsilon)s} = i M \xi^{(\epsilon)s} ~.
\end{equation}
Here $\epsilon = \pm$ specifies the northern or southern patch of the $S^2$ respectively.  The two solutions will be related by a transition function, $\xi^{(+)s} = e^{i p_s \phi} \xi^{(-)s}$, on the overlap.  

This is a classic problem with a completely explicit solution.  (See appendix C of \cite{Moore:2014jfa} for a recent treatment.) The eigenspinors are labeled by three indices, $\upsigma,j,m$, where $\upsigma \in \{+,-,0\}$ and $(j,m)$ are angular momentum quantum numbers.  Let 
\begin{equation}
j_\ast := \half (|p_s| - 1)~.
\end{equation}
Then the eigenspinors with $\upsigma = \pm$ have $j$-values starting at $j_\ast +1$ and increasing integer steps, while $m$ runs from $-j$ to $j$ in integer steps as usual.  They are given by
\begin{equation}
\xi_{\pm,j,m}^{(\epsilon)s}(\theta,\phi) = \frac{1}{\sqrt{2}} N_{m,\frac{1-p_s}{2}}^j e^{i (m + \epsilon p_s/2)\phi} \left( d^{j}_{m, \frac{1-p_s}{2}}(\theta) \mathbbm{1}_2 + i \upsigma \, d^{j}_{m,\frac{-1-p_s}{2}}(\theta) \sigma^1 \right) \xi_{+}^0 ~,
\end{equation}
where $d^{j}_{m,m'}(\theta)$ is a Wigner little $d$ function\footnote{These solutions can also be expressed in terms of spin-weighted spherical harmonics.  The relationship is ${}_{m'} Y_{jm}(\theta,\phi) = \left( \frac{2j+1}{4\pi}\right)^{1/2} e^{im\phi} d^{j}_{m,m'}(\theta)$.} and $\xi_{+}^0 = (1,0)^T$.  The $\upsigma = 0$ spinors correspond to the special value $j = j_\ast$ only, and their form depends on the sign of $p_s$:
\begin{equation}
\xi_{0,j_\ast,m}^{(\epsilon)s}(\theta,\phi) = e^{i (m + \epsilon p_s/2)\phi} \left\{ \begin{array}{l l} N_{m,-j_\ast}^{j_\ast} d^{j_\ast}_{m,-j_\ast}(\theta) \xi_{+}^0 ~, & p_s > 0~, \\[1ex] N_{m,j_\ast}^{j_\ast} d^{j_\ast}_{m,j_\ast}(\theta) \sigma^1 \xi_{+}^0 ~, & p_s < 0~. \end{array} \right.
\end{equation}
Note these solutions only exist when $p_s$ is nonzero; $j_\ast$ takes an unphysical value when $p_s = 0$.  If $p_s = 0$ then the $\upsigma = \pm$ solutions are a complete set with $j \in \{ \half, \frac{3}{2}, \ldots \}$.  The $N^{j}_{m,m'}$ are normalization coefficients:
\begin{equation}\label{spinornorm}
N^{j}_{m,m'} = e^{i \pi |m-m'|/2} \sqrt{ \frac{2j+1}{4\pi}} ~,
\end{equation}
where the choice of phase will be convenient below.  The corresponding eigenvalues are
\begin{equation}
M_{\upsigma,j}^s = \frac{\upsigma}{2} \sqrt{ (2j+1)^2 - p_{s}^2} ~.
\end{equation}
The $\upsigma = 0$ modes are zero modes of $\sigma^\alpha \tilde{\DD}_\alpha$, but this does not mean that they correspond to massless spinors on $AdS_4$ as there are other terms in the equation \eqref{6Dfeqn} that must be taken into account.

To find the four-dimensional spectrum we insert the mode expansion
\begin{equation}\label{6DKKspinor}
\psi^{(\epsilon)s} = \sum_{\upsigma,j,m} \psi_{\upsigma,j,m}^{s}(x^\mu,r) \otimes \xi_{\upsigma,j,m}^{(\epsilon)s}(\theta,\phi)~,
\end{equation}
into the linearized equation \eqref{6Dfeqn}, using \eqref{4plus2split}.  Note that
\begin{equation}
\Sigmabar = \gammabar \otimes \sigma^3~, \qquad B_6 = B_4 \otimes \sigma^3 \sigma^2 = -i B_4 \otimes \sigma^1~.
\end{equation}
Here $B_4$ is the product over the imaginary $\gamma^{\umu,\ur}$ and satisfies $(\gamma^{\mu,r})^\ast = - B_4 \gamma^{\mu,r} B_{4}^{-1}$.  (We also used that it is necessarily the product of an odd number of $\gamma$'s, as charge conjugation reverses chirality for $Spin(1,3)$.)  Hence we'll need the action of $\sigma^3$ and charge conjugation, $\xi \mapsto \sigma^1 \xi^\ast$ on the eigenspinors.  These are found to be
\begin{equation}\label{sigma3action}
\sigma^3 \xi_{\upsigma,j,m}^{(\epsilon),s} = \xi_{-\upsigma,j,m}^{(\epsilon),s}~, \quad \upsigma = \pm ~, \qquad \textrm{and} \quad \sigma^3 \xi_{0,j_\ast,m}^{(\epsilon),s} = \sgn(p_s) \xi_{0,j_\ast,m}^{(\epsilon),s}~,
\end{equation}
and
\begin{equation}
\sigma^1 (\xi_{\upsigma,j_\ast,m}^{(\epsilon),s})^\ast = \upsigma \sgn\left( m + \frac{p_s}{2}\right) \xi_{-\upsigma,j,-m}^{(\epsilon),-s} ~, \quad \upsigma = \pm~, \qquad \textrm{and} \quad \sigma^1 (\xi_{0,j_\ast,m}^{(\epsilon),s})^\ast = \xi_{0,j_\ast,-m}^{(\epsilon),-s} ~.
\end{equation}
In order to obtain the latter one requires the property $d^{j}_{m,m'}(\theta) = (-1)^{m-m'} d^{j}_{-m,-m'}(\theta)$.  The phase of \eqref{spinornorm} was chosen to make the action of charge conjugation as simple as possible.  Remember also that $p_{-s} = - p_{s}$.  See the discussion under \eqref{adaction}.

Using all these facts, we find that \eqref{6Dfeqn} splits into two families of coupled systems for the modes $\psi_{\upsigma,j,m}^{s}$.  The coupled system for the $\upsigma = 0$ modes (which exist when $p_s \neq 0$) is
\begin{align}\label{s0system}
0 =&~ \left\{ \left[  \slashed{D}_4 - \frac{i m_{s,z_3}}{\mu (r^2 + z_{0}^2)^{1/2}} \gammabar \right] \mathbbm{1}_{2} + {\bf M}_0 \right\} \left( \begin{array}{c} \psi_{0,j_\ast,m}^{s} \\[1ex] \BB_4 (\psi_{0,j_\ast,-m}^{-s})^\ast \end{array} \right)~,
\end{align}
where
\begin{align}
{\bf M}_0 =&~ \left[ - M_{\Psi} + \sgn(p_s) \mu (r^2 + z_{0}^2)^{1/2} \left( m_{y,s} - \frac{p_s}{2r} \right) \right] \sigma^3 + \cr
&~ - \frac{\sgn(p_s)}{\mu (r^2 + z_{0}^2)^{1/2}} \left( m_{z_1,s} \sigma^1 + m_{z_2,s} \sigma^2\right) ~,
\end{align}
and $\slashed{D}_4 = \gamma^{\mu} D_\mu + \gamma^r D_r$ is the standard Dirac operator on the asymptotically $AdS_4$ space.  Explicitly, one finds
\begin{align}\label{Dslash4}
\slashed{D}_4 =&~ \mu (r^2 + z_{0}^2)^{1/2} \gamma^{\ur} \pd_r + \frac{1}{\mu (r^2 + z_{0}^2)^{1/2}} \gamma^{\umu} \pd_\mu + \frac{3 \mu r}{2(r^2 + z_{0}^2)^{1/2}} \gamma^{\ur} ~.
\end{align}
Inserting \eqref{Dslash4} into \eqref{s0system} and dividing through by $\mu (r^2 + z_{0}^2)^{1/2}$, we have
\begin{align}\label{s0system2}
\left( \begin{array}{c c} \DD_{+}^{0} & \BB^0 \\[1ex] \BB^{0\ast} & \DD_{-}^0 \end{array} \right) \left( \begin{array}{c} \psi_{0,j_\ast,m}^{s} \\[1ex] B_4 (\psi_{0,j_\ast,-m}^{-s})^\ast \end{array} \right) = 0~,
\end{align}
where
\begin{align}
\DD_{\pm}^0 :=&~ \gamma^{\ur} \left( \pd_r  + \frac{3 r}{2 (r^2 + z_{0}^2)} \right) \pm \left( \sgn(p_s) m_{y,s} - \frac{|p_s|}{2r} - \frac{r}{r^2 + z_{0}^2} \right) + \frac{\gamma^{\umu} \pd_\mu - i m_{z_3,s} \gammabar}{\mu^2 (r^2 + z_{0}^2)} ~, \cr
\BB^0 :=&~ - \sgn(p_s) \frac{(m_{z_1,s} - i m_{z_2,s})}{\mu^2 (r^2 + z_{0}^2)} ~,
\end{align}
which is a more useful form for studying the large $r$ asymptotics of solutions.

At this point we will content ourselves with understanding the $r\to \infty$ behavior of solutions.  Then it is sufficient to expand the matrix operator in \eqref{s0system2} through $O(1/r)$.  To this order it diagonalizes and reduces to
\begin{align}
\left[ \gamma^{\ur} \left( \pd_r  + \frac{3}{2r} \right) + \sgn(p_s) m_{y,s} - \left(1 + \frac{|p_s|}{2} \right) \frac{1}{r} + O(1/r^2) \right] \psi_{0,j_\ast,m}^{s} = 0~,
\end{align}
along with an equivalent equation for the conjugate spinor.  The equation diagonalizes with respect to $\gamma^{\ur}$.  If we decompose $\psi$ into eigenspinors,
\begin{equation}\label{grchirality}
\psi_{0,j_\ast,m}^{s} = \psi_{0,j_\ast,m}^{s,+} + \psi_{0,j_\ast,m}^{s,-}~, \qquad \textrm{with} \quad \gamma^{\ur} \psi_{0,j_\ast,m}^{s,\pm} = \pm \psi_{0,j_\ast,m}^{s,\pm}~,
\end{equation}
then the leading behavior of solutions is
\begin{equation}\label{psi0modes}
\psi_{0,j_\ast,m}^{s,\pm} \propto e^{\mp \sgn(p_s) m_{y,s} r} r^{- \frac{3}{2} \mp m_{0}} \left( 1 + O(1/r) \right)~, \qquad m_{0} := - \left(1 + \frac{|p_s|}{2} \right)~.
\end{equation}
When $m_{y,s} \neq 0$ we have exponential decay or blowup behavior.  When $m_{y,s} = 0$ we have power-law behavior dictated by the mass $m_{0}$, which we have defined in such a way that it can be identified with a standard $AdS_4$ mass for the fermion.  In other words, the asymptotic behavior of solutions to $(\slashed{D}_4 + m)\psi = 0$ on $AdS_4$ is $\psi_{\pm} \propto r^{-\frac{3}{2} \mp m}$.  Since the $m_{0}^s$ are all negative, we see that the normalizable modes in the case $m_{y,s} = 0$ are necessarily associated with $\psi_{0,j_\ast,m}^{s,-}$.  However the normalizable (exponentially decaying) modes when $m_{y,s} \neq 0$ could be associated with either $\psi_{0,j_\ast,m}^{s,\pm}$, depending on the sign of the product $p_s m_{y,s}$.  It will be associated with $\psi_{0,j_\ast,m}^{s,-}$ if this sign is negative.  We will comment further on this below.

Taking similar steps, one finds that the coupled system for the $\upsigma = \pm$ modes can be put in the following form:
\begin{align}\label{spmsystem}
& \left( \begin{array}{c c c c} \DD_+ & \CC & \BB & 0 \\[1ex] \CC^\ast & \DD_+ & 0 & -\BB \\[1ex] \BB^\ast & 0 & \DD_- & \CC \\[1ex] 0 & -\BB^\ast & \CC^\ast & \DD_- \end{array} \right) \left( \begin{array}{c} \psi_{+,j,m}^{s} \\[1ex] \psi_{-,j,m}^{s} \\[1ex] \BB_4 (\psi_{+,j,-m}^{-s})^\ast \\[1ex] \BB_4 (\psi_{-,j,-m}^{-s})^\ast \end{array} \right) = 0~,
\end{align}
where
\begin{align}
\DD_{\pm} =&~ \gamma^{\ur} \left( \pd_r + \frac{3 r}{2 (r^2 + z_{0}^2)} \right) \mp \frac{r}{r^2 + z_{0}^2} + \frac{\gamma^{\umu} \pd_\mu -i m_{z_3,s} \gammabar}{\mu^2 (r^2 + z_{0}^2)}  ~, \cr
\CC =&~ m_{y,s} - \frac{i |M_{\upsigma,j}^{s}| }{r} - \frac{p_s}{2r}~, \cr
\BB =&~  \sgn\left( m+\frac{p_s}{2}\right) \frac{ (m_{z_1,s} - i m_{z_2,s})}{\mu^2 (r^2 + z_{0}^2)} ~.
\end{align}

Henceforth restrict our analysis to the $r \to \infty$ behavior of solutions.  Working through $O(1/r)$ the $\BB$ entries can be dropped and the system reduces to
\begin{equation}\label{spmlarger}
\left[ \left( \begin{array}{c c} \gamma^{\ur} \left( \pd_r + \frac{3}{2r} \right) - \frac{1}{r} & m_{y,s} - \left( \frac{p_s}{2} + i |M_{\upsigma,j}^s| \right) \frac{1}{r} \\[1ex] m_{y,s} - \left( \frac{p_s}{2} - i |M_{\upsigma,j}^s| \right) \frac{1}{r} &  \gamma^{\ur} \left( \pd_r + \frac{3}{2r} \right) - \frac{1}{r} \end{array} \right) + O(1/r^2) \right] \left( \begin{array}{c} \psi_{+,j,m}^{s} \\[1ex] \psi_{-,j,m}^{s} \end{array} \right) = 0~,
\end{equation}
along with an equivalent equation for the conjugates.  Let $\alpha(r)$ denote the phase of $\CC$, $e^{i\alpha} = \CC/|\CC|$, and consider the unitary transformation
\begin{align}\label{chietadef}
\left( \begin{array}{c} \chi_{(j,m)}^s \\[1ex] \eta_{(j,m)}^s \end{array} \right) :=&~ {\bf U} \left( \begin{array}{c} \psi_{+,j,m}^{s} \\[1ex] \psi_{-,j,m}^{s} \end{array} \right) := \frac{1}{\sqrt{2}} \left( \begin{array}{c c} e^{-i \alpha/2} & e^{i\alpha/2} \\[1ex] e^{-i \alpha/2} & - e^{i\alpha/2} \end{array} \right) \left( \begin{array}{c} \psi_{+,j,m}^{s} \\[1ex] \psi_{-,j,m}^{s} \end{array} \right)~.
\end{align}
This transformation diagonalizes \eqref{spmlarger} to the order we are working.  The new variables $\chi,\eta$ satisfy the asymptotic equations
\begin{equation}\label{chietaeqns}
\left[ \gamma^{\ur} \left( \pd_r + \frac{3}{2r} \right) - \frac{1}{r} \pm |\CC(r)| + O(1/r^2) \right] (\chi_{(j,m)}^s,\eta_{(j,m)}^s) = 0~,
\end{equation}
where the $+(-)$ is for $\chi(\eta)$ respectively, and
\begin{align}
|\CC(r)| =&~ \sqrt{ \left(m_{y,s} - \frac{p_s}{2r} \right)^2 + |M_{\upsigma,j}^s|^2} = \sqrt{ m_{y,s}^2 - \frac{p_s m_{y,s}}{r} + \frac{(2j+1)^2}{4 r^2}}  \nonumber \\[1ex]
=&~ \left\{ \begin{array}{l l} |m_{y,s}| - \frac{p_s \sgn(m_{y,s})}{2r} + O(1/r^2)~, & m_{y,s} \neq 0~, \\[1ex] \frac{1}{2r} (2j+1)~, & m_{y,s} = 0~. \end{array} \right.
\end{align}

Let $\chi_{(j,m)}^{\pm,s}$ and $\eta_{(j,m)}^{s,\pm}$ denote the positive and negative chirality components with respect to $\gamma^{\ur}$, as in \eqref{grchirality}.  Then the asymptotic behavior of solutions to \eqref{chietaeqns} is
\begin{align}
\chi_{(j,m)}^{s,\pm} \propto \left\{ \begin{array}{l l} e^{\mp |m_{y,s}| r} r^{-\frac{3}{2} \pm \left(1 + \frac{p_s}{2} \sgn(m_{y,s})\right)}(1 + O(1/r)) ~, & m_{y,s} \neq 0~, \\[1ex ]  r^{- \frac{3}{2} \mp m_{j}^{(\chi)}}(1 + O(1/r)) ~, & m_{y,s} = 0~, \end{array} \right.  \\[2ex]
\eta_{(j,m)}^{s,\pm} \propto \left\{ \begin{array}{l l} e^{\pm |m_{y,s}| r} r^{-\frac{3}{2} \pm \left(1 - \frac{p_s}{2} \sgn(m_{y,s})\right)}(1 + O(1/r)) ~, & m_{y,s} \neq 0~, \\[1ex ]  r^{- \frac{3}{2} \mp m_{j}^{(\eta)}}(1 + O(1/r)) ~, & m_{y,s} = 0~, \end{array} \right.
\end{align}
where the $AdS_4$ masses are
\begin{equation}
m_{j}^{(\chi)} = j - \half ~, \qquad m_{j}^{(\eta)} = - \left( j + \frac{3}{2} \right)~.
\end{equation}
The normalizable modes for $\chi$ are those that have positive $\gamma^{\ur}$ chirality asymptotically, while the normalizable modes of $\eta$ are those that have negative $\gamma^{\ur}$ chiarlity asymptotically.  In both cases the normalizable modes along Lie algebra directions with $m_{y,s} \neq 0$ are exponentially decaying while those along directions with $m_{y,s} = 0$ are power-law decaying.

Recall that $j$ starts at $j_\ast + 1 = \frac{1}{2} (|p_s| + 1)$ for these modes.  However we can view the $\psi_{0,j_\ast,m}^s$ modes as filling in a lower $j = j_\ast$ rung for the $\eta$ tower in the sense that
\begin{equation}
m_{j_\ast}^{(\eta)} = - \left( \frac{|p_s| -1}{2} + \frac{3}{2} \right) = m_0~.
\end{equation}
Also the asymptotic $\gamma^{\ur}$-chiralities match provided $\sgn(p_s) m_{y,s} < 0$ whenever $m_{y,s} \neq 0$.  Assuming this is the case, for the same reasons as discussed under \eqref{lambdageneral}, we can identify 
\begin{equation}
\eta_{(j_\ast,m)}^{s} \equiv \psi_{0,j_\ast,m}^{s}~,
\end{equation}
as the lowest rung of the $\eta$ tower for those $s$ such that $p_s \neq 0$.

Finally we note that the $\gamma^{\ur}$-chirality condition can be translated back to a condition on the six-dimensional $\psi$ or on the ten-dimensional $\Psi$.  First, since the action of $\gamma^{\ur}$ commutes with the rotation ${\bf U}$ relating $\chi,\eta$ to the $\psi_{s,j,m}$, we see that $\psi$ will be an asymptotic eigenspinor of
\begin{equation}
\mathbbm{\gamma^{\ur}} \otimes \mathbbm{1}_2 = -i \Sigma^{\underline{r\theta\phi}}~,
\end{equation}
 when restricted to normalizable modes of $\chi$ or $\eta$ only.  We will have $\psi = -i \Sigma^{\underline{r\theta\phi}} \psi$ asymptotically for the normalizable $\chi$-type modes and $\psi = +i \Sigma^{\underline{r\theta\phi}} \psi$ asymptotically for the normalizable $\eta$-type modes.  One can then show from \eqref{6plus4split} and \eqref{psiNPsi} that
 \begin{equation}\label{6to10chirality}
 \left( \mathbbm{1} \pm i \Sigma^{\underline{r\theta\phi}} \right) \psi = 0 \qquad \iff \qquad \left( \mathbbm{1} \pm \Gamma^{\underline{r\theta\phi y}} \right) \Psi = 0~.
 \end{equation}
Hence positive (negative) $\gamma^{\ur}$ chirality corresponds to negative (positive) $\Gamma^{\underline{r\theta\phi y}}$ chirality.

\section{Boundary supersymmetry}\label{app:bsusy}

In this appendix we provide some of the details of the asymptotic analysis that we quoted in subsection \ref{sec:bndrySUSY}.  We begin with $\BB^{\ur}$ and $\BB^{\rm bndry}$, appearing in \eqref{dSBB}.  From \eqref{Bcurrent},
\begin{align}\label{Bcurrentr}
\BB^{\ur} =&~ \varepsbar \Tr \bigg\{ \bigg[ - \half F_{ab} \Gamma^{ab\ur} + (D_a \Phi_{\um}) \Gamma^{\um} \Gamma^{a\ur} - \half [\Phi_{\um},\Phi_{\un}] \Gamma^{\underline{mn}} \Gamma^{\ur} - M_{\Psi} \Phi_{\um} \Gamma^{\um} \Gamma^{\underline{\theta\phi y}} \Gamma^{\ur}  \bigg] \Psi \bigg\} +   \cr
&~ + \half \Tr \left\{ \Psibar \Gamma^{\ur} \delta_{\vareps} \Psi \right\} ~. \raisetag{20pt}
\end{align}
Meanwhile $\BB^{\rm bndry}$ is defined in terms of the supersymmetry variation of the boundary action, \eqref{Sbndry}, according to \eqref{Bbndrydef}.  Taking the variation of \eqref{Sbndry} with respect to \eqref{SUSY}, we infer
\begin{align}\label{Bbndry}
\BB^{\rm bndry} =&~ \varepsbar \Tr \bigg\{ \bigg[ \left( \mu^2 r \Phi^y + \frac{\mu^2 (r^2 + z_{0}^2)}{r^2 \sin{\theta}} F_{\theta\phi} \right) \Gamma^{\uy} - \frac{r}{r^2 + z_{0}^2} \Phi^{z_i} \Gamma^{\underline{z_i}} + \cr
&~ \qquad \quad + \frac{1}{r} \left( \frac{1}{\sin{\theta}} (D_\phi \Phi^y) \Gamma^{\utheta} - (D_\theta \Phi^y) \Gamma^{\uphi} \right) \bigg] \Psi \bigg\} + \half \Tr \left\{ \Psibar \Gamma^{\underline{\theta\phi y}} \delta_{\vareps} \Psi \right\}  ~. \qquad \quad
\end{align}
Since the boundary measure in \eqref{dSBB} is $O(r^3)$ as $r \to \infty$, we must work through $O(r^{-3})$ in the large $r$ expansion of $\BB^{\ur} +\BB^{\rm bndry}$.

For the moment we set aside the last terms of \eqref{Bcurrentr} and \eqref{Bbndry} involving the variation of the fermion, and we focus on the remaining terms.  Since $\vareps = O(r^{1/2})$ and $\Psi = O(r^{-3/2})$, we must compute the terms in square-brackets through $O(r^{-2})$, utilizing the field asymptotics \eqref{fieldas1}.  All terms can contribute at this order.  We expand out, plug in vielbein factors, and collect terms together as follows:
\begin{align}\label{BrplusBbndry}
\BB^{\ur} + \BB^{\rm bndry} =& \, \varepsbar \Tr \bigg\{ \bigg[ \left( \mu^2 r \Phi^y + \frac{\mu^2 (r^2 + z_{0}^2)}{r^2 \sin{\theta}} F_{\theta\phi} \right) (\Gamma^{\uy} - \Gamma^{\underline{r\theta\phi}}) - \frac{r}{r^2 + z_{0}^2} \Phi^{z_i} \Gamma^{\underline{z_i}} (\mathbbm{1} - \Gamma^{\underline{r\theta\phi y}}) + \cr
&~ \qquad + \frac{\mu^2 (r^2 + z_{0}^2)}{r} D_\theta \Phi^y (\Gamma^{\underline{y\theta r}} - \Gamma^{\underline{\phi}}) + \frac{\mu^2 (r^2 + z_{0}^2)}{r \sin{\theta}} D_\phi \Phi^y ( \Gamma^{\underline{y \phi r}} + \Gamma^{\underline{\theta}}) + \cr
&~ \qquad + \bigg( - \frac{1}{r} F_{\mu\theta} \Gamma^{\underline{\mu\theta r}} - \frac{1}{r \sin{\theta}} F_{\mu\phi} \Gamma^{\underline{\mu \phi r}} - [ \Phi^{z_i}, \Phi^{y}] \Gamma^{\underline{z_i y r}} + (D_\mu \Phi^{y}) \Gamma^{\underline{y \mu r}} + \cr
&~ \qquad  \qquad +  \frac{1}{r} (D_\theta\Phi^{z_i}) \Gamma^{\underline{z_i \theta r}} + \frac{1}{r \sin{\theta}} (D_\phi \Phi^{z_i}) \Gamma^{\underline{z_i \phi r}}   \bigg)  + \cr
&~ \qquad - \frac{1}{\mu^2 (r^2 + z_{0}^2)} \left(\half F_{\mu\nu} \Gamma^{\underline{\mu\nu}} + D_\mu \Phi^{z_i} \Gamma^{\underline{\mu z_i}} + \half [\Phi^{z_i}, \Phi^{z_j}] \Gamma^{\underline{z_i z_j}} \right) \Gamma^{\ur} \bigg] \Psi \bigg\}  + \cr
&~ +  \half \Tr \left\{ \Psibar \Gamma^{\underline{r}} \left(\mathbbm{1} + \Gamma^{\underline{r\theta\phi y}} \right) \delta_{\vareps} \Psi \right\} ~. \raisetag{22pt}
\end{align}

The first four sets of terms are proportional to the projector $\half (\mathbbm{1} - \Gamma^{\underline{r\theta\phi y}})$ acting to the left.  In the case of the $\Phi^{z_i}$ term, the relevant spinor bilinear is $\varepsbar^- \Psi^- = O(1/r^2)$.  However, recall that if the vev $\phi_{({\rm nn})}^{z_i}$ is nonzero, then we must set the superconformal generators $\eta_0$ to zero, which implies $\vareps^- = 0$.  Hence, we get an extra order of suppression from the fact that the leading $\phi_{({\rm nn})}^{z_i}$ part of $\Phi^{z_i}$ does not contribute, and therefore this term can be neglected.  The same reasoning applies to the $\Phi_{\infty}^y$ part of $\Phi^y$ in the first term.   The remaining terms in this set involve the spinor bilinear $\varepsbar^- \Psi^+$, which is $O(1/r^2)$.  Thus we need to evaluate them through $O(1/r)$, which corresponds precisely to the contribution from $\vec{\XX}$ in $\Phi^y,A_\theta,A_\phi$.  Specifically,
\begin{align} 
\mu^2 r \Phi^y + \frac{\mu^2 (r^2 + z_{0}^2)}{r^2 \sin{\theta}} F_{\theta\phi} \to &~ -\frac{1}{r} \hat{r} \cdot \vec{\XX}_{({\rm n})} + \cdots ~, \cr
\frac{\mu^2 (r^2 + z_{0}^2)}{r} D_\theta \Phi^y \to &~ \frac{1}{r} \hat{\theta} \cdot \vec{\XX}_{({\rm n})} + \cdots ~, \cr
\frac{\mu^2 (r^2 + z_{0}^2)}{r \sin{\theta}} D_\phi \Phi^y \to &~ \frac{1}{r} \hat{\phi} \cdot \vec{\XX}_{({\rm n})} + \cdots ~.  \cr
\end{align}
Hence the relevant combination is
\begin{align}
\frac{1}{r} \left(-\Gamma^{\uy} \hat{r} - \Gamma^{\uphi} \hat{\theta} + \Gamma^{\utheta} \hat{\phi} \right) \cdot \vec{\XX}_{({\rm n})} (1 + \Gamma^{\underline{r\theta\phi y}}) \Psi =&~  \frac{2}{r} \Gamma^{\underline{ry}} \left(\Gamma^{\ur} \hat{r} + \Gamma^{\utheta} \hat{\theta} + \Gamma^{\uphi} \hat{\phi} \right)\cdot \vec{\XX}_{({\rm n})}  \Psi^+ \cr
=&~ \frac{2}{r} \Gamma^{\underline{ry}} (\vec{\Gamma}_{(r)} \cdot \vec{\XX}_{({\rm n})}) \Psi^+ ~,
\end{align}
where we used $\Gamma^{\underline{r\theta\phi y}} \Psi^+ = \Psi^+$ and $\hat{r} \Gamma^{\ur} + \hat{\theta} \Gamma^{\utheta} + \hat{\phi} \Gamma^{\uphi} = (\Gamma^{\underline{r_1}},\Gamma^{\underline{r_2}},\Gamma^{\underline{r_3}}) \equiv \vec{\Gamma}_{(r)}$.  (See \eqref{CtoS}.)

Next consider the set of six terms inside the large round brackets of \eqref{BrplusBbndry}.  It follows from the field asymptotics \eqref{fieldas1} that all of these terms start at $O(1/r^2)$.  Furthermore all of the gamma matrix structures associated with these terms \emph{commute} with $\Gamma^{\underline{r\theta\phi y}}$.  Hence they involve $\varepsbar^+ \Psi^+ = O(1/r)$ and $\varepsbar^- \Psi^- = O(1/r^2)$, and we only need to worry about the former.  The order $O(1/r^2)$ terms in the round brackets are all of the form $D_{\mu}^{({\rm nn})}$ or $\ad(\phi_{({\rm nn})}^{z_i})$ acting on $\vec{\XX}$, where $D_{\mu}^{({\rm nn})} = \pd_\mu + \ad(a_{\mu}^{({\rm nn})})$.  Specifically, the relevant combination of terms is
\begin{align}
& \frac{1}{\mu^2 r^2}\left( (\Gamma^{\underline{\mu \theta r}} \hat{\phi} - \Gamma^{\underline{\mu \phi r}} \hat{\theta} + \Gamma^{\underline{y \mu r}} \hat{r} ) \cdot D_{\mu}^{({\rm nn})}\vec{\XX}_{({\rm n})} + (\Gamma^{\underline{z_i \theta r}} \hat{\phi} - \Gamma^{\underline{z_i \phi r}} \hat{\theta} -\Gamma^{\underline{z_i y r}} \hat{r} ) \cdot [\phi_{({\rm nn})}^{z_i}, \vec{\XX}_{({\rm n})}]  \right) \Psi^+ \cr
& \qquad = \frac{1}{\mu^2 r^2} \Gamma^{\uy} \left( \Gamma^{\umu} D_{\mu}^{({\rm nn})} + \Gamma^{\underline{z_i}} \ad(\phi_{({\rm nn})}^{z_i}) \right) (\vec{\Gamma}_{(r)} \cdot \vec{\XX}_{({\rm n})}) \Psi^+ ~.  \raisetag{18pt}
\end{align}

The remaining terms in the square brackets of \eqref{BrplusBbndry} start at $O(1/r^2)$ and anti-commute with $\Gamma^{\underline{r\theta\phi y}}$.  Hence they involve the couplings $\varepsbar^+ \Psi^-$ and $\varepsbar^- \Psi^+$, and we only need to keep the former.  One simply needs to evaluate $(A_\mu, \Phi^{z_i})$ on their leading behavior, $(a_{\mu}^{({\rm nn})}, \phi_{({\rm nn})}^{z_i})$.  Collecting results, we have
\begin{align}\label{BBsum2}
\BB^{\ur} + \BB^{\rm bndry} =&~ \frac{2}{r} \varepsbar^- \Tr \left\{ \Gamma^{\underline{ry}} (\vec{\Gamma}_{(r)} \cdot \vec{\XX}_{({\rm n})}) \Psi^+ \right\} + \cr
& + \frac{1}{\mu^2 r^2} \varepsbar^+ \Tr \left\{ \Gamma^{\uy} \left( \Gamma^{\umu} D_{\mu}^{({\rm nn})} + \Gamma^{\underline{z_i}} \ad(\phi_{({\rm nn})}^{z_i}) \right) (\vec{\Gamma}_{(r)} \cdot \vec{\XX}_{({\rm n})}) \Psi^+ \right\} +  \cr
& - \frac{1}{\mu^2 r^2} \varepsbar^+ \Tr \left\{ \Gamma^{\ur} \left( \half f_{\mu\nu}^{({\rm nn})} \Gamma^{\underline{\mu\nu}} + D_{\mu}^{({\rm nn})} \phi_{({\rm nn})}^{z_i} \Gamma^{\underline{\mu z_i}} + \half [\phi_{({\rm nn})}^{z_i}, \phi_{({\rm nn})}^{z_j}] \Gamma^{\underline{z_i z_j}} \right) \Psi^- \right\} + \cr
& +   \half \Tr \left\{ \Psibar \Gamma^{\underline{r}} \left(\mathbbm{1} + \Gamma^{\underline{r\theta\phi y}} \right) \delta_{\vareps} \Psi \right\} + O(r^{-7/2})~.
\end{align}
Plugging in \eqref{braneKScart} and \eqref{fieldasPsi} leads to the result in the text, \eqref{BBsum}.

The next step is to analyze the asymptotics of $\delta_{\vareps} \Psi$, as given in \eqref{SUSY}.  Our goal will be to compute $(\delta_{\vareps} \Psi)^+$ through $O(r^{-3/2})$ since this is the only order that can contribute to \eqref{BBsum2}, given the asymptotics of $\Psi^{-}$, \eqref{fieldas1}.  We note that the $O(r^{-3/2})$ terms of $(\delta_{\vareps} \Psi)^+$ give the supersymmetry variation of the non-normalizable mode, $\uppsi_{0}^{({\rm nn})}$.  Even if we choose to set this field to zero, its variation need not be zero.  The reason is that we are allowing certain non-normalizable modes of the bosonic fields---namely $(a_{\mu}^{({\rm nn})},\phi_{({\rm nn})}^{z_i}, \vec{\XX}_{({\rm nn})})$---to be turned on, and they can source the supersymmetry variation of the non-normalizable fermion modes.

We expand out \eqref{SUSY} and collect terms as follows:
\begin{align}
\delta_{\vareps} \Psi =&~ \left\{ M_r + M_\phi + M_\phi + M_{\rm rest} \right\} \vareps~,
\end{align}
where
\begin{align}
M_r =&~ \mu (r^2 + z_{0}^2)^{1/2} D_r \Phi^{\uy} \Gamma^{\underline{ry}} + \frac{\mu r}{(r^2 + z_{0}^2)^{1/2}} \Phi^{\uy} \Gamma^{\underline{\theta\phi}} + \frac{\mu^2 (r^2 + z_{0}^2)}{r^2 \sin{\theta}} F_{\theta\phi} \Gamma^{\underline{\theta\phi}} ~, \cr
M_{\theta} =&~ \frac{\mu (r^2 + z_{0}^2)^{1/2}}{r} D_\theta \Phi^{\uy} \Gamma^{\underline{\theta y}} + \frac{\mu^2 (r^2 + z_{0}^2)}{r\sin{\theta}} F_{r\phi} \Gamma^{\underline{r\phi}} ~, \cr
M_\phi =&~  \frac{\mu (r^2 + z_{0}^2)^{1/2}}{r \sin{\theta}} D_\phi \Phi^{\uy} \Gamma^{\underline{\phi y}} + \frac{\mu^2 (r^2 + z_{0}^2)}{r} F_{r\theta} \Gamma^{\underline{r\theta}} ~,
\end{align}
and
\begin{align}
M_{\rm rest} =&~  \left( \mu (r^2 + z_{0}^2)^{1/2} D_r \Phi^{\underline{z_i}} \Gamma^{\underline{r z_i}} + \frac{\mu r}{(r^2 + z_{0}^2)^{1/2}} \Phi^{\underline{z_i}} \Gamma^{\underline{\theta\phi y z_i}} \right) + \cr
&~ + \bigg[ F_{\mu r} \Gamma^{\underline{\mu r}} + \frac{1}{r} F_{\mu \theta} \Gamma^{\underline{\mu\theta}} + \frac{1}{r\sin{\theta}} F_{\mu\phi} \Gamma^{\underline{\mu\phi}} + \frac{1}{\mu (r^2 + z_{0}^2)^{1/2}} D_\mu \Phi^{\uy} \Gamma^{\underline{\mu y}} + \cr
&~ \qquad + \frac{\mu (r^2 + z_{0}^2)^{1/2}}{r} \left( D_\theta \Phi^{\underline{z_i}} \Gamma^{\underline{\theta z_i}} + \frac{1}{\sin{\theta}} D_\phi \Phi^{\underline{z_i}} \Gamma^{\underline{\phi z_i}} \right) + [\Phi^{\underline{y}}, \Phi^{\underline{z_i}}] \Gamma^{\underline{y z_i}} \bigg] + \cr
&~ + \frac{1}{\mu^2 (r^2 + z_{0}^2)} \left(\half F_{\mu\nu} \Gamma^{\underline{\mu\nu}} + D_\mu \Phi^{z_i} \Gamma^{\underline{\mu z_i}} + \half [\Phi^{z_i}, \Phi^{z_j}] \Gamma^{\underline{z_i z_j}} \right) ~.
\end{align}

Let's start with $M_{\rm rest}$.  It follows from the field asymptotics that all seven terms in the big square-brackets are $O(1/r^2)$.  Furthermore the gamma matrix structure of each of these terms is such that it maps the $(\pm)$-chirality eigenspace of $\Gamma^{\underline{r\theta\phi y}}$ to the $(\mp)$-chirality eigenspace.  Hence, these terms acting on $\vareps^+$ give an $O(r^{-3/2})$ contribution to $(\delta_{\vareps} \Psi)^-$, while these terms acting on $\vareps^-$ give an $O(r^{-5/2})$ contribution to $(\delta_{\vareps} \Psi)^+$.  Therefore these terms can be neglected to the order we are working.  In contrast the terms in the last line preserve the chirality and so acting on $\vareps^+$ they give a contribution to $(\delta_{\vareps} \Psi)^+$ that is $O(r^{-3/2})$ that must be kept.  Finally, consider the first two terms of $M_{\rm rest}$.  Using \eqref{canonscalar} one finds
\begin{align}
& \mu (r^2 + z_{0}^2)^{1/2} D_r \Phi^{\underline{z_i}} \Gamma^{\underline{r z_i}} + \frac{\mu r}{(r^2 + z_{0}^2)^{1/2}} \Phi^{\underline{z_i}} \Gamma^{\underline{\theta\phi y z_i}} = \cr
& \qquad \qquad \qquad \qquad = D_r \Phi^{z_i} \Gamma^{\underline{r z_i}} + \frac{r}{r^2 + z_{0}^2} \Phi^{z_i} \Gamma^{\underline{z_i r}} \left( \mathbbm{1} - \Gamma^{\underline{r\theta\phi y}} \right)~.
\end{align}
The $D_{r} \Phi^{z_i}$ term is $O(1/r^2)$ and exchanges $\Gamma^{\underline{r\theta\phi y}}$ chiralities.  If $\phi_{({\rm nn})}^{z_i}$ is nonzero then the projector annihilates $\vareps$, so the last term is also effectively $O(1/r^2)$ and exchanges chiralities.  Hence these terms are on the same footing as the square-bracketed terms and can be neglected.  In summary,
\begin{align}\label{Mrestoneps}
(M_{\rm rest} \vareps)^+ =&~ \frac{1}{\mu^2 r^2} \left( \half f_{\mu\nu}^{({\rm nn})} \Gamma^{\underline{\mu\nu}} + D_{\mu}^{({\rm nn})} \phi_{({\rm nn})}^{z_i} \Gamma^{\underline{\mu z_i}} + \half [\phi_{({\rm nn})}^{z_i}, \phi_{({\rm nn})}^{z_j}] \Gamma^{\underline{z_i z_j}} \right) \vareps^+  + O(r^{-5/2})~, \cr
(M_{\rm rest} \vareps)^- =&~ O(r^{-3/2})~. \raisetag{20pt}
\end{align}

Now consider $M^y$.  Plugging in \eqref{canonscalar} we have
\begin{align}
M_r =&~  \mu^2 (r^2 + z_{0}^2) D_r \Phi^y \Gamma^{\underline{r y}} + \mu^2 r \Phi^y ( \Gamma^{\underline{ry}} + \Gamma^{\underline{\theta\phi}}) + \frac{\mu^2 ( r^2 + z_{0}^2)}{r^2 \sin{\theta}} F_{\theta\phi} \Gamma^{\underline{\theta\phi}} \cr
=&~ \left[ \mu^2 (r^2 + z_{0}^2) D_r \Phi^y + \mu^2 r \Phi^y \right] \left( \Gamma^{\underline{ry}} + \Gamma^{\underline{\theta\phi}}\right) - \mu^2 (r^2 + z_{0}^2) \left[ D_r \Phi^y - \frac{1}{r^2 \sin{\theta}} F_{\theta\phi} \right] \Gamma^{\underline{\theta\phi}} \cr
=&~ \left[ \mu^2 r \Phi_{\infty}^y + O(1/r) \right] \Gamma^{\underline{r y}} \left(\mathbbm{1} - \Gamma^{\underline{r\theta\phi y}} \right) - \frac{1}{\mu^4 r^2 (r^2 + z_{0}^2)} \PP^y \Gamma^{\underline{\theta\phi}} ~,
\end{align}
where in the last step we recalled the definition, \eqref{boundaryPs}.  The first term will drop out of \eqref{BBsum2} since it involves the opposite projector.  The large $r$ expansion of $\PP^y$ was determined in \eqref{Pylarger}.  Using that result here gives
\begin{equation}\label{Mrresult}
M_r = O(1/r) \cdot \Gamma^{\underline{ry}} \left( \mathbbm{1} - \Gamma^{\underline{r\theta\phi y}}\right) - \frac{1}{\mu^2 r^2} \left( \hat{r} \cdot \vec{\XX}_{({\rm nn})} + O(1/r) \right) \Gamma^{\underline{\theta\phi}} ~.
\end{equation}
Similar manipulations lead to
\begin{align}
M_\theta =&~ \frac{\mu^2 (r^2 + z_{0}^2)}{r} D_\theta \Phi^y \Gamma^{\underline{\theta y}} \left( \mathbbm{1} - \Gamma^{\underline{r\theta\phi y}}\right) + \frac{\sin{\theta}}{\mu^4 r (r^2 + z_{0}^2)} \PP^\phi \Gamma^{\underline{r\phi}} \cr
=&~  O(1/r) \cdot \Gamma^{\underline{\theta y}} \left( \mathbbm{1} - \Gamma^{\underline{r\theta\phi y}} \right) + \frac{1}{\mu^2 r^2} \left( \hat{\theta} \cdot \vec{\XX}_{({\rm nn})} + O(1/r) \right) \Gamma^{\underline{r\phi}} ~,
\end{align}
and
\begin{align}
M_\phi =&~ \frac{\mu^2 (r^2 + z_{0}^2)}{r \sin{\theta}} D_\phi \Phi^y \Gamma^{\underline{\phi y}} \left( \mathbbm{1} - \Gamma^{\underline{r\theta\phi y}} \right) + \frac{1}{\mu^4 r (r^2 + z_{0}^2)} \PP^\theta \Gamma^{\underline{r\theta}} \cr
=&~ O(1/r) \cdot \Gamma^{\underline{\phi y}} \left( \mathbbm{1} - \Gamma^{\underline{r\theta\phi y}} \right) - \frac{1}{\mu^2 r^2} \left( \hat{\phi} \cdot \vec{\XX}_{({\rm nn})} + O(1/r)\right) \Gamma^{\underline{r\theta}} ~.
\end{align}

Thus we have
\begin{align}\label{Mrtponeps}
\left( (M_r + M_\theta + M_\phi) \vareps\right)^+ =&~ -\frac{1}{\mu^2 r^2} \vec{\XX}_{({\rm nn})} \cdot \left( \hat{r} \Gamma^{\underline{\theta\phi}} - \hat{\theta} \Gamma^{\underline{r\phi}} + \hat{\phi} \Gamma^{\underline{r\theta}} \right) \vareps^+ + O(r^{-5/2}) \cr
=&~  -\frac{1}{\mu^2 r^2} \Gamma^{\uy} \left( \hat{r} \Gamma^{\ur} + \hat{\theta} \Gamma^{\utheta} + \hat{\phi} \Gamma^{\uphi} \right) \cdot \vec{\XX}_{({\rm nn})} \, \vareps^+ + O(r^{-5/2}) \cr
=&~ -\frac{1}{\mu^2 r^2} \Gamma^{\uy} \vec{\Gamma}_{(r)} \cdot \vec{\XX}_{({\rm nn})} \, \vareps^+ + O(r^{-5/2})~, \cr
\left( (M_r + M_\theta + M_\phi) \vareps\right)^- =&~ O(r^{-3/2})~,
\end{align}
Combining \eqref{Mrestoneps} and \eqref{Mrtponeps} leads to the result quoted in the text, \eqref{Psivaras}.

Our final goal is to derive the asymptotics of $\Psi$ due to the massless $AdS_4$ fermions, as given in \eqref{Psi2uppsi} with \eqref{Psisolcart}.  The leading behavior of these modes as $r \to \infty$ is $O(r^{-3/2})$ and the first subleading behavior is $O(r^{-5/2})$.  They are solutions to the fermion equation of motion
\begin{equation}
0 = \bigg\{ \Gamma^\mu D_\mu + \Gamma^r D_r + \Gamma^\alpha D_\alpha + M_{\Psi} \Gamma^{\underline{\theta\phi y}} + \Gamma^{\underline{z_i}} \ad(\Phi^{\underline{z_i}}) + \Gamma^{\uy} \ad(\Phi^{\uy}) \bigg\} \Psi ~.
\end{equation}
Our analysis in appendix \ref{app:modesf} shows that the massless modes are in the simultaneous kernel of $\ad(\Phi_{\infty}^y)$ and $\ad(P)$.  Taking this into account with respect to the field asymptotics \eqref{fieldas1}, the large $r$ form of the equation of motion is
\begin{align}\label{Psieomlarger}
0 =&~ \bigg\{ \mu r \Gamma^{\ur} \left( \pd_r + \frac{3}{2r} \right) + \mu \tilde{\Gamma}^{\alpha} \tilde{D}_\alpha + \mu \Gamma^{\underline{\theta\phi y}} + \frac{1}{\mu r} \left[ \Gamma^{\umu} D_{\mu}^{({\rm nn})} + \Gamma^{\underline{z_i}} \ad(\phi_{({\rm nn})}^{z_i}) \right] + \cr
&~ \quad + \frac{1}{\mu r} \Gamma^{\ur} \ad(a_{r}^{({\rm n})}) + \frac{1}{\mu r} \left[ - \Gamma^{\utheta} \hat{\phi} + \Gamma^{\uphi} \hat{\theta} + \Gamma^{\uy} \hat{r} \right] \cdot \ad(\vec{\XX}_{({\rm n})}) + O(1/r^2) \bigg\} \Psi~,
\end{align}
where $\tilde{\Gamma}^\alpha \tilde{D}_\alpha$ is (the 10D embedding of) the standard Dirac operator on the two-sphere.  The first three terms give the leading order equation of motion while the remaining terms give $O(1/r)$ corrections.

Note that this equation only involves the asymptotics of the bosonic modes that we keep in the truncation, \eqref{truncation}, and therefore the asymptotics of the solution to the order we need will be the same as in the truncated theory.  Hence we will derive the equations of motion for the fermion in the truncated theory, which we quoted in \eqref{4Deoms}, and then consider the asymptotics of it. 

\subsection{The massless fermion modes}\label{app:fermsol}

We first use results from appendix \ref{app:modesf} to determine the form of the 10D fermion, $\Psi$, restricted to the massless $AdS_4$ modes.  These are the $j=1/2$ doublet $\chi_{(\half,m)}(x^\mu,r)$.  They satisfy \eqref{chietaeqns} with the plus sign, and since $m_{y,s} = p_s = 0$ for these modes, we have $|\mathcal{C}(r)| = |M_{\upsigma,\half}^s | /r = 1/r$.  Hence
\begin{equation}\label{chidoubletas}
\chi_{(\half,m)}(x^\mu,r) = (\mu r)^{-3/2} \tilde{\chi}_{(\half,m)}(x^\mu) \left( 1 + O(1/r) \right)~.
\end{equation}
The boundary data $\tilde{\chi}_{(\half,m)}$ can be decomposed into eigenspinors of $\gamma^{\ur}$, $\tilde{\chi}_{(\half,m)}^{\pm} = \pm \gamma^{\ur} \tilde{\chi}_{(\half,m)}^{\pm}$, and we will see that $\tilde{\chi}_{(\half,m)}^{-}$ corresponds to the normalizable modes and $\tilde{\chi}_{(\half,m)}^{+}$ to the non-normalizable modes.

The phase of $\CC(r)$ that appears in the unitary transformation of \eqref{chietadef} is $\alpha = -\pi/2$, and therefore the corresponding $\psi_{\upsigma,\half,m}$ modes are $\psi_{\pm, \half,m} = \frac{1}{\sqrt{2}} e^{\mp i \pi/4} \chi_{(\half,m)}$.  Hence the 6D spinor, \eqref{6DKKspinor}, restricted to these modes, which we will denote by $\psi_{j=1/2}^{(\chi)}$, takes the form
\begin{align}
\psi_{j=1/2}^{(\chi)} =&~ \sum_{m = \pm 1/2}  \chi_{(\half,m)}(x^\mu,r) \otimes \frac{1}{\sqrt{2}}  \left( e^{-i \pi/4} \xi_{+,\half,m} + e^{i\pi/4} \xi_{-,\half,m} \right) ~,
\end{align}
where the $\xi$ are given by 
\begin{equation}
\xi_{\pm, \half,m} = N^{1/2}_{m,1/2} e^{i m \phi} \left( \begin{array}{c} d^{1/2}_{m,1/2}(\theta) \\[1ex] \pm i d^{1/2}_{m,-1/2}(\theta) \end{array} \right)~.  
\end{equation}
Hence
\begin{align}
\psi_{j=1/2}^{(\chi)} =&~ \sum_{m}  \chi_{(\half,m)}(x^\mu,r) \otimes N^{1/2}_{m,1/2} e^{im\phi} \left( \begin{array}{c} d^{1/2}_{m,1/2}(\theta) \\[1ex]  d^{1/2}_{m,-1/2}(\theta) \end{array}\right) ~.
\end{align}
Now, using $d^{1/2}_{1/2,1/2} = d^{1/2}_{-1/2,-1/2} = \cos{\frac{\theta}{2}}$, $d^{1/2}_{-1/2,1/2} = -d^{1/2}_{1/2,-1/2} = \sin{\frac{\theta}{2}}$, and $N^{1/2}_{1/2,-1/2} = i N^{1/2}_{1/2,1/2} \equiv i N_{1/2}$, one finds that this spinor can be expressed in the form
\begin{align}\label{6Dlead1} 
\psi_{j=1/2}^{(\chi)} =& N_{1/2} \left( \begin{array}{c c} \cos{\frac{\theta}{2}} & \sin{\frac{\theta}{2}} \\[1ex] - \sin{\frac{\theta}{2}} & \cos{\frac{\theta}{2}} \end{array} \right) e^{i \sigma^3 \phi/2} \left( \begin{array}{c c} \chi_{(\half,\half)}(x^\mu,r) \\[1ex] i \chi_{(\half,-\half)}(x^\mu,r) \end{array} \right) \cr
=&~ \exp \left( i \Sigma^{\uphi} \frac{\theta}{2}\right) \exp\left(\Sigma^{\underline{\theta\phi}} \frac{\phi}{2}\right) \uppsi_{\rm 6D}(x^\mu,r) ~, 
\end{align}
where in the last step we introduced the 6D spinor
\begin{equation}
\uppsi_{\rm 6D}(x^\mu,r) := N_{1/2} \left( \begin{array}{c c} \chi_{(\half,\half)}(x^\mu,r) \\[1ex] i \chi_{(\half,-\half)}(x^\mu,r) \end{array} \right)~,
\end{equation}
and wrote the expression in 6D notation with the definitions \eqref{4plus2split}.  

$\uppsi_{\rm 6D}$ has a large $r$ expansion starting at $O(r^{-3/2})$ with the leading behavior given in terms of the boundary spinors $\tilde{\chi}$, \eqref{chidoubletas}.  If one restricts the 4D spinors $\chi$ to $\gamma^{\ur}$ eigenspaces, $\chi^{\pm}$, this corresponds to restricting $\uppsi_{\rm 6D}$ to $\uppsi_{\rm 6D}^{\pm}$ defined by
\begin{equation}
\mp i \Sigma^{\underline{r\theta\phi}} \uppsi_{\rm 6D}^{\pm} = (\pm \gamma^{\ur} \otimes \mathbbm{1}_2) \uppsi_{\rm 6D}^{\pm} = \uppsi_{\rm 6D}^{\pm} ~.
\end{equation}
We use this to express $\psi_{j=1/2}^{(\chi)}$ in the form
\begin{equation}
\psi_{j=1/2}^{(\chi),\pm} = \exp \left( \Sigma^{\underline{r\theta}} \frac{\theta}{2} \right) \exp \left( \Sigma^{\underline{\theta\phi}} \frac{\phi}{2} \right) \uppsi_{\rm 6D}^+(x^\mu,r) + \exp \left( -\Sigma^{\underline{r\theta}} \frac{\theta}{2} \right) \exp \left( \Sigma^{\underline{\theta\phi}} \frac{\phi}{2} \right) \uppsi_{\rm 6D}^-(x^\mu,r) ~.
\end{equation}
This result is straightforwardly expressed in 10D notation via \eqref{6plus4split}.  We find
\begin{align}\label{leading10Dsol}
\Psi_{j=1/2}^{(\chi)} =&~ h_{S^2}(\theta, \phi) \uppsi^+(x^\mu,r) + h_{S^2}(-\theta,\phi) \uppsi^-(x^\mu,r)~,
\end{align}
where we made use of \eqref{hS2def}, $\uppsi = \uppsi^+ + \uppsi^-$ is defined in terms of $\uppsi_{\rm 6D}$ via \eqref{psiNPsi}, and $\uppsi^{\pm}$ satisfy
\begin{equation}\label{uppsiproj}
\Gamma^{\underline{r\theta\phi y}} \uppsi^{\pm} = \pm \uppsi^{\pm}~.
\end{equation}
This is \eqref{Psi2uppsiS2}, which is given in a natural basis with respect to the $S^2$ frame in which  $\Gamma^{\ur}, \Gamma^{\utheta},\Gamma^{\uphi}$ are constant.  Indeed, this was assumed throughout the analysis in appendix \ref{app:modesf}.

This is to be plugged into the full fermion equation of motion, 
\begin{equation}
E_{\Psi} := \left( \Gamma^a D_a + M_{\Psi} \Gamma^{\underline{\theta\phi y}} \right) \Psi + \Gamma_m [\Phi^m, \Psi] = 0~,
\end{equation}
with the bosonic fields restricted to \eqref{truncation} as well.  The basic idea it to pull the factors of $h_{S^2}(\pm \theta,\phi)$ through to the left and collect the terms that are proportional to each.  We expand the Dirac operator,
\begin{align}
\Gamma^a D_a =&~ \Gamma^\mu D_\mu + \Gamma^r D_r + \frac{\mu (r^2 + z_{0}^2)^{1/2}}{r} \left( \tilde{\slashed{D}}_{S^2} + \Gamma^{\utheta} \ad(A_\theta) + \frac{1}{\sin{\theta}} \Gamma^{\uphi} \ad(A_\phi) \right)~,
\end{align}
with $\tilde{\slashed{D}}_{S^2}$ the standard Dirac operator on the unit $S^2$.  Then we make use of the following identities:
\begin{equation}
\tilde{\slashed{D}}_{S^2} h(\pm \theta,\phi) =  \mp h(\mp \theta,\phi) \Gamma^{\underline{r}}~, \qquad \Gamma^{\ur} h(\pm \theta,\phi) = h(\mp\theta,\phi) \Gamma^{\ur} ~,
\end{equation}
\begin{equation}
\Gamma^{\underline{\theta\phi y}} h(\pm \theta,\phi) \uppsi^{\pm} = \Gamma^{\ur} \Gamma^{\underline{r\theta\phi y}} h(\pm \theta,\phi) \uppsi^{\pm} = \pm h(\mp \theta,\phi) \Gamma^{\ur}~,
\end{equation}
and
\begin{align}\label{Gammahcom}
\Gamma^{\utheta} h_{S^2}(\pm \theta,\phi) \uppsi^{\pm} =&~ - h_{S^2}(\mp \theta,\phi) \, \hat{\phi} \cdot (\Gamma^{\uphi},-\Gamma^{\utheta},\Gamma^{\uy}) \uppsi^{\pm}~, \cr
\Gamma^{\uphi} h_{S^2}(\pm \theta,\phi)\uppsi^{\pm} =&~  h_{S^2}(\mp \theta,\phi) \, \hat{\theta} \cdot (\Gamma^{\uphi},-\Gamma^{\utheta},\Gamma^{\uy}) \uppsi^{\pm}~, \cr
\Gamma^{\uy} h_{S^2}(\pm \theta,\phi) \Psi^{\pm} =&~  h_{S^2}(\mp \theta,\phi) \, \hat{r} \cdot (\Gamma^{\uphi},-\Gamma^{\utheta},\Gamma^{\uy}) \uppsi^{\pm}~.
\end{align}
Note for these last three we are employing \eqref{uppsiproj} as well.  Then we find
\begin{align}\label{EPsiexpand}
E_{\Psi} =&~ h_{S^2}(\theta,\phi) \bigg\{  \Gamma^\mu D_\mu \uppsi^+ + \Gamma^r D_r \uppsi^- + \left(\frac{1}{r} - \frac{r}{r^2 + z_{0}^2}\right) \Gamma^r \uppsi^- + \Gamma_{z_i} [\tilde{\Phi}^{z_i}, \uppsi^+]  \cr 
&~ \qquad \qquad + \mu (r^2 + z_{0}^2)^{1/2} (\Gamma^{\uphi}, -\Gamma^{\utheta}, \Gamma^{\uy}) \cdot \left[ -\frac{1}{r} \hat{\phi} A_{\theta} + \frac{1}{r\sin{\theta}}\hat{\theta} A_\phi + \hat{r} \Phi^y, \uppsi^- \right]  \bigg\}  + \cr
&~ + h_{S^2}(-\theta,\phi) \bigg\{  \Gamma^\mu D_\mu \uppsi^- + \Gamma^r D_r \uppsi^+ - \left(\frac{1}{r} - \frac{r}{r^2 + z_{0}^2}\right) \Gamma^r \uppsi^+ + \Gamma_{z_i} [\tilde{\Phi}^{z_i}, \uppsi^-]  \cr 
&~ \qquad \qquad + \mu (r^2 + z_{0}^2)^{1/2} (\Gamma^{\uphi}, -\Gamma^{\utheta}, \Gamma^{\uy}) \cdot \left[ -\frac{1}{r} \hat{\phi} A_{\theta} + \frac{1}{r\sin{\theta}}\hat{\theta} A_\phi + \hat{r} \Phi^y, \uppsi^+ \right]  \bigg\} ~. \cr
\end{align}

The mass-like term
\begin{align}
\pm \left(\frac{1}{r} - \frac{r}{r^2 + z_{0}^2}\right) \Gamma^r \uppsi^{\mp} =&~ - \frac{z_{0}^2}{r (r^2 +  z_{0}^2)} \Gamma^r \Gamma^{\underline{r\theta\phi y}} \uppsi^{\mp} = - \frac{\mu z_{0}^2}{r (r^2 + z_{0}^2)^{1/2}} \Gamma^{\underline{\theta\phi y}} \uppsi^{\mp} \cr
=&~ - m_{\uppsi} \Gamma^{\underline{h_1 h_2 h_3}} \uppsi^{\mp}~,
\end{align}
where we used \eqref{4Dfermimass} and \eqref{hyperGtrip}, vanishes for the $AdS_4$ background where $z_0 = 0$, and in general the $r$-dependent mass vanishes asymptotically like $O(1/r^2)$.  Plugging in the truncation ansatz \eqref{truncation} for the bosonic modes, observe that
\begin{equation}
-\frac{1}{r} \hat{\phi} A_{\theta} + \frac{1}{r\sin{\theta}}\hat{\theta} A_\phi + \hat{r} \Phi^y = \frac{1}{\mu^2 r^2} (\hat{\phi} \hat{\phi} \cdot \vec{\XX} + \hat{\theta} \hat{\theta} \cdot \vec{\XX} + \hat{r} \hat{r} \cdot \vec{\XX}) = \frac{1}{\mu^2 r^2} \vec{\XX}~.
\end{equation}
Hence the quantities in curly brackets in \eqref{EPsiexpand} are independent of $\theta,\phi$ on this ansatz.  After introducing the triplet notation \eqref{hyperGtrip} and the metric \eqref{hypermet}, we obtain the result quoted in the text, \eqref{Etoe} and \eqref{4Deoms}:
\begin{align}
e_{\uppsi} =&~ \left(\Gamma^\mu D_\mu  + \Gamma^r D_r - \frac{\mu z_{0}^2}{r (r^2 + z_{0}^2)^{1/2}} \Gamma^{\underline{h_1 h_2 h_3}} \right) \uppsi + \Gamma_{z_i} [\phi^{z_i}, \uppsi]  + \Gamma_{h_i} [\XX^i, \uppsi] = 0~.
\end{align}

Now we analyze the large $r$ asymptotics of this equation.  Keeping terms through $O(1/r)$ in the operator acting on $\uppsi$, one finds
\begin{align}\label{epsilarger}
0 =&~ \bigg\{ \mu r \Gamma^{\ur} \left( \pd_r + \frac{3}{2r} \right) + \frac{1}{\mu r} \left[ \Gamma^{\umu} D_{\mu}^{({\rm nn})} + \Gamma^{\underline{z_i}} \ad(\phi_{({\rm nn})}^{z_i}) \right] + \cr
&~ \qquad \qquad \qquad + \frac{1}{\mu r} \left[ \Gamma^{\ur} \ad(a_{r}^{({\rm n})}) + \Gamma^{\underline{h_i}} \ad(\XX_{({\rm n})}^i) \right] + O(1/r^2) \bigg\} \uppsi~.
\end{align}
The asymptotics of $\uppsi^{\pm}$ are
\begin{align}\label{uppsiexp}
\uppsi^{+}(x^\mu,r) =&~ \frac{1}{(\mu r)^{3/2}} \uppsi_{0}^{({\rm nn})}(x^\mu) + \frac{1}{(\mu r)^{5/2}} \uppsi_{1}^+(x^\mu) + O(r^{-7/2})~, \cr
\uppsi^-(x^\mu,r) =&~  \frac{1}{(\mu r)^{3/2}} \Gamma^{\ur} \uppsi_{0}^{({\rm n})}(x^\mu) + \frac{1}{(\mu r)^{5/2}} \uppsi_{1}^{-}(x^\mu) + O(r^{-7/2})~. 
\end{align}
This is consistent with \eqref{fieldasPsi}, remembering that $(\Gamma^{\underline{r_3}})_{\rm cart} = (\Gamma^{\ur})_{S^2}$.  The $\uppsi_{1}^{\pm}$ are found by plugging this expansion back into \eqref{epsilarger} and solving it at the first subleading order.  We find
\begin{align}
\mu \uppsi_{1}^+ =& - \left[ \Gamma^{\umu} D_{\mu}^{({\rm nn})} + \Gamma^{\underline{z_i}} \ad(\phi_{({\rm nn})}^{z_i}) \right] \uppsi_{0}^{({\rm n})} +  \left[ \ad(a_{r}^{({\rm nn})}) + \Gamma^{\ur} \Gamma^{\underline{h_i}} \ad(\XX_{({\rm n})}^i) \right] \psi_{0}^{({\rm nn})} ~, \cr
\mu\uppsi_{1}^- =&~ \Gamma^{\ur} \left[ \Gamma^{\umu} D_{\mu}^{({\rm nn})} + \Gamma^{\underline{z_i}} \ad(\phi_{({\rm nn})}^{z_i}) \right]\uppsi_{0}^{({\rm nn})} + \Gamma^{\ur} \left[ \ad(a_{r}^{({\rm nn})}) +  \Gamma^{\underline{h_i}} \Gamma^{\ur}\ad(\XX_{({\rm n})}^i) \right]  \psi_{0}^{({\rm n})} ~. \qquad
\end{align}
This can be expressed in terms of Cartesian frame quantities using $(\Gamma^{\ur})_{S^2} = (\Gamma^{\underline{r_3}})_{\rm cart}$ and
\begin{align}
& (\Gamma^{\ur} \vec{\Gamma}_{(h)})_{S^2} \uppsi^+ =  (\Gamma^{\underline{r\phi}},-\Gamma^{\underline{r\theta}},\Gamma^{\underline{ry}})_{S^2} \uppsi^+ = (\Gamma^{\underline{\theta y}}, \Gamma^{\underline{\phi y}}, \Gamma^{\underline{r y}})_{S^2} \uppsi^+  = - \Gamma^{\uy} (\vec{\Gamma}_{(r)})_{\rm cart} \uppsi^+ ~,
\end{align}
which leads to the results for \eqref{uppsiexp} quoted in \eqref{Psisolcart}.

\section{Some details on the truncation}\label{app:trunc}

Here we collect expressions for the components of the non-abelian fieldstrength and covariant derivatives evaluated on the truncation ansatz \eqref{truncation}.  We use a 10D notation $\hat{A}_M$ for the gauge field and Higgs fields in which we identify $(\hat{A}_{z_i},\hat{A}_{y}) \equiv (\Phi^{z_i}, \Phi^y)$ and, for example, $\hat{F}_{\mu z_i} = D_\mu \Phi^{z_i}$.  There is nothing to say about $F_{\mu\nu}, F_{\mu r}, \hat{F}_{\mu z_i}, \hat{F}_{r z_i}, \hat{F}_{z_i z_j}$.  For the remaining ones we have
\begin{align}\label{Fprthetaphi}
& F_{\mu \theta} ~\xrightarrow{\textrm{trnc}}~ - \frac{1}{\mu^2 r} \hat{\phi} \cdot D_\mu \vec{\XX}~, \cr
&  F_{\mu\phi} ~\xrightarrow{\textrm{trnc}}~ \frac{\sin{\theta}}{\mu^2 r} \hat{\theta} \cdot D_\mu \vec{\XX}~, \cr
& \hat{F}_{\mu y} ~\xrightarrow{\textrm{trnc}}~ \frac{1}{\mu^2 r^2} \hat{r} \cdot D_\mu \vec{\XX}~,
\end{align}
\begin{align}
& \hat{F}_{z_i \theta} ~\xrightarrow{\textrm{trnc}}~ -\frac{1}{\mu^2 r} \hat{\phi} \cdot [\Phi^{z_i}, \vec{\XX}] ~, \cr
& \hat{F}_{z_i \phi} ~\xrightarrow{\textrm{trnc}}~ \frac{\sin{\theta}}{\mu^2 r} \hat{\theta} \cdot [\Phi^{z_i}, \vec{\XX}]~, \cr
& \hat{F}_{z_i y} ~\xrightarrow{\textrm{trnc}}~ \frac{1}{\mu^2 r^2} \hat{r} \cdot [\Phi^{z_i}, \vec{\XX}]~,
\end{align}
and
\begin{align}\label{Frthetaphiy}
F_{r\theta} ~\xrightarrow{\textrm{trnc}}~&~ \frac{1}{\mu^2 r^2} \hat{\phi} \cdot \vec{\XX} - \frac{1}{\mu^2 r} \hat{\phi} \cdot D_r \vec{\XX}~, \cr
F_{r\phi} ~\xrightarrow{\textrm{trnc}}~&~ -\frac{\sin{\theta}}{\mu^2 r^2} \hat{\theta} \cdot \vec{\XX} + \frac{\sin{\theta}}{\mu^2 r} \hat{\theta} \cdot D_r \vec{\XX}~, \cr
\hat{F}_{ry} ~\xrightarrow{\textrm{trnc}}~&~ \frac{P}{2 r^2} -\frac{2}{\mu^2 r^3} \hat{r} \cdot \vec{\XX} + \frac{1}{\mu^2 r^2} \hat{r} \cdot D_r \vec{\XX} ~, \cr
F_{\theta\phi} ~\xrightarrow{\textrm{trnc}}~&~ \frac{P}{2} \sin{\theta} - \frac{\sin{\theta}}{\mu^2 r} \hat{r} \cdot \left( 2 \vec{\XX} - \frac{1}{2\mu^2 r} [\vec{\XX}, \times \vec{\XX}] \right)~, \cr
\hat{F}_{\theta y} ~\xrightarrow{\textrm{trnc}}~&~ \frac{1}{\mu^2 r^2} \hat{\theta} \cdot \left( \vec{\XX} - \frac{1}{2\mu^2 r} [\vec{\XX}, \times \vec{\XX}] \right) ~, \cr
\hat{F}_{\phi y} ~\xrightarrow{\textrm{trnc}}~&~ \frac{\sin{\theta}}{\mu^2 r^2} \hat{\phi} \cdot \left( \vec{\XX} - \frac{1}{2\mu^2 r} [\vec{\XX}, \times \vec{\XX}] \right)~.
\end{align}

We also list some formulae that are used in subsection \ref{sec:reducedBPS} for the reduction of the BPS equations.  From \eqref{Frthetaphiy} one finds that
\begin{align}
\frac{1}{r^2 \sin{\theta}} F_{\theta\phi} - D_r \Phi^y  ~\xrightarrow{\textrm{trnc}}~ &~ - \frac{\sin{\theta}}{\mu^2 r^2} \hat{r} \cdot \left( D_r \vec{\XX} - \frac{1}{2\mu^2 r^2} [\vec{\XX}, \times \vec{\XX}] \right)~,  \cr
 F_{r\theta} - \frac{1}{\sin{\theta}} D_\phi \Phi^y ~\xrightarrow{\textrm{trnc}}~&~  - \frac{1}{\mu^2 r} \hat{\phi} \cdot \left( D_r \vec{\XX} - \frac{1}{2\mu^2 r^2} [\vec{\XX}, \times \vec{\XX}] \right)~,    \cr
 \frac{1}{\sin{\theta}} F_{r\phi} + D_\theta \Phi^y  ~\xrightarrow{\textrm{trnc}}~&~  \frac{1}{\mu^2 r} \hat{\theta} \cdot  \left( D_r \vec{\XX} - \frac{1}{2\mu^2 r^2} [\vec{\XX}, \times \vec{\XX}] \right)~,
\end{align}
and converting to the Cartesian coordinate system results in
\begin{align}
F_{r_1 r_2} - D_{r_3} \Phi^y ~\xrightarrow{\textrm{trnc}}~&~ -\frac{1}{\mu^2 r^2} \left( D_r \XX^3 - \frac{1}{\mu^2 r^2} [\XX^1,\XX^2] \right) ~, \cr
F_{r_2 r_3} - D_{r_2} \Phi^y ~\xrightarrow{\textrm{trnc}}~&~ -\frac{1}{\mu^2 r^2} \left( D_r \XX^1 - \frac{1}{\mu^2 r^2} [\XX^2,\XX^3] \right) ~, \cr
F_{r_3 r_1} - D_{r_2} \Phi^y ~\xrightarrow{\textrm{trnc}}~&~ -\frac{1}{\mu^2 r^2} \left( D_r \XX^2 - \frac{1}{\mu^2 r^2} [\XX^3,\XX^1] \right) ~.
\end{align}

Likewise, converting from $F_{pr},F_{p\theta},F_{p\phi}$, to the Cartesian frame $F_{pr_i}$ results in
\begin{align}
F_{pr_1} ~\xrightarrow{\textrm{trnc}}~&~ \frac{1}{\mu^2 r^2} \left( \sin{\theta} \cos{\phi} (\mu^2 r^2 F_{pr}) + \sin{\theta} \sin{\phi} D_p \XX^3 - \cos{\theta} D_p \XX^2 \right)  ~, \cr
F_{p r_2} ~\xrightarrow{\textrm{trnc}}~&~ \frac{1}{\mu^2 r^2} \left( - \sin{\theta} \cos{\phi} D_p \XX^3 + \sin{\theta} \sin{\phi} (\mu^2 r^2 F_{pr}) + \cos{\theta} D_p \XX^1 \right)  ~, \cr
F_{p r_3} ~\xrightarrow{\textrm{trnc}}~&~ \frac{1}{\mu^2 r^2} \left( \sin{\theta} \cos{\phi}D_p \XX^2 - \sin{\theta} \sin{\phi} D_p \XX^1 + \cos{\theta} (\mu^2 r^2 F_{pr}) \right)  ~,
\end{align}
while
\begin{equation}
\hat{F}_{py} ~\xrightarrow{\textrm{trnc}}~ \frac{1}{\mu^2 r^2} \left( \sin{\theta} \cos{\phi} D_p \XX^1 + \sin{\theta} \sin{\phi} D_p \XX^2 + \cos{\theta} D_p \XX^3 \right)~.
\end{equation}
Identical expressions hold for the $\hat{F}_{z_p r_i}$ and $\hat{F}_{z_p y}$ upon replacing $D_p \to \ad(\Phi^{z_p})$.

\section{The BPS energy}\label{app:BPSEnergy}

In this appendix we show how one obtains \eqref{dwBPSbound2} from \eqref{dwBPSbound1}, and as a special case, \eqref{BPSbound2} from \eqref{BPSbound1}.  

First we introduce some notation that exposes the structure of $\Omega_{4}'$.  Let $x^{\tilde{p}} = (x^1,x^2,\hat{z}_1,\hat{z}_2)$ parameterize $\mathbbm{R}^4$ with the standard orientation.  Introduce a basis of self-dual two-forms, 
\begin{equation}\label{twoformbasis}
\upomega^1 = \ed x^2 \ed \hat{z}_2 - \ed x^1 \ed \hat{z}_1~, \quad \upomega^2 = \ed x^2 \ed \hat{z}_1 + \ed x^1 \ed \hat{z}_2 ~, \quad \upomega^3 = \ed x^1 \ed x^2 + \ed \hat{z}_1 \ed \hat{z}_2 ~.
\end{equation}
These can be expressed in terms of 't Hooft matrices,
\begin{equation}
\upomega^i := \half \eta_{\tp\tq}^i \ed x^{\tp} \ed x^{\tq}~.
\end{equation}
where our conventions are
\begin{equation}
\eta^1 = \left( \begin{array}{c c c c} 0 & 0 & -1 & 0 \\ 0 & 0 & 0 & 1 \\ 1 & 0 & 0 & 0 \\ 0 & -1 & 0 & 0 \end{array}\right)~, \qquad \eta^2 = \left( \begin{array}{c c c c} 0 & 0 & 0 & 1 \\ 0 & 0 & 1 & 0 \\ 0 & -1 & 0 & 0 \\ -1 & 0 & 0 & 0 \end{array}\right)~, \qquad \eta^3 = \left( \begin{array}{c c c c} 0 & 1 & 0 & 0 \\ -1 & 0 & 0 & 0 \\ 0 & 0 & 0 & 1 \\ 0 & 0 & -1 & 0 \end{array}\right)~.
\end{equation}
Note this is a slightly different convention than the standard one given in \cite{tHooft:1976snw} in that 
\begin{equation}
(\eta^1,\eta^2,\eta^3)_{\rm here} = (\eta^2,\eta^1,\eta^3)_{\rm standard} ~.
\end{equation}
With our convention matrix multiplication gives the quaternion algebra, $\eta^i \eta^j = - \delta^{ij} + \epsilon^{ij}_{\phantom{ij}k} \eta^k$, with a plus sign in front of the $\epsilon$ rather than a minus.

Then, in terms of the two-forms \eqref{twoformbasis}, one has
\begin{align}
\Omega_{4}' =&~ \frac{1}{\mu^4 (r^2 + z_{0}^2)^2} \ed \hat{y} \ed r_1 \ed r_2 \ed r_3 + (\ed \hat{y} \ed r_1 + \ed r_2 \ed r_3 ) \wedge \upomega^1 +  \cr
&~ + (\ed \hat{y} \ed r_2 + \ed r_3 \ed r_1) \wedge \upomega^2 + ( \ed \hat{y} \ed r_3 + \ed r_1 \ed r_2) \wedge \upomega^3 ~.
\end{align}
Dropping the $\upomega^{1}$ and $\upomega^2$ terms gives $\omega_{4}'$.  

Converting to spherical coordinates, $(r,\theta,\phi)$, results in
\begin{align}
\Omega_{4}' =&~ \left( r^2 \sin{\theta} \ed \theta \ed \phi \, \hat{r}_i + r \ed y \ed \theta \, \hat{\theta}_i + r \sin{\theta} \ed y \ed \phi \, \hat{\phi}_i \right) \wedge \upomega^i + ~ \textrm{$\ed r$ terms}~.
\end{align}
Here we have suppressed terms that have a leg along the radial direction since they will not contribute to the boundary integral.  It follows that
\begin{equation}\label{magdensity}
(\Omega_{4}' \wedge \omega_{\rm CS})_{12\theta\phi \hat{z}_1 \hat{z}_2 \hat{y}} = \half (\eta^i)^{\tp \tq} \,\omega_{\rm CS} \left( \hat{A}_y \, r^2 \sin{\theta} \, \hat{r}_i + \hat{A}_{\phi} \, r \, \hat{\theta}_i - \hat{A}_\theta \, r \sin{\theta} \, \hat{\phi}_i ~, ~ \hat{A}_{\tp} ~, ~\hat{A}_{\tq} \right)~,
\end{equation}
where we are using the notation $\omega_{\rm CS}(\hat{A}_A,\hat{A}_B,\hat{A}_C) \equiv (\omega_{\rm CS})_{ABC}$ for the components of the Chern--Simons form.  If we want $\omega_{4}' \wedge \omega_{\rm CS}$ instead, then we drop the $i=1,2$ terms.  

These expressions integrated against $\ed x^1 \ed x^2 \ed\theta\ed\phi \ed \hat{z}_1 \ed \hat{z}_2 \ed \hat{y}$ at the boundary $r \to \infty$.  Hence we need the large $r$ limit of \eqref{magdensity}.  The leading behavior of the $\hat{A}_{\tp}$ is $O(1)$ and given by the non-normalizable $S^2$ singlet modes.  Therefore the furthest we need to go in the subleading asymptotics of $(\Phi^y, A_{\theta,\phi})$ is the $\vec{\XX}_{(\rm n)}$ terms, which will yield a finite contribution to \eqref{magdensity} as $r \to \infty$.  In fact, if one restricts to the $\vec{\XX}_{(\rm n)}$ terms, the first factor in $\omega_{\rm CS}$ collapses nicely:
\begin{equation}
\hat{A}_y \, r^2 \sin{\theta} \, \hat{r} + \hat{A}_{\phi} \, r \, \hat{\theta} - \hat{A}_\theta \, r \sin{\theta} \, \hat{\phi}  \to \frac{1}{\mu^2}\sin{\theta} \, \vec{\XX}_{({\rm n})}~.
\end{equation}
One might worry that the $\Phi_{\infty}^y$ and 't Hooft charge terms in the asymptotics of $(\Phi^y, A_{\theta,\phi})$ will lead to a divergence, but this is not the case.  The 't Hooft charge drops out of \eqref{magdensity}.  The $\Phi_{\infty}^y$ term can contribute, but integration over the two-sphere will pick out subleading behavior in the $\hat{A}_{\tp}$ factors such that the result is finite.  (The integration over $S^2$ should be carried out before the $r\to\infty$ limit is taken.)  We thus have
\begin{align}\label{magenint}
\int_{\pd \hat{M}_8} \Omega_{4}' \wedge \omega_{\rm CS} =&~ \lim_{r\to \infty} \half  \int_{\mathbbm{R}^2} \ed^2 x \int_{S^2} \ed \theta \ed\phi \sin{\theta} \times \cr
&~ \times  \vec{\eta}\,{}^{\tp\tq} \cdot \left\{ \frac{1}{\mu^2} \omega_{\rm CS}(\vec{\XX}_{({\rm n})}, \hat{A}_{\tp}, \hat{A}_{\tq}) + r^2 \hat{r} \, \omega_{\rm CS}(\Phi_{\infty}^y, \hat{A}_{\tp}, \hat{A}_{\tq}) \right\}~. \qquad
\end{align}

Both Chern--Simons terms are of a similar structure in that they involve an adjoint-valued scalar in one of the factors.  When this is the case, one can obtain the following equivalent expression, starting from the definition \eqref{CS3form}:
\begin{align}
\omega_{\rm CS}(\vec{\XX}_{({\rm n})}, \hat{A}_{\tp}, \hat{A}_{\tq}) =&~ 2 \Tr \left\{ \vec{\XX}_{({\rm n})} \hat{F}_{\tp\tq} \right\} + \pd_{\tq} \left[ \Tr \{ \vec{\XX}_{({\rm n})} \hat{A}_{\tp} \} \right] - \pd_{\tp} \left[ \Tr \{ \vec{\XX}_{({\rm n})} \hat{A}_{\tq} \}\right] ~.
\end{align}
Here it should be understood that the total derivative term is only present when $\tp, \tq = 1,2$.  An analogous expression holds with $\vec{\XX}_{({\rm n})} \to \Phi_{\infty}^y$.  However in this case we can use that $\Phi_{\infty}^y$ is constant and that any power-law modes of $\hat{A}_{\tp}$ commute with $\Phi_{\infty}^y$ to observe that the total derivative terms just subtract off half of the first term, resulting in:
\begin{equation}
\lim_{r \to \infty} \int_{S^2} \ed\Omega \, r^2 \hat{r}_i \, \omega_{\rm CS}(\Phi_{\infty}^y, \hat{A}_{\tp}, \hat{A}_{\tq}) = \lim_{r \to \infty} \int_{S^2} \ed\Omega \, r^2 \hat{r}_i \, \Tr \left\{ \Phi_{\infty}^y \hat{F}_{\tp \tq} \right\}~.
\end{equation}

Now let us recall the mode expansion of $\hat{A}_{\tp} = (A_{p}, \Phi^{z_p})$.  The terms we need are
\begin{equation}
\hat{A}_{\tp}(x^\mu,r,\theta,\phi) = a_{\tp}(x^\mu,r) + \cdots + \sum_{m=-1}^1 a_{\tp,(1,m)}(x^\mu,r) Y_{1m}(\theta,\phi) + \cdots~,
\end{equation}
where $a_{\tp}(x^\mu, r) = a_{\tp}^{({\rm nn})}(x^\mu) + O(r^{-1})$ as usual, and we introduce the triplet notation, $\vec{a}_{\tp}$, such that
\begin{equation}
\sum_{m=-1}^1 a_{\tp,(1,m)}(x^\mu,r) Y_{1m}(\theta,\phi) = - \frac{\sqrt{3}}{\mu^2 r^2} \hat{r} \cdot \vec{a}_{\tp}^{\,({\rm n})}(x^\mu) + O(r^{-3}) ~.
\end{equation}
Here the normalization convention is consistent with the one taken in \eqref{XXtripdef}.  Then \eqref{magenint} is equivalent to 
\begin{align}\label{dsmagenint}
\half \int_{\pd \hat{M}_8} \Omega_{4}' \wedge \omega_{\rm CS} =&~ \frac{\pi}{\mu^2} \int_{\mathbbm{R}^2} \ed^2 x \, \vec{\eta}\,{}^{\tp\tq} \cdot \bigg\{ 2 \Tr \{ \vec{\XX}_{({\rm n})} f_{\tp \tq}^{({\rm nn})} \} - \frac{1}{\sqrt{3}} \Tr \{ \Phi_{\infty}^y \vec{f}_{\tp\tq}^{\,({\rm n})} \} +  \cr
&~ \qquad \qquad \qquad \qquad +  \pd_{\tq} \left[ \Tr \{ \vec{\XX}_{({\rm n})} a_{\tp}^{({\rm nn})} \} \right] - \pd_{\tp} \left[ \Tr \{ \vec{\XX}_{({\rm n})} a_{\tq}^{({\rm nn})} \}\right] \bigg\}~, \qquad \qquad
\end{align}
where $f_{\tp\tq}^{({\rm nn})} = 2 \pd_{[\tp} a_{\tq]}^{({\rm nn})} + [a_{\tp}^{({\rm nn})}, a_{\tq}^{({\rm nn})}]$ and $\vec{f}_{\tq\tq}^{\,({\rm n})} = 2 \pd_{[\tp} \vec{a}_{\tq]}^{\, ({\rm n})}$, and we used the integral
\begin{equation}
\int_{S^2} \ed \Omega \hat{r}_i \hat{r}_j = \frac{4\pi}{3} \delta_{ij}~.
\end{equation}
This reproduces the magnetic contribution to the energy bound given in \eqref{dwBPSbound2}.

Dropping the terms proportional to the first two 't Hooft symbols will give the result for $\Omega_{4}' \to \omega_{4}'$.  Furthermore there are some simplifications if we plug in the explicit form of $\eta^{3}_{\tp\tq}$:
\begin{align}\label{magenint2}
\half \int_{\pd \hat{M}_8} \omega_{4}' \wedge \omega_{\rm CS} =&~ \frac{2\pi}{\mu^2} \int_{\mathbbm{R}^2} \ed^2 x \bigg\{ 2 \Tr \left\{ \XX_{({\rm n})}^3 \left(f_{12}^{({\rm nn})} + [\phi_{({\rm nn})}^{z_1}, \phi_{({\rm nn})}^{z_2}] \right) \right\} - \frac{1}{\sqrt{3}} \Tr \{ \Phi_{\infty}^y f_{12}^{3 ({\rm n})} \} +  \cr
&~ \qquad \qquad \qquad +  \pd_{2} \left[ \Tr \{ \XX_{({\rm n})}^3 a_{1}^{({\rm nn})} \} \right] - \pd_{1} \left[ \Tr \{ \XX_{({\rm n})}^3 a_{2}^{({\rm nn})} \}\right] \bigg\} ~. \qquad \raisetag{20pt}
\end{align}
For the second term we can pull $\Phi_{\infty}^y$ out of the integral.  Then we are simply computing the total magnetic flux of the third component of the normalizable mode of the gauge field triplet.  (See \eqref{fluxtrip}.)  Meanwhile by Stokes' theorem the last two terms give a line integral around the circle at infinity:
\begin{equation}
 \int_{\mathbbm{R}^2} \ed^2 x \left\{   \pd_{2} \left[ \Tr \{ \XX_{({\rm n})}^3 a_{1}^{({\rm nn})} \} \right] - \pd_{1} \left[ \Tr \{ \XX_{({\rm n})}^3 a_{2}^{({\rm nn})} \}\right] \right\} = - \oint_{S_{\infty}^1} \Tr \left\{ \XX_{({\rm n})}^{3} a^{({\rm nn})} \right\}~. \end{equation}
Taking these facts into account one finds that \eqref{magenint2} reproduces the magnetic energy contribution to \eqref{BPSbound2}.

For the electric energy contribution, we first note that
\begin{equation}
(\star E)_{12\theta\phi \hat{z}_1 \hat{z}_2 \hat{y}} = r^2 \sin{\theta} g^{rr} F_{r0}~.
\end{equation}
Then using $g^{00} = - (g^{rr})^{-1}$ and the definition \eqref{asEflux}, one quickly finds the remaining terms in \eqref{BPSbound2} and \eqref{dwBPSbound2}.


\bibliographystyle{utphys}
\bibliography{FBH0bib}

\end{document}